\documentclass[twocolumn,prd,aps,floatfix,preprintnumbers,superscriptaddress,bibnotes,nofootinbib]{revtex4-1}

\usepackage{amsmath}
\usepackage{latexsym}
\usepackage[psamsfonts]{amssymb}
\usepackage{graphicx}
\usepackage{longtable}
\usepackage{ulem}
\usepackage{longtable}
\usepackage{epstopdf}
\usepackage{bm}
\usepackage{color}
\usepackage{hyperref}
\usepackage{multirow}

\newcommand{\be}{\begin{equation}}
\newcommand{\ee}{\end{equation}}
\newcommand{\ben}{\begin{eqnarray}}
\newcommand{\een}{\end{eqnarray}}
\newcommand{\bi}{\begin{itemize}}
\newcommand{\ei}{\end{itemize}}

\newcommand{\la}{\langle}
\newcommand{\ra}{\rangle}

\begin{document}

\preprint{UTHEP-725, UTCCS-P-116, HUPD-1808}


\title{Nucleon form factors and root-mean-square radii \\
on a (10.8 fm$)^4$ lattice at the physical point}

\author{Eigo Shintani\:}
\email{shintani@riken.jp} 
\affiliation{RIKEN Center for Computational Science, Kobe, Hyogo 650-0047, Japan}
\author{Ken-Ichi Ishikawa\:}
\affiliation{Graduate School of Science, Hiroshima University, Higashi-Hiroshima, 
Hiroshima 739-8526, Japan}
\author{Yoshinobu Kuramashi\:}
\email{kuramasi@het.ph.tsukuba.ac.jp}
\affiliation{Center for Computational Sciences, University of Tsukuba, Tsukuba, Ibaraki 305-8577, Japan}
\author{Shoichi Sasaki\:}
\email{ssasaki@nucl.phys.tohoku.ac.jp} 
\affiliation{Department of Physics, Tohoku University, Sendai 980-8578, Japan}
\author{Takeshi Yamazaki\:}
\email{yamazaki@het.ph.tsukuba.ac.jp}
\affiliation{Faculty of Pure and Applied Sciences, University of Tsukuba, Tsukuba, Ibaraki, 305-8571, Japan}
\affiliation{Center for Computational Sciences, University of Tsukuba, Tsukuba, Ibaraki 305-8577, Japan}

\collaboration{PACS Collaboration}

\date{\today}
\begin{abstract}
We present the nucleon form factors and root-mean-square (RMS) radii measured on a (10.8 fm$)^4$ lattice at the physical point. We compute the form factors at small momentum transfer region in $q^2\le 0.102$ GeV$^2$ with the standard plateau method choosing four source-sink separation times $t_{\rm sep}$ from 0.84 to 1.35 fm to examine the possible excited state contamination. We obtain the electric and magnetic form factors and their RMS radii for not only the isovector channel but also the proton and neutron ones without the disconnected diagram. We also obtain the axial-vector coupling and the axial radius from the axial-vector form factor. We find that these three form factors do not show large $t_{\rm sep}$ dependence in our lattice setup, and those RMS radii are consistent with the experimental values. On the other hand, the induced pseudoscalar and pseudoscalar form factors show the clear effects of the excited state contamination, which affect the generalized Goldberger-Treiman relation.

\end{abstract}

\maketitle

 
\section{Introduction}
\label{sec:1}

For the deep understanding of the nucleon and nucleus structures, a precise determination of structure functions is an essential ingredient. Recently an unknown effect for the proton charge radius has been revealed as a significant discrepancy between different approaches~\cite{Mohr:2015ccw} in the $ep$ scattering~\cite{Mohr:2015ccw} process and the muonic hydrogen spectroscopy~\cite{Antognini:1900ns}, in which 5.6-$\sigma$ deviation appears as the so-called ``Proton radius puzzle.'' The measurement of the atomic spectroscopy~\cite{Mohr:2015ccw} has also agreed with the value from the $ep$ scattering, while a recent measurement of the regular hydrogen spectroscopy~\cite{Beyer:2017} agrees with the value from the muonic hydrogen spectroscopy. Under such a confusing circumstance, the theoretical estimation is demanded as an independent test.

Similarly, the axial-vector form factor and the axial radius are important inputs for the weak process associated with the neutrino-nucleus scattering. The $q^2$ dependence of the axial-vector form factor can be used to estimate the neutrino properties such as the neutrino mass and mixing angle~\cite{Hill:2017wgb}. Furthermore, the axial-vector coupling $g_A$ obtained from the cross section of the muon-nucleus scattering measured in the muon capture experiment serves as an independent test to check consistency with the high-precision data of $g_A$ from the neutron beta decay. Having achived the three times higher precision from the current measurement in the muon capture experiment~\cite{Andreev:2012fj}, it would also provide $g_A$ comparable with that from the neutron beta decay, which is expected to be less than 1\% level, using an input of accurate axial radius~\cite{Hill:2017wgb}. The current experimental value of the axial radius~\cite{Bodek:2007ym} is 3\% accuracy from the dipole fit of the neutrino-deuteron scattering and the pion electroproduction experiment, whereas, according to the argument in Refs.~\cite{Bhattacharya:2011ah,Hill:2017wgb}, this error may be underestimated by the model-dependent analysis. It means that the theoretical value of the axial radius is desired to use as an input for the analysis in both muon capture and neutrino scattering experiments.

Lattice QCD (LQCD) is able to nonperturbatively determine the QCD values of the nucleon form factors and RMS radii from the first principles. The recent developments of computational techniques and incredible growth of computational resources make it possible to perform a realistic LQCD computation at the physical point with the light (degenerate up and down) and strange quark flavors, even for the baryonic system in which the systematics involved are more complicated than the mesonic system, e.g., the finite volume effect etc.. Although the results for the nucleon form factors at the physical point have been made available by several LQCD groups~\cite{Bhattacharya:2016zcn,Rajan:2017lxk,Alexandrou:2017ypw,Alexandrou:2017hac,Hasan:2017wwt,Chang:2018uxx,Gupta:2018qil,Ishikawa:2018rew}, the precision has not been enough to be comparable with the experimental values. This is due to the exponential growth of the statistical noise as the light quark mass gets close to the physical point, besides the possible systematic uncertainties of the finite volume (FV) effects, the excited state contamination and the lattice cutoff effects we should take into account. Some LQCD groups~\cite{Bhattacharya:2016zcn,Rajan:2017lxk,Chang:2018uxx,Gupta:2018qil} have tried to subtract the excited state contamination by introducing the ``2-state ansatz'' \cite{Bhattacharya:2016zcn} and the simultaneous fit of the data off the physical point with the use of the ansatz, e.g., chiral perturbation theory~\cite{Chang:2018uxx} or polynomial functions, to remove the FV effects and the lattice cutoff effects. This approach, instead, is an introduction of the other systematic uncertainties originating from the model dependence. For the purpose of high precision to a few \% level and below, much effort to remove the systematic uncertainties in LQCD simulations is needed. We think the most reliable way is the direct simulation at the physical point on sufficiently large volume, which is the critical importance for LQCD computation of the nucleon form factors and RMS radii to directly compare experimental values and theoretically verify the prediction in effective models~\cite{Alarcon:2017ivh,Alarcon:2018irp,Sick:2018fzn,Alarcon:2018zbz}. 

This work is an extension of our earlier study~\cite{Ishikawa:2018rew}. In the previous work, one of the authors analyzed the isovector electric ($G_E$) and magnetic ($G_M$) form factors, and obtained the isovector electric RMS charge radius $\sqrt{\langle r_E^2 \rangle}$ in a model-independent way with the $z$-expansion method. Their results were consistent with two experimental values of $ep$ scattering~\cite{Mohr:2015ccw} and $\mu$H atom spectroscopy~\cite{Antognini:1900ns} within 1-$\sigma$ statistical error, although statistical error was much larger than experimental ones. It was then difficult to argue which experimental values can be favored in LQCD. For the magnetic moment $\mu_v$, LQCD results were also in agreement with experimental value although its statistical precision was almost 15\%. On the other hand, somewhat puzzling situation in the axial-vector ($F_A$) and induced pseudoscalar ($F_P$) form factors in the nucleon axial-vector matrix element occurred. $F_A$ was barely consistent with the experimental results in the low $q^2$ region of $0\le q^2 \le 0.2$ GeV$^2$ and the axial charge $g_A=F_A(q^2=0)$ was slightly underestimated in comparison with the experimental value. Furthermore, $F_P$ had an apparent discrepancy from the experimental expectation at very low $q^2$, which was a consequence of the violation of the generalized Goldberger-Treiman relation.

The purpose of this paper is further reduction of the statistical and systematic errors for the nucleon form factors and understanding of the issues associated with $F_A$ and $F_P$ raised in the previous work. We have made several essential improvements from the previous work;\\
(i) The lattice size is enlarged from  (8.1 fm$)^4$ to (10.8 fm$)^4$ employing the stout-smeared $O(a)$-improved Wilson-clover quark action and the Iwasaki gauge action at $\beta$=1.82 \cite{Ishikawa:2018jee}, which are exactly same as in the previous work.  We expect that the spatial extent 10.8 fm has a strong advantage for both suppression of the finite volume effects and reduction of the minimum value of the momentum transfer to $q^2=0.013$ GeV$^2$ which is a half of the previous work \cite{Ishikawa:2018rew}.\\
(ii) The quark masses are carefully tuned to the physical point~\cite{Ishikawa:2018jee}. The slight deviation from the physical point with $m_\pi = 146$ MeV in the previous work~\cite{Ishikawa:2018jee} is removed by adjusting the pion mass to 135 MeV.\\
(iii) Using the variation of the source-sink separation as $t_{\rm sep}/a=t_{\rm sink}/a-t_{\rm src}/a=10,12,14,16$, where the largest one is about 1.35 fm, we can examine the possible excited state contributions, which has not been studied in the previous work~\cite{Ishikawa:2018jee}, where a single choice of $t_{\rm sep}/a=15$ was used.\\
(iv) Significant reduction of the computational cost is possible to utilize the all-mode-averaging (AMA) method~\cite{Blum:2012uh,Shintani:2014vja,vonHippel:2016wid} optimized by the deflation technique \cite{Luscher:2007se}. 
  
This paper is organized as follows: Section~\ref{sec:nff_def} presents the definition of nucleon form factors to fix the notations in this paper. The general features of the nucleon form factors have been already explained in Sec.~II of Ref.~\cite{Ishikawa:2018rew}. In Sec.~\ref{sec:simulation} we first present a brief description of gauge configurations, which are a partial set of ``PACS10'' configurations generated by the PACS Collaboration~\cite{Ishikawa:2018jee}. We also explain the error reduction technique employed in this study. The results for the nucleon form factors are presented in Sec.~\ref{sec:nff_results_gema}. We investigate $t_{\rm sep}$ dependence for the nucleon form factors and the corresponding RMS radii. We also discuss the violation of the generalized GT relation associated with the form factors $F_A$, $F_P$, and $G_P$. Section~\ref{sec:summary} is devoted to summary and outlook. 

In this paper the matrix elements are given in the Euclidean metric convention. $\gamma_5$ is defined by $\gamma_5\equiv \gamma_1 \gamma_2 \gamma_3 \gamma_4 =-\gamma_5^{M}$, which has the opposite sign relative to that in the Minkowski convention ($\vec{\gamma}^M=i\vec{\gamma}$ and $\gamma_0^{M}=\gamma_4$) adopted in the particle data group~\cite{Patrignani:2016xqp}. The sign of all the form factors is chosen to be positive. The Euclidean four-momentum squared $q^2$ corresponds to the spacelike momentum squared as $q_M^2=-q^2<0$ in Minkowski space. 

\section{LQCD computation of nucleon form factors}
\label{sec:nff_def}
\subsection{Definition of nucleon form factors}\label{subsec:define_ff}
We present our convention of the nucleon form factors. We measure the electric and magnetic Sachs form factors, $G_E(q^2)$ and $G_M(q^2)$, which are relevant for the experimental data of elastic electron-nucleon scattering. They are linear combinations of the Dirac and Pauli form factors, $F_{1}^N,\,F_{2}^N$ $(N=p,n)$, as 
%
%
\ben
G^{N}_E(q^2)&=&F_1^N(q^2)-\frac{q^2}{4M_N^2}F_2^N(q^2),
\label{Eq:GE}\\
G^{N}_M(q^2)&=&F_1^N(q^2)+F_2^N(q^2),
\label{Eq:GM}
\een
extracted from the nucleon vector matrix elements,
%
%
\begin{multline}
\langle N(P^\prime)|j_{\alpha}^{\rm em}|N(P)\rangle\cr
{}=\bar{u}_N(P^\prime)\left(
\gamma_{\alpha}F^N_1(q^2)+i\sigma_{\alpha \beta}\frac{q_{\beta}}{2M_N}F^N_2(q^2)
\right)u_N(P)
\end{multline}
with $\sigma_{\alpha \beta}=\frac{1}{2}[\gamma_\alpha, \gamma_\beta]$. The momentum transfer between the nucleon initial state ($P$) to the final state ($P^\prime$) is defined as $q=P-P^\prime$. The electromagnetic current $j_{\alpha}^{\rm em}$ is expressed in terms of the flavor-diagonal vector current:
%
%
\ben
j_{\alpha}^{\rm em}=\sum_{q} Q_q \bar{q}\gamma_\alpha q=\frac{2}{3}\bar{u}\gamma_\alpha u -\frac{1}{3}\bar{d}\gamma_\alpha d + \cdots,
\label{Eq:EMcur}
\een
where $Q_q$ denotes the charge (in units of proton charge $e$) for quark flavor $q$, and the ellipsis denote terms for strange and heavier quarks. Here we rewrite Eq.~(\ref{Eq:EMcur}) into the following form for the later discussion, 
%
%
\ben
j_{\alpha}^{\rm em}=\frac{1}{6}j_{\alpha}^{s}+\frac{1}{2}j_{\alpha}^{v}+\cdots
\een
with the isoscalar and isovector currents: 
$j_{\alpha}^{s}=\bar{u}\gamma_{\alpha}u+\bar{d}\gamma_{\alpha}d$ and $j_{\alpha}^{v}=\bar{u}\gamma_{\alpha}u-\bar{d}\gamma_{\alpha}d$.
Recall that in the case of $m_u=m_d$ (SU(2) isospin symmetry), 
the nucleon three-point correlation function for the isovector current does not 
receive any contribution from the disconnected diagram of all quark flavors since 
they are canceled each other.  Therefore, the isovector part of nucleon electromagnetic form factors
can be determined only by the connected-type contribution, whose numerical evaluation is much easier in lattice simulations. 

On the other hand, the isoscalar component receives the full contribution including the disconnected diagrams even under the exact isospin symmetry. Nevertheless, all disconnected-type contributions from the light and heavier quark flavors are known to be relatively small in comparison to the connected-type contributions especially in the proton (see for example Ref.~\cite{Sufian:2017osl}), whereas it will not in the neutron. Here we simply evaluate individual proton (neutron) form factors in the vector channel, which is extracted from the matrix element with the isoscalar and isovector parts of the electromagnetic current, only by the connected-type contributions, since a computation of disconnected diagram is much costly and beyond our scope of this study. Later we will show the numerical evidence of the appearance of missing effect of disconnected-type contribution to the electromagnetic form factor in both proton and neutron. 

The isovector nucleon form factors can be related to the isovector combination of the proton's and neutron's form factors assuming the SU(2) isospin symmetry
%
%
\ben
G_{l}^v(q^2)&=&G_{l}^p(q^2)-G_{l}^n(q^2),\quad l=\{E,M\},
\label{Eq:CVC}
\een
where the normalization of the above form factors at $q^2=0$ are given by the proton/neutron electric charge and the magnetic moment: $G_E^p(0)=1$ and $G_M^p(0)=\mu_p=+2.79285$ for the proton and $G_E^n(0)=0$ and $G_M^n(0)=\mu_n=-1.91304$ for the neutron. Therefore, one finds
%
%
\ben
G_E^v(0)=1,\quad G_M^v(0)-1&=3.70589.
\een

The nucleon axial-vector matrix element is represented with the axial-vector form factor $F_A(q^2)$ and the induced pseudoscalar form factor $F_P(q^2)$ as 
%
%
\begin{multline}
\langle p(P^\prime)|A_{\alpha}^{+}(x)|n(P)\rangle\cr
{}=\bar{u}_p(P^\prime)\left(
\gamma_{\alpha}\gamma_5 F_A(q^2)+iq_{\alpha}\gamma_5F_P(q^2)
\right)u_n(P)e^{iq\cdot x}
\end{multline}
with the axial-vector current, $A^+_\alpha=\bar{u}\gamma_5\gamma_\alpha d$. The axial-vector coupling $g_A=F_A(q^2=0)$, which governs the life time of the neutron beta decay, is experimentally determined as $g_A=1.2724(23)$\cite{Tanabashi:2018oca}, and the $q^2$ dependence of $F_A(q^2)$ is well described by the dipole form $F_A(q^2)=F_A(0)/(1+q^2/M_A^2)^2$ below $q^2\approx 1$ $({\rm GeV})^2$~\cite{Bernard:2001rs,Bodek:2007ym}. On the other hand, the properties of the induced pseudoscalar form factor  $F_P(q^2)$ is less clear in the experiments~\cite{Choi:1993vt,Gorringe:2002xx}.

On the theoretical side, $F_A(q^2)$ and $F_P(q^2)$ are related through the generalized Goldberger-Treiman  (GT) relation~\cite{Weisberger:1966ip,Sasaki:2007gw}: 
%
%
\begin{align}
  2M_NF_A(q^2)-q^2F_P(q^2)=2\hat{m}G_P(q^2)
  \label{Eq:GTrelation}
\end{align}
with a degenerate up and down quark mass $\hat m=m_u=m_d$, derived from the axial Ward-Takahashi (AWT) identity: $\partial_\alpha A^+_\alpha(x)=2\hat{m}P^{+}(x)$. Here, the additional form factor $G_P(q^2)$ is defined in the nucleon pseudoscalar matrix element
%
%
\begin{align}
\langle p(P^\prime)|P^{+}(x)|n(P)\rangle
=\bar{u}_p(P^\prime)\left(
\gamma_5 G_P(q^2)
\right)u_n(P)e^{iq\cdot x}
  \label{Eq:PSFF}
\end{align}
with the local pseudoscalar density $P^{+}= \bar{u}\gamma_5d$. In order to test the GT relation in LQCD, the simultaneous investigation of three form factors $F_A(q^2)$, $F_P(q^2)$ and $G_P(q^2)$ is essential. 

The RMS radius $R_l=\sqrt{\langle r^2_l\rangle}$ is defined from the expansion of the normalized form factor $G_l(q^2)$ in the powers of $q^2$:
%
%
\begin{eqnarray}
G_l(q^2)&=&G_l(0)\left(1-\frac{1}{6}\langle r_l^2\rangle q^2+\frac{1}{120}\langle
r_l^4\rangle q^4 + \cdots\right),\nonumber\\
&&l=\{E,M,A\},
\end{eqnarray}
which measures a typical size in the coordinate space. Here we define $G_A\equiv F_A$, and the RMS radius is computed from the derivative of nucleon form factor with respect to $q^2$ at $q^2=0$, 
%
%
\ben
\la r^2_l\ra&=&-\left.\frac{6}{G_l(0)}\frac{dG_l(q^2)}{dq^2}\right\vert_{q^2=0}.
\een
When the form factors can be described by the dipole form,
%
%
\begin{equation}
  G_l(q^2) = \frac{G_l(0)}{\big(1+\frac{q^2}{\Lambda_l^2}\big)^2}
  \label{Eq:dipole}
\end{equation}
with the dipole mass parameter $\Lambda$, the RMS radius $R$ is obtained as $R_l=\sqrt{12}/{\Lambda_l}$. In the $z$-expansion method~\cite{Boyd:1995cf,Hill:2010yb}, whose convergence has been carefully studied in Ref.~\cite{Ishikawa:2018rew}, the form factors can be described by a convergent Taylor series in a new variable $z$ which is conformally mapped from $q^2$:
%
%
\begin{eqnarray}
  &&G_l(z) = \sum_{k=0}^{k_{\rm max}} c_k z(q^2)^k,  
  \label{Eq:z-exp}\\
  &&z(q^2)=\frac{\sqrt{t_{\rm cut}+q^2}-\sqrt{t_{\rm cut}}}{\sqrt{t_{\rm cut}+q^2}+\sqrt{t_{\rm cut}}},
  \label{Eq:z-value}
\end{eqnarray}
where the expansion is truncated at the $k_{\rm max}$-th order with $t_{\rm cut}=4m_{\pi}^2$ for $G_E$ and $G_M$, or with $t_{\rm cut}=9m_{\pi}^2$ for $F_A$, and the RMS radius is then determined by $\sqrt{\la r^2_l\ra}=\sqrt{-6(c_1/c_0)/(4t_{\rm cut})}$ for $l=\{ E, M, A\}$.

For a precise determination of the RMS radii $\sqrt{\la r^2_l\ra}$ in LQCD, the data of low $q^2$ variation is important. The systematic uncertainty due to fitting LQCD data at low $q^2$ with extrapolation ansatz into $q^2=0$ can be reduced by using small $q^2$ data we can obtain. It means that large spatial extent is advantageous to this study since the accessible minimum value of $q^2$ is essentially determined by the spatial extent of the lattice. Our spatial lattice volume of (10.8 fm$)^3$ allows $q^2_{\rm min}=0.013$ GeV$^2$, and it is then 1.8 times smaller $q^2$ data we can use than (8.1 fm$)^3$ box employed in the previous work~\cite{Ishikawa:2018rew}. Furthermore, LQCD analysis concentrating on the small $q^2$ region allows us to avoid the additional systematic uncertainty of lattice cutoff effect stemming from $O((aq)^2)$. 

\subsection{Nucleon two- and three-point functions for form factors}
\label{subsec:3pt}

We first define the nucleon (proton) interpolating operator as
%
%
\begin{multline}
  N_X(t,{\bm p})\\
  {}=\sum_{{\bm x},{\bm x}_{1},{\bm x}_{2},{\bm x}_{3}}e^{-i{\bm p}\cdot{\bm x}}\varepsilon_{abc}[u_a^T({\bm x}_1,t)C\gamma_5d_b({\bm x}_2, t)]u_c({\bm x}_3, t)\cr
  {}\times\phi_X({\bm x}_1-{\bm x})\phi_X({\bm x}_2-{\bm x})\phi_X({\bm x}_3-{\bm x}),
  \label{eq:nuc_ope}
\end{multline}
where the superscript $T$ denotes a transposition and $C$ is the charge conjugation matrix defined as $C=\gamma_4\gamma_2$. The indices $a,b,c$ and $u,d$ label the color and the flavor, respectively. The function $\phi_X$ $(X=L,S)$ represents two types of the smearing functions employed in this study: local type (L) given by $\phi_L({\bm x}_i-{\bm x})=\delta({\bm x}_i -{\bm x})$ and exponentially smeared one (S) by $\phi_S({\bm x}_i-{\bm x})=A\exp\left(-B|{\bm x}_i-{\bm x}|\right)$.

The nucleon two-point functions are constructed with the source and sink operators located at $t_{\rm src}$ and $t_{\rm sink}$, respectively:
\begin{multline}
  C_{XS}(t_{\rm sink}-t_{\rm src}, {\bm p})\\
  {}=\frac{1}{4}{\rm Tr}\left\{{\cal P_+}\langle N_X(t_{\rm sink},{\bm p}){\bar N}_S(t_{\rm src},-{\bm p})\rangle
  \right\}, \label{eq:2pt}
\end{multline}
where the smeared operator is employed at the source and at the sink we use both the smeared ($X=S$) and local ($X=L$) operators. The lattice momentum is defined as ${\bm p}=2\pi/(N_{\rm s}a)\times {\bm n}$ with a vector of integers ${\bm n}\in Z^3$ and $N_{\rm s}$ the number of the spatial lattice sites. A projection operator ${\cal P}_+=\frac{1+\gamma_4}{2}$ is applied to extract the contributions from the positive-parity state for $|{\bm p}|=0$~\cite{Sasaki:2001nf,Sasaki:2005ug}.

In our study, the nucleon form factor is extracted from the nucleon three-point function, 
\begin{multline}
C_{{\cal O},\alpha}^{{\cal P}_k}(t,{\bm p}^\prime,{\bm p})\cr
{}=\frac{1}{4}{\rm Tr}\left\{{\cal P}_k\langle N_S(t_{\rm sink},{\bm p}^\prime)
J^{\cal O}_\alpha(t,{\bm q}){\bar N}_S(t_{\rm src},-{\bm p})\rangle
\right\}, \label{eq:3pt_J}
\end{multline}
using the local currents $J^{{\cal O}}_\alpha(x)=\bar{q}(x)\Gamma^{\cal O}_{\alpha} q(x)$ of $\Gamma^P_\alpha=Z_P\gamma_5$, $\Gamma^V_\alpha=Z_V\gamma_\mu$, $\Gamma^A_\alpha=Z_A\gamma_\mu\gamma_5$ with the renormalization factor $Z_{\cal{O}}$. In the above equation, ${\bm q}={\bm p}-{\bm p}^\prime$ represents the three-dimensional momentum transfer, and ${\cal P}_k$ denotes the projection operator to extract the form factors for unpolarized case, ${\cal P}_k ={\cal P}_t\equiv {\cal P}_+\gamma_4$ and polarized case (in $z$ direction) ${\cal P}_k={\cal P}_{5z}\equiv{\cal P}_+\gamma_5\gamma_3$. In a conventional way to remove the unwanted nucleon wavefunction, we use the following ratio, 
\begin{widetext}
\be
   {\cal R}^{k}_{{\cal O},\alpha}(t,{\bm p}^\prime,{\bm p})= 
   \frac{C_{{\cal O},\alpha}^{{\cal P}_k}(t,{\bm p}^\prime,{\bm p})}{C_{SS}(t_{\rm sink}-t_{\rm src}, {\bm p}^\prime)}
   \sqrt{
     \frac{C_{LS}(t_{\rm sink}-t, {\bm p})C_{SS}(t-t_{\rm src}, {\bm p}^\prime)C_{LS}(t_{\rm sink}-t_{\rm src}, {\bm p}^\prime)}
          {C_{LS}(t_{\rm sink}-t, {\bm p}^\prime)C_{SS}(t-t_{\rm src}, {\bm p})C_{LS}(t_{\rm sink}-t_{\rm src}, {\bm p})}
},
   \label{Eq:RatioQ}
\ee
\end{widetext}
as a function of initial and final nucleon momenta, $\bm p$ and $\bm p'$, and the temporal position of local current $t$ in the fixed temporal position of source and sink nucleon interpolation operator $t_{\rm src}$ and $t_{\rm sink}$. In this study we use various nucleon source-sink separations as $t_{\rm sep}/a=t_{\rm sink}/a-t_{\rm src}/a=10,12,14,16$ to examine a possible excited state contamination. Here, by restricting the kinematics of the nucleon final state at rest, where $q^2=2M_N(E_N({\bm q})-M_N)$ with ${\bm p}^\prime={\bm 0}$, the above ratio can be represented as ${\cal R}_{{\cal O},\alpha}(t,{\bm q})$. 

In the electromagnetic vector channel, ${\cal O}=j^{\rm em}$, the ratio of Eq.~(\ref{Eq:RatioQ}) is supposed to give the following asymptotic form~\cite{Sasaki:2007gw}: 
\begin{eqnarray}
  &&{\cal R}^{t,N}_{j^{\rm em},4}(t,{\bm q})\rightarrow \sqrt{\frac{E_N+M_N}{2E_N}}G^N_E(q^2),
  \label{Eq:R_GE}\\
  &&{\cal R}^{5z,N}_{j^{\rm em},i}(t,{\bm q})\rightarrow \frac{-i\varepsilon_{ij3}q_j}{\sqrt{2E_N(E_N+M_N)}}G^N_M(q^2).
  \label{Eq:R_GM}
\end{eqnarray}
in the limit of $t_{\rm sink}\gg t \gg t_{\rm src}$ 
with $N = p, n$.

Similarly in the axial-vector current and pseudoscalar cases, ${\cal O}=A^+,P^+$, we obtain
\begin{eqnarray}
  &&{\cal R}^{5z}_{A^+,i}(t,{\bm q}) \cr
  &&{}\rightarrow\sqrt{\frac{E_N+M_N}{2E_N}}\left[F_A(q^2)\delta_{i3}-\frac{q_iq_3}{E_N+M_N}F_P(q^2)\right],
  \label{Eq:R_FA_FP}\\
  &&{\cal R}^{5z}_{P^+}(t,{\bm q})\rightarrow \frac{iq_3}{\sqrt{2E_N(E_N+M_N)}}G_P(q^2).
\label{Eq:R_GP}
\end{eqnarray}

The three-point correlation functions of Eq.~(\ref{eq:3pt_J}) are calculated by the sequential source method with $t_{\rm sink}$ and $t_{\rm src}$ fixed~\cite{Martinelli:1988rr,Sasaki:2003jh}, which requires to solve the sequential quark propagators individually in the four choices of $t_{\rm sep}$ and the projection operators ${\cal P}={\cal P}_t,{\cal P}_{5z}$. 

\section{Simulation details}
\label{sec:simulation}

\subsection{PACS10 configurations}
\label{subsec:config}

In this work we have used a partial set of PACS10 configurations~\cite{Ishikawa:2018jee}. We briefly present relevant points to make the paper self-contained (see in Ref.~\cite{Ishikawa:2018jee} for the detailed description).
  
The gauge configurations in 2+1 flavor QCD with the stout-smeared $O(a)$-improved Wilson-clover quark action and the Iwasaki gauge action~\cite{Iwasaki:2011np} on an $N_{\rm s}^3\times N_{\rm t}=128^3\times 128$ lattice at $\beta=1.82$, which corresponds to a (10.8 fm$)^4$ physical space-time with a lattice cutoff of $a^{-1}= 2.333(18)$ GeV ($a = 0.08457(67)$ fm)~\cite{Ishikawa:2015rho} have been generated by PACS Collaboration. The Schr{\"o}dinger functional (SF) scheme is employed to determine the nonperturbative improvement coefficient $c_{\rm SW}=1.11$~\cite{Taniguchi:2012kk}. Since the improvement factor for the axial-vector current $c_A$ is consistent with zero within the statistical error~\cite{Taniguchi:2012kk}, we do not take account of the ${\cal O}(a)$ improvement of the quark bilinear currents. The hopping parameters of $(\kappa_{ud}, \kappa_{s})$ = (0.126117, 0.124902) are carefully chosen to be at the physical point. We use 20 gauge configurations separated by 10 trajectories. The statistical error is estimated by the single elimination jackknife method.

\subsection{Utilization of all-mode-averaging technique}
\label{subsec:ama}

Here we employ the AMA technique to efficiently implement the LQCD computation of two- and three-point functions. For the implementation of AMA, we compute the combination of correlator with high-precision $O^{\rm (org)}$ and low-precision $O^{\rm (approx)}$ as
%
%
\begin{eqnarray}
  O^{\rm (ama)}&=&\frac{1}{N_{\rm org}}\sum^{N_{\rm org}}_{f\in G}\big(O^{{\rm (org)}\,f} - O^{{\rm (approx)}\,f}\big)\nonumber\\
  &&+ \frac{1}{N_G}\sum^{N_G}_{g\in G}O^{{\rm (approx)}\,g},
  \label{Eq:ama}
\end{eqnarray}
where the superscript $f,\,g$ denotes the transformation under the lattice symmetry $G$, for instance translational symmetry. 
$N_{\rm org}$ and $N_G$ are the number of such a transformed observable for $O^{\rm (org)}$ and $O^{\rm (approx)}$ respectively. To achieve the high performance of AMA, we need to set $N_{\rm org}\ll N_{\rm G}$ satisfying the strong correlation $r$ between $O^{\rm org}$ and $O^{\rm (approx)}$, as $2(1-r)<1/N_G$~\cite{Shintani:2014vja,vonHippel:2016wid}. Following Refs.~\cite{vonHippel:2016wid,Izubuchi:2018tdd} we employ the optimized AMA which adopts the deflated Schwartz Alternative Procedure (SAP)~\cite{Luscher:2003qa} and Generalized Conjugate Residual (GCR) \cite{Luscher:2007se} in the computation of both high-precision $O^{\rm (org)}$ and low-precision $O^{\rm (approx)}$. As demonstrated by the performance test in Ref.~\cite{Izubuchi:2018tdd}, the utilization of deflated SAP-GCR can significantly save the computational cost compared to the low-mode deflation originally suggested in Refs.~\cite{Blum:2012uh,Shintani:2014vja}.

\subsection{LQCD parameters}
\label{subsec:lqcd_param}

First we tune the parameters for the source and sink smearing function as $A=1.2$ and $B=0.16$ in Eq.~(\ref{eq:nuc_ope}). The smearing parameters are slightly different from previous work~\cite{Ishikawa:2018rew} to gain a better overlap with the ground state in three-point function. As mentioned in Sec.~\ref{subsec:define_ff}, to avoid the considerable lattice cutoff effect, we choose the eight lowest variations of $q^2$ listed in Table~\ref{tab:q}, up to $q^2=0.11$ GeV$^2$, in our analysis. 

\begin{table*}[t]
\caption{Choices for the nonzero spatial momenta: ${\bm q}=\pi/(64a)\times {\bm n}$, and corresponding nucleon energy $E_N$ measured by global fitting of two-point function with the same range as in Q0 (also see in a text). The degeneracy of $\vert{\bm n}\vert^2$ due to the permutation symmetry between $\pm x$, $\pm y$, $\pm z$ directions is listed in the bottom raw.
\label{tab:q}
}
\begin{ruledtabular}
\begin{tabular}{c|cccccccccc} 
  label & Q0 & Q1 & Q2 & Q3 & Q4 & Q5 & Q6 & Q7 \\ \hline
  ${\bm n}$ & (0,0,0) & (1,0,0) & (1,1,0) & (1,1,1) & (2,0,0) & (2,1,0) & (2,1,1) & (2,2,0)\\
  $\vert{\bm n}\vert^2$ & 0 & 1 & 2 & 3 & 4 & 5 & 6 & 8\\
  $aE_N$ & 0.4041(47) & 0.4073(47) & 0.4105(48) & 0.4137(49) & 0.4159(49) & 0.4192(49) & 0.4222(50) & 0.4278(51)\\
 $q^2\,[{\rm GeV}^2]$ & 0 & 0.013 & 0.026 & 0.039 & 0.052 & 0.064 & 0.077 & 0.102\\
  degeneracy & 1 & 6 & 12 & 8 & 6 & 24 & 24 & 12\\ 
\end{tabular}
\end{ruledtabular} 
\end{table*} 

The renormalization factors $Z_{\cal O}$ $({\cal O}=V,A)$ are obtained  by the SF scheme at the vanishing quark mass as $Z_V=0.95153(76)(1487)$, $Z_A=0.9650(68)(95)$~\cite{Ishikawa:2015fzw}, where the first error is statistical one and the second is a systematic error coming from the difference of two volumes. In our analysis this systematic error is regarded as negligible since we here choose the larger volume. 

We compute the three-point function of Eq.~(\ref{eq:3pt_J}) with four variations of $t_{\rm sep}/a=10,\,12,\,14,\,16$ to examine the excited state contamination. Since the LQCD calculation with large $t_{\rm sep}$ suffers from the large statistical noise,  we increase the $N_G$ as shown in Table~\ref{tab:meas} to keep the signal-to-noise ratio as $t_{\rm sep}$ becomes large. As for the AMA tuning parameter, the approximation is obtained by 5 GCR iteration using $8^4$ SAP domain size with 50 deflation fields. $O^{(\rm org)}$ is given in the stopping criteria of residual norm less than $10^{-8}$. 

\begin{table}
  \caption{We present $N_{\rm org}$, $N_G$ and total number of measurements ($N_G\times N_{\rm conf}$) at each $t_{\rm sep}$. Fit range for the ratio of Eq.~(\ref{Eq:RatioQ}) to extract $G_E$, $G_M$, $F_A$, $F_P$, and $G_P$ is also listed.}\label{tab:meas}
  \begin{ruledtabular}
    \begin{tabular}{rrrrr}
      $t_{\rm sep}/a$ & $N_{\rm org}$ & $N_G$ & $\#$meas & fit range\\
      \hline
      10 & 1 & 128 & 2,560 & [3,7]\\
      12 & 1& 256 & 5,120 & [4,8]\\
      14 & 2 & 320 & 6,400 & [5,9]\\
      16 & 4 & 512 & 10,218 & [6,10]\\
    \end{tabular}    
  \end{ruledtabular}
\end{table}

\subsection{Nucleon effective mass}

\begin{figure}[ht!]
\includegraphics[width=80mm,keepaspectratio,clip]{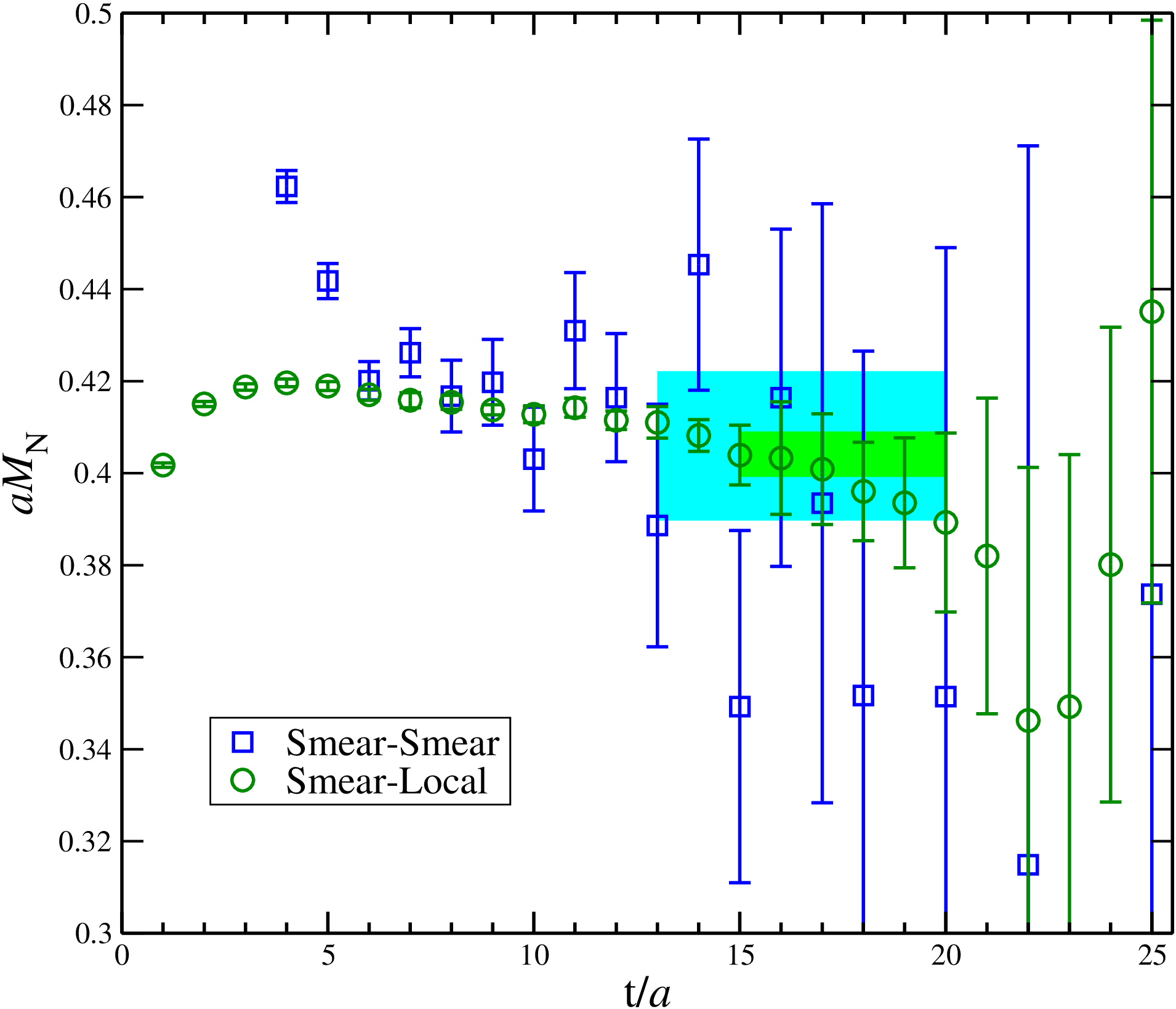}
\caption{Nucleon effective mass plot for the smeared-smeared (square symbol) and smeared-local (circle symbol) operators in the nucleon two-point functions. Horizontal band with green (cyan) color denotes the fitting range and its statistical error for the smeared-local (smeared-smeared) function.}
\label{fig:effm_n}
\end{figure}

In Fig.~\ref{fig:effm_n} we show the nucleon effective mass plot with the smeared-smeared and smeared-local operators. The single exponential function is used in the correlated fit with the range of $t/a=$15--20 for the smeared-local and 13--20 for the smeared-smeared cases, and those are consistent with each other. The he nucleon mass from the smeared-local case is obtained as 
\begin{equation}
  aM_N = 0.4041(47), \quad M_N = 0.9416(110)\,{\rm GeV},
\end{equation}
where the error is statistical. This is consistent with the experimental value  $M_N^{\rm exp}=0.93891874$ GeV obtained by averaging the proton and neutron masses. For the extraction of the form factors in Eqs.~(\ref{Eq:R_GE}), (\ref{Eq:R_GM}), (\ref{Eq:R_FA_FP}) and (\ref{Eq:R_GP}), we use the central values of the nucleon mass and the energy with finite momenta determined from the smeared-local case. In fact, even if we input the statistically fluctuating mass and energy onto those equations, the variation of extracted form factors is negligibly small compared to statistical fluctuation of three-point function. We have summarized the measured nucleon mass and energy in Table~\ref{tab:q} together with values of $q^2$.

\section{Results for nucleon electromagnetic form factors and axial form factor}
\label{sec:nff_results_gema}

\subsection{Electric form factor and electric charge radius}
\label{subse:nff_ge}

\begin{figure*}[ht!]
  \includegraphics[width=8.0cm,keepaspectratio,clip]{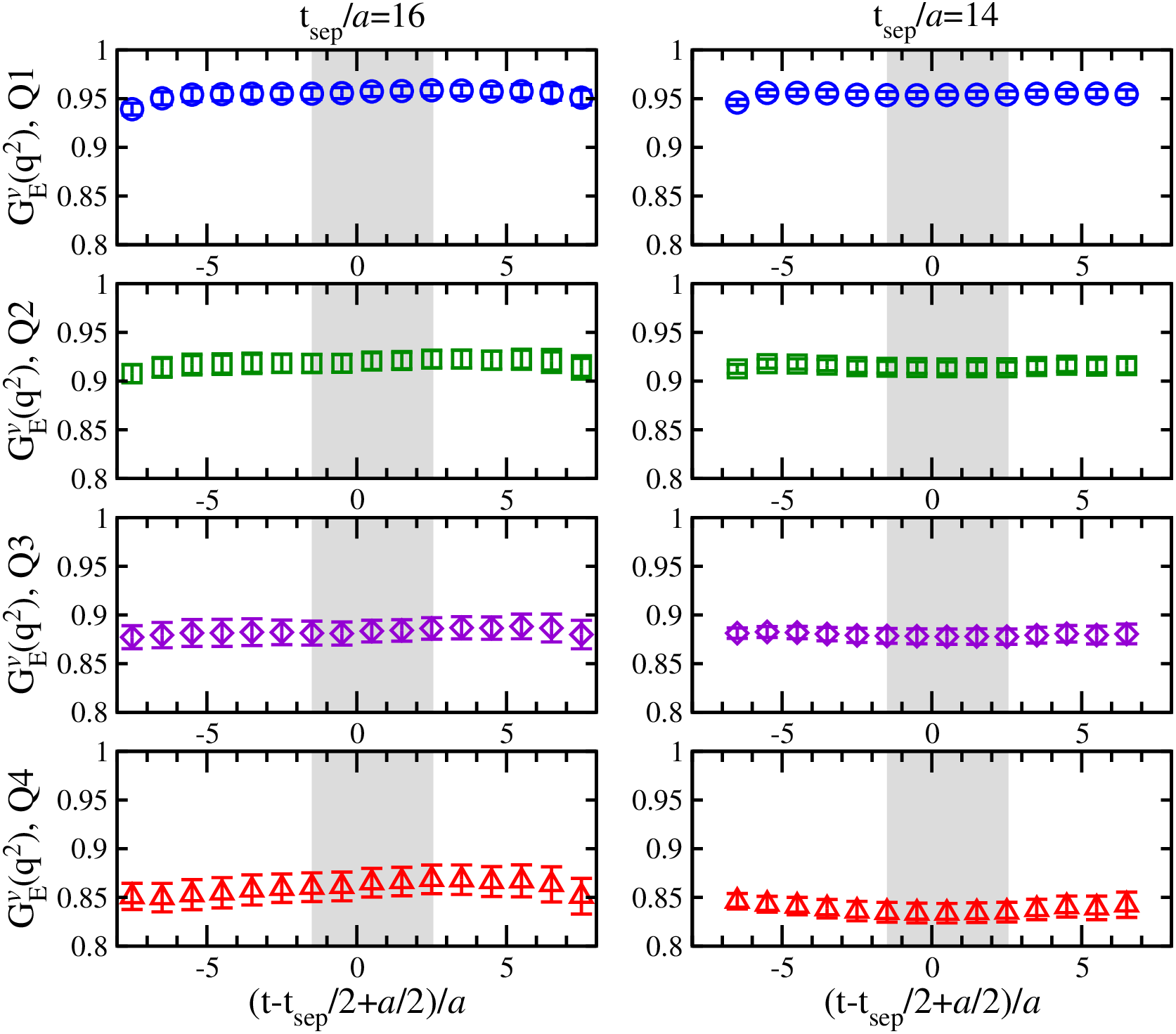}
  \hspace{3mm}
  \includegraphics[width=8.0cm,keepaspectratio,clip]{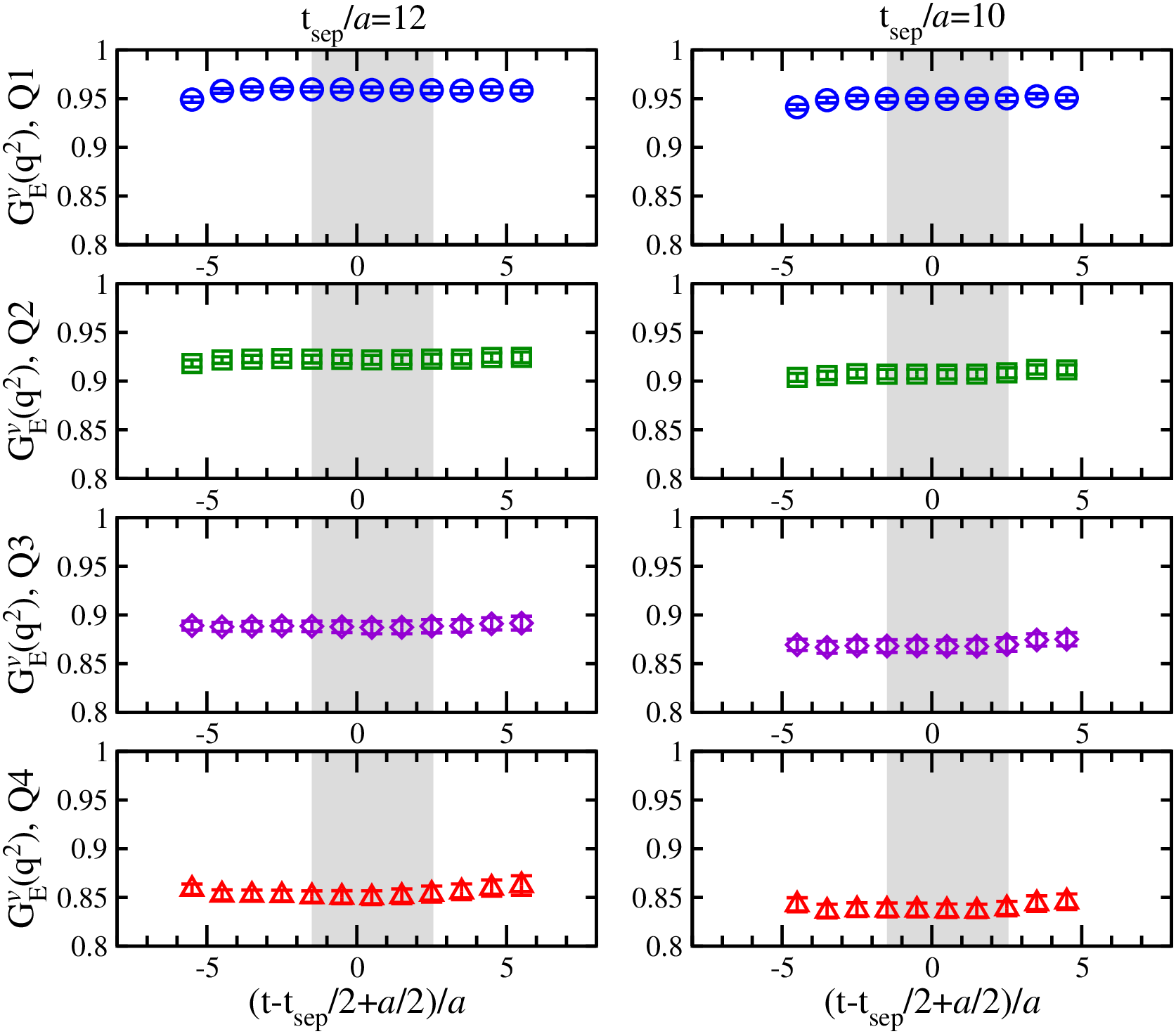}
\caption{Isovector electric form factor $G_E^v(q^2)$, which is extracted from the ratios of three- to two-point functions of Eq.~(\ref{Eq:R_GE}), for $t_{\rm sep}/a=10,12,14,16$ with four lowest nonzero momentum transfers. Gray-shaded area denotes the fit range in each panel.}
\label{fig:ratio_ge}
\end{figure*}

Figure~\ref{fig:ratio_ge} shows $t$ dependence of the ratio ${\cal R}^{t,N}_{j^{\rm em},4}(t,{\bm q})$ of Eq.~(\ref{Eq:R_GE}) for the isovector electric form factor $G_E^v(q^2)$ at $t_{\rm sep}/a=10,12,14,16$ in the smallest four nonzero momenta corresponding to Q1, Q2, Q3, and Q4 (see Table~\ref{tab:q}). We observe clear plateau for all the cases of $t_{\rm sep}$ and $|{\bm n}|^2$ thanks to our elaborate tuning of the smearing parameter. The $G_E^v(q^2)$ is extracted by the constant fit with the fit range listed in Table~\ref{tab:meas}. 

\begin{figure}[ht!]
  \includegraphics[width=8.0cm,keepaspectratio,clip]{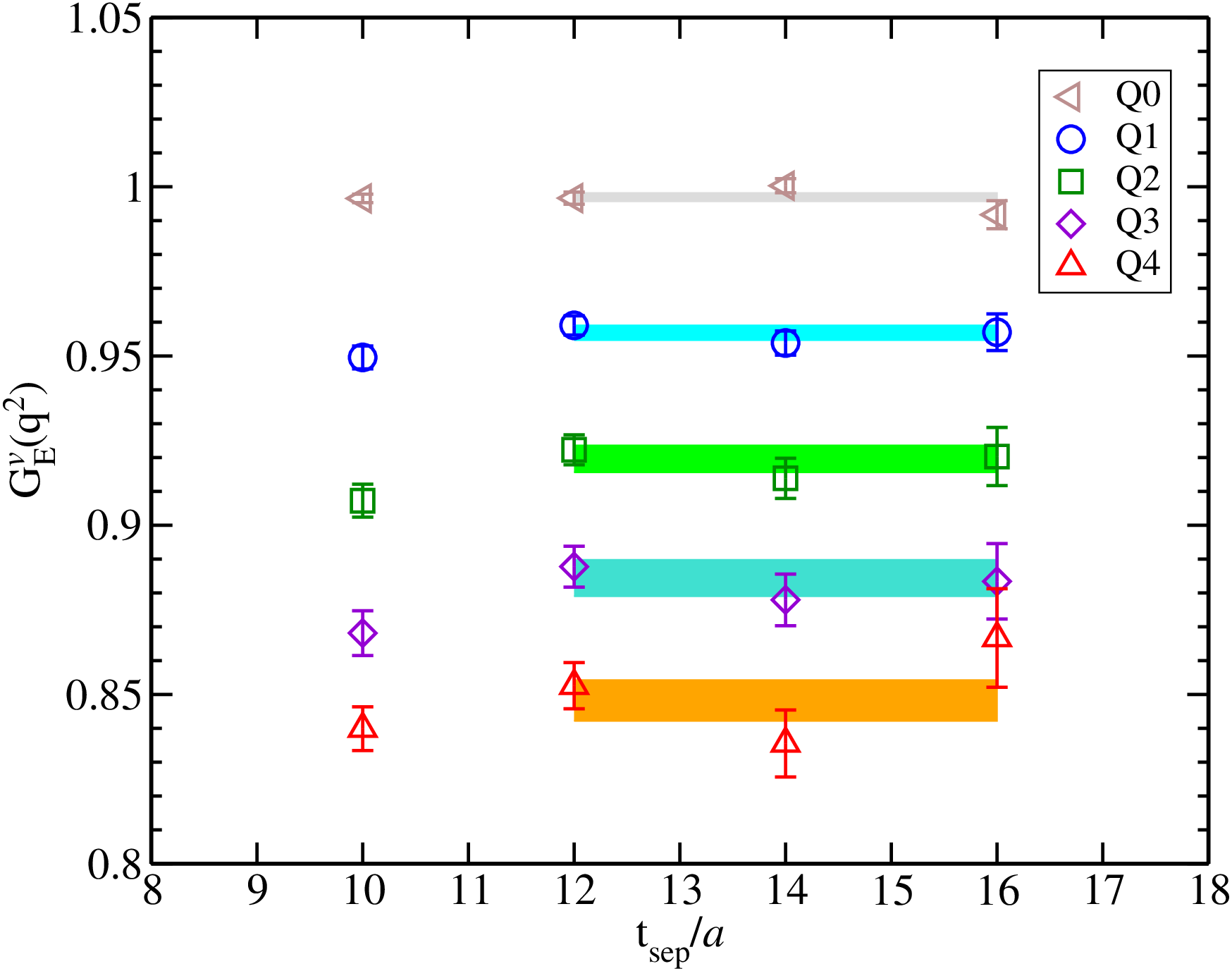}
\caption{$t_{\rm sep}$ dependence of the isovector electric form factor $G_E^v(q^2)$ with five lowest momentum transfers. Horizontal band represents the fit result of $G_E^v(q^2)$ at $t_{\rm sep}/a=12,14,16$ for each $q^2$.}
\label{fig:ge_ts}
\end{figure}

In Fig.~\ref{fig:ge_ts} we plot the $t_{\rm sep}$ dependence of $G_E^v(q^2)$ for the smallest five values of $\vert {\bm n}\vert^2$. One can see the data at $t_{\rm sep}/a=12,14,16$ is statistically consistent within the error, which means there is negligibly small $t_{\rm sep}$ dependence, while the data at $t_{\rm sep}/a=10$ differs from others at smaller nonzero $q^2$. This observation allows us to use two possible combined values with $t_{\rm sep}/a=\{12,14,16\}$ and $t_{\rm sep}/a=\{14,16\}$ to obtain $G_E^v(q^2)$ without considerable excited state contamination. 

\begin{figure}[ht!]
    \includegraphics[width=8.0cm,keepaspectratio,clip]{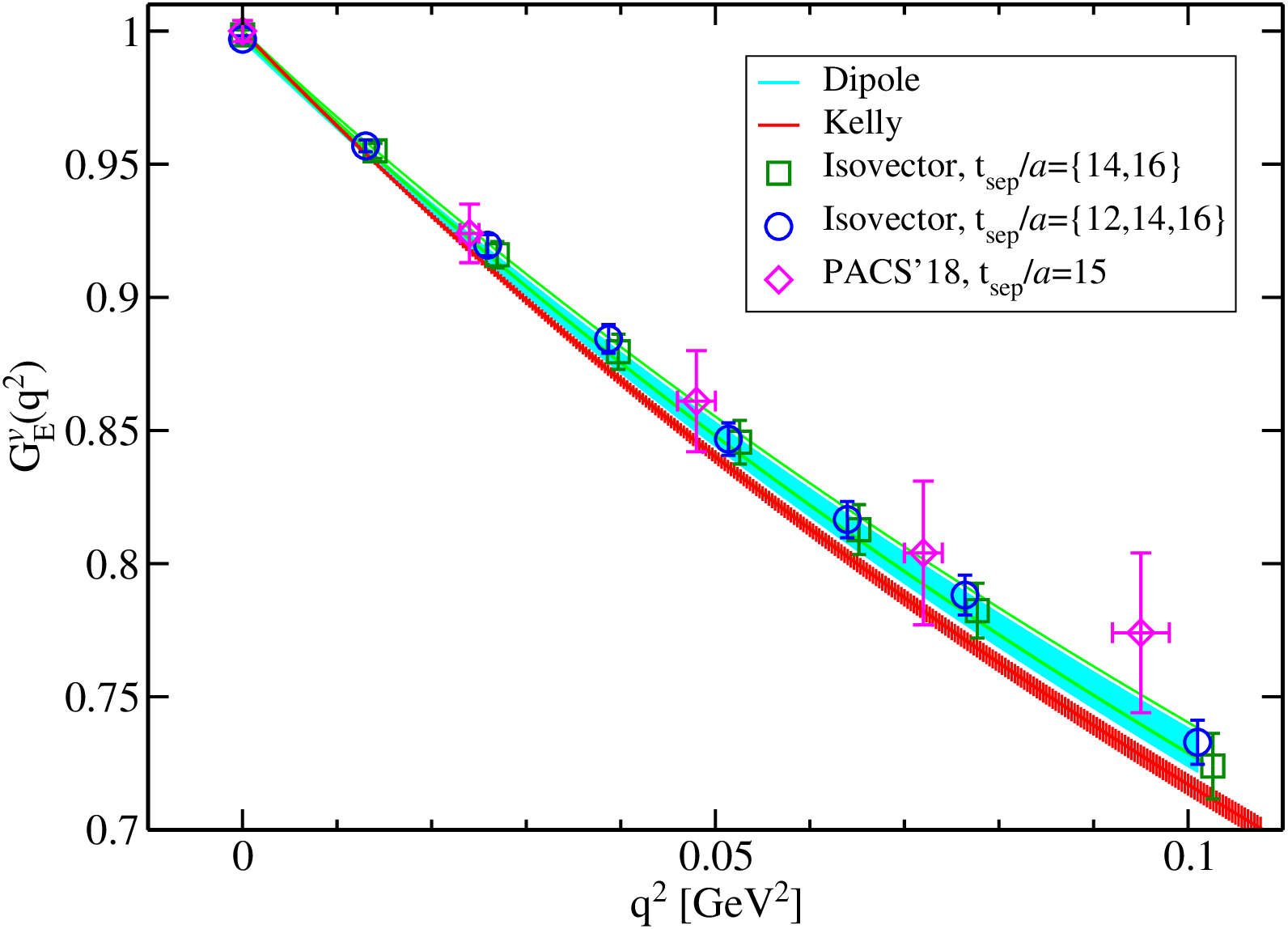}
\caption{$q^2$ dependence of the isovector electric form factor $G_E^v(q^2)$ obtained by the combined analysis of the results at $t_{\rm sep}/a=\{12,14,16\}$ (circle) and $t_{\rm sep}/a=\{14,16\}$ (square). Diamond symbols, which are obtained with $t_{\rm sep}/a=15$ on a $96^4$ lattice at $m_\pi = 146$ MeV in Ref.~\cite{Ishikawa:2018rew}, are also plotted for comparison.}
\label{fig:ge_q2}
\end{figure}

Figure~\ref{fig:ge_q2} shows the $q^2$ dependence of $G_E^v(q^2)$ together with the Kelly's fit~\cite{Kelly:2004hm}. The combined results with $t_{\rm sep}/a=\{12,14,16\}$ and $t_{\rm sep}/a=\{14,16\}$ are consistent with each other in all $q^2$. In the small $q^2$ region the LQCD data closely follow the Kelly's fit, while in the large $q^2$ region it is slightly but systematically above that. The figure also shows that our error is much smaller than the one of our previous results at $m_\pi = 0.146$ GeV on (8.1 fm)$^4$ in Ref.~\cite{Ishikawa:2018rew} thanks to the AMA technique described in Sec.~\ref{subsec:ama} and tuning the smearing parameters.

\begin{figure*}[ht!]
  \includegraphics[width=8.0cm,keepaspectratio,clip]{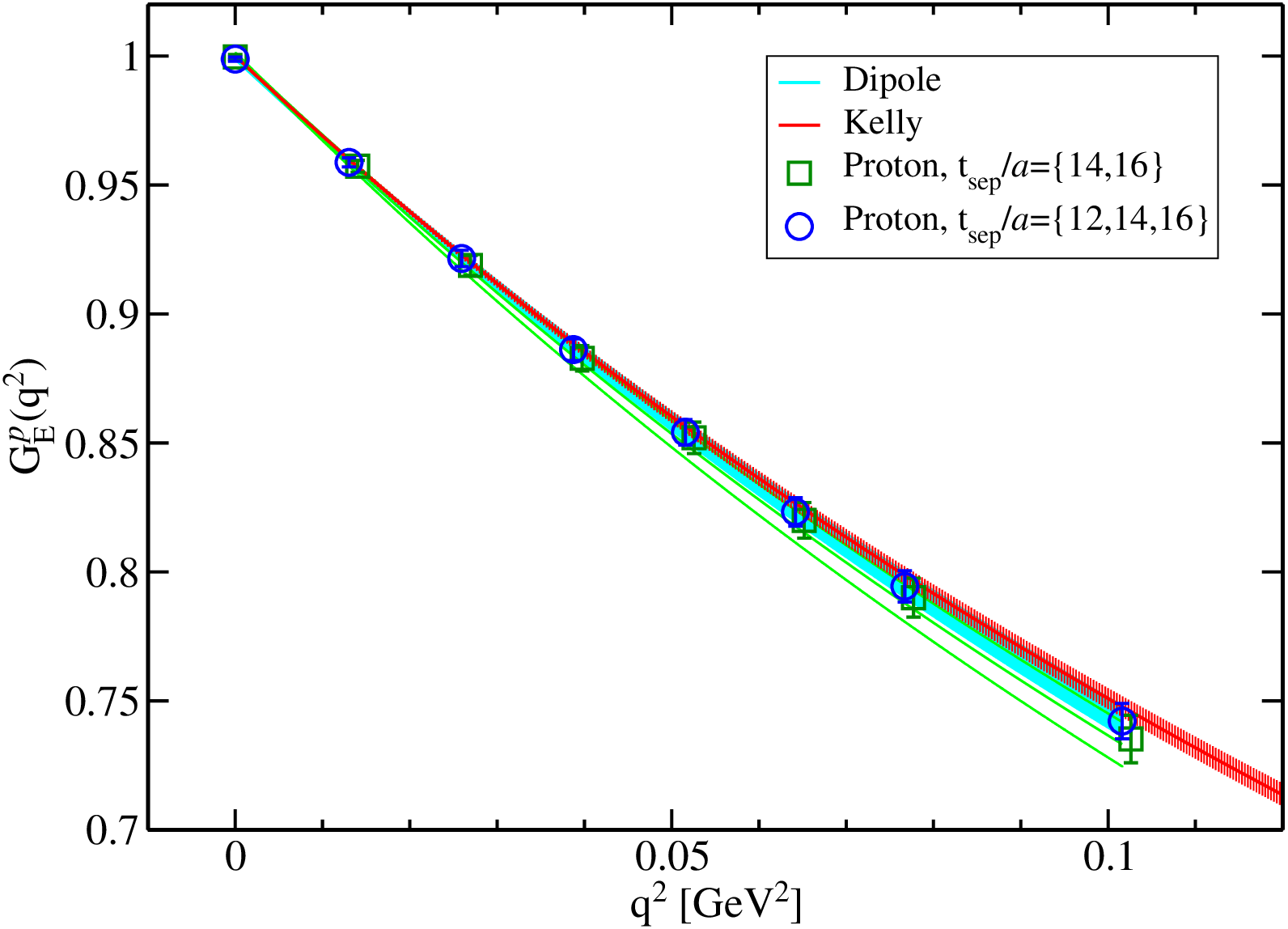}
  \hspace{3mm}
  \includegraphics[width=8.0cm,keepaspectratio,clip]{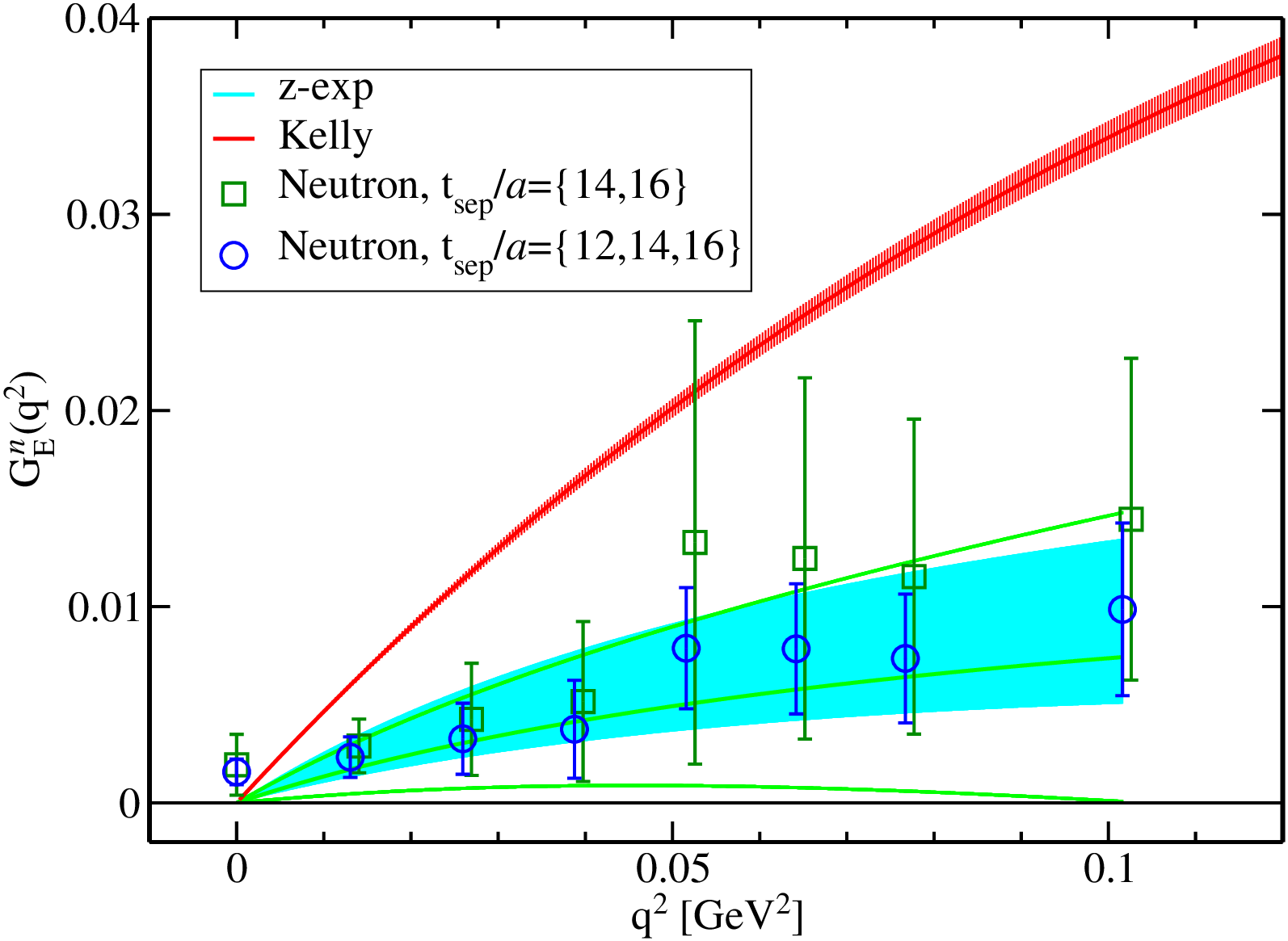}
\caption{Same as Figure~\ref{fig:ge_q2} for the proton (left) and neutron (right). Results are obtained without the disconnected diagram.}
\label{fig:ge_q2_p_n}
\end{figure*}

\begin{figure*}[ht!]
  \includegraphics[width=8.0cm,keepaspectratio,clip]{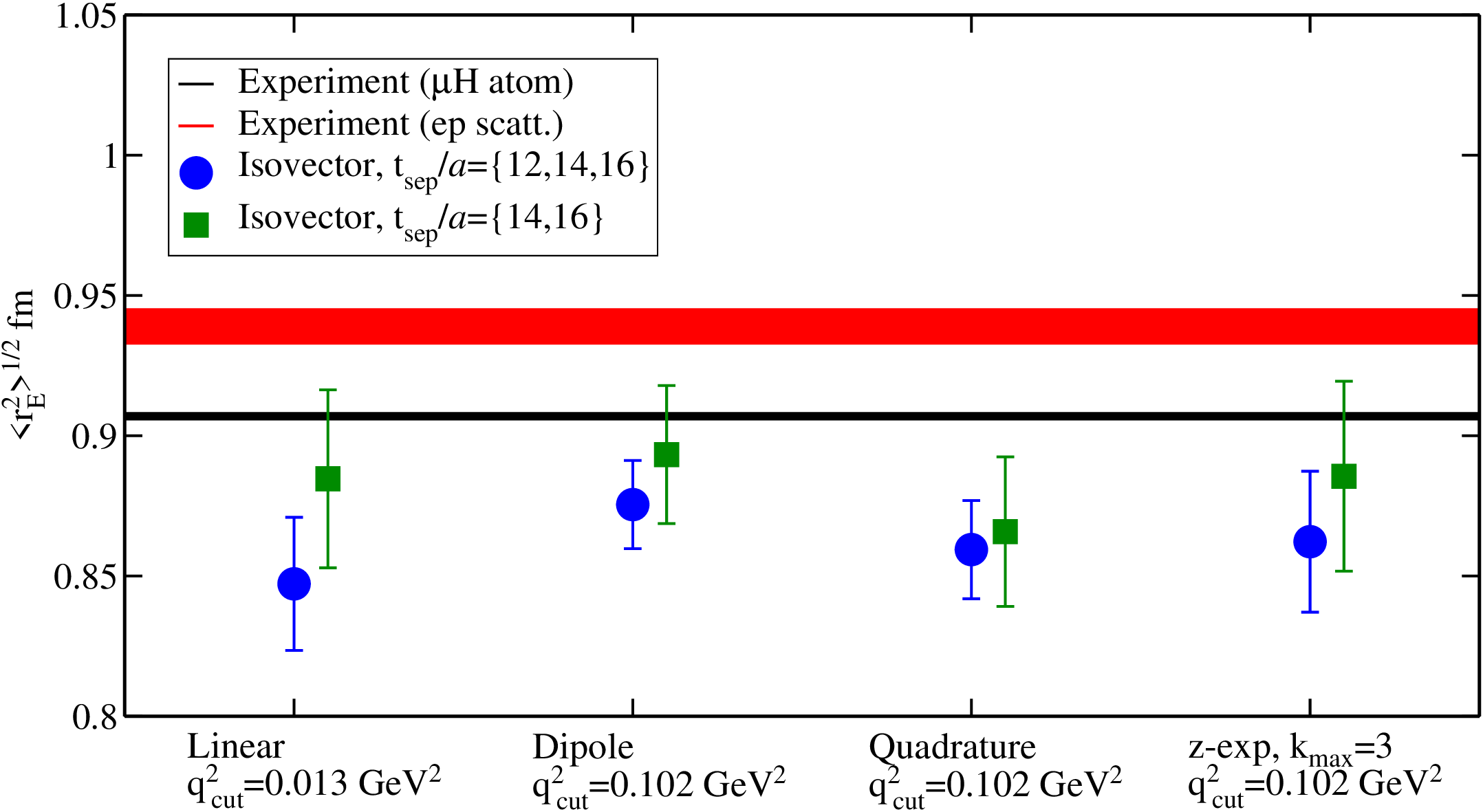}
  \hspace{3mm}
  \includegraphics[width=8.0cm,keepaspectratio,clip]{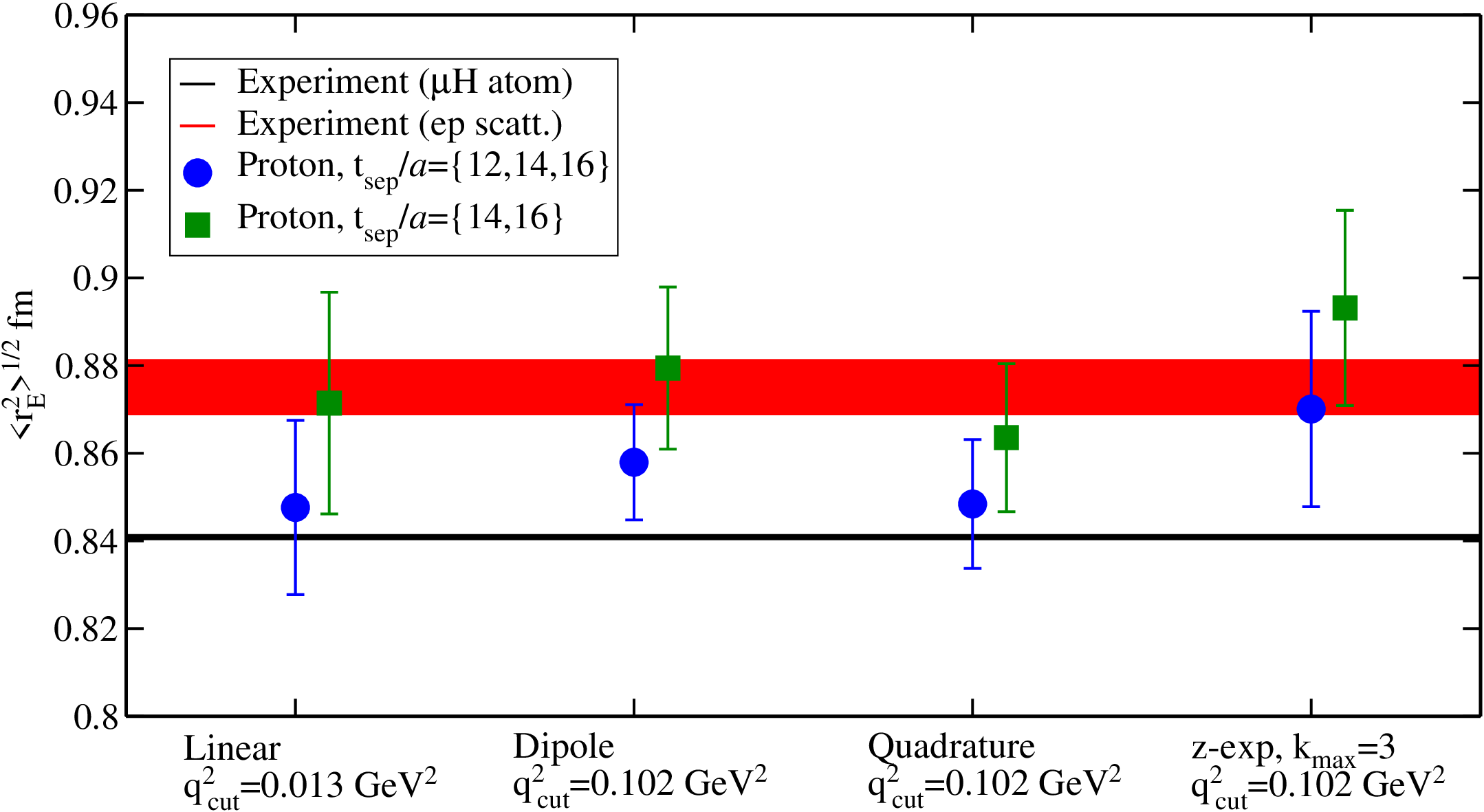}
\caption{Electric RMS radius $\sqrt{\la r^2_E\ra}$ for the isovector (left) and proton (right) obtained by linear, dipole, quadratic and $z$-expansion fits for the combined data. Horizontal bands represent the experimental results from $ep$ scattering (upper) and $\mu$H spectroscopy (lower). The results for the proton channel is obtained without the disconnected diagram.}
\label{fig:re_fitdep}
\end{figure*}

\begin{table*}[ht!]
\begin{ruledtabular}
\caption{
  Results for the electric RMS charge radius $\sqrt{\la r^2_{E}\ra}$ in the isovector, proton and neutron channels. In the row of ``This work'' we present our best estimates, where the first error is statistical and the second one is systematic as explained in the text. Results for the proton and neutron are obtained without the disconnected diagram. Our previous work was performed on a $96^4$ lattice at $m_\pi = 146$ MeV in Ref.~\cite{Ishikawa:2018rew}, where only the statistical errors are presented. \label{tab:re}}
\begin{tabular}{ccccccccccc}
  & & & \multicolumn{2}{c}{Isovector} & \multicolumn{2}{c}{Proton} & \multicolumn{2}{c}{Neutron}\\
  \hline
  Fit type & $q^2_{\rm cut}$ [GeV$^2$] & $t_{\rm sep}/a$ & $\sqrt{\la r^2_{E}\ra}$ [fm] & $\chi^2$/dof & $\sqrt{\la r^2_{E}\ra}$ [fm] & $\chi^2$/dof & $\la r^2_{E}\ra$ [fm$^2$] & $\chi^2$/dof\cr
  \hline
  \multirow{2}{*}{Linear}     & \multirow{2}{*}{0.013} & $\{12,14,16\}$ & { 0.847(23) } & {$-$} & { 0.848(19) } & {$-$} & {$-$}  & {$-$}\\
                              &                        & $\{14,16\}$    & { 0.885(31) } & {$-$} & { 0.871(25) } & {$-$} & {$-$}  & {$-$}\\
  \hline
  \multirow{2}{*}{Dipole}     & \multirow{2}{*}{0.102} & $\{12,14,16\}$ & { 0.875(15) } & { 0.8(6)}& { 0.858(13) } & { 1.1(8)}   & {$-$} & {$-$}\\
                              &                        & $\{14,16\}$    & { 0.893(24) } & { 0.5(5)}& { 0.879(18) } & { 0.8(7)}   & {$-$} & {$-$}\\
  \hline
  \multirow{2}{*}{Quadrature} & \multirow{2}{*}{0.102} & $\{12,14,16\}$ & { 0.859(17) } & { 0.6(6)}& { 0.848(14) } & { 1.2(1.0)} & {$-$0.037(18)} & {2.2(1.9)}\\
                              &                        & $\{14,16\}$    & { 0.866(26) } & { 0.7(7)}& { 0.864(16) } & { 1.4(1.1)} & {$-$0.029(23)} & {2.6(2.2)}\\
  \hline
  z-exp      & \multirow{2}{*}{0.102} & $\{12,14,16\}$ & { 0.862(25) } & { 0.9(8)}& { 0.870(22) } & { 1.1(9)}   & {$-$0.047(20)} & {1.8(1.7)}\\
  ($k_{\rm max}=3$)             &                        & $\{14,16\}$    & { 0.886(33) } & { 0.5(6)}& { 0.893(22) } & { 0.7(7)}   & {$-$0.035(25)} & {2.4(2.1)}\\
  \hline
  \multicolumn{2}{ c }{This work} & & 0.875(15)(28) & & 0.858(13)(35) & & $-$0.047(20)(18)\\
  \hline\hline
  \multicolumn{2}{ c }{PACS'18~\cite{Ishikawa:2018rew}} \cr
  \hline
  Dipole    & 0.215 & 15 & 0.822(63) & $-$ & $-$ & $-$ & $-$ & $-$ \cr
  \hline
  Quadratic & 0.215 & 15 & 0.851(62) & $-$ & $-$ & $-$ & $-$ & $-$  \cr
  \hline
  z-exp     & \multirow{2}{*}{0.215} & \multirow{2}{*}{15} & \multirow{2}{*}{0.914(101)} & \multirow{2}{*}{$-$} & \multirow{2}{*}{$-$} & \multirow{2}{*}{$-$}  & \multirow{2}{*}{$-$} & \multirow{2}{*}{$-$} \cr
  ($k_{\rm max}=3$)\\
  \hline\hline
  \multicolumn{2}{ c }{Experimental value}  &  &  \cr
  \hline
  \multicolumn{2}{ c }{$ep$ scattering} & & 0.939(6) & & 0.875(6) & & $-$0.1161(22) \cr
  \multicolumn{2}{ c }{$\mu$H atom} & & 0.907(1) & & 0.8409(4) & & $-$\cr
\end{tabular}
\end{ruledtabular}
\end{table*}

We also calculate $G_E^p(q^2)$ and $G_E^n(q^2)$ separately without the disconnected diagram.
They have similar properties to $G_E^v(q^2)$:
tiny $t_{\rm sep}$ dependence and good plateau in ${\cal R}^{t,p}_{j_{em},4}(t,{\bm q})$ and ${\cal R}^{t,n}_{j_{em},4}(t,{\bm q})$.
The combined results with $t_{\rm sep}/a = \{12,14,16\}$ and $\{14,16\}$
for $G_E^p(q^2)$ and $G_E^n(q^2)$ are summarized in Appendix~\ref{app:tab}
together with the ones of $G_E^v(q^2)$.
The results for $G_E^p(q^2)$ and $G_E^n(q^2)$
are compared with the Kelly's fit in Fig.~\ref{fig:ge_q2_p_n}.
We observe that $G_E^p(q^2)$ is closer to Kelly's fit rather than 
$G_E^v(q^2)$.
On the other hand, $G_E^n(q^2)$ is much smaller than the Kelly's fit. One possible reason is an uncertainty due to the missing disconnected diagram in the isoscalar channel, which could affect $G_E^p(q^2)$ and $G_E^n(q^2)$. 
Actually, a recent work using the mixed actions with the domain-wall sea quarks and the overlap valence ones implies the contribution of the disconnected diagram to $G_E^p(q^2)$ and $G_E^n(q^2)$ amounts to $\sim$ 0.005 with rather large statistical errors\footnote{This is just the value in the light quark flavor since the strange quark contribution is negligible.} at $q^2\approx0.05$ GeV$^2$ in $m_\pi=135$ MeV~\cite{Sufian:2017osl}, whose magnitude is comparable to the difference between our $G_E^n(q^2)$ and the experimental value. To completely resolve the problem we need to evaluate the isoscalar vector form factor in the future. 

With the use of the correlated fit procedure, we compare four types of fitting functions to examine the uncertainty in the extrapolation of the slope to $q^2=0$: linear function $G_E(q^2)  = d_0+d_1q^{2}$, dipole form of Eq.~(\ref{Eq:dipole}), quadratic function $G_E(q^2)  = d_0+d_1q^{2}+d_2q^{4}$ and the model-independent $z$-expansion method with Eq.~(\ref{Eq:z-value}) with $k_{\rm max} = 3$. In Figs.~\ref{fig:ge_q2} and \ref{fig:ge_q2_p_n} we find that the dipole form well describes the LQCD results for $G_E^v(q^2)$ and $G^p_E(q^2)$ up to the maximum fitting range of $q^2_{\rm cut}=0.102$ GeV$^2$. We plot the fit form dependence of $\sqrt{\la r^2_{E}\ra}$ in Fig.~\ref{fig:re_fitdep}, where the upper and lower bands denote the experimental results of the $ep$ scattering  and the spectroscopy of the muonic hydrogen ($\mu$H) atom, respectively. The numerical results are summarized in Table~\ref{tab:re} together with the experimental values.  We observe that all the fit procedures show good consistency within the error bars both for the isovector and proton channels with a reasonable $\chi^2$/dof, which is evaluated by jackknife estimator in correlated fit. We also find that the combined results with $t_{\rm sep}/a=\{12,14,16\}$ are consistent with those with $t_{\rm sep}/a=\{14,16\}$ within the error bars, which indicates that the excited state contamination in $G_E(q^2)$ is under control. Note that, in the case of neutron, one can find a clear deviation from the experimental value due to the lack of the disconnected diagram as already mentioned above. 

As shown in this section, the LQCD calculation at the low $q^2$ region up to 0.11 GeV$^2$ allows us to successfully reduce the uncertainties stemming from the choice of the fitting procedures. Their central values, however, slightly fluctuate depending on each fitting procedure and choice of $t_{\rm sep}$ range. We then take a result of $\sqrt{\la r^2_{E}\ra}$ with the dipole form at $t_{\rm sep}/a=\{12,14,16\}$ as our best estimate of the central value and its statistical error. The maximum difference between the central value in the dipole fit with $t_{\rm sep}/a=\{12,14,16\}$ and those in other fitting procedures with two choices of the combined $t_{\rm sep}$ ranges is taken as the systematic error (see Table~\ref{tab:re}).

Although our result of the isovector channel stays around the value of the $\mu$H experiment rather than that of the $ep$ scattering experiment, it may be too early to conclude its preference at this stage because of relatively large error bars. For the proton case, on the other hand, LQCD value stays amid those experimental values. For the definite conclusion we need more precise calculation including the disconnected diagram. This is, however, an encouraging situation indicating a possibility that LQCD can distinguish the two experimental results in near future.

\subsection{Magnetic form factor and magnetic RMS radius}
\label{subse:nff_gm}

\begin{figure*}[ht!]
  \includegraphics[width=8.0cm,keepaspectratio,clip]{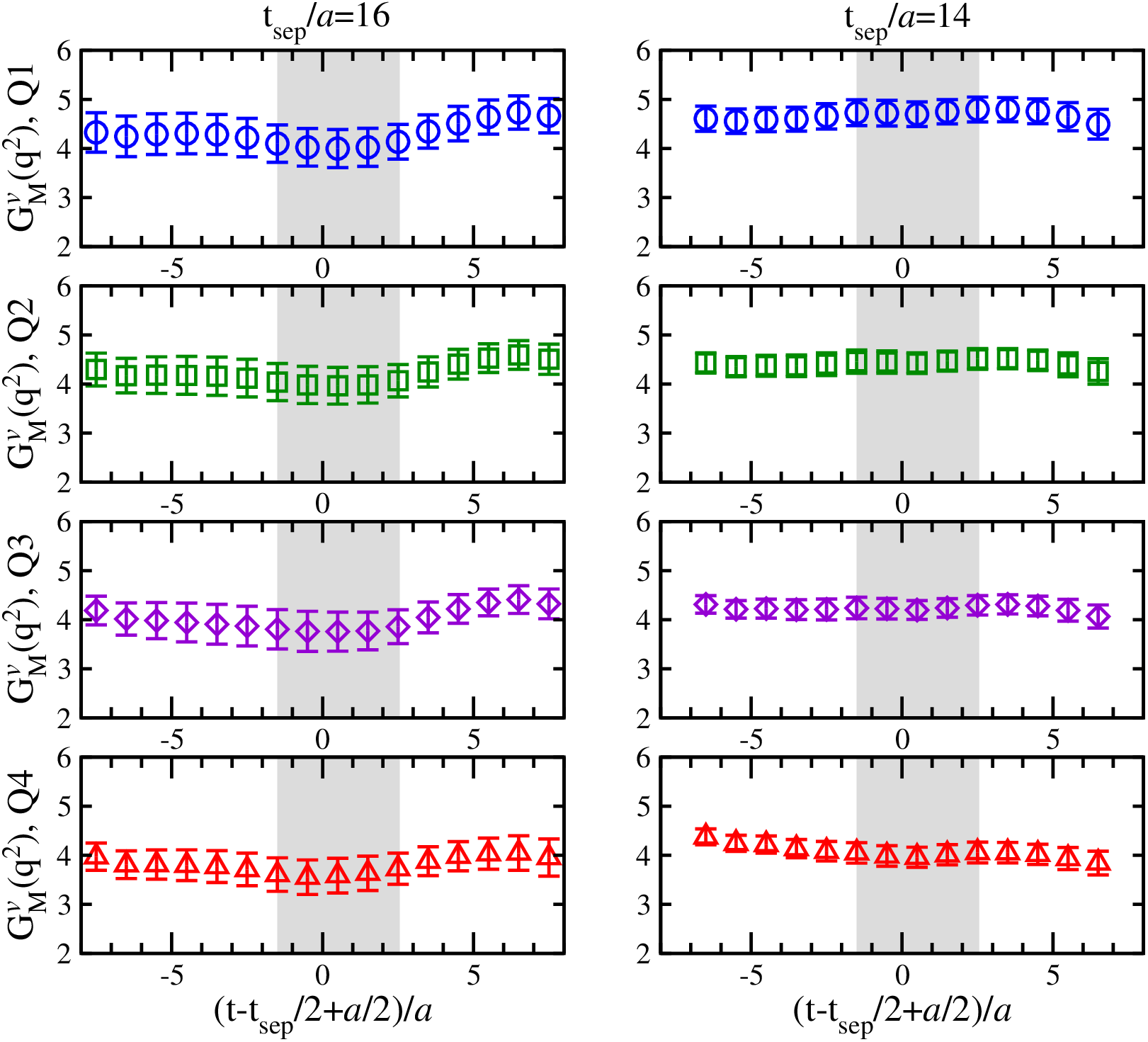}
  \hspace{3mm}
  \includegraphics[width=8.0cm,keepaspectratio,clip]{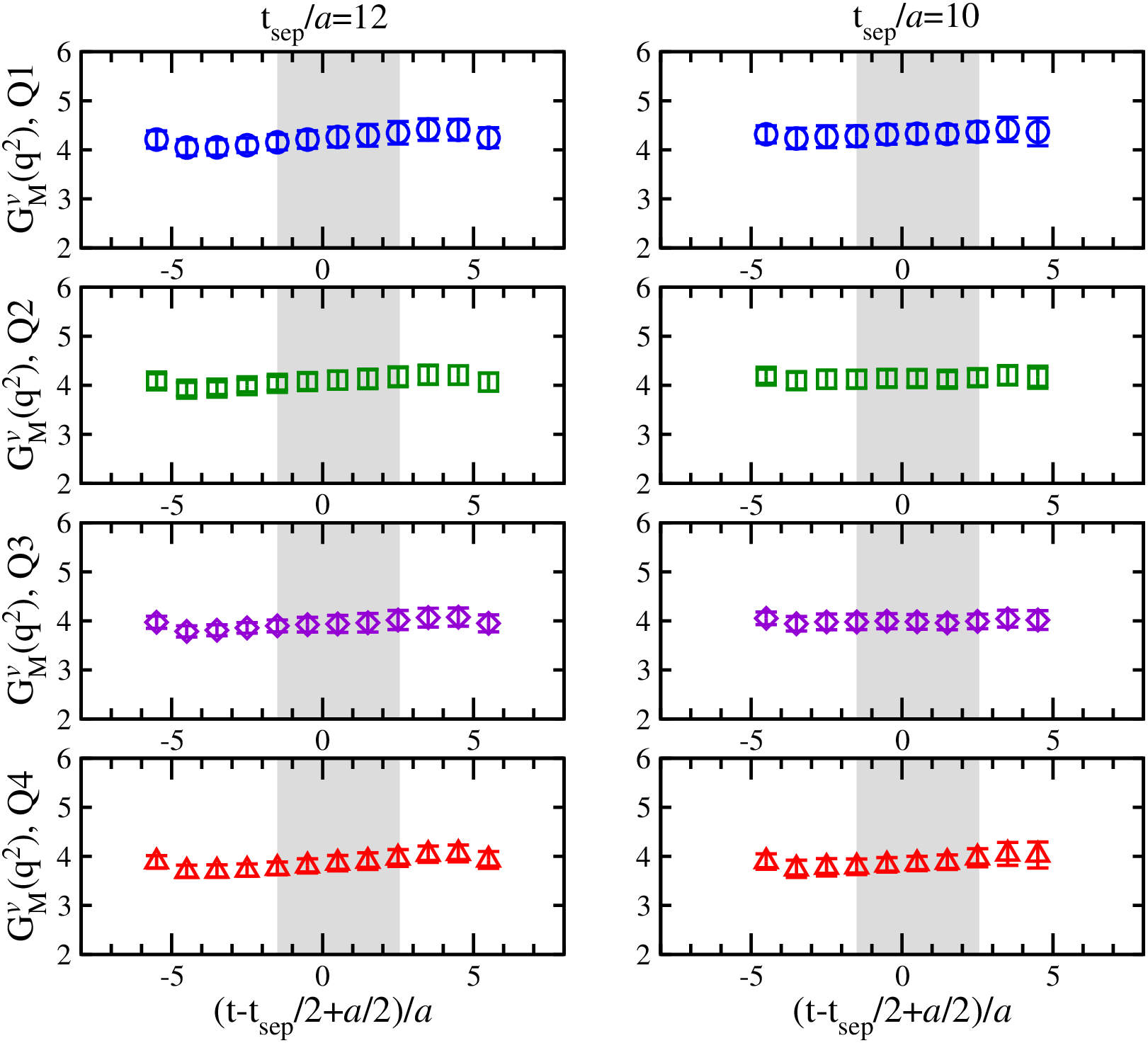}
\caption{Same as Fig.~\ref{fig:ratio_ge} for the ratio of Eq.~(\ref{Eq:R_GM}) to extract the isovector magnetic form factor $G_M^v(q^2)$.}
\label{fig:ratio_gm}
\end{figure*}

\begin{figure}[ht!]
  \includegraphics[width=8.0cm,keepaspectratio,clip]{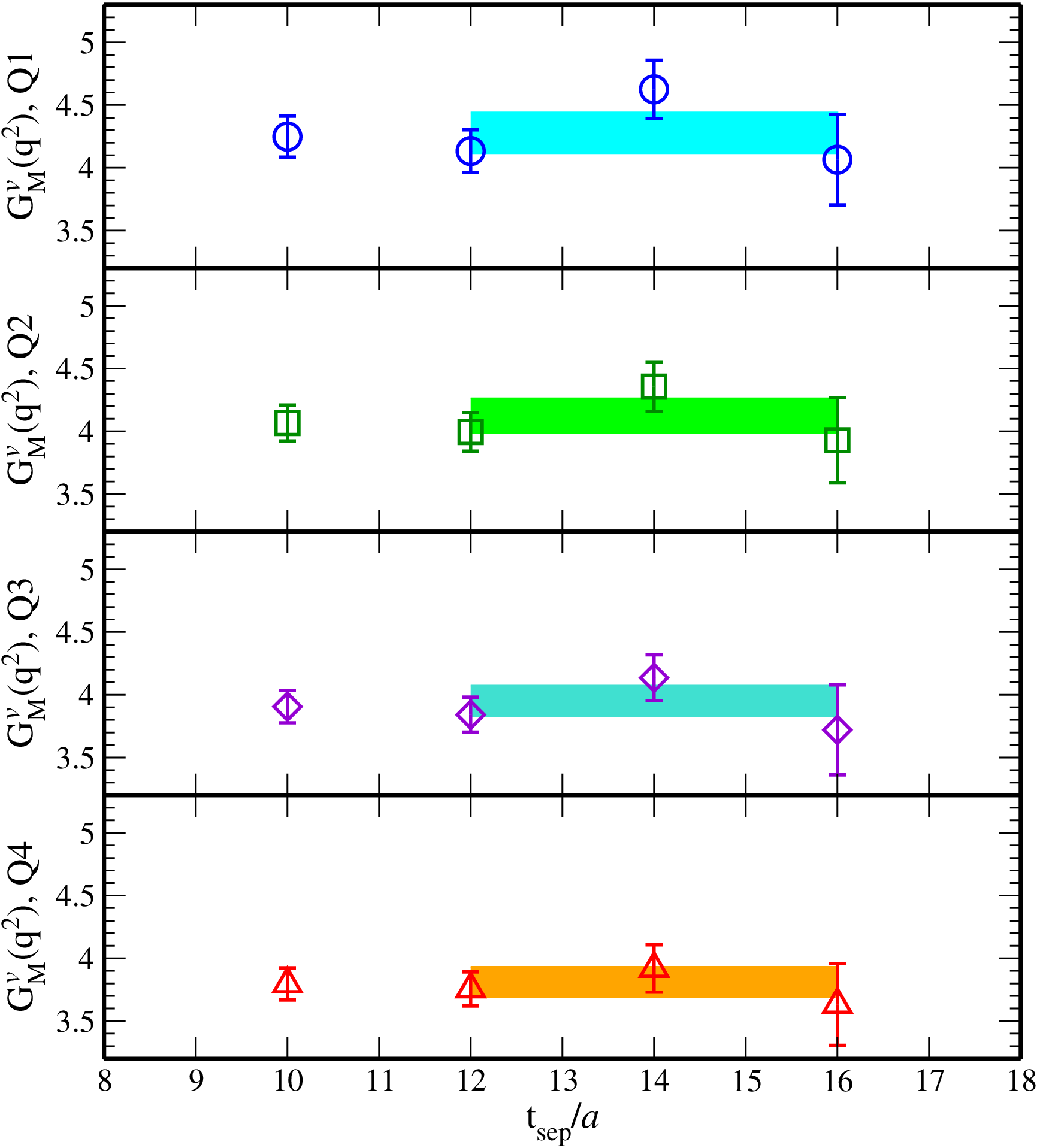}
  \caption{Same as Fig.~\ref{fig:ge_ts} for the isovector magnetic form factor $G_M^v(q^2)$ with four lowest nonzero momentum transfers.}
\label{fig:gm_ts}
\end{figure}

The isovector magnetic form factor $G_M^v(q^2)$ is extracted from the ratio ${\cal R}^{5z,N}_{j_{em},i}(t,{\bm q})$ of Eq.~(\ref{Eq:R_GM}). The analysis of $G_M^v(q^2)$ is performed in parallel with $G_E^v(q^2)$. We first plot the $t$ dependence of the ratio with $\vert {\bm n}\vert^2=1,2,3,4$ for $t_{\rm sep}/a=10,12,14,16$ in Fig.~\ref{fig:ratio_gm}, which show good plateau for all the cases of $\vert {\bm n}\vert^2$ and $t_{\rm sep}/a$. We extract $G_M^v(q^2)$ with the constant fit employing the same fitting range as in the $G_E^v(q^2)$ case. Figure~\ref{fig:gm_ts} shows that the results for $t_{\rm sep}/a=10,12, 14, 16$ agree with each other within 1-$\sigma$ error bars, though the statistical fluctuation is much larger than the $G_E^v(q^2)$ case. We evaluate $G_M^p(q^2)$ and $G_M^n(q^2)$ separately from each ${\cal R}^{5z,N}_{j^{\rm em},i}(t,{\bm q})$ for $N=p,n$, where we omit the disconnected diagram. As in $G_M^v(q^2)$, all the ratios of ${\cal R}^{5z,N}_{j^{\rm em},i}(t,{\bm q})$ have reasonable plateaus, and those values are consistent in the four $t_{\rm sep}$ cases. At each $q^2$ we take two combined values obtained by the constant fit in the two ranges of $t_{\rm sep}/a=\{12,14,16\}$ and $t_{\rm sep}/a=\{14,16\}$ for $G_M^v(q^2)$, $G_M^p(q^2)$, and $G_M^n(q^2)$. 
Those values are summarized in Appendix~\ref{app:tab}.

\begin{figure}[ht!]
  \includegraphics[width=8.0cm,keepaspectratio,clip]{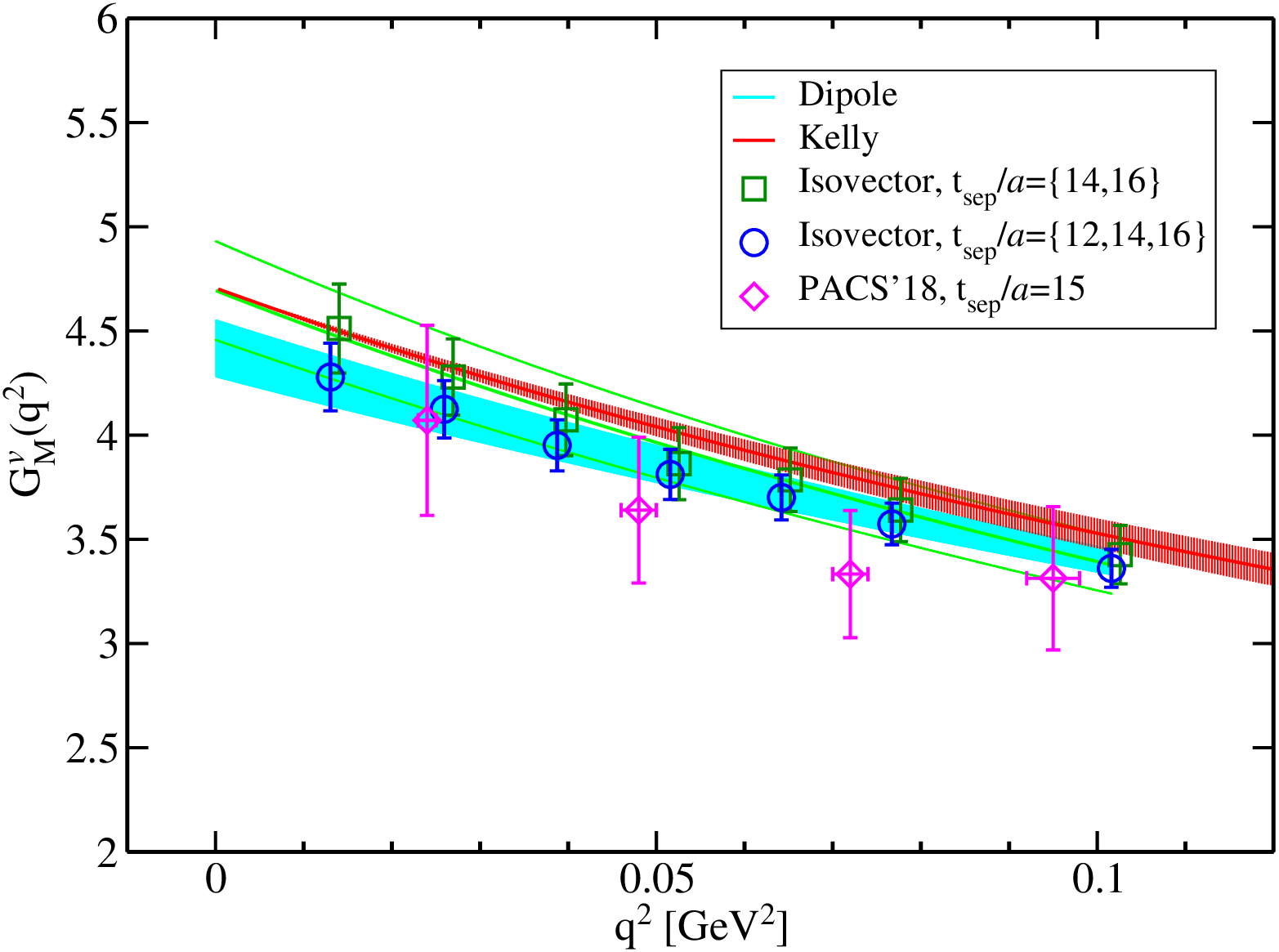}
\caption{Same as Fig.~\ref{fig:ge_q2} for the isovector magnetic form factor $G_M^v(q^2)$.}
\label{fig:gm_q2}
\end{figure}

\begin{figure*}[ht!]
  \includegraphics[width=8.0cm,keepaspectratio,clip]{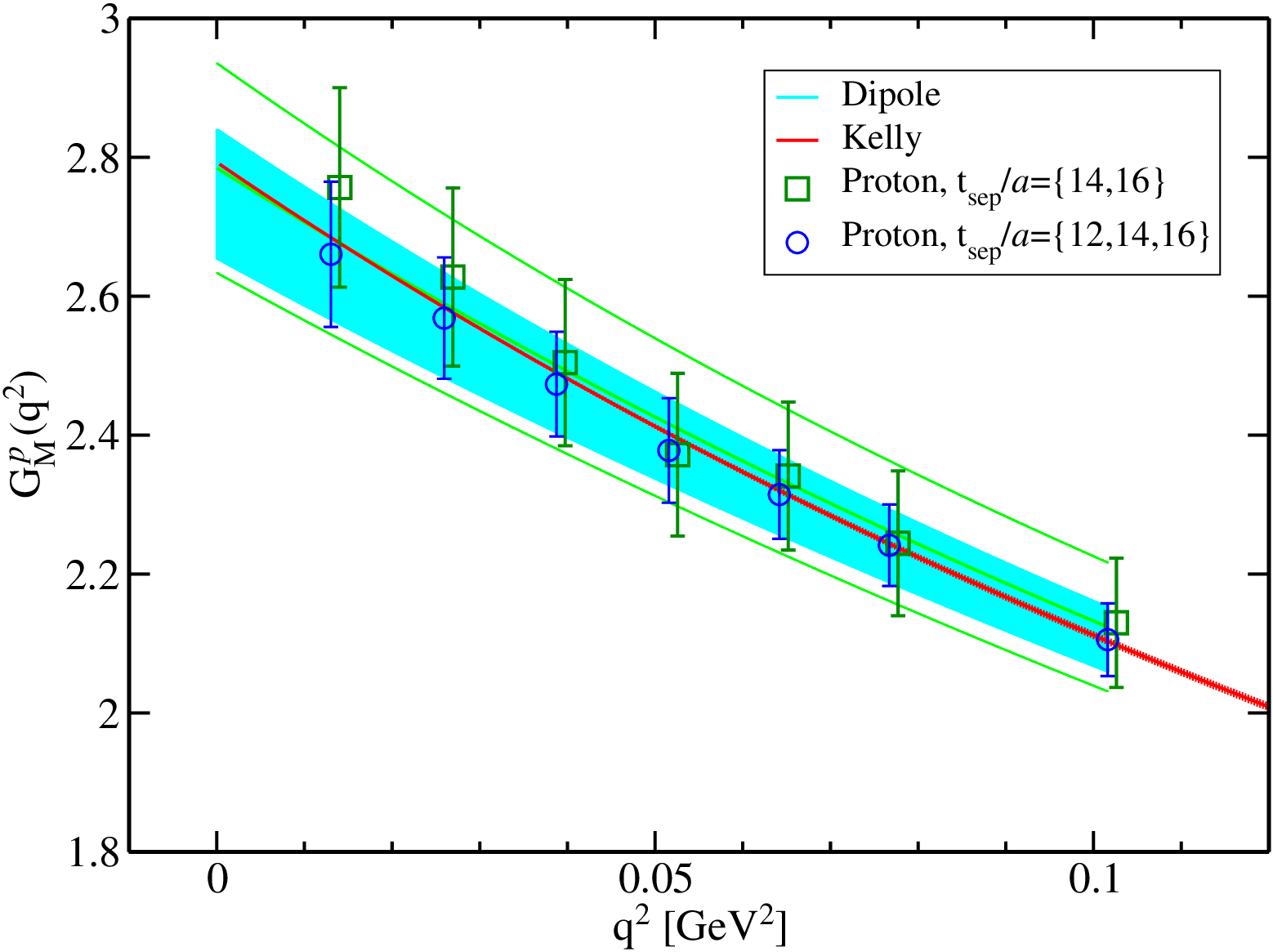}
  \hspace{3mm}
  \includegraphics[width=8.0cm,keepaspectratio,clip]{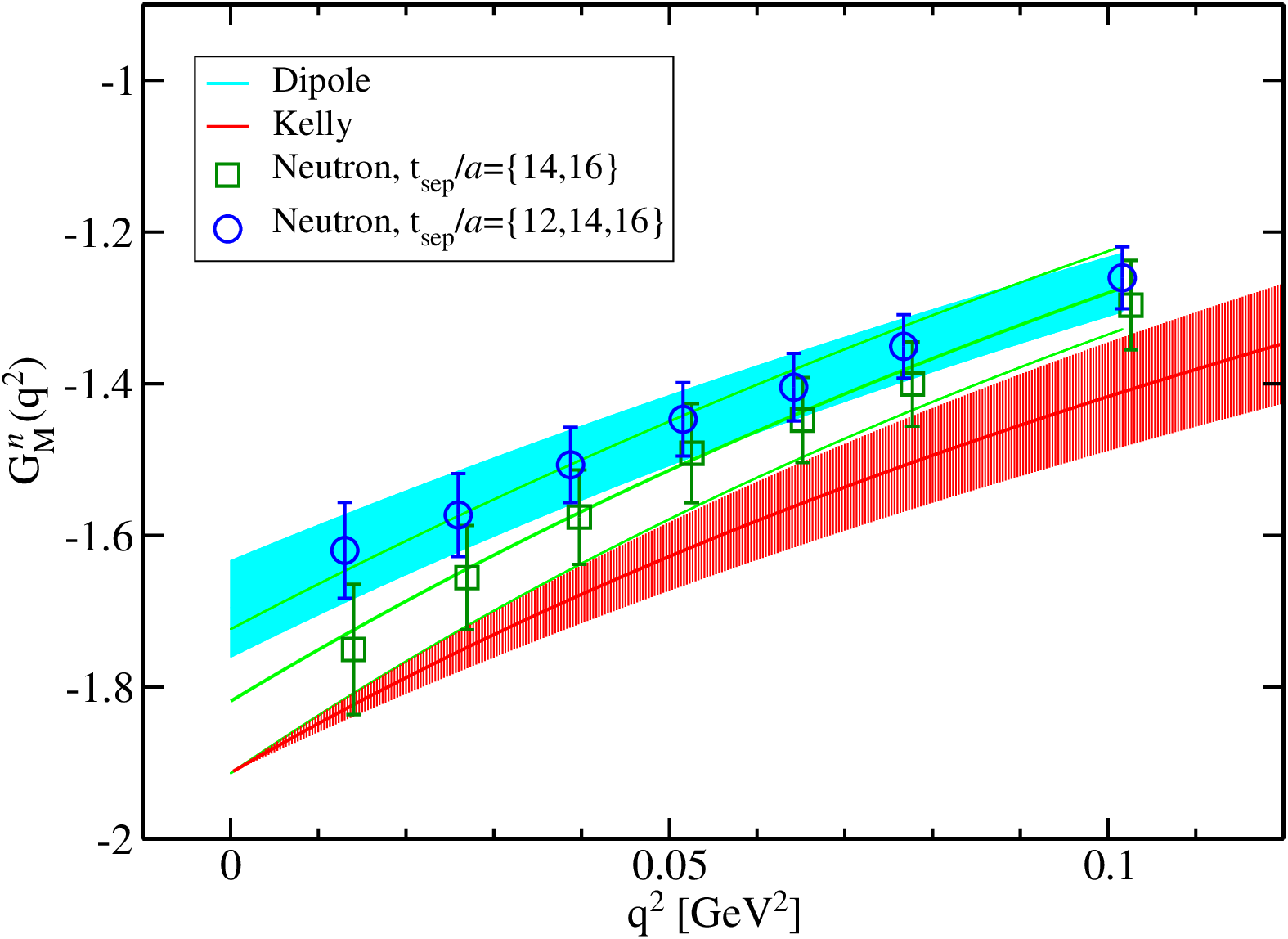}
\caption{Same as Fig.~\ref{fig:ge_q2} for the proton (left) and neutron (right) magnetic form factor $G_M(q^2)$. The
results are obtained without the disconnected diagram.}
\label{fig:gm_q2_p_n}
\end{figure*}

Figure~\ref{fig:gm_q2} shows that the results from the two combined $t_{\rm sep}$ ranges are consistent with each other. These results are compared with that of our previous calculation~\cite{Ishikawa:2018rew}. Our current result has much smaller error than the previous one, and closer to the Kelly's fit.
In Figs.~\ref{fig:gm_q2} and \ref{fig:gm_q2_p_n} we observe that the $q^2$ dependence of $G_M^{v}(q^2)$ and $G_M^p(q^2)$ is consistent with the Kelly's fit within the 1.5-$\sigma$ error, though $G_M^n(q^2)$ for $t_{\rm sep}=\{12,14,16\}$ in the smaller $q^2$ region shows slight deviation from the Kelly's fit. This could be due to the lack of the disconnected contribution as well as $G_E^n(q^2)$ case. Note that the negative disconnected contribution of $\sim-0.03$ at $q^2\approx0.05$ GeV$^2$ in $m_\pi=135$ MeV implied in Ref.~\cite{Sufian:2017osl} could make our result of $G_M^n(q^2)$ closer to the experimental value. 

\begin{figure}[ht!]
  \includegraphics[width=7.0cm,keepaspectratio,clip]{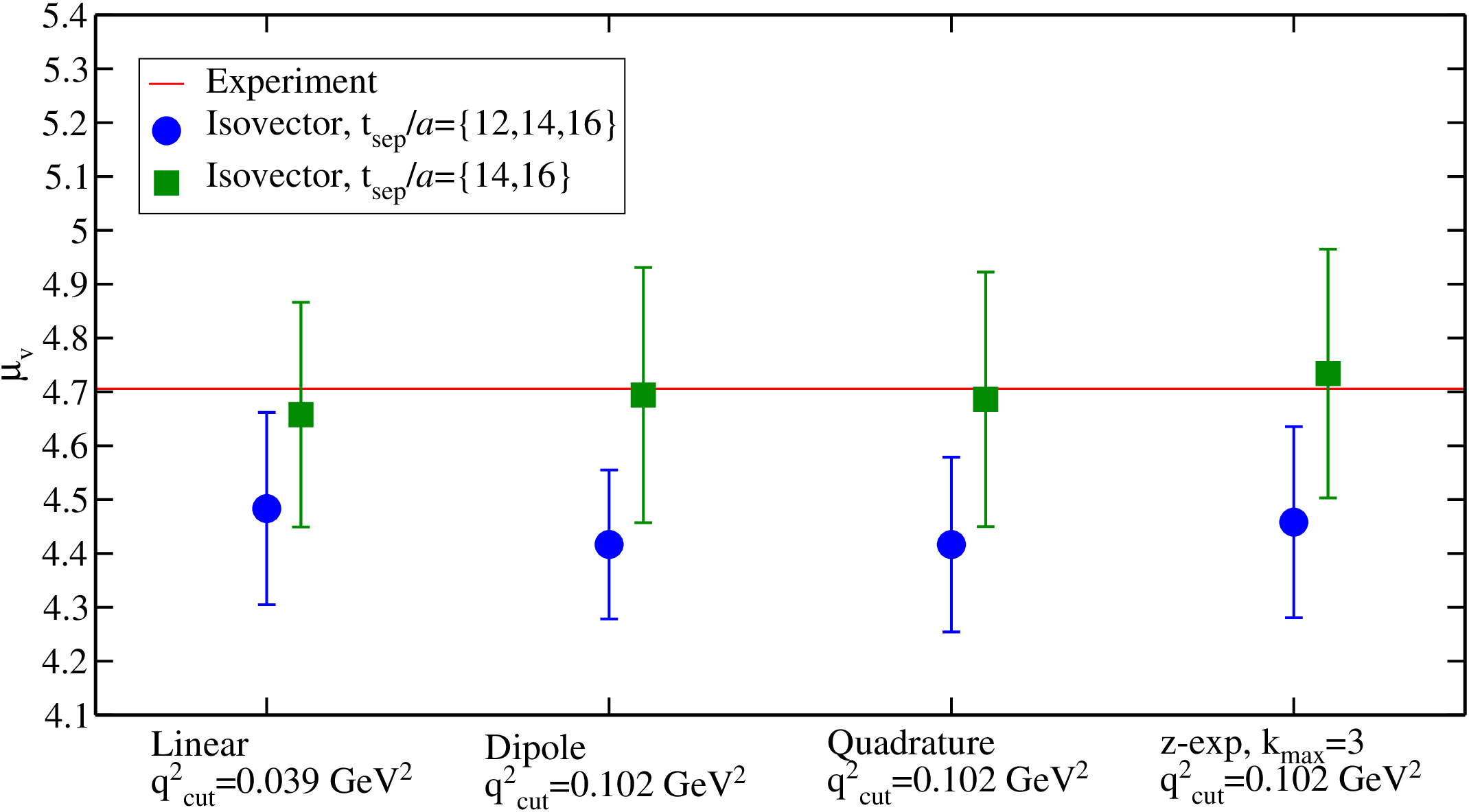}
  \vskip 3mm
  \includegraphics[width=7.0cm,keepaspectratio,clip]{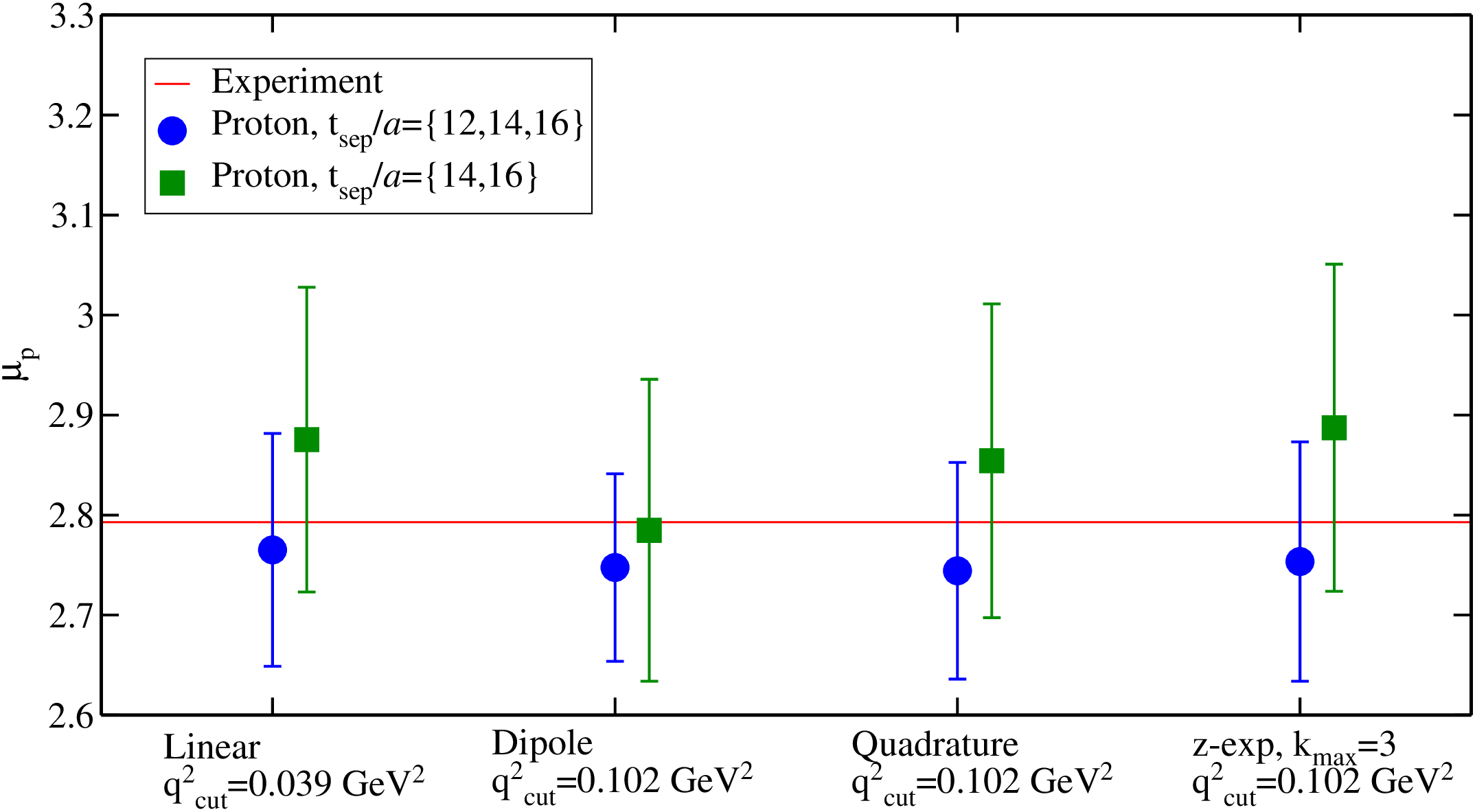}
  \vskip 3mm
  \includegraphics[width=7.0cm,keepaspectratio,clip]{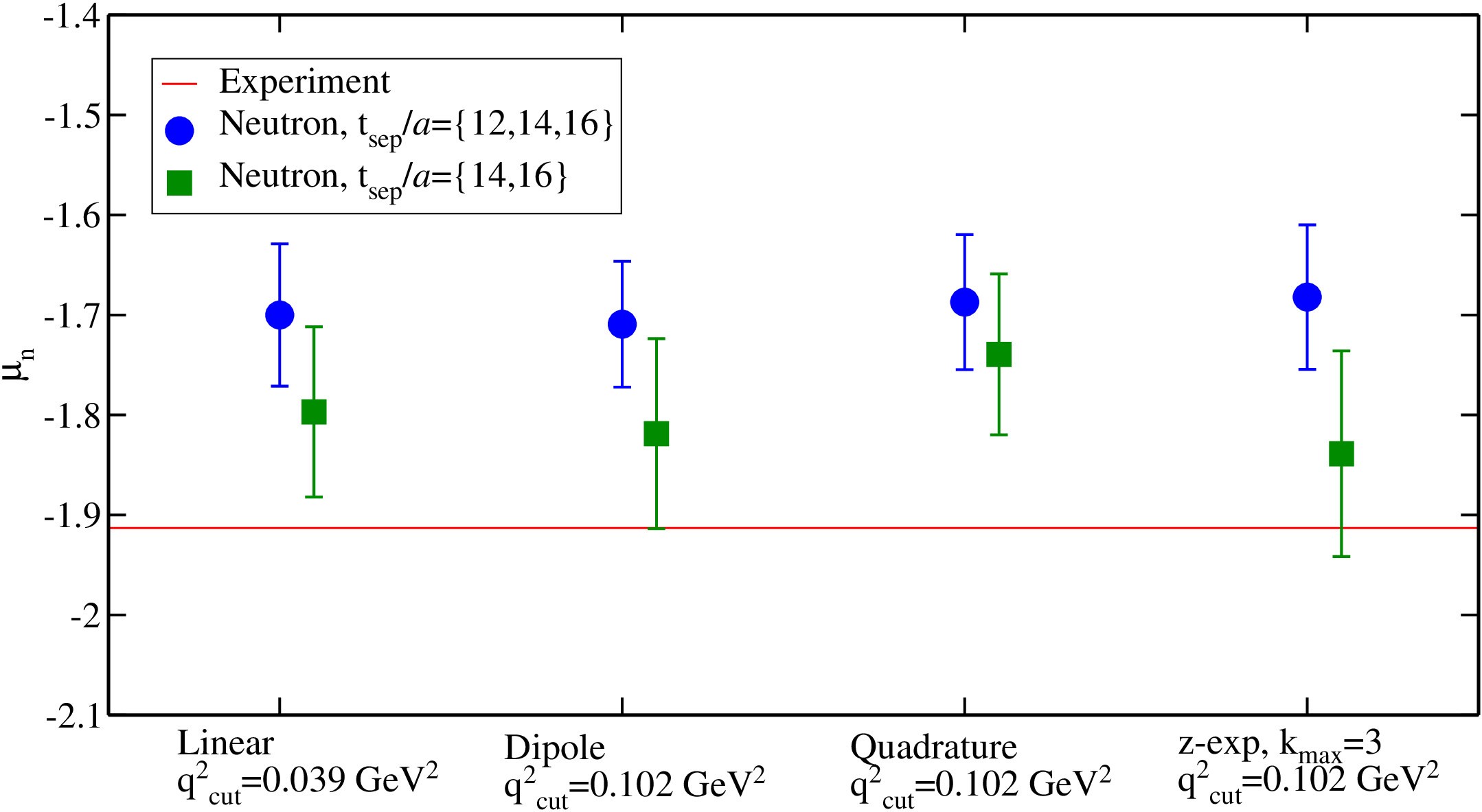}  
\caption{Magnetic moment $\mu$ for the isovector (top), proton (middle) and neutron (bottom) channels obtained by the fitting with the linear, dipole, quadratic forms and the $z$-expansion method for the combined data. Horizontal bands represent the experimental results. Two types of symbols denote the results with two choices of the combined $t_{\rm sep}$ ranges. Results for the proton and neutron channels are obtained without the disconnected diagram.}
\label{fig:mm_fit}
\end{figure}

\begin{figure}[ht!]
  \includegraphics[width=7.0cm,keepaspectratio,clip]{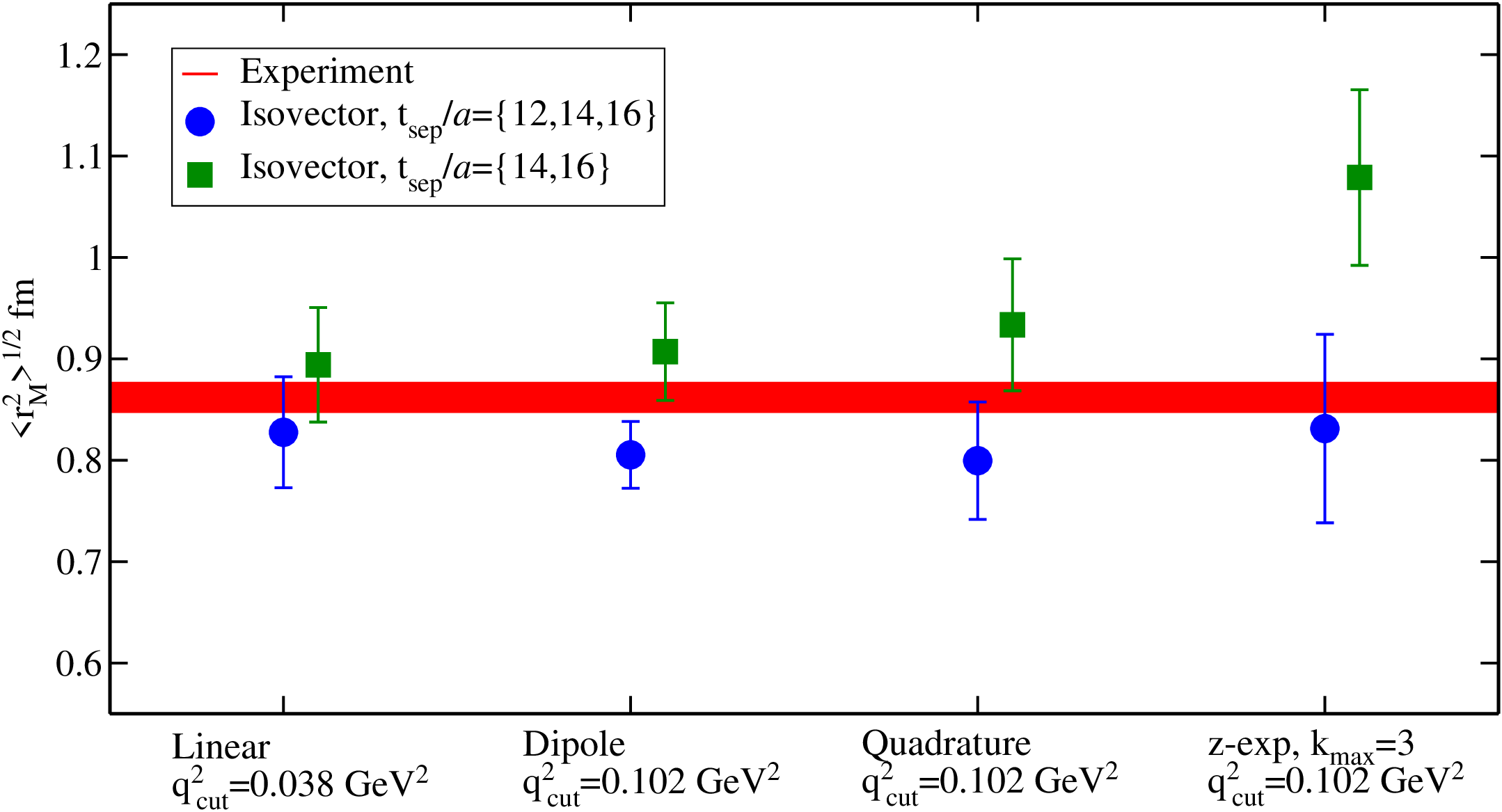}
  \vskip 3mm
  \includegraphics[width=7.0cm,keepaspectratio,clip]{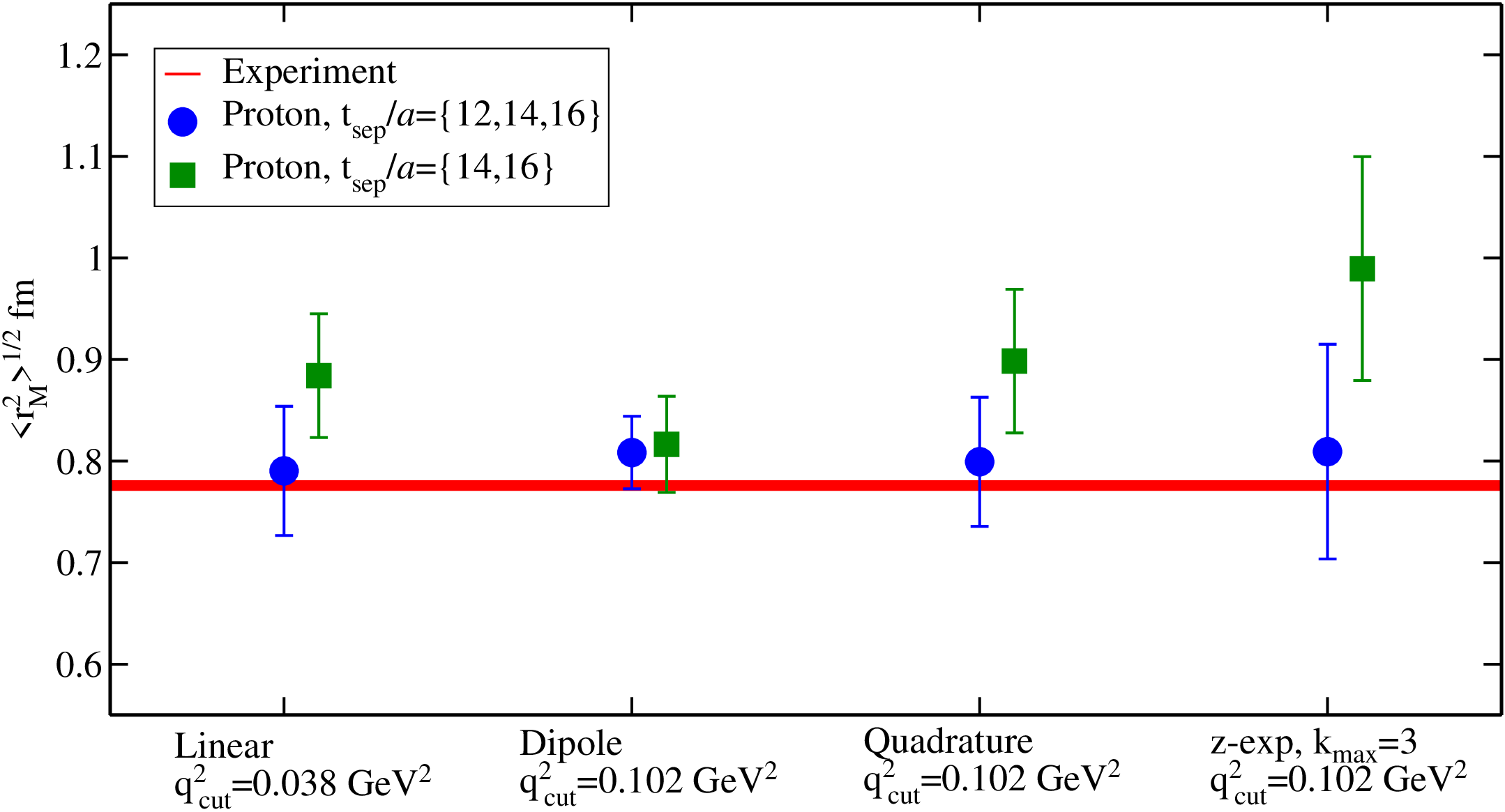}
    \vskip 3mm
  \includegraphics[width=7.0cm,keepaspectratio,clip]{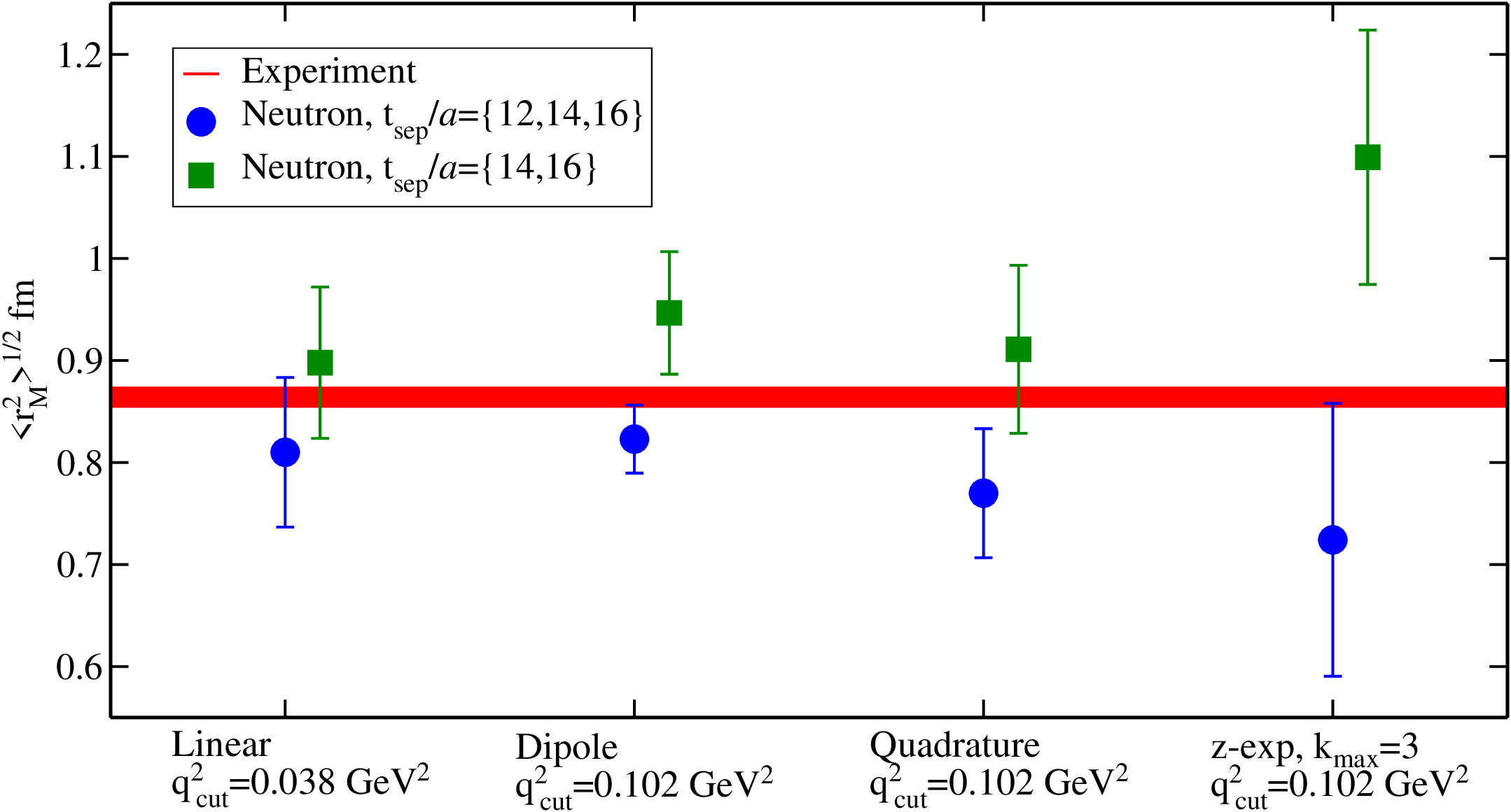}
\caption{Same as Fig.~\ref{fig:mm_fit} for the magnetic RMS radius $\sqrt{\la r^2_M\ra}$.}
\label{fig:rm_fit}
\end{figure}

\begin{table*}[ht!]
\begin{ruledtabular}
\caption{
Results for the magnetic moments $\mu$ and magnetic RMS radius
$\sqrt{\la r^2_{M}\ra}$ for the isovector, proton and neutron channels. In the row of ``This work'' we present our best estimates, where the first error is statistical and the second one is systematic as explained in the text. Results for the proton and neutron are obtained without the disconnected diagram. Previous work was performed on a $96^4$ lattice at $m_\pi = 146$ MeV in Ref.~\cite{Ishikawa:2018rew}, where only the statistical errors are presented.
\label{tab:murm}}
\begin{tabular}{ccccccccccccc}
  & & & \multicolumn{3}{c}{Isovector}\\
  \hline
  Fit type & $q^2_{\rm cut}$ [GeV$^2$] & $t_{\rm sep}/a$ & $\mu_v$ & $\sqrt{\la r^2_{M}\ra}$ [fm]& $\chi^2$/dof \\
  \hline
  \multirow{2}{*}{Linear}     & \multirow{2}{*}{0.039} & $\{12,14,16\}$ & { 4.483(178) } & { 0.828(54) } & { 0.6(1.7)}\\
                              &                        & $\{14,16\}$    & { 4.658(208) } & { 0.894(56) } & { 0.4(1.3)}\\
  \hline
  \multirow{2}{*}{Dipole}     & \multirow{2}{*}{0.102} & $\{12,14,16\}$ & { 4.417(138) } & { 0.805(32) } & { 0.9(7)}\\
                              &                        & $\{14,16\}$    & { 4.694(236) } & { 0.907(48) } & { 3.1(3.6)}\\
  \hline
  \multirow{2}{*}{Quadrature} & \multirow{2}{*}{0.102} & $\{12,14,16\}$ & { 4.417(162) } & { 0.800(57) } & { 1.2(1.1)}\\
                              &                        & $\{14,16\}$    & { 4.546(201) } & { 0.938(59) } & { 1.6(1.2)}\\
  \hline
  z-exp      & \multirow{2}{*}{0.102} & $\{12,14,16\}$ & { 4.458(177) } & { 0.831(92) } & { 1.1(1.0)}\\
  ($k_{\rm max}=3$) &                        & $\{14,16\}$    & { 4.734(231) } &{ 1.079(86) }  & { 0.9(9)}\\
  \hline
  \multicolumn{2}{ c }{This work} & & 4.417(138)(317) & 0.805(32)(274) & \\
  \hline\hline
  \multicolumn{2}{ c }{PACS'18~\cite{Ishikawa:2018rew}} \cr
  \hline
  Dipole    & 0.215 & 15 & 3.96(46) & 0.656(133) & $-$ &  \cr
  \hline
  Quadratic & 0.215 & 15 & 4.24(52) & 0.852(130) & $-$ \cr
  \hline
  z-exp     & \multirow{2}{*}{0.215} & \multirow{2}{*}{15} & \multirow{2}{*}{4.86(82)} & \multirow{2}{*}{1.495(437)} & $-$ \cr
  ($k_{\rm max}=3$)\\
  \hline\hline
  \multicolumn{2}{ c }{Experimental value} \cr
  \hline
  & & & 4.70589 & 0.862(14) \\
  \hline\hline\\
 \hline\hline
  & & & \multicolumn{3}{c}{Proton} & \multicolumn{3}{c}{Neutron}\\
  \hline
  Fit type & $q^2_{\rm cut}$ [GeV$^2$] & $t_{\rm sep}/a$ & $\mu_p$ & $\sqrt{\la r^2_{M}\ra}$ [fm]& $\chi^2$/dof & $\mu_n$ & $\sqrt{\la r^2_{M}\ra}$ [fm]& $\chi^2$/dof \\
  \hline
  \multirow{2}{*}{Linear}     & \multirow{2}{*}{0.039} & $\{12,14,16\}$ & { 2.765(116) } & { 0.790(63) }  & { 0.1(7)}   & { $-$1.700(71) }  & { 0.810(73) } & { 2.2(2.9)}\\
                              &                        & $\{14,16\}$    & { 2.875(152) } & { 0.884(60) }  & { 0.1(5)}   & { $-$1.797(85) }  & { 0.898(74) } & { 0.6(1.7)}\\
  \hline
  \multirow{2}{*}{Dipole}     & \multirow{2}{*}{0.102} & $\{12,14,16\}$ & { 2.748(93) }  & { 0.808(35) }  & { 0.3(4)}   & { $-$1.709(62) }  & { 0.823(33) } & { 0.9(7)}\\
                              &                        & $\{14,16\}$    & { 2.785(150) } & { 0.816(47) }  & { 1.5(1.7)} & { $-$1.819(95) }  & { 0.947(60) } & { 1.1(2.3)}\\
  \hline
  \multirow{2}{*}{Quadrature} & \multirow{2}{*}{0.102} & $\{12,14,16\}$ & { 2.744(108) } & { 0.799(63) }  & { 0.3(5)}   & { $-$1.687(67) }  & { 0.770(63) } & { 1.0(9)}\\
                              &                        & $\{14,16\}$    & { 2.816(149) } & { 0.931(58) }  & { 2.8(1.7)} & { $-$1.739(80) }  & { 0.911(82) } & { 1.3(1.2)}\\
  \hline
  z-exp      & \multirow{2}{*}{0.102} & $\{12,14,16\}$ & { 2.753(119) } & { 0.809(105) } & { 0.3(5)}   & { $-$1.682(72) }  & { 0.724(133) } & { 1.0(9)}\\
  ($k_{\rm max}=3$)          &                        & $\{14,16\}$    & { 2.887(163) } & { 0.990(110) } & { 0.6(1.5)} & { $-$1.839(102) } & { 1.099(124) } & { 0.7(8)}\\
  \hline
  \multicolumn{2}{ c }{This work} & & 2.748(93)(139) & 0.808(35)(182) & & $-$1.709(62)(130) & 0.823(33)(276)\\
  \hline\hline
  \multicolumn{2}{ c }{Experimental value} \cr
  \hline
  & & & 2.79285 & 0.776(38) & & $-$1.91304 & 0.864(9)\cr  
\end{tabular}
\end{ruledtabular}
\end{table*}

We obtain the magnetic RMS radius $\sqrt{\la r^2_{M}\ra}$ together with the magnetic moment $\mu=G_M(0)$ with four types of fitting functions as in the electric case. The numerical values of $\mu$ and $\sqrt{\la r^2_{M} \ra}$ for the isovector, proton and neutron channels are summarized in Table~\ref{tab:murm} with the linear, dipole, quadratic forms and the z-expansion method with $k_{\rm max}=3$. In Figs.~\ref{fig:gm_q2} and \ref{fig:gm_q2_p_n}, one can see that the dipole form up to $q_{\rm cut}^2=0.102$ GeV$^2$ can well describe the LQCD data for both choices of the combined $t_{\rm sep}$ ranges, and they show good consistency with each other. Figure~\ref{fig:mm_fit} illustrates a comparison between four types of the fit procedures for the magnetic moment $\mu$, which shows good consistency within 1-$\sigma$ error as well as the case of the electric RMS radius. The situation is similar to the magnetic RMS radius $\sqrt{\la r^2_{M}\ra}$ in  Fig.~\ref{fig:rm_fit}, though the $z$-expansion method at $t_{\rm sep}/a=\{14,16\}$ gives the deviation beyond 1-$\sigma$ error between two choices of the combined $t_{\rm sep}$ ranges. We take the result of the dipole form with $t_{\rm sep}/a=\{12,14,16\}$ as our best estimate of the central value and the statistical error for $\mu$ and $\sqrt{\la r^2_{M}\ra}$, and take the maximum difference between the central value in the dipole fit with $t_{\rm sep}/a=\{12,14,16\}$ and those in other fitting procedures with two choices of the combined $t_{\rm sep}$ range as the systematic error (see Table~\ref{tab:murm}). Compared to the previous work \cite{Ishikawa:2018rew}, the statistical precision is significantly improved with less discrepancies between four types of the fit procedures. The results of the magnetic moment and RMS radius show consistency with the experimental values within 1-$\sigma$ error bars of the isovector, proton, and neutron channels, though the systematic uncertainty in a choice of the combined $t_{\rm sep}$ range is relatively large due to the excited state contamination.

\subsection{Axial-vector coupling, axial-vector form factor, and axial radius} 
\label{subsec:nff_fa}

\subsubsection{Axial-vector coupling}
\label{subsubsec:nff_ga}

\begin{figure}[ht!]
\includegraphics[width=8.0cm,keepaspectratio,clip]{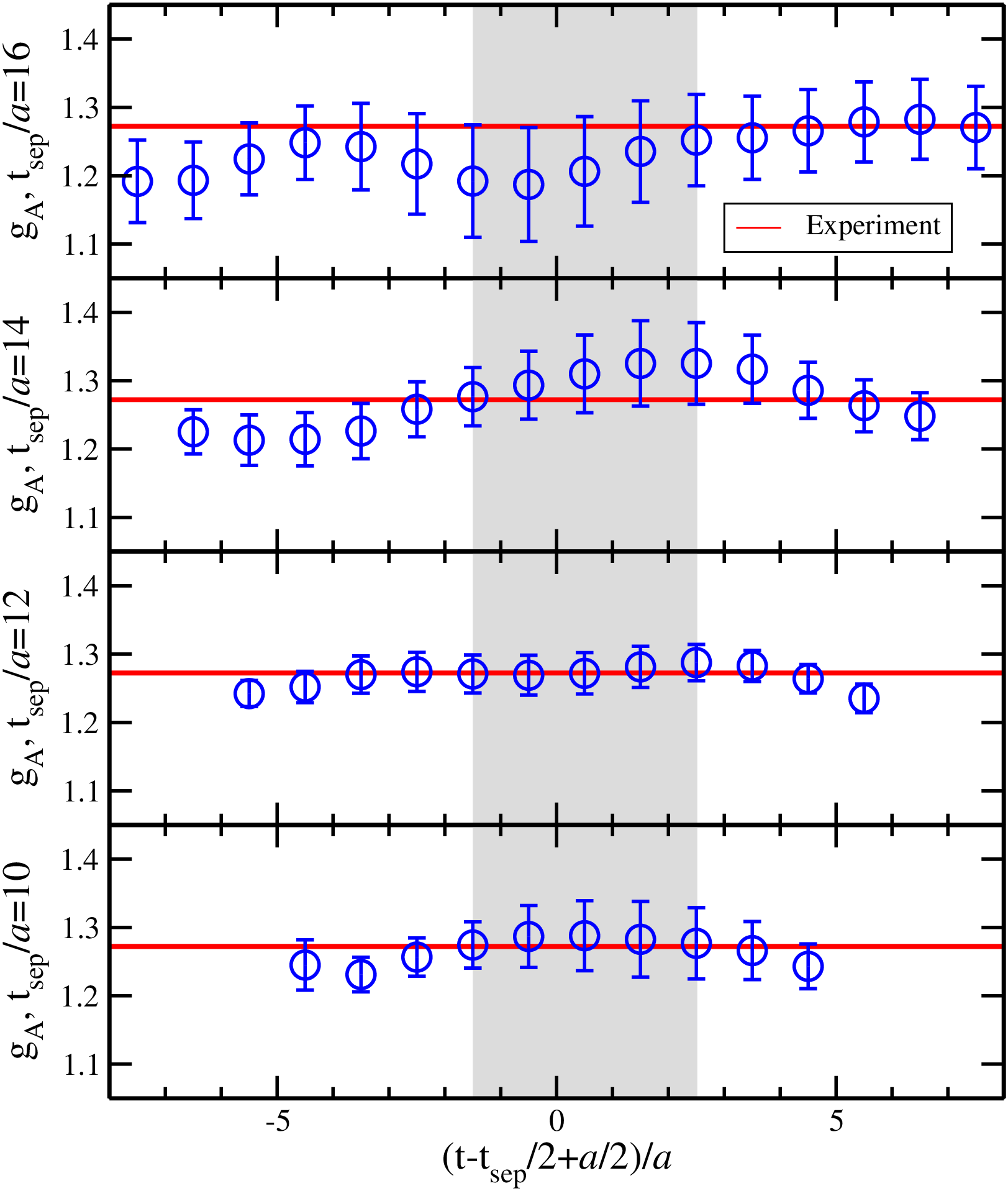}
\caption{Same as Fig.~\ref{fig:ratio_ge} for the axial-vector coupling $g_A=F_A(0)$ extracted from the ratio of Eq.~(\ref{Eq:R_FA_FP}) at the zero momentum transfer. Red band denotes the experimental result~\cite{Tanabashi:2018oca}.}
\label{fig:ga_p0_ts}
\end{figure}

\begin{figure}[ht!]
\includegraphics[width=8.0cm,keepaspectratio,clip]{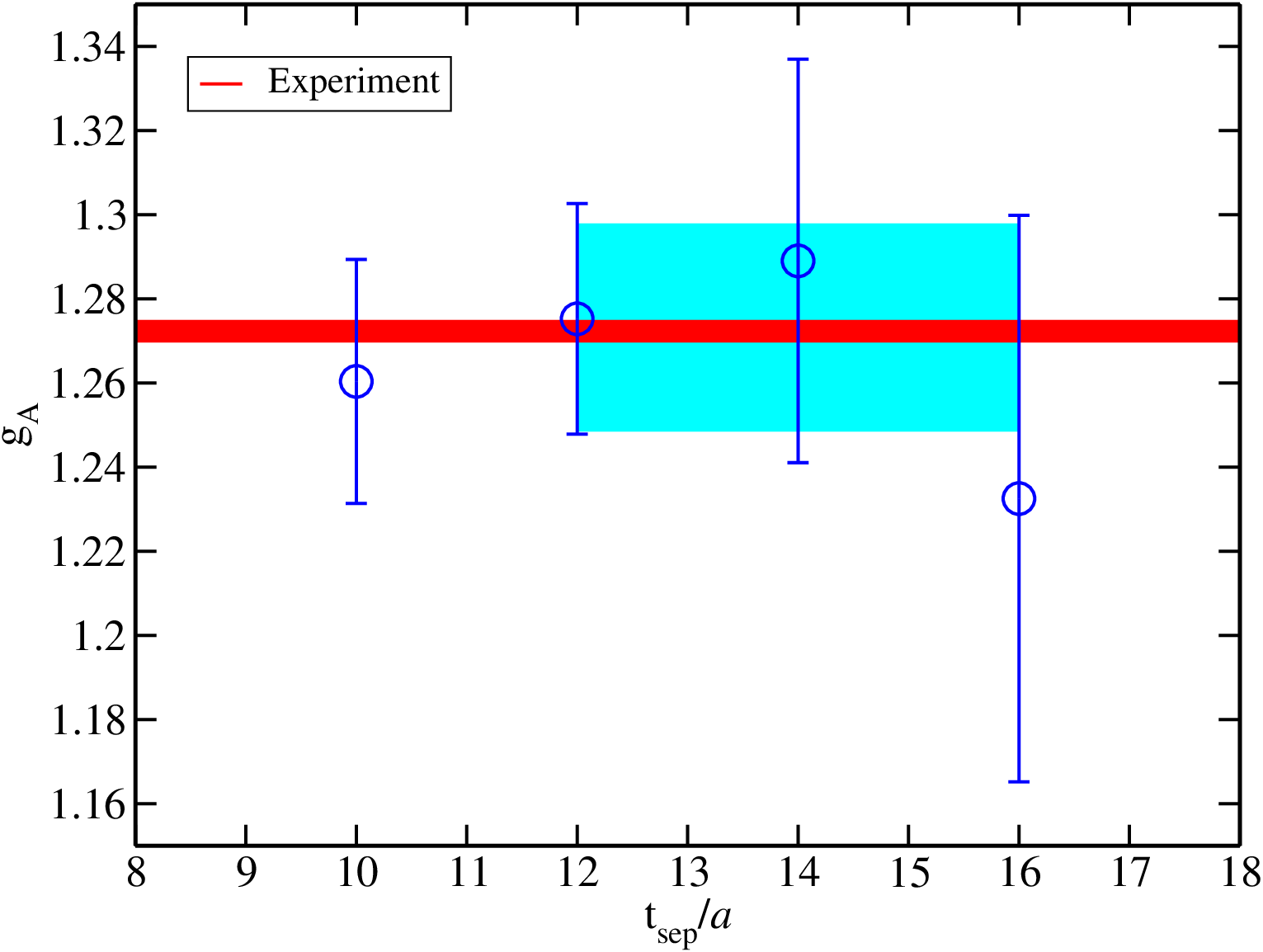}
\caption{Same as Fig.~\ref{fig:ge_ts} for the axial-vector coupling $g_A$. Red band denotes the experimental result~\cite{Tanabashi:2018oca}.}
\label{fig:ga_ts}
\end{figure}

The axial-vector coupling $g_A=F_A(0)$ has been extensively calculated with LQCD by various groups (see Ref.~\cite{Ishikawa:2018rew} and references therein). We first show the $t$ dependence of $g_A$ extracted from Eq.~(\ref{Eq:R_FA_FP}) with Q0 for $t_{\rm sep}/a=10,12,14,16$ in Fig.~\ref{fig:ga_p0_ts}. We observe reasonable plateau for all the cases of $t_{\rm sep}$. Figure~\ref{fig:ga_ts} shows that our results of $g_A$ in all the $t_{\rm sep}$ cases agree with the experimental value, 1.2724(23)~\cite{Tanabashi:2018oca}. 

We determine the central value of $g_A$ from the combined value with $t_{\rm sep}/a=\{12,14,16\}$ presented in the figure and a difference from the $t_{\rm sep}=\{14,16\}$ case is regarded as the systematic error. The result is summarized in Table~\ref{tab:gara}. Our best estimate of $g_A$ in this work is
\begin{equation}
g_A = 1.273(24)(5)(9),
\end{equation}
where we also include a systematic error stemming from the error of $Z_A$\footnote{The error of $Z_A$ coming from the difference of two volumes is not included, because we choose the larger volume in Ref.~\cite{Ishikawa:2015fzw} to set the physical scale.} as the third one. This result entirely agrees with the experiment.

Here it may be useful to present the up- and down-quark spin component in the nucleon spin, which can be obtained by decomposing the axial-vector coupling into the up- and down-quark contributions: $g_A^u=0.967(30)(16)(7)$ and $g_A^d=-0.306(19)(21)(2)$, in which the first error is statistical one and, the second and third ones are systematic errors due to the excited state contamination and uncertainty of $Z_A$, respectively. Again note that these results are obtained without the disconnected diagram.

\subsubsection{Axial-vector form factor and axial radius}
\label{subsubsec:nff_fa}

\begin{figure*}[ht!]
  \includegraphics[width=8.0cm,keepaspectratio,clip]{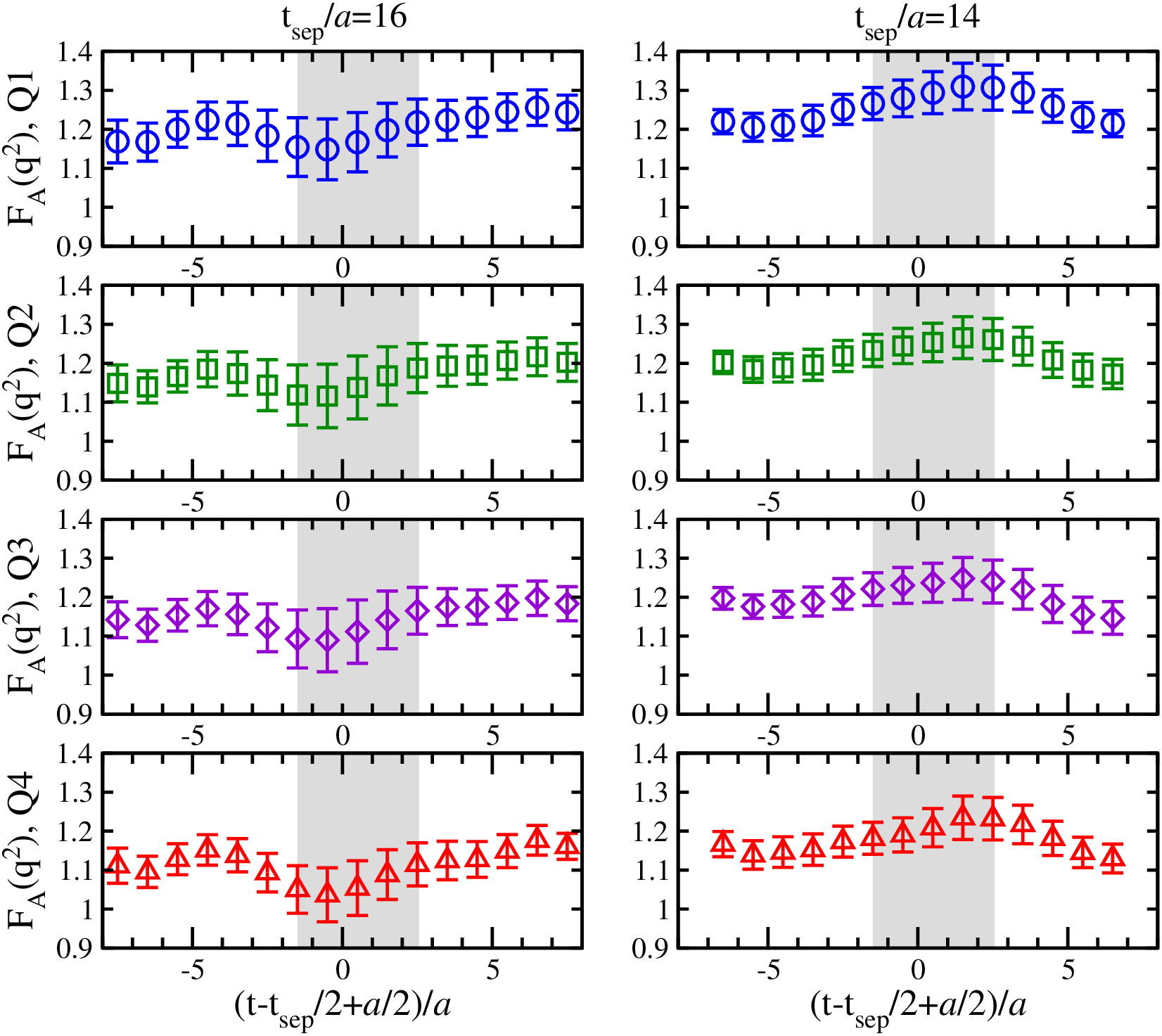}
  \hspace{3mm}
  \includegraphics[width=8.0cm,keepaspectratio,clip]{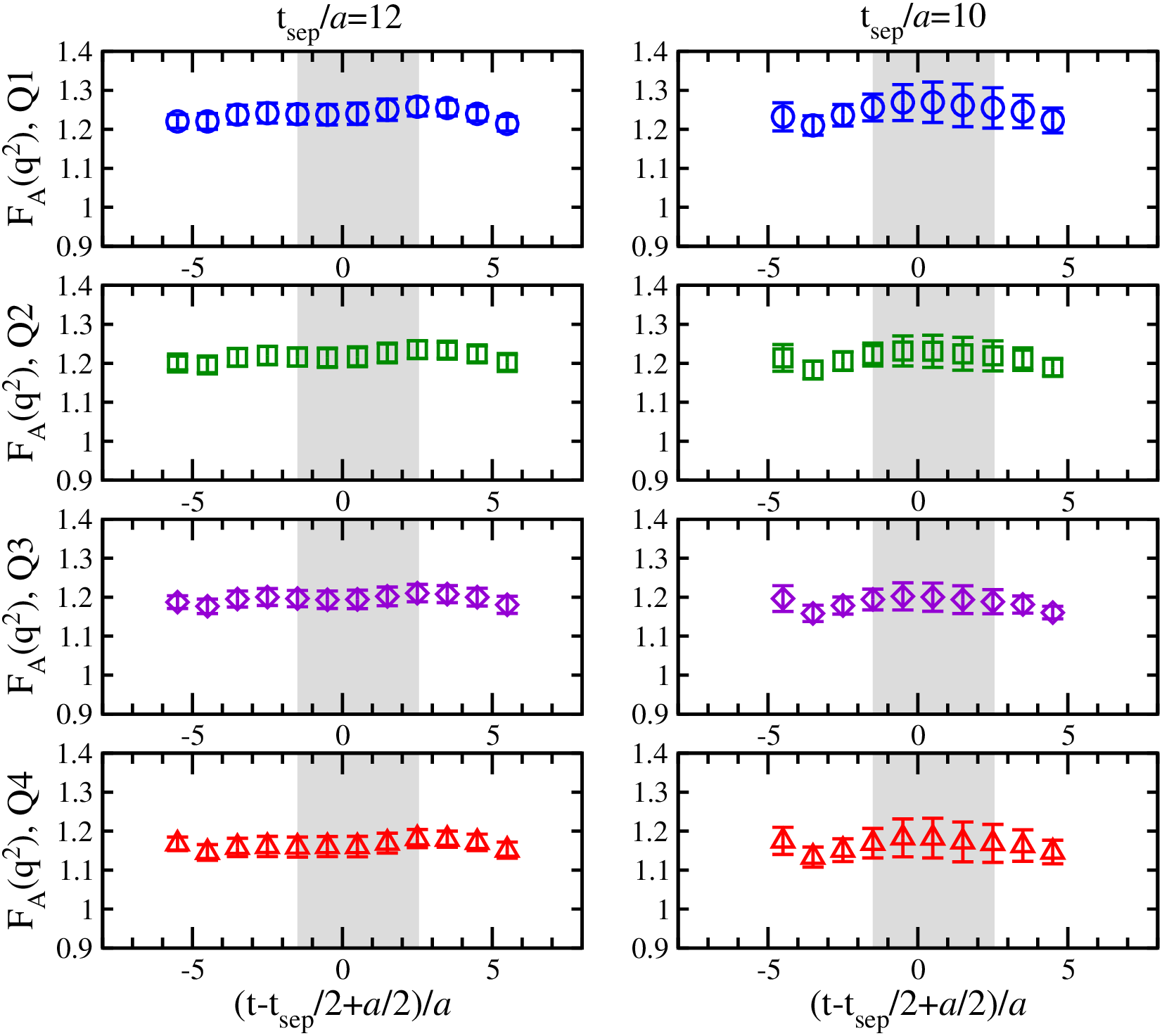}
\caption{Same as Fig.~\ref{fig:ratio_ge} for the axial-vector form factor $F_A(q^2)$ extracted from the ratio of Eq.~(\ref{Eq:R_FA_FP}).}
\label{fig:ratio_fa}
\end{figure*}

\begin{figure}[ht!]
\includegraphics[width=8.0cm,keepaspectratio,clip]{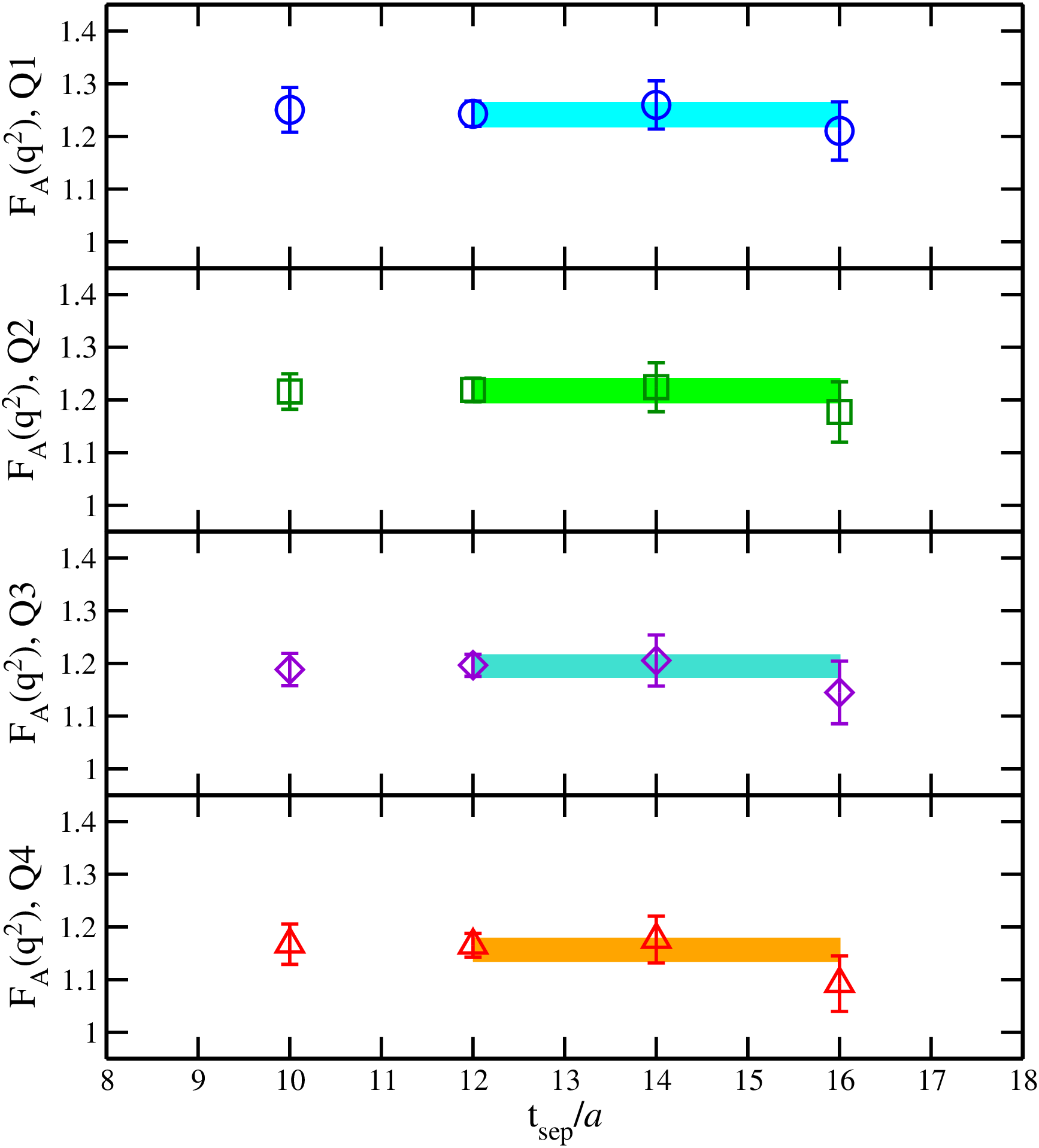}
\caption{Same as Fig.~\ref{fig:ge_ts} for $F_A(q^2)$ with four lowest nonzero momentum transfers.}
\label{fig:fa_ts}
\end{figure}

We next show the $t$ dependence of $F_A$ extracted from Eq.~(\ref{Eq:R_FA_FP}) with Q1, Q2, Q3, and Q4 for $t_{\rm sep}/a=10,12,14,16$ in Fig.~\ref{fig:ratio_fa}. We observe reasonable plateau for all the cases of $t_{\rm sep}$ and 
finite ${\bm n}$. As in the case of $G_M(q)$, we do not observe a significant $t_{\rm sep}$ dependence in Fig.~\ref{fig:fa_ts}. We employ two combined values obtained by applying the constant fit to the data in the two ranges of $t_{\rm sep}/a=\{12,14,16\}$ and $t_{\rm sep}/a=\{14,16\}$, which are used for the investigation of $q^2$ dependence of $F_A(q^2)$. The two combined values for $F_A(q^2)$ are summarized in Appendix~\ref{app:tab}.

\begin{figure}[ht!]
  \includegraphics[width=8.0cm,keepaspectratio,clip]{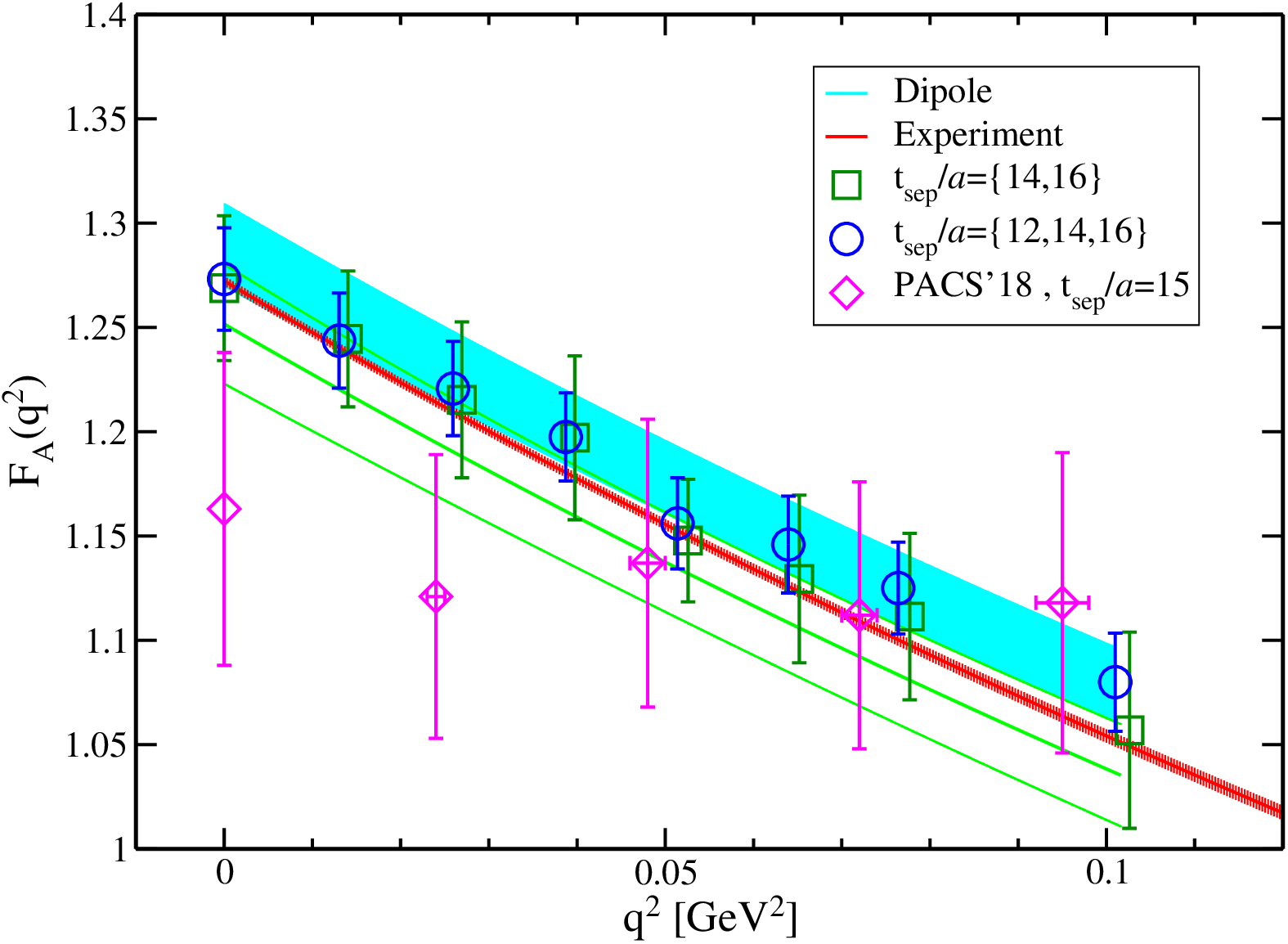}
\caption{Same as Fig.~\ref{fig:ge_q2} for the axial-vector form factor $F_A(q^2)$. Experimental line is obtained by the dipole form with the value of dipole mass~\cite{Bodek:2007ym,Bernard:2001rs} and $g_A$~\cite{Tanabashi:2018oca}.}
\label{fig:fa_q2}
\end{figure}

We plot the $q^2$ dependence of $F_A(q^2)$ in Fig.~\ref{fig:fa_q2}, where any strong curvature is not observed in terms of $q^2$. We also find that our two combined results show good agreement with each other and both of them are consistent with the experimental values~\cite{Bodek:2007ym} within 1-$\sigma$ error bars. The isovector axial-vector coupling $F_A(0)$ and the axial RMS radius $\sqrt{\la r^2_{A}\ra}$ are obtained from several types of fitting procedure with the linear, dipole, quadratic forms and the $z$-expansion method with $k_{\rm max}=3$. Note that we employ $t_{\rm cut}=9m_{\pi}^2$ in Eq.~(\ref{Eq:z-value}) for the $z$-expansion method.

\begin{figure*}[ht!]
  \includegraphics[width=8.0cm,keepaspectratio,clip]{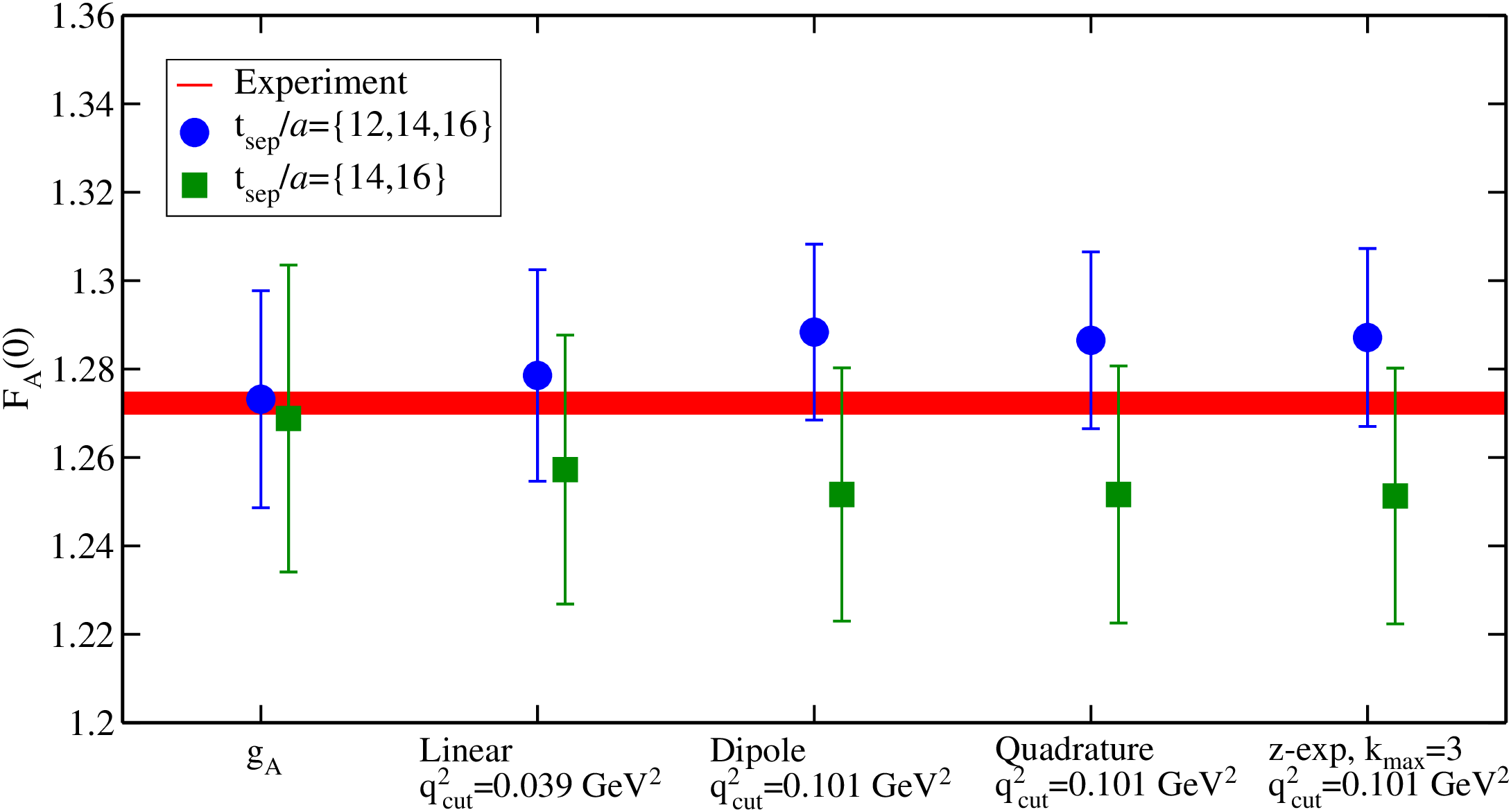}
  \hspace{3mm}
  \includegraphics[width=8.0cm,keepaspectratio,clip]{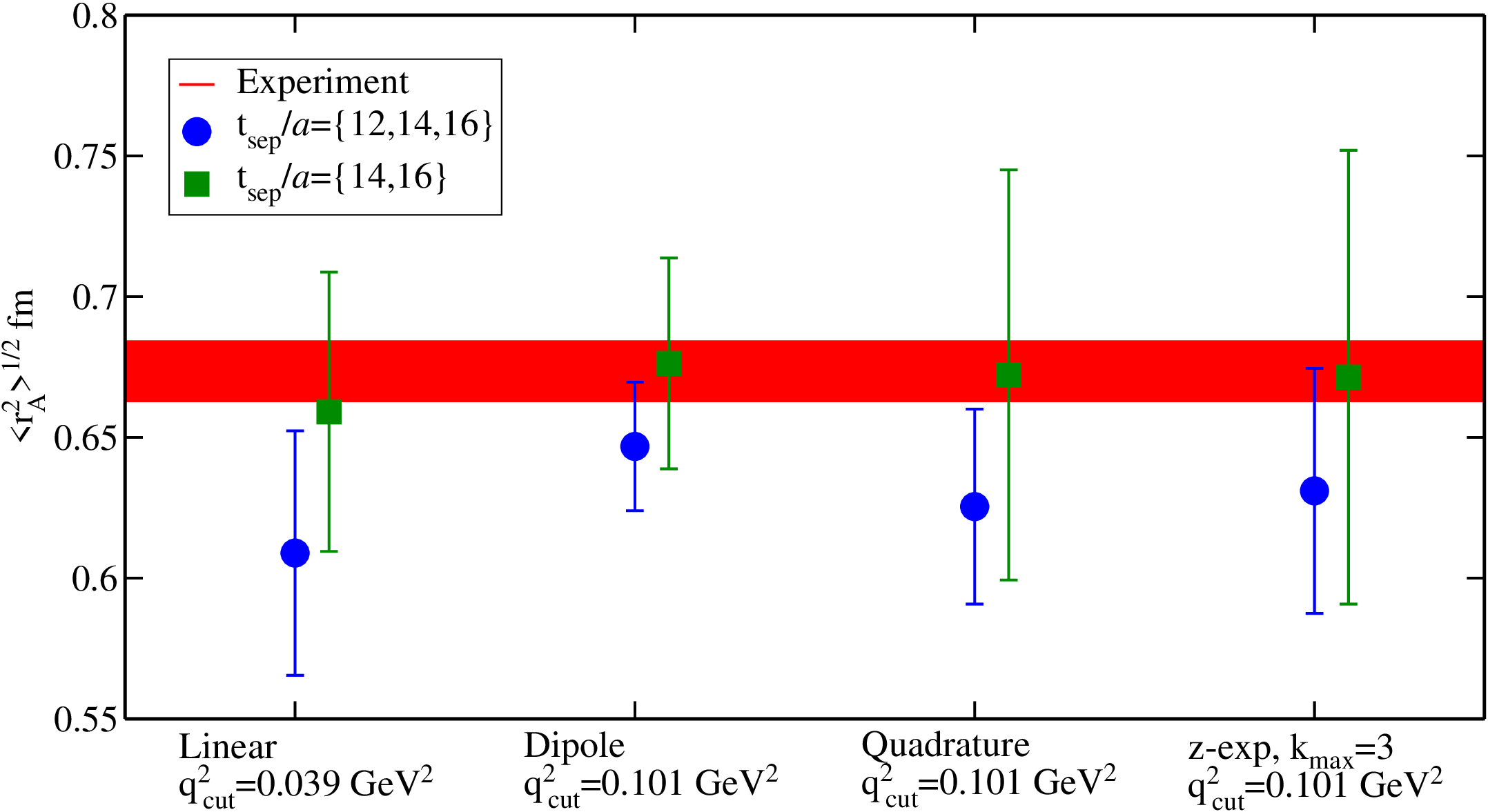}
\caption{Axial-vector coupling $g_A=F_A(0)$ (left) and the axial-vector RMS radius $\sqrt{\la r^2_A\ra}$ (right) obtained by the fit for the combined data with the linear, dipole, quadratic forms and the $z$-expansion method. Horizontal bands represent the experimental values.}
\label{fig:ra_fit}
\end{figure*}

The dipole form fits for two combined data with different choices of $t_{\rm sep}$ range are presented in Fig.~\ref{fig:fa_q2}. The fit results up to $q^2_{\rm cut}=0.102$ GeV$^2$ well describe our data. As shown in Fig.~\ref{fig:fa_q2}, the fitted curve for $t_{\rm sep}/a=\{14,16\}$ appears slightly below their respective data points. This could be due to a poor determination of the covariance matrix in the correlated fit for the highly correlated data among different $q^2$ points. The results from four types of fit form are compared graphically in Fig.~\ref{fig:ra_fit} and numerically in Table~\ref{tab:gara}. The fit results for both $F_A(0)$ and $\sqrt{\la r^2_{A}\ra}$ show good consistency among all four types of fitting procedures, and they are also in agreement with the experimental values. 

Following the analysis in $G_E(q^2)$ and $G_M(q^2)$, we take the result of the dipole form with $t_{\rm sep}/a=\{12,14,16\}$ as our best estimate of the central value and the statistical error for $\sqrt{\la r^2_{A}\ra}$, and take the maximum difference between the dipole fit with $t_{\rm sep}/a=\{12,14,16\}$ and other fitting with two $t_{\rm sep}$ ranges as the systematic error (see Table~\ref{tab:gara}).

\begin{table*}[ht!]
\begin{ruledtabular}
\caption{Results for the axial-vector coupling $g_A=F_A(0)$ and axial-vector RMS radius $\sqrt{\la r^2_A\ra}$.  In the row of ``This work'' we present our best estimates, where the first error is statistical and the second one is systematic as explained in the text. Results for the proton and neutron are obtained without the disconnected diagram. Previous work was performed on a $96^4$ lattice at $m_\pi = 146$ MeV in Ref.~\cite{Ishikawa:2018rew}, where only the statistical errors are presented.
\label{tab:gara}}
\begin{tabular}{ccccccc}
  Fit type & $q^2_{\rm cut}$ GeV$^2$ & $t_{\rm sep}/a$ & $g_A$ & $F_A(0)$ & $\sqrt{\la r^2_A\ra}$ [fm]& $\chi^2$/dof   \cr
  \hline
  \multirow{2}{*}{Linear}     & {0.039} & $\{12,14,16\}$ & & { 1.279(23) } & { 0.609(43) } & { 1.0(1.4)}\\
                              & {0.052} & $\{14,16\}$    & & { 1.257(30) } & { 0.659(49) } & { 0.5(1.6)}\\
  \hline
  \multirow{2}{*}{Dipole}     & {0.102} & $\{12,14,16\}$ & & { 1.288(19) } & { 0.647(22) } & { 1.1(8)}\\
                              & {0.077} & $\{14,16\}$    & & { 1.252(28) } & { 0.676(37) } & { 0.5(9)}\\
  \hline
  \multirow{2}{*}{Quadrature} & {0.102} & $\{12,14,16\}$ & & { 1.287(19) } & { 0.625(34) } & { 1.2(9)}\\
                              & {0.077} & $\{14,16\}$    & & { 1.252(29) } & { 0.672(72) } & { 0.9(1.9)}\\
  \hline
  z-exp                       & {0.102} & $\{12,14,16\}$ & & { 1.287(20) } & { 0.631(43) } & { 1.2(9)}\\
  ($k_{\rm max}=3$)           & {0.077} & $\{14,16\}$    & & { 1.251(28) } & { 0.671(80) } & { 0.9(1.9)}\\
  \hline
  \multirow{2}{*}{$g_A$}    & $-$  & $\{12,14,16\}$ & {1.273(24)} \\
                            & $-$  & $\{14,16\}$    & {1.268(35)} \\
  \hline
  \multicolumn{2}{ c }{This work} & & 1.273(24)(5) & & 0.647(22)(38)\\
  \hline\hline
  \multicolumn{2}{ c }{PACS'18~\cite{Ishikawa:2018rew}} \cr
  \hline
  Dipole    & 0.215 & 15 & & $-$ & 0.40(12) & $-$ \cr
  \hline
  Quadratic & 0.215 & 15 & & $-$ & 0.22(49) & $-$ \cr
  \hline
  z-exp     & \multirow{2}{*}{0.215} & \multirow{2}{*}{15} & & \multirow{2}{*}{$-$} & \multirow{2}{*}{0.46(11)} & \multirow{2}{*}{$-$} \cr
  ($k_{\rm max}=3$)\\
  \hline
  $g_A$  & 0.215 & 15 & 1.163(75) \\
  \hline\hline
  \multicolumn{2}{ c }{Experimental value}  &  & & \cr
  \hline
  & & & 1.2724(23) & & 0.67(1) &\cr
\end{tabular}
\end{ruledtabular}
\end{table*}

\subsection{Induced pseudoscalar form factor} 
\label{subsec:nff_fp}

\begin{figure*}[ht!]
  \includegraphics[width=8.0cm,keepaspectratio,clip]{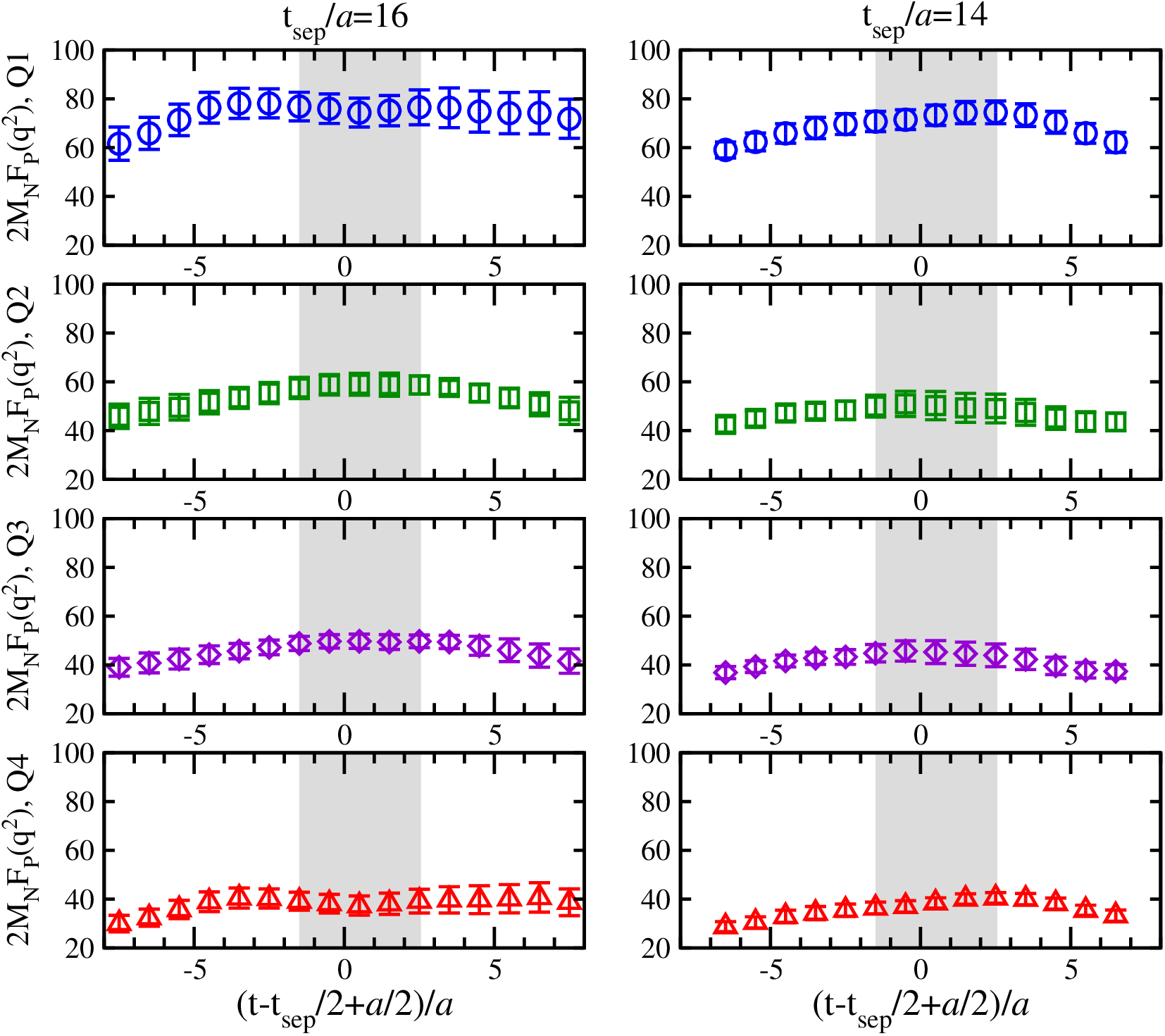}
  \hspace{3mm}
  \includegraphics[width=8.0cm,keepaspectratio,clip]{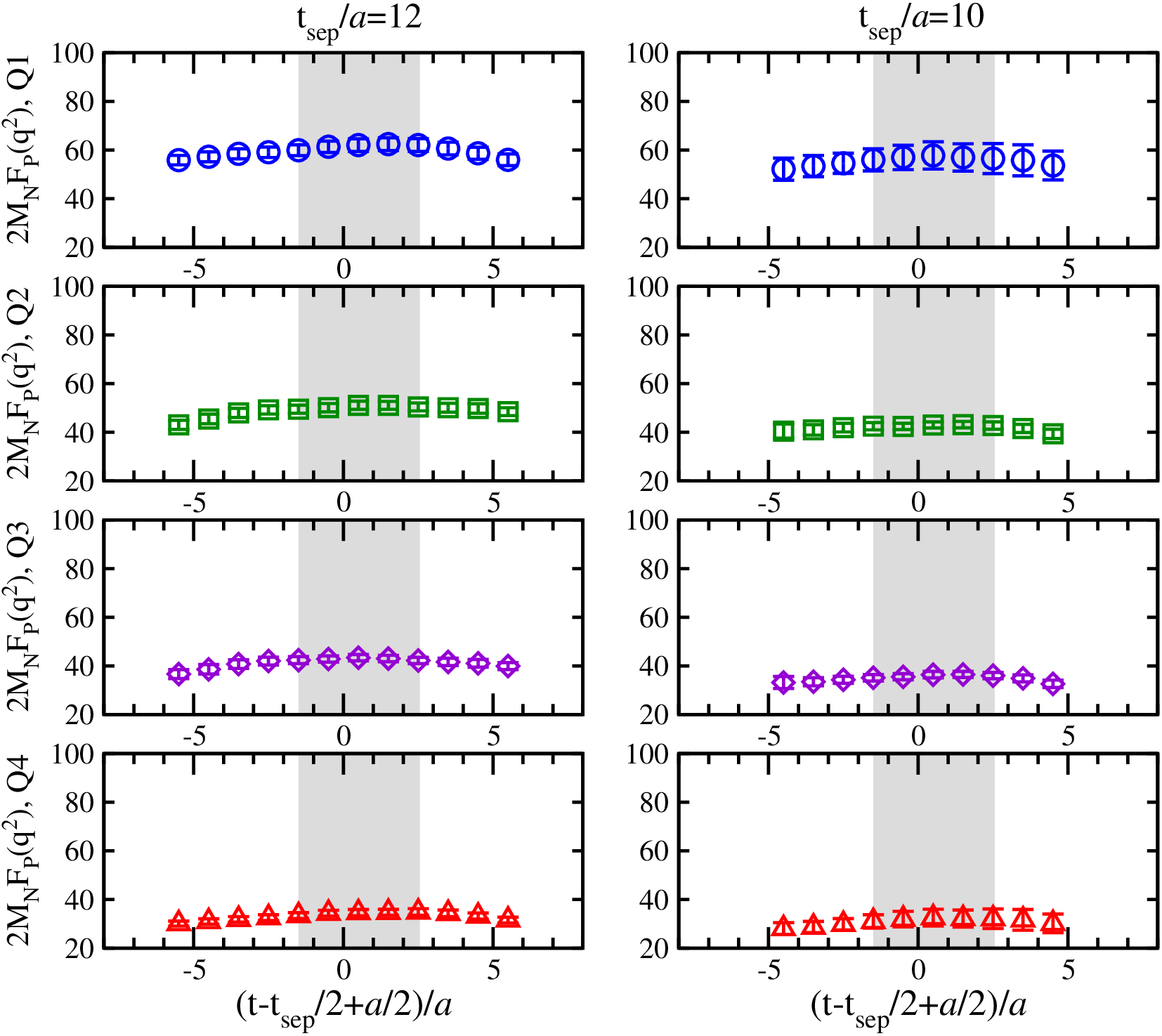}
\caption{Same as Fig.~\ref{fig:ratio_ge} for the induced pseudoscalar form factor $F_P(q^2)$ extracted from the ratio of Eq.~(\ref{Eq:R_FA_FP}).}
\label{fig:ratio_fp}
\end{figure*}

\begin{figure}[ht!]
  \includegraphics[width=8.0cm,keepaspectratio,clip]{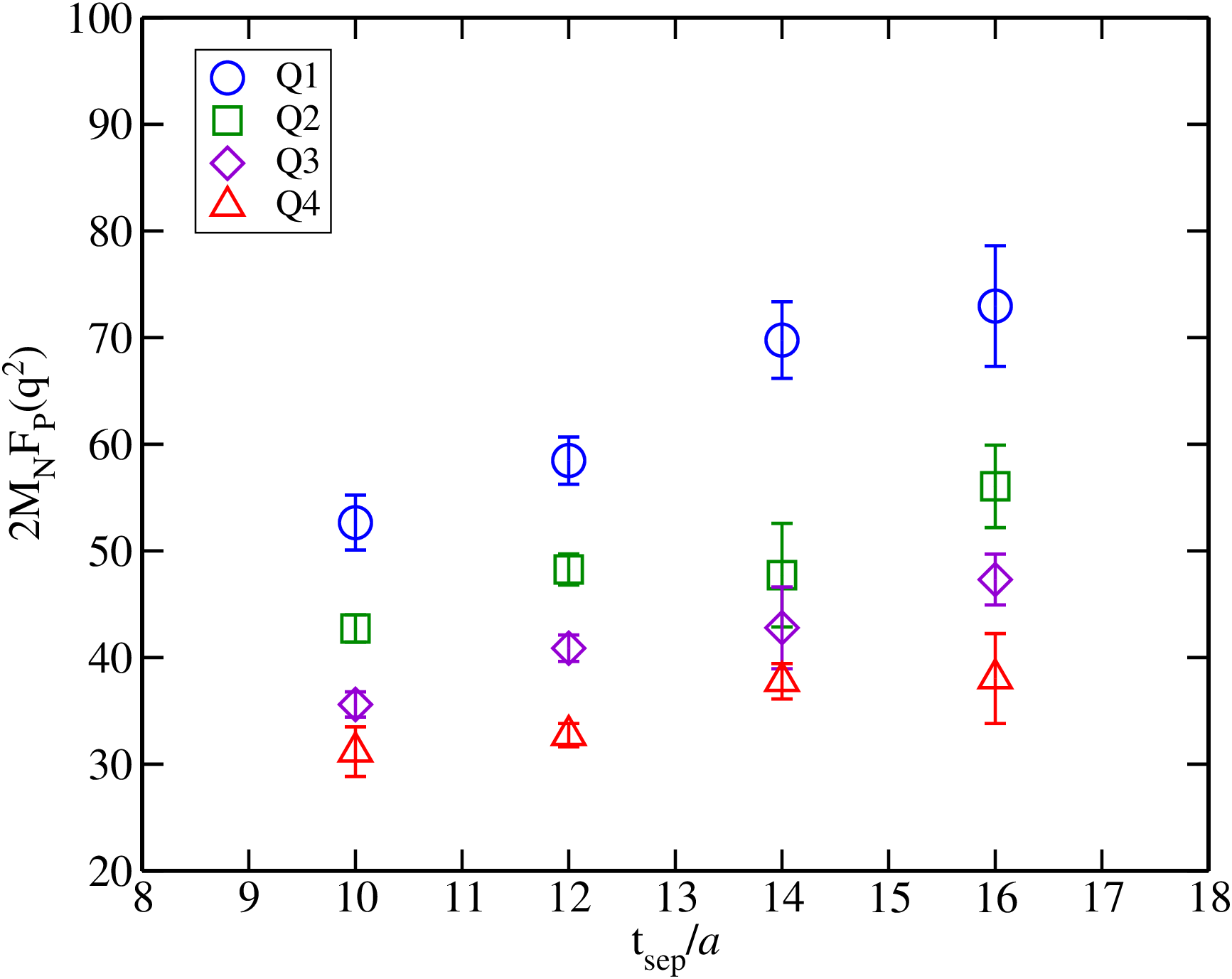}
\caption{Same as Fig.~\ref{fig:ge_ts} for the normalized induced pseudoscalar form factor $2M_NF_P(q^2)$ with four lowest nonzero momentum transfers.}
\label{fig:fp_ts}
\end{figure}

In Fig.~\ref{fig:ratio_fp} we plot the $t$ dependence of $F_P(q^2)$ extracted from Eq.~(\ref{Eq:R_FA_FP}) with Q1, Q2, Q3, and Q4 for $t_{\rm sep}/a=10,12,14,16$. One can observe that its dependence has slight convex shape for all the cases in contrast to the form factors $G_E(q^2)$, $G_M(q^2)$, and $F_A(q^2)$ discussed above. Inside the fitting range of $t$, however, the data points are overlapping within 1$\sigma$ statistical error, so that employing a constant fit to obtain $F_P(q^2)$ is appropriate. Figure~\ref{fig:fp_ts} shows the $t_{\rm sep}$ dependence of $F_P(q^2)$ at the smallest three values of $q^2$. We find that $F_P(q^2)$ clearly increases as  $t_{\rm sep}$ increases. This indicates that the significant contribution from the excited states is involved in the $F_P(q^2)$ case. In fact, the previous work~\cite{Ishikawa:2018rew} on a $96^4$ lattice at the 146 MeV pion with $t_{\rm sep}/a=15$ gives $F_{P}(q^2)$ close to $t_{\rm sep}/a=14$ in our case. 

\begin{figure}[ht!]
  \includegraphics[width=8.0cm,keepaspectratio,clip]{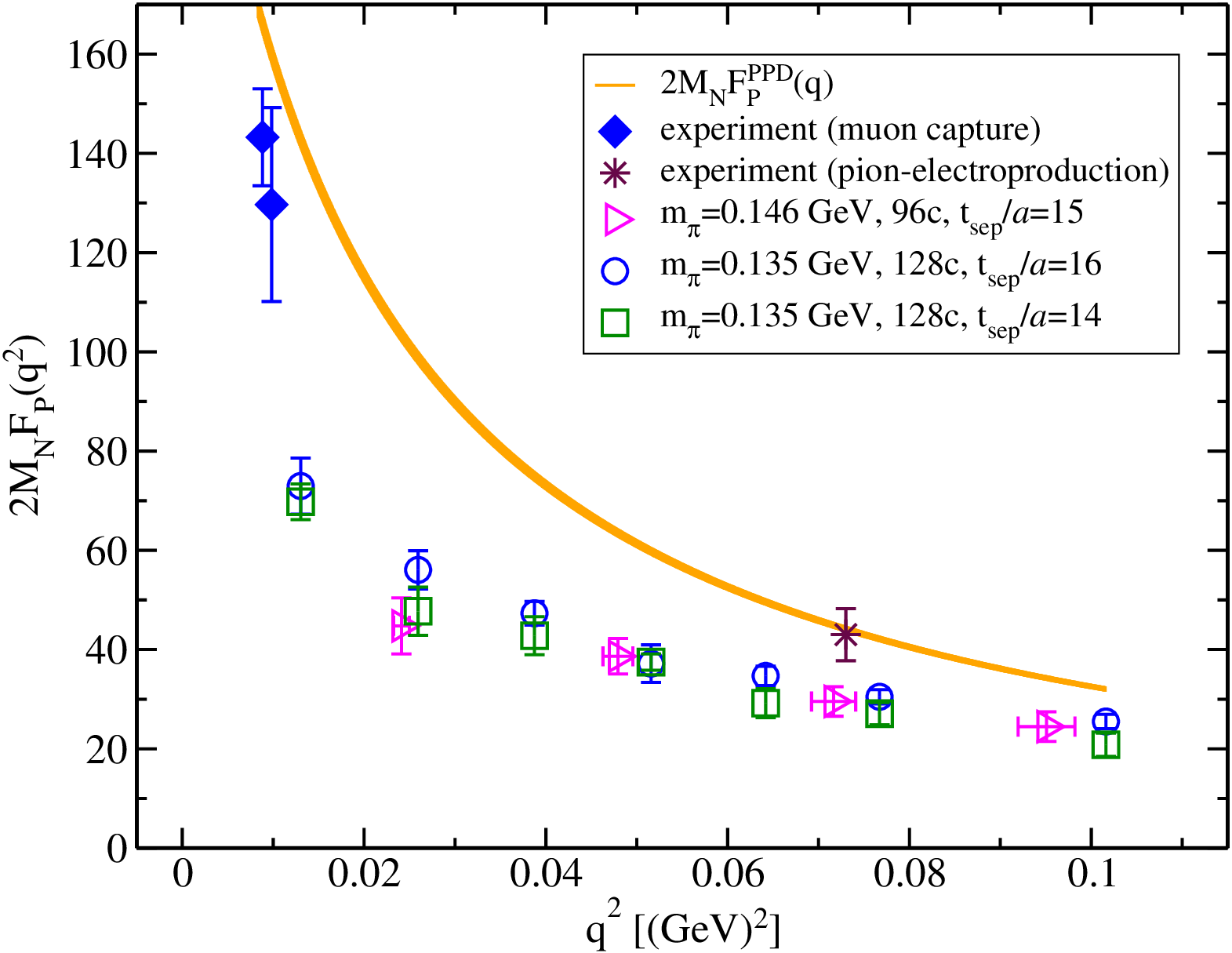}
  \caption{$q^2$ dependence of the normalized induced pseudoscalar form factor $2M_N F_P(q^2)$. We also plot the results in the previous work~\cite{Ishikawa:2018rew} for comparison. The colored band shows the function of the pion pole dominance model using the fit result of $F_A(q^2)$ with the dipole form in Fig.~\ref{fig:fa_q2}.}
\label{fig:fp_q2}
\end{figure}

We plot the $q^2$ dependence of the normalized induced pseudoscalar form factor $2M_N F_P(q^2)$ for $t_{\rm sep}/a=14$ and $16$ compared to the previous work~\cite{Ishikawa:2018rew} in Fig.~\ref{fig:fp_q2}. The colored curve denotes a prediction of the pion-pole dominance (PPD) model with the measured values of $m_\pi$, $M_N$ and the global fit result of $F_A(q^2)$ in the dipole form:
\be
F_P^{\rm PPD}(q^2)=2M_N F_A(q^2)/(q^2+m_\pi^2),
\label{Eq:PPD_FP}
\ee
which successfully describes two experimental results of the muon capture \cite{Andreev:2007wg} and the pion-electroproduction \cite{Choi:1993vt}. The $t_{\rm sep}$ dependence of $2M_N F_P(q^2)$ found in Fig.~\ref{fig:fp_ts}, where the significant change of form factor as $t_{\rm sep}$ increases appears, gives an important hint to explain a discrepancy from experimental values and PPD model prediction. 

For more careful verification of the excited state contamination in the induced pseudoscalar form factor, we need more accurate data at $t_{\rm sep}/a=16$ and an additional calculation with at least one more large $t_{\rm sep}$, which may help to extrapolate $F_P$ in the infinite $t_{\rm sep}$ limit. We note that the baryon chiral perturbation theory suggests the aforementioned discrepancy as the contamination of the $\pi$-$N$ excited states in the standard plateau method~\cite{Bar:2018akl,Bar:2018xyi}. It is also noted that this contamination is recently investigated using a proper projection~\cite{Bali:2018qus}. More detailed comparison is also interesting for the future work.

\section{Pseudoscalar form factor and Goldberger-Treiman relation} 
\label{sec:nff_gp}

In the previous section, we have found the relatively large excited state contamination in $F_P(q^2)$ compared to $F_A(q^2)$, which may one of the reasons for the considerable discrepancy between the LQCD result of $2M_NF_P(q^2)$ and the experimental values. We expect that the generalized GT relation of Eq.~(\ref{Eq:GTrelation}), which is associated with the AWT identity, may also suffer from the serious effects of the excited state contamination. 

\begin{figure*}[ht!]
  \includegraphics[width=8.0cm,keepaspectratio,clip]{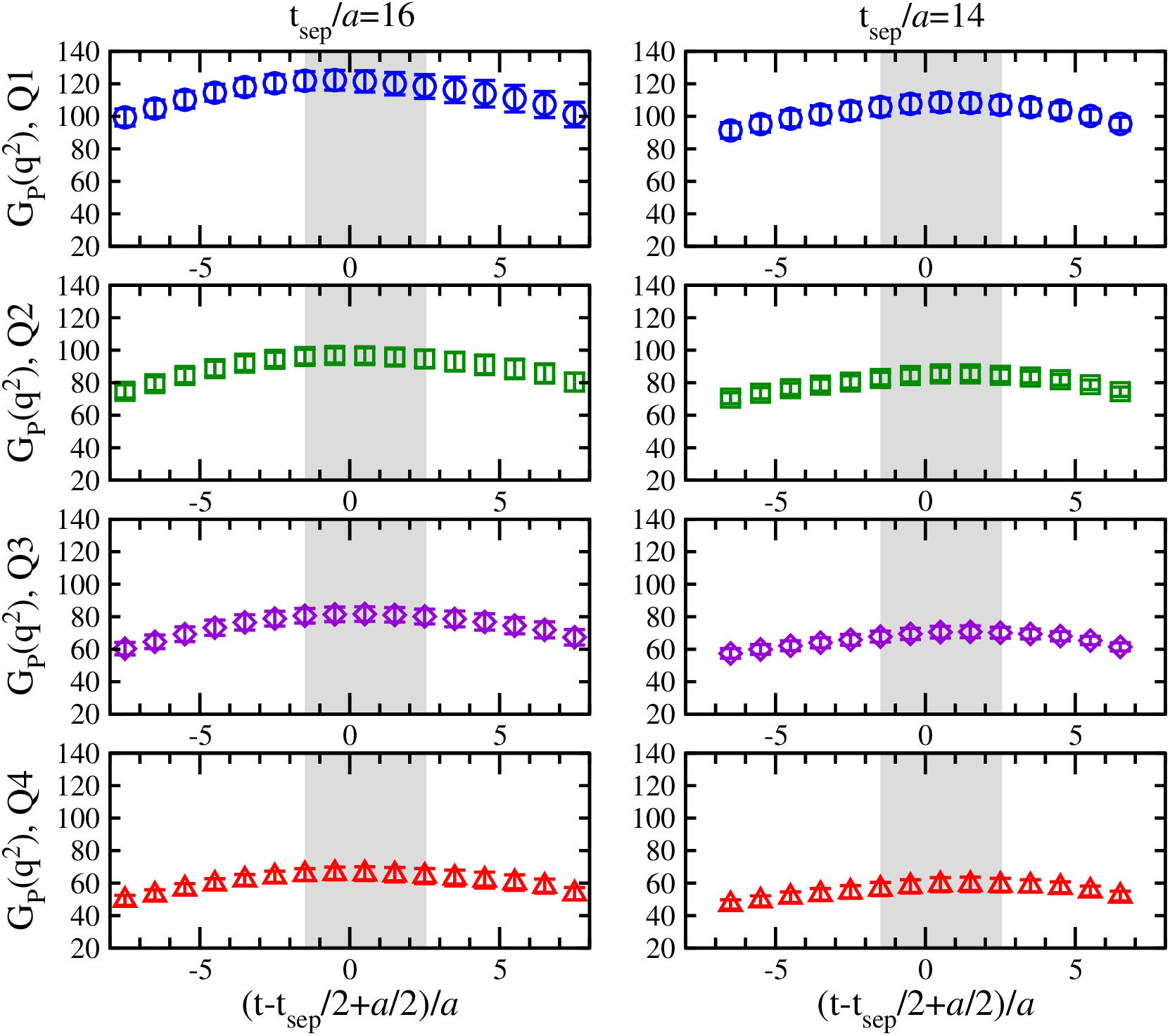}
  \hspace{3mm}
  \includegraphics[width=8.0cm,keepaspectratio,clip]{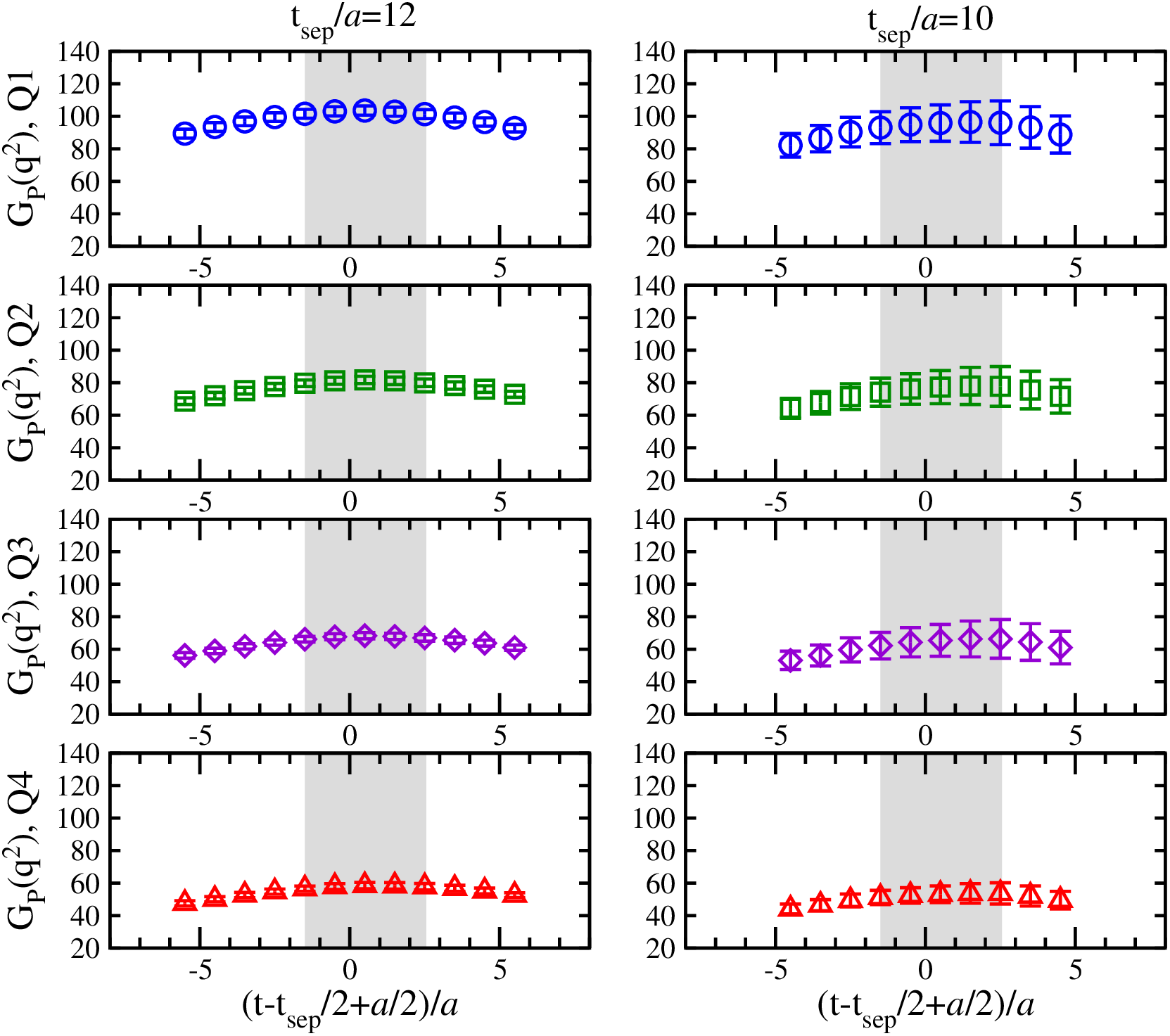}
\caption{Same as Fig.~\ref{fig:ratio_ge} for the pseudoscalar form factor $G_P(q^2)$ extracted from the ratio of Eq.~(\ref{Eq:R_GP}).}
\label{fig:ratio_gp}
\end{figure*}

The pseudoscalar form factor $G_P(q^2)$ is defined by Eq.~(\ref{Eq:PSFF}) and extracted from the ratio ${\cal R}^{5z}_{P}(t,{\bm q})$ of Eq.~(\ref{Eq:R_GP}). Figure~\ref{fig:ratio_gp} shows the $t$ dependence of the ratio with $\vert {\bm n}\vert^2=1,2,3,4$ for $t_{\rm sep}/a=10,12,14,16$. We observe that the convex shape is much clearer than the $F_P$ case and its top value increases for larger $t_{\rm sep}$. One can see that the pseudoscalar form factor is also strongly affected by the excited state contributions. 

\begin{figure}[ht!]
  \includegraphics[width=8.0cm,keepaspectratio,clip]{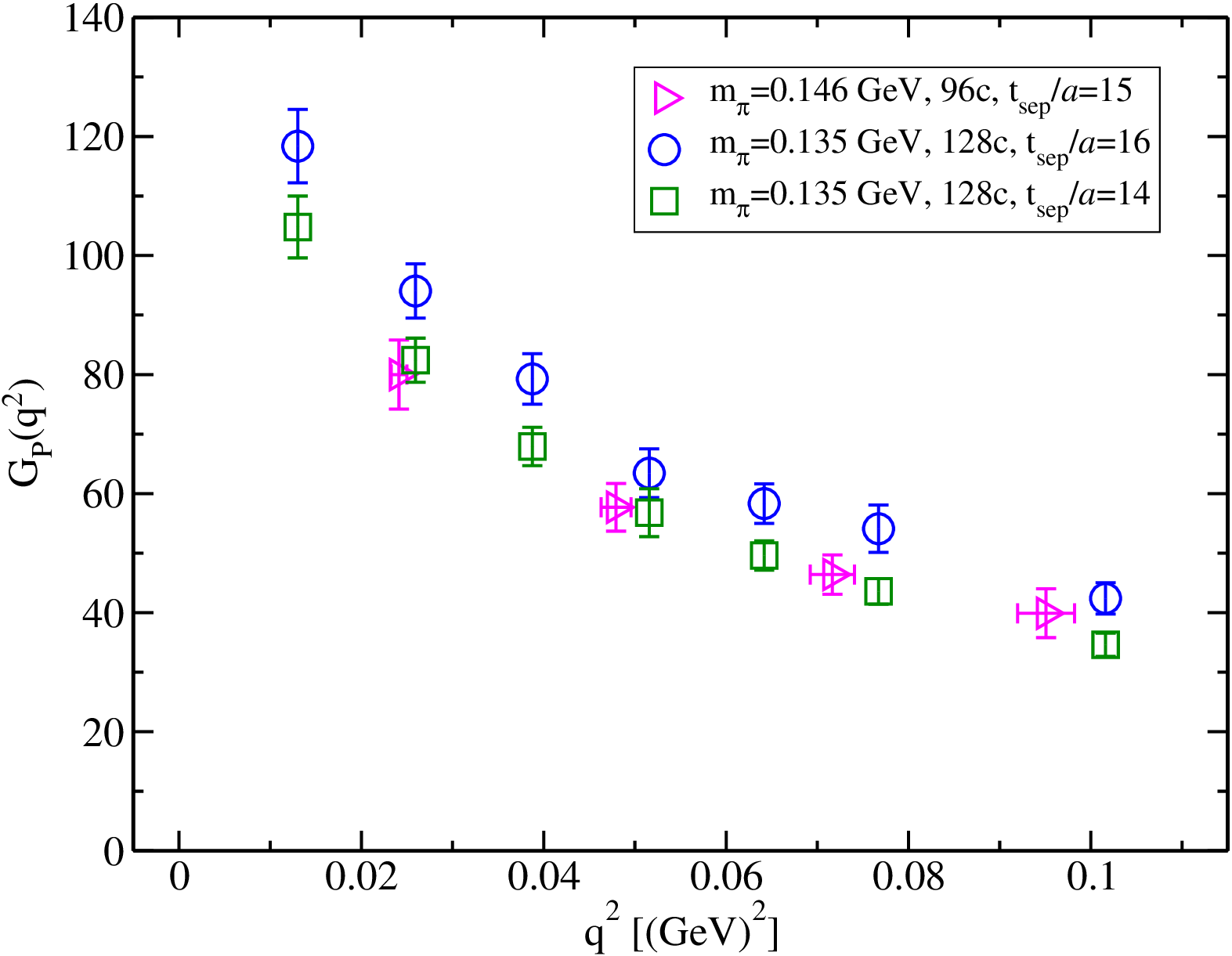}
\caption{$q^2$ dependence for the pseudoscalar form factor $G_P(q^2)$ at $t_{\rm sep}/a=14,\,16$. We also plot the results in the previous work~\cite{Ishikawa:2018rew} for comparison.}
\label{fig:gp_q2}
\end{figure}

In Fig.~\ref{fig:gp_q2}, we plot our values of $G_P(q^2)$ as a function of $q^2$, which are obtained by the constant fit of data for both $t_{\rm sep}/a=14$ and 16 choosing the same fit range as the other form factors. The stronger curvature appears around $q^2=0$ as $t_{\rm sep}$ increases. This behavior is found to be similar to $F_P(q^2)$.
Although data points of $G_P(q^2)$ for $t_{\rm sep}/a=14$ are comparable with the previous result~\cite{Ishikawa:2018rew}, where $t_{\rm sep}/a=15$ was chosen, the magnitude of $G_P(q^2)$ for $t_{\rm sep}/a=16$ becomes about 10 percent larger than that of $t_{\rm sep}/a=14$ at all the simulated $q^2$ points. The $t_{\rm sep}$ dependence of $G_P(q^2)$ is much more prominent than that of $F_P(q^2)$. 

\begin{figure}[ht!]
  \includegraphics[width=8.0cm,keepaspectratio,clip]{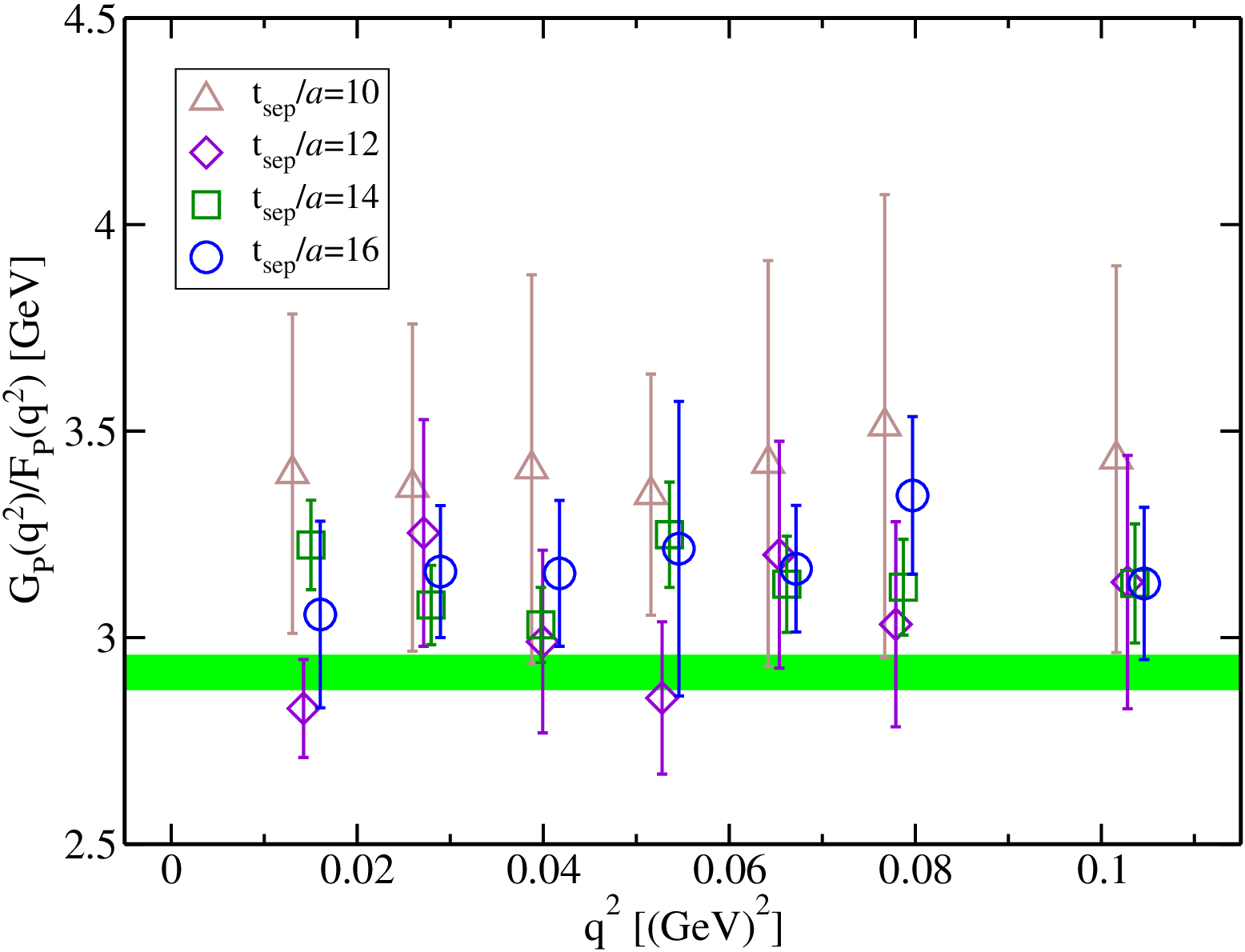}
\caption{Ratio of the pseudoscalar form factor $G_P(q^2)$ over the induced one $F_P(q^2)$ as a function of $q^2$ at each $t_{\rm sep}$.}
\label{fig:gp_q2_ratio}
\end{figure}

According to the PPD model or the generalized GT relation associated with the AWT identity, $F_P(q^2)$ and $G_P(q^2)$ are supposed to share the same pion-pole structure, {\it i.e.}, $\propto 1/(q^2+m_\pi^2)$, at lower $q^2$. In the previous work~\cite{Ishikawa:2018rew}, it was indeed observed that the ratio of $G_P(q^2)/F_P(q^2)$ exhibited a flat $q^2$ dependence at lower $q^2$ and was in good agreement with the bare value of the low-energy constant $B_0=m_{\pi}^2/(2\hat{m})$ with the simulated pion mass $m_{\pi}$ and the PCAC quark mass  $\hat{m}=m^{\rm PCAC}_{\rm AWTI}$. In order to test whether this feature holds against the variation of $t_{\rm sep}$, we plot the ratios of $G_P(q^2)/F_P(q^2)$ for all the case of $t_{\rm sep}=10,12,14,16$ in Fig.~\ref{fig:gp_q2_ratio}. Each ratio of $G_P(q^2)/F_P(q^2)$ does not depend on $q^2$ and those are in good agreement with the bare value of the low-energy constant $B_0$ as illustrated by the green band. This strongly indicates that the individual effects of the excited state contamination on the $G_P(q^2)$ and $F_P(q^2)$ form factors are canceled in the ratio of $G_P(q^2)/F_P(q^2)$. 

\begin{figure*}[ht!]
  \includegraphics[width=7.0cm,keepaspectratio,clip]{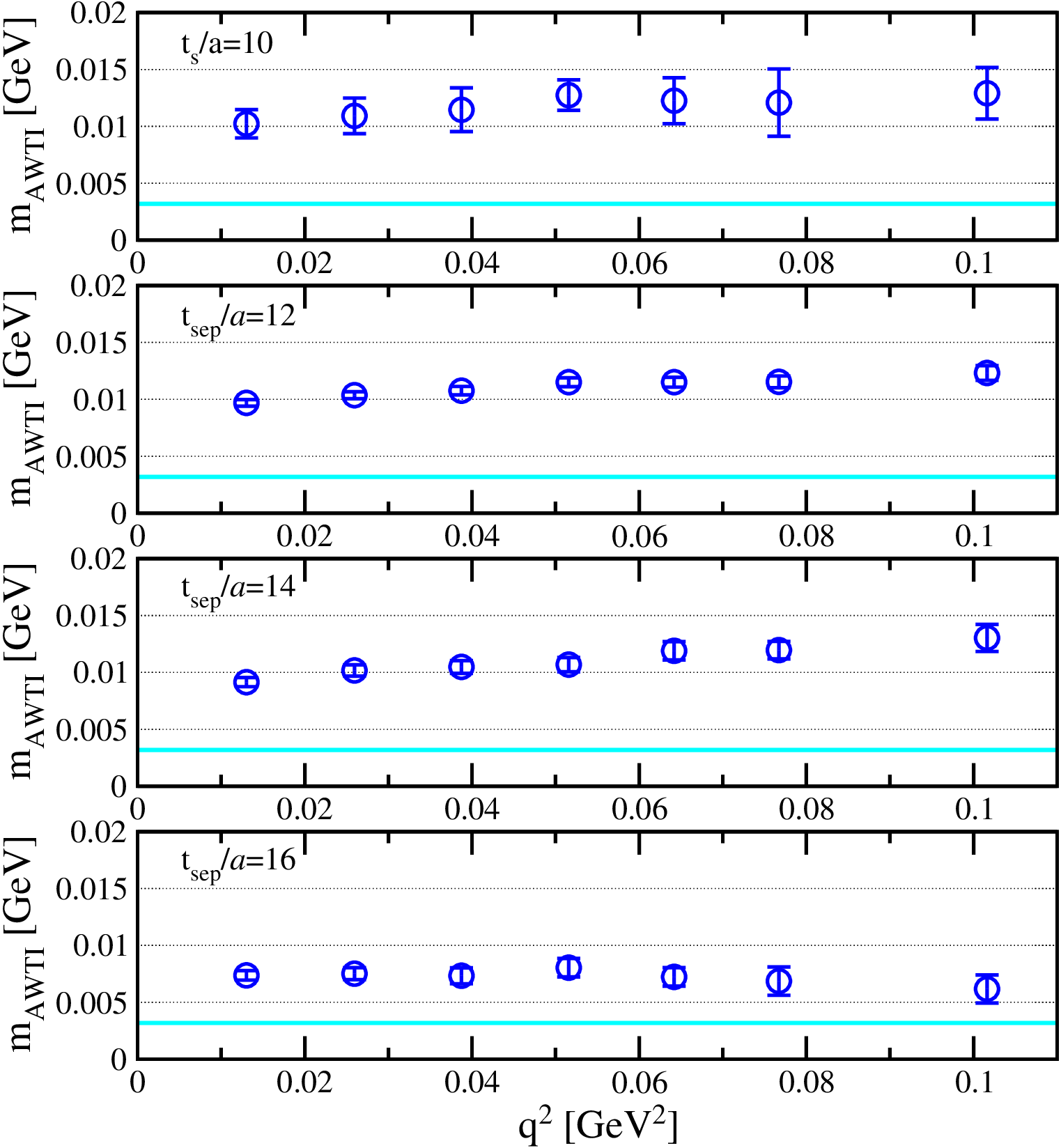}
  \hspace{3mm}
  \includegraphics[width=8.0cm,keepaspectratio,clip]{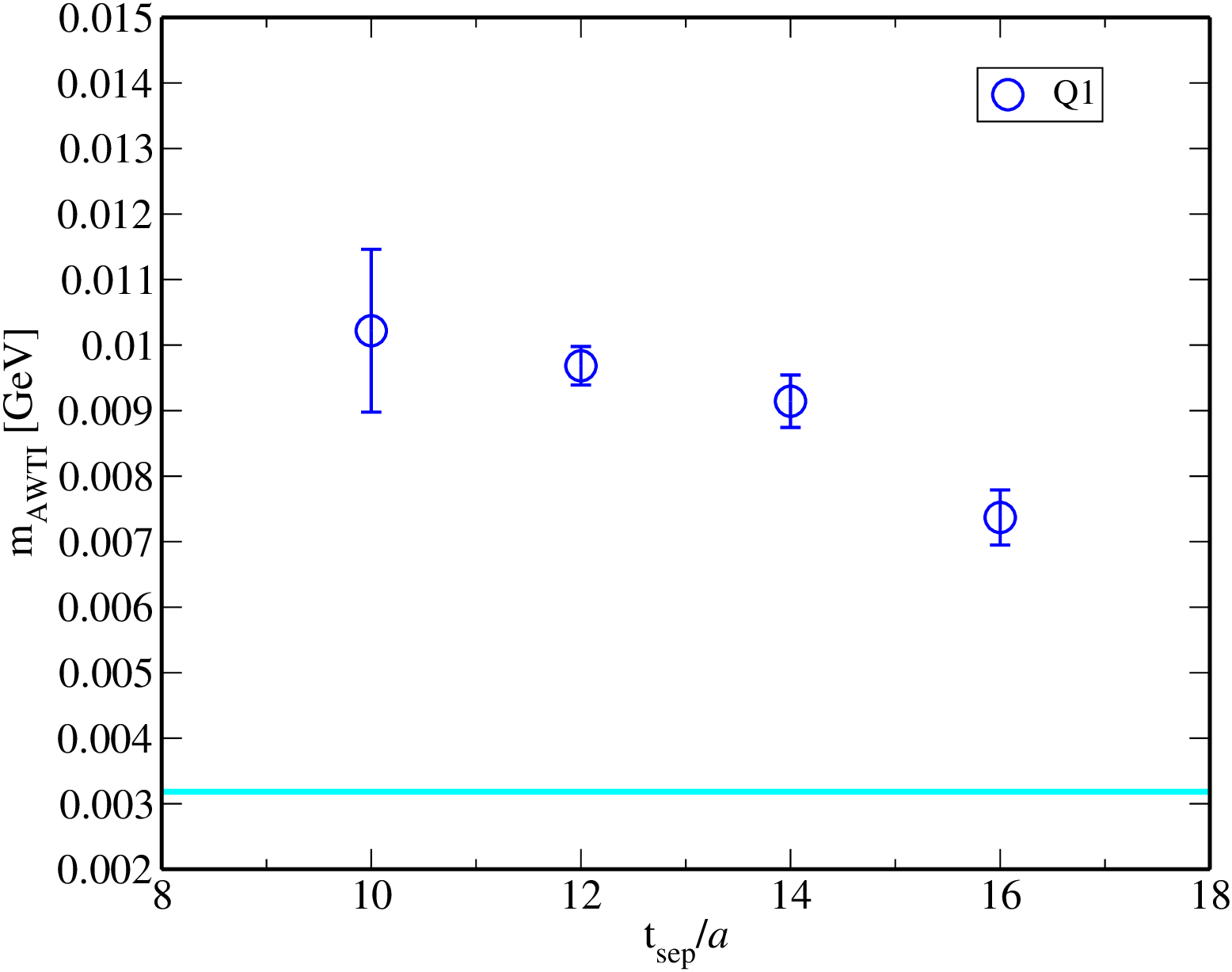}
\caption{$q^2$ dependence (left) and $t_{\rm sep}$ dependence (right) of the quark mass defined by the generalized GT relation of Eq.~(\ref{Eq:GTrelation}). The data in the right panel is obtained at Q1 corresponding to $|{\bm n}|^2=1$. Horizontal band denotes the quark mass obtained from the PCAC relation in Ref.~\cite{Ishikawa:2018jee}.}
\label{fig:mq_awti}
\end{figure*}

In order to test the generalized GT relation of Eq.~(\ref{Eq:GTrelation}), it is convenient to define the following (bare) quark mass as in Ref.~\cite{Ishikawa:2018rew}:
\be
m^{\rm GT}_{\rm AWTI} = \frac{2M_NF_A(q^2) - q^2 F_P(q^2)}{2G_P(q^2)}.
\label{Eq:MAWTI}
\ee
Since the generalized GT relation is an expression of the AWT identity in terms of the nucleon matrix elements, the above quark mass should coincide with the PCAC quark mass $m^{\rm PCAC}_{\rm AWTI}$ extracted from the pseudoscalar matrix elements. In Fig.~\ref{fig:mq_awti} we plot the quark mass $m^{\rm GT}_{\rm AWTI}$ as a function of $q^2$ for all the cases of $t_{\rm sep}/a=10,12,14,16$. The results do not show any strong $q^2$ dependence but they are systematically decreased when $t_{\rm sep}$ increases. Compared to the measured $m^{\rm PCAC}_{\rm AWTI}$~\cite{Ishikawa:2018jee}  from the pion propagator on the same gauge ensembles, $m^{\rm GT}_{\rm AWTI}$ approaches to $m^{\rm PCAC}_{\rm AWTI}$ as $t_{\rm sep}/a$ is increased: $m^{\rm GT}_{\rm AWTI}/m^{\rm PCAC}_{\rm AWTI}=3.0(1)$ at $t_{\rm sep}/a=12$ to 2.3(1) at $t_{\rm sep}/a=16$, quoted from the value of Fig.~\ref{fig:mq_awti} at the minimum $q^2=0.013$ GeV$^2$, with only statistical error. This tendency provides a hint to resolve an issue of ``distortion of pion-pole structure'' discussed in Ref.~\cite{Ishikawa:2018rew}.

\section{Summary and discussion} 
\label{sec:summary}

We have calculated the nucleon electric and magnetic form factors, $G_E(q^2)$ and $G_M(q^2)$, for not only the isovector channel but also the individual form factors of proton and neutron without the disconnected diagram on a (10.8 fm$)^4$ lattice at the physical point in 2+1 flavor QCD. We have also measured the axial-vector form factor $F_A(q^2)$ together with the axial-vector coupling $g_A$ and the axial radius and the induced pseudoscalar form factor $F_P(q^2)$. Utilizing the optimized all-mode-averaging (AMA) technique with the Wilson-clover fermion, we have investigated the effects of the excited state contamination by varying $t_{\rm sep}$ from $0.85$ fm to 1.35 fm with $t_{\rm sep}/a=10,12,14,16$ in the plateau method, which has not been studied in the previous work~\cite{Ishikawa:2018rew}. After elaborate tuning of the sink and source functions, we can obtain clear signal of the nucleon asymptotic state for $G_E(q^2)$, $G_M(q^2)$, and $F_A(q^2)$ without significantly large excited state contamination. Taking account of the uncertainties with the extrapolation onto $q^2=0$ and the excited state contamination, our best estimates for the RMS radii
and magnetic moments are obtained as follows: 
\ben
&&\begin{array}{cc}
\sqrt{\la r^2_{E}\ra}=0.875(15)(28)\,[{\rm fm}]& \textrm{(isovector)}\\
\sqrt{\la r^2_{E}\ra}=0.858(13)(35)\,[{\rm fm}]& \textrm{(proton)}\\
\la r^2_{E}\ra=\,-0.047(20)(18)[{\rm fm^2}]& \textrm{(neutron)}\\
\end{array}
,\label{Eq:final_rns_ge}\\
&&\begin{array}{cc}
\sqrt{\la r^2_{M}\ra}=0.805(32)(274)\,[{\rm fm}]& \textrm{(isovector)}\\
\sqrt{\la r^2_{M}\ra}=0.808(35)(182)\,[{\rm fm}]& \textrm{(proton)}\\
\sqrt{\la r^2_{M}\ra}=0.823(33)(276)\,[{\rm fm}]& \textrm{(neutron)}\\
\end{array}
,\label{Eq:final_rns_gm}\\
&&\begin{array}{cc}
  \mu_v = 4.417(138)(317)& \textrm{(isovector)}\\
  \mu_p = 2.748(93)(139)& \textrm{(proton)}\\
  \mu_n = -1.709(62)(130)& \textrm{(neutron)}\\
\end{array}
,\label{Eq:final_mm}
\een 
where the first error is statistical, the second one is systematic error of which the uncertainty of possible excited state contamination and the fit dependence for extrapolation to $q^2=0$ (see Secs.~\ref{subse:nff_ge} and \ref{subse:nff_gm} for the details) are included. They are comparable with the experimental values of $\sqrt{\la r^2_{E}\ra}$, which are given by 0.939(6) fm (isovector) and 0.875(6) fm (proton) for the $ep$ scattering, and $0.907(1)$ fm (isovector) and 0.8409(4) fm (proton) for the $\mu$H spectroscopy. The experimental values of $\sqrt{\la r^2_{M}\ra}$ is given by 0.862(14) fm (isovector), 0.776(38) fm (proton) and 0.864(9) fm (neutron), and those for the magnetic moment are $\mu_v=4.70589$, $\mu_p=2.79285$ and $\mu_n=-1.91304$ quoted from PDG'18~\cite{Tanabashi:2018oca}. Although our results for the electric RMS charge radius in the isovector channel seem to favor the experimental result of the $\mu$H spectroscopy within 1-$\sigma$ error, it is still too early to draw any definitive conclusion because of rather large error bars of 4\% level. For the proton and neutron channels we leave the inclusion of the disconnected diagram to future work.

\begin{figure*}[ht!]
  \includegraphics[width=16cm,keepaspectratio,clip]{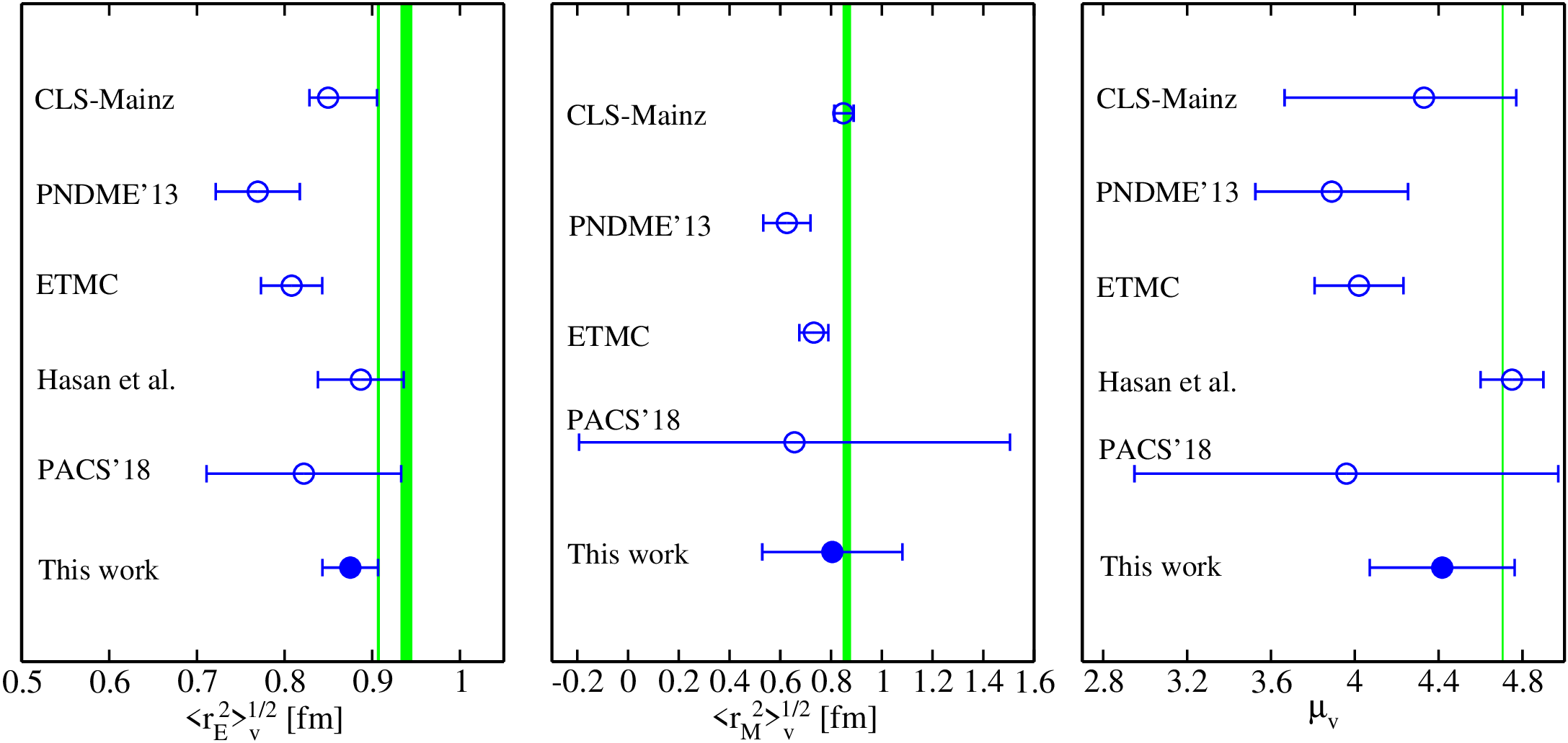}
  \vskip 3mm
  \includegraphics[width=11cm,keepaspectratio,clip]{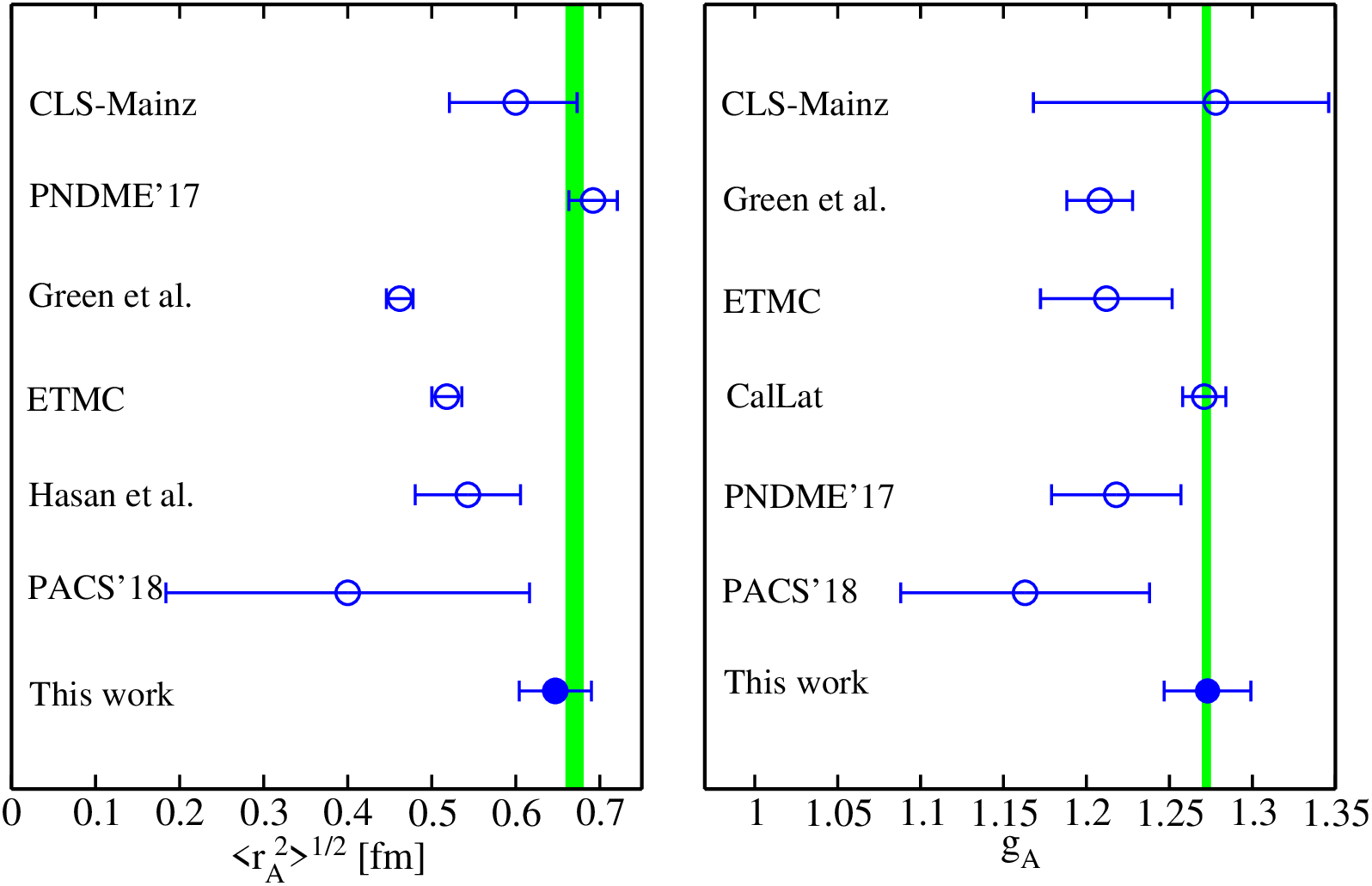}
\caption{(Top) Comparison of recent LQCD results for the isovector nucleon electric and the magnetic RMS radii, and magnetic moment obtained by CLS-Mainz~\cite{Capitani:2015sba}, PNDME~\cite{Bhattacharya:2013ehc}, ETMC~\cite{Alexandrou:2017ypw}, Hasan et al.~\cite{Hasan:2017wwt} and PACS~\cite{Ishikawa:2018rew}. (Bottom) Same as top panels for the axial RMS radius and the axial-vector coupling obtained by CLS-Mainz~\cite{Capitani:2017qpc}, Green et al.~\cite{Green:2017keo}, PNDME~\cite{Rajan:2017lxk,Gupta:2018qil}, ETMC~\cite{Alexandrou:2017hac}, CalLat~\cite{Chang:2018uxx}, and PACS~\cite{Ishikawa:2018rew}. Those errors are total one combined with statistical and systematic errors in the quadrature, except that $g_A$ for PACS'18 has only statistical error. Vertical bands denote experimental values.}
\label{fig:summary_iso}
\end{figure*}

\begin{figure*}[ht!]
  \includegraphics[width=16cm,keepaspectratio,clip]{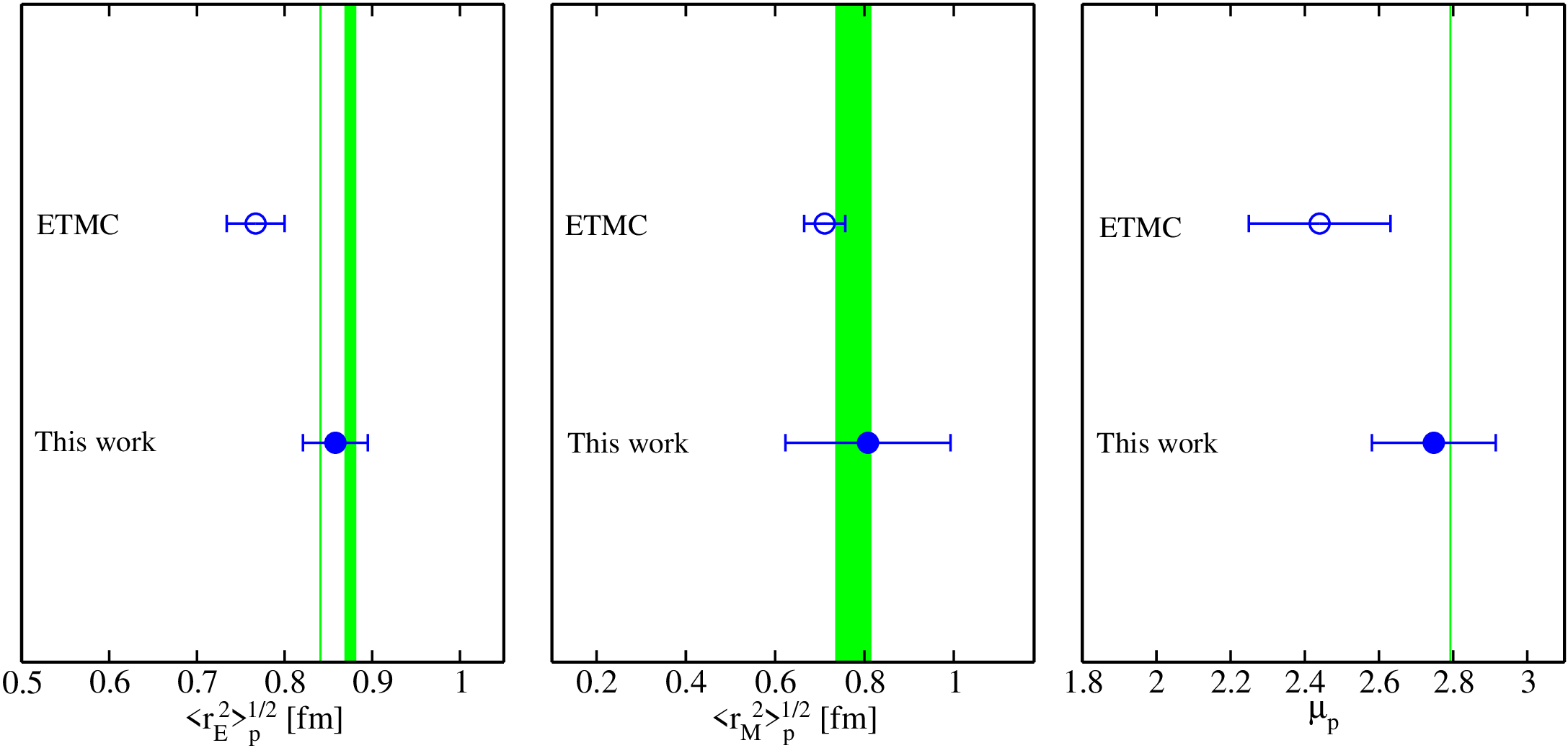}
  \vskip 3mm
  \includegraphics[width=11cm,keepaspectratio,clip]{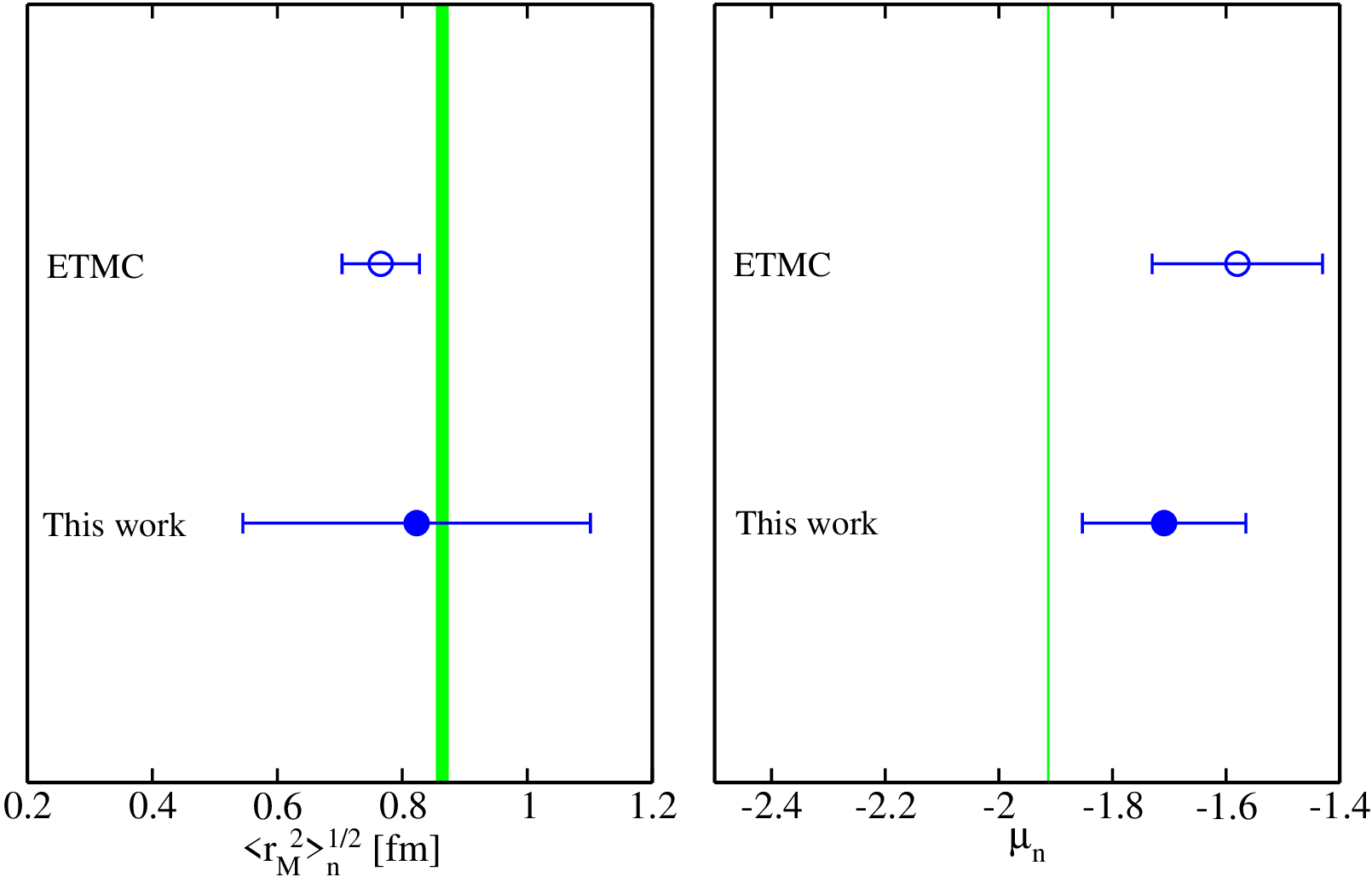}
\caption{Summary plot for the proton electric and magnetic RMS radii and magnetic moment (top) and the neutron electric and magnetic RMS radii and magnetic moment (bottom). We also plot the results of the ETM Collaboration~\cite{Alexandrou:2017ypw} for comparison. Vertical bands denote experimental values. Our results are obtained without the disconnected diagram.}
\label{fig:summary_p_n}
\end{figure*}

In Fig.~\ref{fig:summary_iso} we compare our results with those obtained by previous LQCD calculations for the isovector channel (see Table~\ref{tab:comp_previous_studies} for the simulation parameters). Those errors are combined with statistical and systematic errors in the quadrature, except that $g_A$ for PACS'18 has only statistical error. The electric RMS charge radius given by the PNDME~\cite{Bhattacharya:2013ehc} and ETM~\cite{Alexandrou:2017ypw} Collaborations are below the experimental values beyond 1-$\sigma$ error. In spite of that the other LQCD calculation, {\it e.g.}, the ETM Collaboration, has also been at the physical pion, their values have differed from our results and experimental values. This may be due to some sorts of finite volume effect on their results~\cite{Alexandrou:2017ypw}. The spatial extent of $10.8$ fm in our case allows $q^2=0.013$ GeV$^2$ as the minimum momentum transfer, which is 6 times smaller than that of the ETM Collaboration\cite{Alexandrou:2017ypw}\footnote{Recently, the updated results of electromagnetic form factor in ETM Collaboration appear in Ref.~\cite{Alexandrou:2018sjm}. Their results of electric RMS radius still has a large discrepancy from experimental values. As we have argued in this paper, this may be due to finite volume effect on their relatively small spatial size, which is up to 6 fm ($N_f=2$), compared to our calculation on 10.8 fm.}, who have employed gauge configurations with $N_f=2$ twisted mass fermion on a (4.5 fm$)^3$ box at the physical pion mass. It means that our simulation on large volume is a strong advantage for the determination of a slope at $q^2=0$ to correctly obtain RMS radius. Actually the deficit of the charge radius observed in the ETMC results is consistent with the theoretical expectation discussed in Refs.~\cite{Sick:2018fzn,Ishikawa:2018jee}. For the magnetic RMS radius and the magnetic moment, our results are consistent with the experimental values, though we find relatively large error bars compared to other LQCD results. Our results indicate that $G_M^v$ is sensitive to the source-sink separation $t_{\rm sep}$ rather than $G_E^v$. In Eq.~(\ref{Eq:final_rns_gm}) our large systematic error takes into account such an uncertainty due to the excited state contamination. We find similar stories for the proton and neutron charge radii and the magnetic moment as shown in Fig.~\ref{fig:summary_p_n}, even though our result is obtained from only the connected diagram. As discussed in Secs.~\ref{subse:nff_ge} and \ref{subse:nff_gm} we expect that the disconnected contribution may compensate the difference between our results and the experimental values.

Our best estimates for the axial vector coupling and the axial radius
are obtained as
\begin{eqnarray}
\sqrt{\la r^2_{A}\ra}&=&  0.647(22)(38)\,[{\rm fm}],\\
g_A                  &=& 1.273(24)(5)(9),\\
g_A^u &=& 0.967(30)(16),\\
g_A^d &=& -0.306(19)(21),
\end{eqnarray}
in which the first error is statistical and the second one is systematic explained in Secs.~\ref{subsubsec:nff_ga} and \ref{subsubsec:nff_fa}.
They are comparable with experimental values, $g_A=1.2724(23)$~\cite{Tanabashi:2018oca} and $\sqrt{\la r^2_{A}\ra}=0.67(1)$~\cite{Bernard:2001rs,Bodek:2007ym}. Our result of the axial RMS radius
is consistent with the experimental value, while the 7\% precision is 4.5 times larger than the experimental one. For the axial-vector coupling, our value is also consistent with the experimental one, though the 2\% precision of the former is an order-of-magnitude larger than that of the latter. We also present the results for $g_A^u$ and $g_A^d$ neglecting the disconnected contribution, which are in agreement with other LQCD results \cite{Alexandrou:2017hac,Green:2017keo} within 2-$\sigma$ error. In comparison with other LQCD results as shown in Fig.~\ref{fig:summary_iso}, our results show consistency with experimental values within comparable magnitude of error bars to other groups. For the axial radius and the axial-vector coupling our results are significantly improved from the ETMC's results~\cite{Alexandrou:2017hac}. Here again the spatial lattice size may play a crucial role. Our results on a (10.8 fm$)^3$ spatial box, which is about 14 times larger than a $\sim$(4.5 fm$)^3$ spatial box employed by the other groups {\it e.g.}, the ETM Collaboration, clearly show consistency of $G_A(q^2)$ with the Kelly's fit (see Fig.~\ref{fig:fa_q2}), and we observe less excited state contamination effects. 

On the other hand, the induced pseudoscalar form factor $F_P$ in the axial-vector channel shows clear $t_{\rm sep}$ dependence and considerable deficit from the experimental value in very low $q^2$ region. Investigation of the generalized GT relation associated with $F_A$, $F_P$, and $G_P$ strongly suggests a sizable amount of the excited state contributions to the determination of $F_A$ and $F_P$ in the plateau method. More dominant excited state contamination compared to the other form factors could be a resolution of ``distortion of pion-pole structure''~\cite{Ishikawa:2018rew}, and it would be solvable once the high-precision data at larger $t_{\rm sep}$ is available. Note that, thanks to our spatial size more than 10 fm, we can first obtain the low-$q^2$ LQCD data of induced pseudoscalar form factor, which is close to $q^2$ in MuCap experiment~\cite{Andreev:2012fj}, at the physical point.

This is the first lattice QCD calculation that succeeds in simultaneously reproducing the experimental values for $\sqrt{\la r^2_{E}\ra}$, $\mu$, $\sqrt{\la r^2_{M}\ra}$,  $g_A$, and $\sqrt{\la r^2_{A}\ra}$, and makes an important step for the LQCD calculations to successfully improve its precision to be comparable with the experimental results. In order to assure the reliability of the results, a next step would be further reduction of both statistical and systematic errors such as the cutoff effects and the isospin breaking effects including quark disconnected diagrams.

%
\begin{table*}[ht] 
\caption{Summary of simulation parameters in recent LQCD calculations for the nucleon form factors. 
$N_f$ denotes the number of dynamical quark flavors. In the column ``Fermion,'' ``TM-Clover'' stands for the twisted mass clover-improved Wilson-Dirac operator, ``ST-Clover'' denotes the stout smeared Wilson-clover fermion, and ``HEX-Clover'' denotes the HEX smeared Wilson-clover fermion.  ``MDWF'' denotes M\"obius domain-wall fermion. In the column of ``Method,'' ``R,'' ``S," ``TSF," ``D," and ``FH" stand for the standard plateau (ratio) method, the summation method, the two-state fit method, the derivative method and the method based on the Feynman-Hellmann theorem.
\label{tab:comp_previous_studies}
}
\begin{ruledtabular}
\begin{tabular}{lccccccccccccc} \hline
&
&
&
&
&
&
&
&
&\multicolumn{5}{c}{Observables}\\
\cline{10-14}
\multicolumn{1}{l}{Publication}
& \multicolumn{1}{c}{$N_f$} 
& \multicolumn{1}{c}{Type} 
& \multicolumn{1}{c}{Fermion} 
& \multicolumn{1}{c}{$m_{\pi}$ [MeV]} 
& \multicolumn{1}{c}{$a$ [fm]} 
& \multicolumn{1}{c}{$La$ [fm]} 
& \multicolumn{1}{c}{$t_{\rm sep}$ [fm]}
& \multicolumn{1}{c}{Method}
& \multicolumn{1}{c}{$\langle r^2_{E}\rangle$}
& \multicolumn{1}{c}{$\mu_v$}
& \multicolumn{1}{c}{$\langle r^2_{M}\rangle$}
& \multicolumn{1}{c}{$g_A$}
& \multicolumn{1}{c}{$\langle r^2_A\rangle$}
\\ \hline
CLS-Mainz~\cite{Capitani:2015sba,Capitani:2017qpc} &2 & Full & Clover & $\ge261$ &  0.050 & $4.0$ 
& $\le 1.1$ & R, S, TSF & $\bigcirc$ & $\bigcirc$ & $\bigcirc$ & $\bigcirc$ & $\bigcirc$ \\
&  &  &  & $\ge193$ &  0.063 &  $4.0$ 
& $\le 1.1$ & R, S, TSF & $\bigcirc$ & $\bigcirc$ & $\bigcirc$ & $\bigcirc$ & $\bigcirc$ \\
&  &  &  & $\ge268$ &  0.079 &  $4.0$ 
& $\le 1.26$ & R, S, TSF & $\bigcirc$ & $\bigcirc$ & $\bigcirc$ & $\bigcirc$ & $\bigcirc$\\
ETMC~\cite{Alexandrou:2017ypw,Alexandrou:2017hac} &2 & Full & TM-Clover& 130 & 0.094 & 4.5  
& $\le 1.69$~\footnotemark[1]\footnotetext[1]{The electric form factor determined 
with the projection operator ${\cal P}_t$ is evaluated up to $t_{\rm sep}/a=18$ ($t_{\rm sep}=1.69$ [fm]), 
while the magnetic, axial-vector and pseudoscalar form factors determined with 
the projection operator ${\cal P}_{5z}$ are evaluated only up to $t_{\rm sep}/a=14$
($t_{\rm sep}=1.32$ [fm]).}  & R, S, TSF & $\bigcirc$ & $\bigcirc$ & $\bigcirc$ & $\bigcirc$ & $\bigcirc$\\
PNDME'13~\cite{Bhattacharya:2013ehc} &2+1+1 &Hybrid~\footnotemark[2]\footnotetext[2]{Clover fermions on highly improved staggered quarks (HISQ) ensembles}  & Clover  &220 & 0.12 & 3.8  
&$\le 1.44$  & R, TSF & $\bigcirc$ & $\bigcirc$ & $\bigcirc$ & --- & ---\\
             & & & Clover &310 & 0.12 & 2.9 
             & $\le 1.44$ & R, TSF & $\bigcirc$ & $\bigcirc$ & $\bigcirc$ &  ---  & ---\\
PNDME'17~\cite{Rajan:2017lxk,Gupta:2018qil} 
&2+1+1 &Hybrid~\footnotemark[2] & Clover  & $\ge 135$ & 0.06 & 5.5  
&$\le 1.25$  & R, TSF & --- & --- & --- & $\bigcirc$ & $\bigcirc$\\
             & & & Clover & $\ge 130$ & 0.09 & 5.6  
             & $\le 1.44$ & R, TSF & --- & --- & --- & $\bigcirc$ & $\bigcirc$\\
             & & & Clover & $\ge 220$ & 0.12 & 4.8  
             & $\le 1.66$ & R, TSF & --- & --- & --- & $\bigcirc$ & $\bigcirc$\\
             & & & Clover & $310$ & 0.15 & 2.4  
             & 1.35 & R, TSF & --- & --- & --- & $\bigcirc$ & $\bigcirc$\\             
CalLat~\cite{Chang:2018uxx} &2+1+1 & Hybrid~\footnotemark[3]\footnotetext[3]{M\"obius domain-wall fermions on HISQ ensembles} & MDWF & $\ge 220$ & 0.09 & 4.3
 &  & FH & --- & --- & --- & $\bigcirc$ & ---\\
& & & MDWF & $\ge 130$ & 0.12 & 5.8 
& & FH & --- & --- & --- & $\bigcirc$ & ---\\
& & & MDWF & $\ge 130$ & 0.15 & 4.8 
& & FH & --- & --- & --- & $\bigcirc$ & ---\\
Hasen {\it et al.}~\cite{Hasan:2017wwt} & 2+1 & Full & ST-Clover & 135 & 0.093 & 5.9 
& $\le 1.49$  & R, S, D & $\bigcirc$ & $\bigcirc$ & --- & --- & $\bigcirc$\\             
Green {\it et al.}~\cite{Green:2017keo} & 2+1 & Full & HEX-Clover & 317 & 0.114 & 3.6  
& $\le 1.60$ & R, S & --- & --- & --- & $\bigcirc$ & $\bigcirc$\\
PACS'18~\cite{Ishikawa:2018rew} &2+1&Full  & ST-Clover & 146 & 0.085 & 8.1 
& 1.27 & R & $\bigcirc$ & $\bigcirc$ & $\bigcirc$ & $\bigcirc$ & $\bigcirc$\\
This work &2+1&Full  & ST-Clover & 135 & 0.085 & 10.8 
& $\le 1.36$ & R & $\bigcirc$ & $\bigcirc$ & $\bigcirc$ & $\bigcirc$ & $\bigcirc$\\
\hline

\end{tabular}
\end{ruledtabular} 
\end{table*} 
%

\begin{acknowledgments}
We thank Toshimi Suda and members of the PACS Collaboration for useful discussions. We originally developed the computation code based on Columbia Physics System(CPS) in which the tuned OpenQCD system\footnote{http://luscher.web.cern.ch/luscher/openQCD/} is embedded. Numerical calculations for the present work have been carried out on Oakforest-PACS in Joint Center for advanced high performance computing, HOKUSAI GreatWave at Advanced Center for Computing and Communication (ACCC) of RIKEN, the K computer in RIKEN Center for Computational Science (CCS) and XC40 at YITP at Kyoto University. This research used computational resources of the HPCI system provided by the Information Technology Center of the University of Tokyo, RIKEN CCS through the HPCI System Research Project (Project ID: hp140155, hp150135, hp160125, hp170022, hp180051, hp180072, hp180126). 
This work is supported by Grants-in-Aid for Scientific Research from the Ministry of Education, Culture, Sports, Science and Technology (No. 16H06002), and a Grant-in-Aid for Scientific Research (C) (No. 18K03605), in part by the U.S.-Japan Science and Technology Cooperation Program in High Energy Physics for FY2018, Interdisciplinary Computational Science Program of Center for Computational Sciences in University of Tsukuba, and general use No.~G18001 at ACCC of RIKEN.
\end{acknowledgments}
\appendix
\section{Table of nucleon form factor}
\label{app:tab}

The results for the three isovector form factors $G_E^v(q^2)$, $G_M^v(q^2)$, and $F_A(q^2)$
obtained with $t_{\rm sep}/a=\{12,14,16\}$ and $\{14,16\}$ are summarized in Table~\ref{tab:ff_q2_comb}.
The electric and magnetic form factors for the proton and neutron, $G_E^p(q^2)$, $G_E^n(q^2)$, $G_M^p(q^2)$, and $G_M^n(q^2)$, are presented in Table~\ref{tab:ff_p_n_q2_comb}.

\begin{table*}[hb!]
\caption{$q^2$ dependence of the isovector form factors obtained by the constant fit for $t_{\rm sep}/a=\{12,14,16\}$ and $t_{\rm sep}/a=\{14,16\}$. In the previous work~\cite{Ishikawa:2018rew} the results are obtained with $t_{\rm sep}/a=15$ on a $96^4$ lattice at $m_\pi = 146$ MeV.
\label{tab:ff_q2_comb}
}
\begin{ruledtabular}
\begin{tabular}{c|cccccc} 
\multirow{2}{*}{$q^2\,[{\rm GeV}^2]$}  & \multicolumn{3}{c}{$t_{\rm sep}/a=\{12,14,16\}$} & \multicolumn{3}{c}{$t_{\rm sep}/a=\{14,16\}$} \\
  \cline{2-4}\cline{5-7}
  & $G_E^v(q^2)$ & $G_M^v(q^2)$ & $F_A(q^2)$ & $G_E^v(q^2)$ & $G_M^v(q^2)$ & $F_A(q^2)$\\  
  \hline
0.000 & 0.997(1) & --- & 1.273(24) &  0.999(1) & --- & 1.269(34)\\
0.013 & 0.957(2) & 4.279(162) & 1.244(22)& 0.955(2)&4.511(213)&1.244(32)\\
0.026 & 0.920(3) & 4.124(137) & 1.221(22)&  0.916(4)&4.279(182)&1.215(37)\\
0.039 & 0.885(5) & 3.951(122) & 1.197(20)&  0.880(6)&4.074(172)&1.197(39)\\
0.052 & 0.848(6) & 3.812(120) & 1.158(21)&  0.846(8)&3.863(173)&1.148(29)\\
0.064 & 0.818(6) & 3.701(108) & 1.147(22)&  0.813(9)&3.786(152)&1.129(40)\\
0.077 & 0.789(7) & 3.574(99) & 1.126(21)&  0.782(10)&3.641(151)&1.111(39)\\
0.102 & 0.735(8) & 3.360(90) & 1.082(23)&  0.724(12)&3.427(140)&1.057(47)\\
\hline\hline
\multirow{2}{*}{$q^2\,[{\rm GeV}^2]$} & \multicolumn{3}{c}{PACS'18~\cite{Ishikawa:2018rew} $t_{\rm sep}/a=15$}  \\
  \cline{2-4}
& $G_E^v(q^2)$ & $G_M^v(q^2)$ & $F_A(q^2)$\\  
  \hline
0.000 & 1.000(4) & --- & 1.163(75)\\
0.024 & 0.924(11) & 4.071(456) & 1.121(68)\\
0.048 & 0.861(19) & 3.640(350) & 1.137(69)\\
0.072 & 0.804(27) & 3.333(305) & 1.112(64)\\
0.095 & 0.774(30) & 3.313(344) & 1.118(72)\\
\end{tabular}
\end{ruledtabular} 
\end{table*}

\begin{table*}[hb!]
\caption{$q^2$ dependence of the proton and neutron form factors obtained by the constant fit for $t_{\rm sep}/a=\{12,14,16\}$ and $t_{\rm sep}/a=\{14,16\}$. 
\label{tab:ff_p_n_q2_comb}
}
\begin{ruledtabular}
  \begin{tabular}{c|cccccccc}
\multirow{3}{*}{$q^2\,[{\rm GeV}^2]$}  & \multicolumn{4}{c}{$t_{\rm sep}/a=\{12,14,16\}$} & \multicolumn{4}{c}{$t_{\rm sep}/a=\{14,16\}$} \\
\cline{2-5}\cline{6-9}
  & \multicolumn{2}{c}{Proton} & \multicolumn{2}{c}{Neutron} & \multicolumn{2}{c}{Proton} & \multicolumn{2}{c}{Neutron} \\
  \cline{2-5}\cline{6-9}
  & $G_E^p(q^2)$ & $G_M^n(q^2)$ & $G_E^n(q^2)$ & $G_M^n(q^2)$ & $G_E^p(q^2)$ & $G_M^p(q^2)$ & $G_E^n(q^2)$ & $G_M^n(q^2)$\\  
  \hline
  0.000 & 0.9988(8) & --- &  0.0016(7) & --- & 1.000(0)&---& 0.002(1)&---\\
0.013 & 0.959(1)& 2.660(104)& 0.002(1)& $-$1.620(63)& 0.957(2)&2.757(143)&0.003(1)& $-$1.750(85)\\
0.026 & 0.922(3)& 2.569(87)& 0.003(1)& $-$1.573(54)& 0.919(3)&2.628(128)& 0.004(2)& $-$1.656(68)\\
0.039 & 0.886(4)& 2.473(75)& 0.004(2)& $-$1.507(49)& 0.883(4)& 2.504(119)&0.005(4)& $-$1.576(62)\\
0.052 & 0.854(4)& 2.378(75)& 0.008(3)& $-$1.447(48)& 0.852(6)& 2.372(117)&0.013(11)& $-$1.492(65)\\
0.064 & 0.823(5)& 2.315(63)& 0.008(3)& $-$1.404(44)& 0.820(6)& 2.341(106)&0.012(9)& $-$1.448(56)\\
0.077 & 0.794(6)& 2.241(58)& 0.007(3)& $-$1.351(41)& 0.790(7)& 2.244(104)&0.012(8)& $-$1.400(55)\\
0.102 & 0.742(6)& 2.106(52)& 0.010(4)& $-$1.260(40)& 0.735(9)& 2.130(93)& 0.014(8)& $-$1.296(58)\\
\end{tabular}
\end{ruledtabular} 
\end{table*}

\bibliographystyle{apsrev4-1}
\bibliography{ref}

\pagebreak
\clearpage
\maxdeadcycles=200

\onecolumngrid
\begin{center}
  \textbf{\large Erratum: Nucleon form factors and root-mean-square radii \\
on a (10.8 fm$)^4$ lattice at the physical point [Phys. Rev. D 99, 014510 (2019)]}\\[.2cm]
  Eigo Shintani,$^{1}$ Ken-Ichi Ishikawa,$^{2}$ Yoshinobu Kuramashi,$^3$ Shoichi Sasaki,$^{4}$ and Takeshi Yamazaki, $^{5,3}$\\[.1cm]
  (PACS Collaboration)\\[.1cm]
  {\itshape ${}^1$RIKEN Center for Computational Science, Kobe, Hyogo 650-0047, Japan\\
    ${}^2$Graduate School of Science, Hiroshima University, Higashi-Hiroshima, Hiroshima 739-8526, Japan\\
    ${}^3$Center for Computational Sciences, University of Tsukuba, Tsukuba, Ibaraki 305-8577, Japan\\
    ${}^4$Department of Physics, Tohoku University, Sendai 980-8578, Japan\\
    ${}^5$Faculty of Pure and Applied Sciences, University of Tsukuba, Tsukuba, Ibaraki, 305-8571, Japan\\
  }
(Dated: \today)\\[1cm]
\end{center}
\twocolumngrid

\setcounter{equation}{0}
\setcounter{figure}{0}
\setcounter{table}{0}
\setcounter{page}{1}
\renewcommand{\theequation}{S\arabic{equation}}
\renewcommand{\thefigure}{S\arabic{figure}}

In our analysis we employed different normalization factors from Eqs.~(20)-(23)
in the paper to evaluate the electric, magnetic, axial-vector, 
induced pseudoscalar, and pseudoscalar form factors in the nonzero $q^2$.
In some parts of the normalization factors, $M_N$ was used in stead of $E_N$.
We corrected the normalization factors in our analysis code, and then 
recalculated the form factors with the correct normalization factors
as in the equations.

\begin{figure}[ht!]
    \includegraphics[width=8.0cm,keepaspectratio,clip]{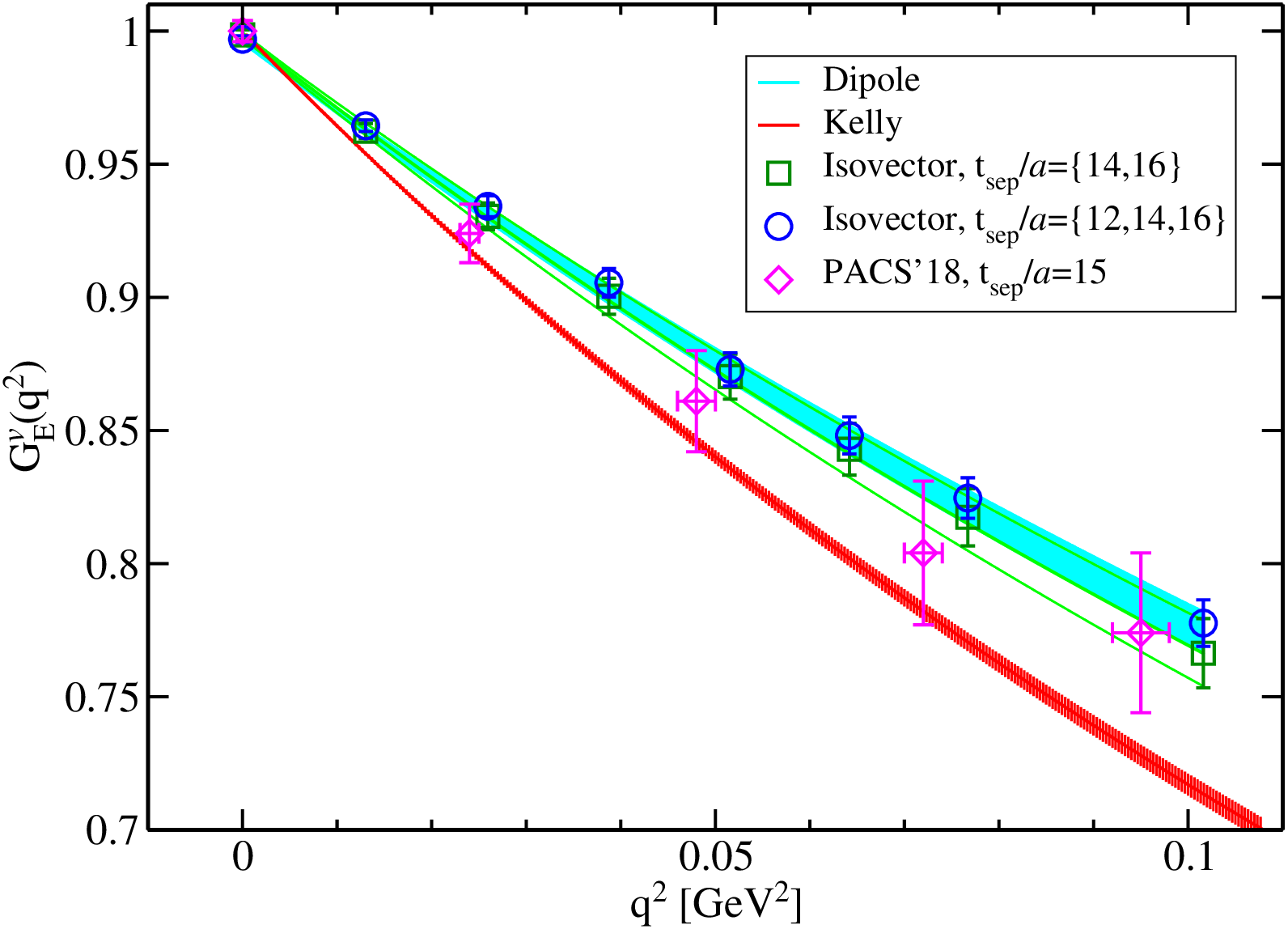}
\caption{$q^2$ dependence of the isovector electric form factor $G_E^v(q^2)$ obtained by the combined analysis of the results at $t_{\rm sep}/a=\{12,14,16\}$ (circle) and $t_{\rm sep}/a=\{14,16\}$ (square). Diamond symbols, which are obtained with $t_{\rm sep}/a=15$ on a $96^4$ lattice at $m_\pi = 146$ MeV in Ref.~\cite{Ishikawa:2018rew}, are also plotted for comparison. This figure replaces Fig.~4 in the paper.}
\label{fig_new:ge_q2}
\end{figure}

\begin{figure*}[ht!]
  \includegraphics[width=8.0cm,keepaspectratio,clip]{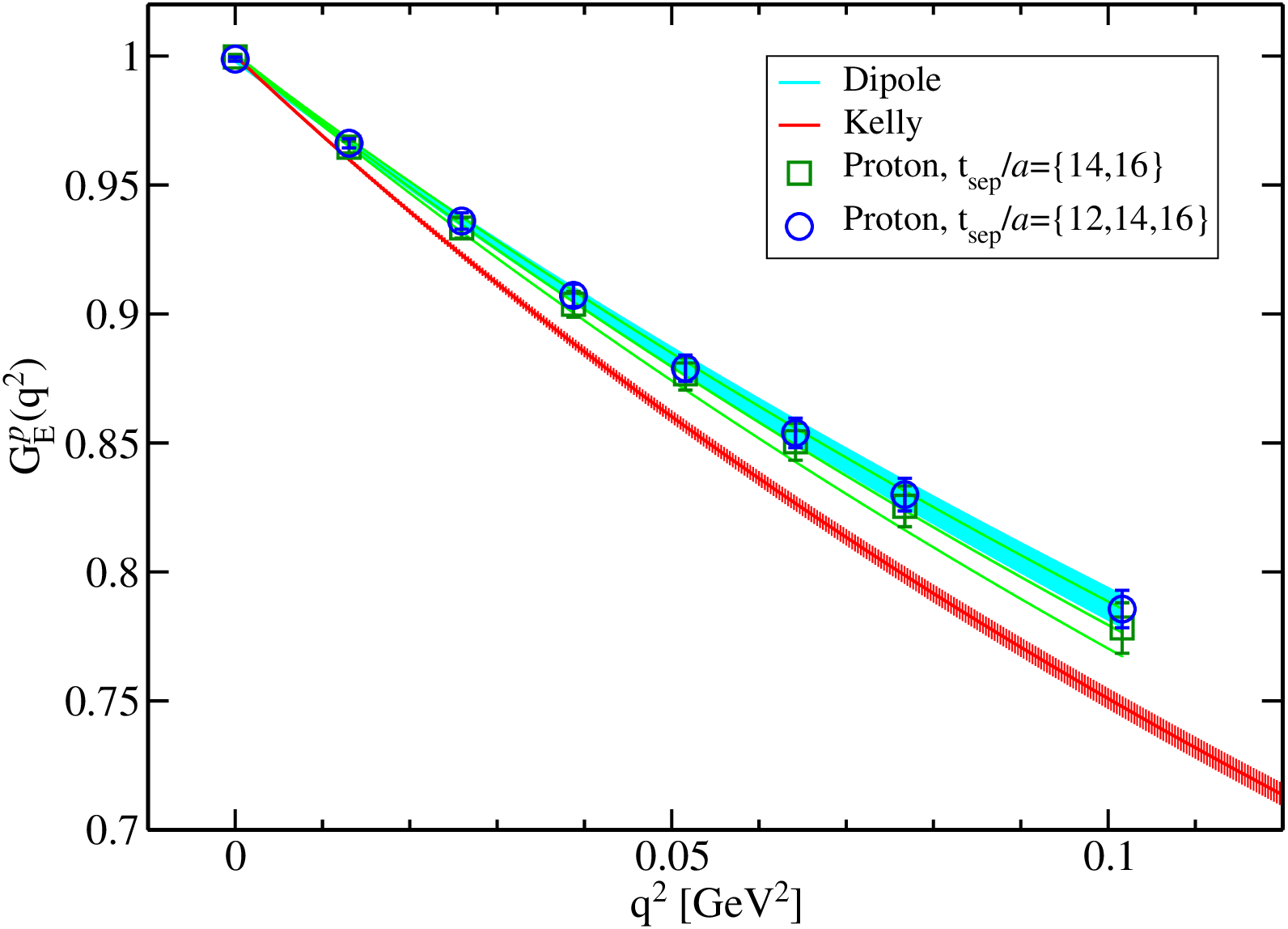}
  \hspace{3mm}
  \includegraphics[width=8.0cm,keepaspectratio,clip]{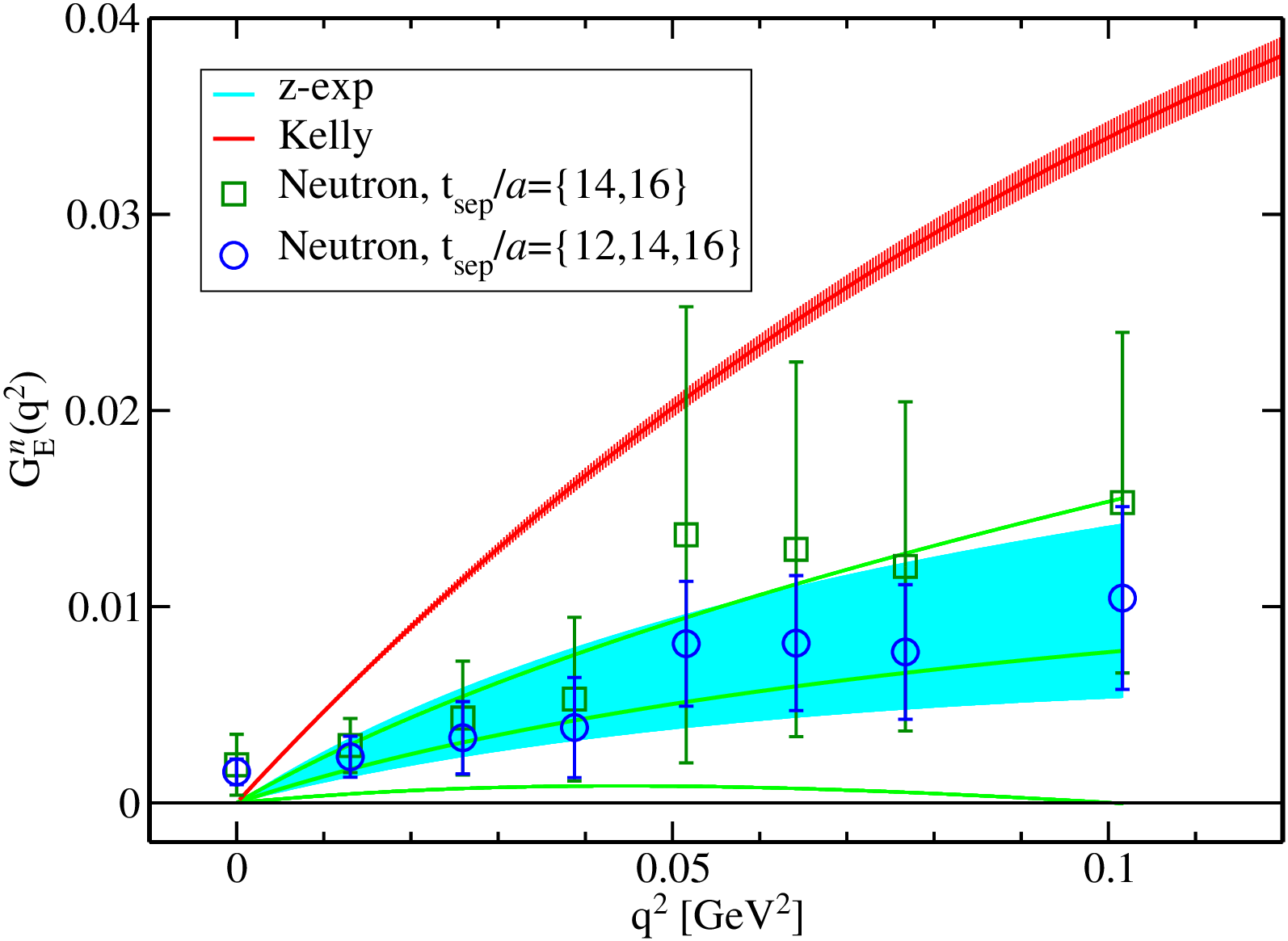}
\caption{Same as Figure~\ref{fig_new:ge_q2} for the proton (left) and neutron (right). Results are obtained without the disconnected diagram. This figure replaces Fig.~5 in the paper.}
\label{fig_new:ge_q2_p_n}
\end{figure*}

\begin{figure*}[ht!]
  \includegraphics[width=8.0cm,keepaspectratio,clip]{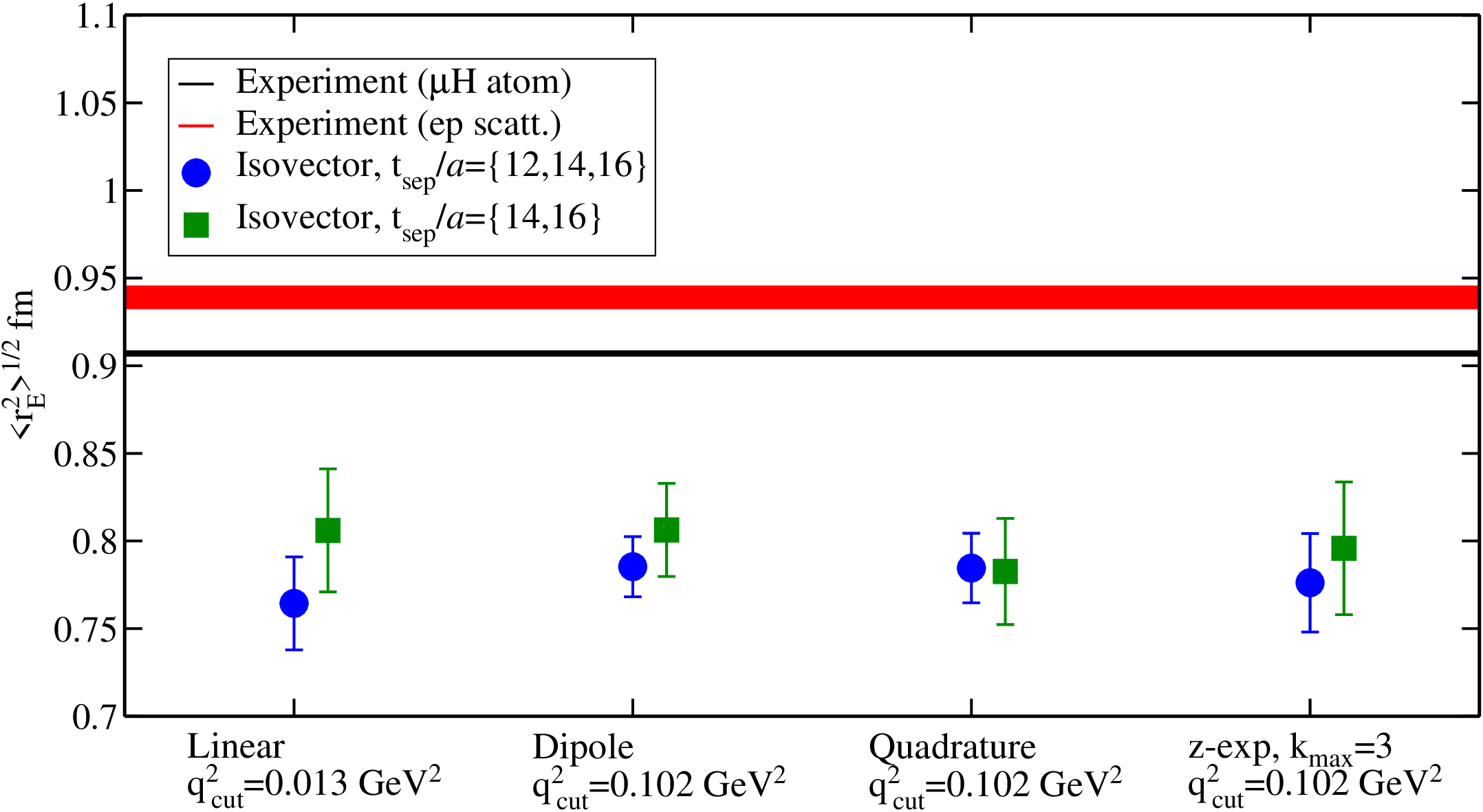}
  \hspace{3mm}
  \includegraphics[width=8.0cm,keepaspectratio,clip]{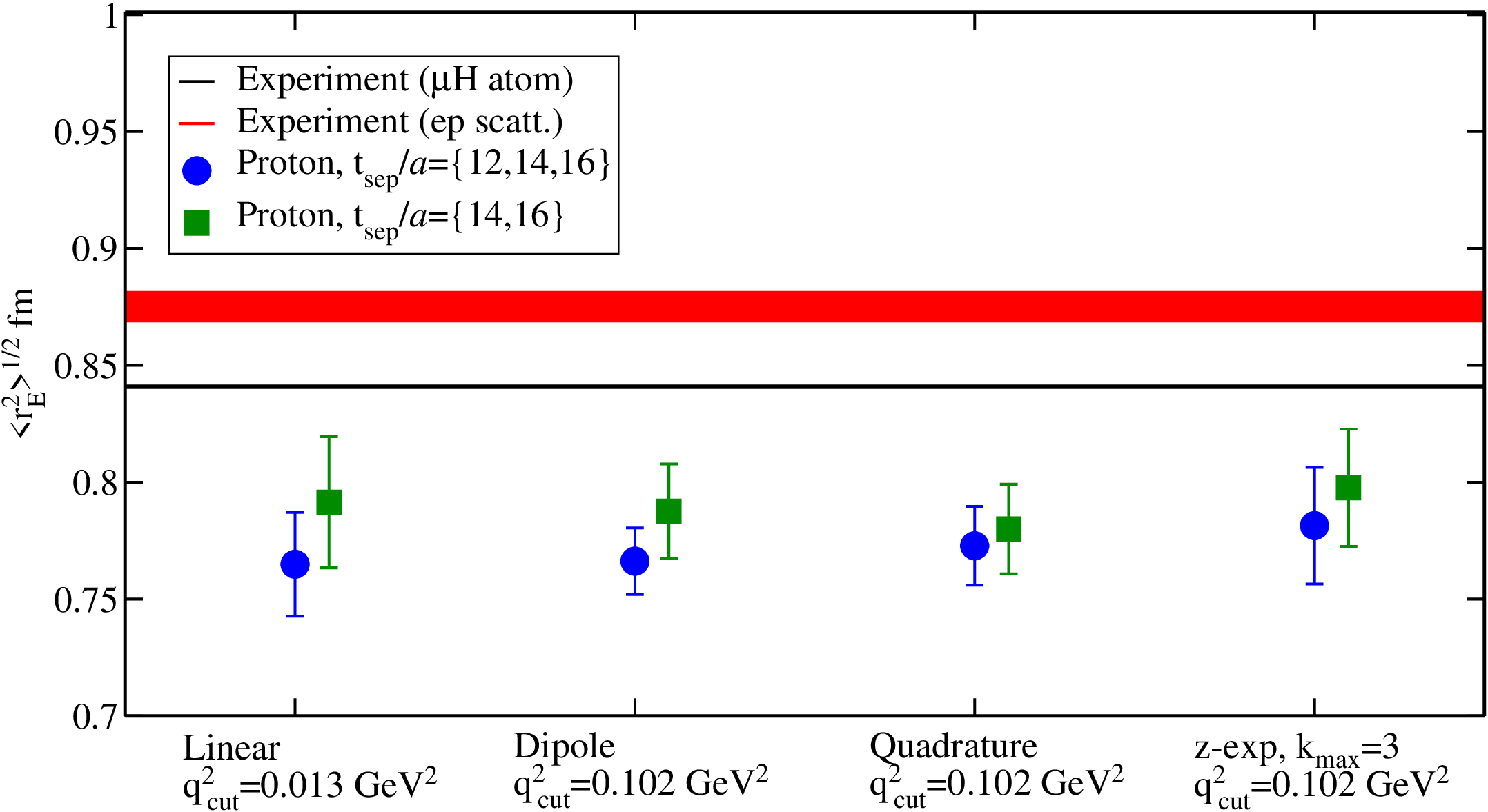}
\caption{Electric RMS radius $\sqrt{\la r^2_E\ra}$ for the isovector (left) and proton (right) obtained by linear, dipole, quadratic and $z$-expansion fits for the combined data. Horizontal bands represent the experimental results from $ep$ scattering (upper) and $\mu$H spectroscopy (lower). The results for the proton channel is obtained without the disconnected diagram. This figure replaces Fig.~6 in the paper.}
\label{fig_new:re_fitdep}
\end{figure*}

\begin{figure}[ht!]
  \includegraphics[width=8.0cm,keepaspectratio,clip]{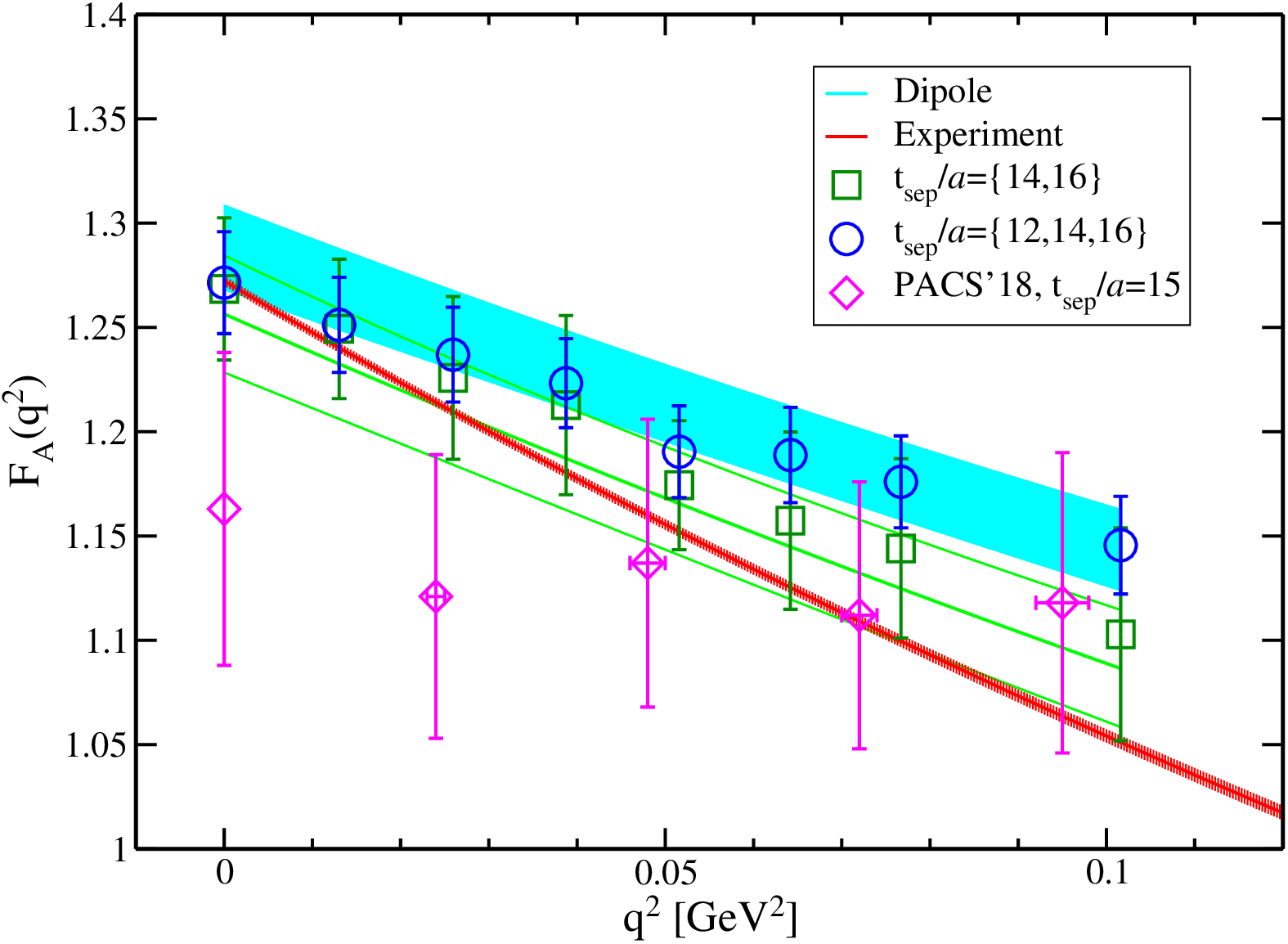}
\caption{Same as Fig.~\ref{fig_new:ge_q2} for the axial-vector form factor $F_A(q^2)$. Experimental line is obtained by the dipole form with the value of dipole mass~\cite{Bodek:2007ym,Bernard:2001rs} and $g_A$~\cite{Tanabashi:2018oca}. This figure replaces Fig.~17 in the paper.}
\label{fig_new:fa_q2}
\end{figure}

\begin{figure*}[ht!]
  \includegraphics[width=8.0cm,keepaspectratio,clip]{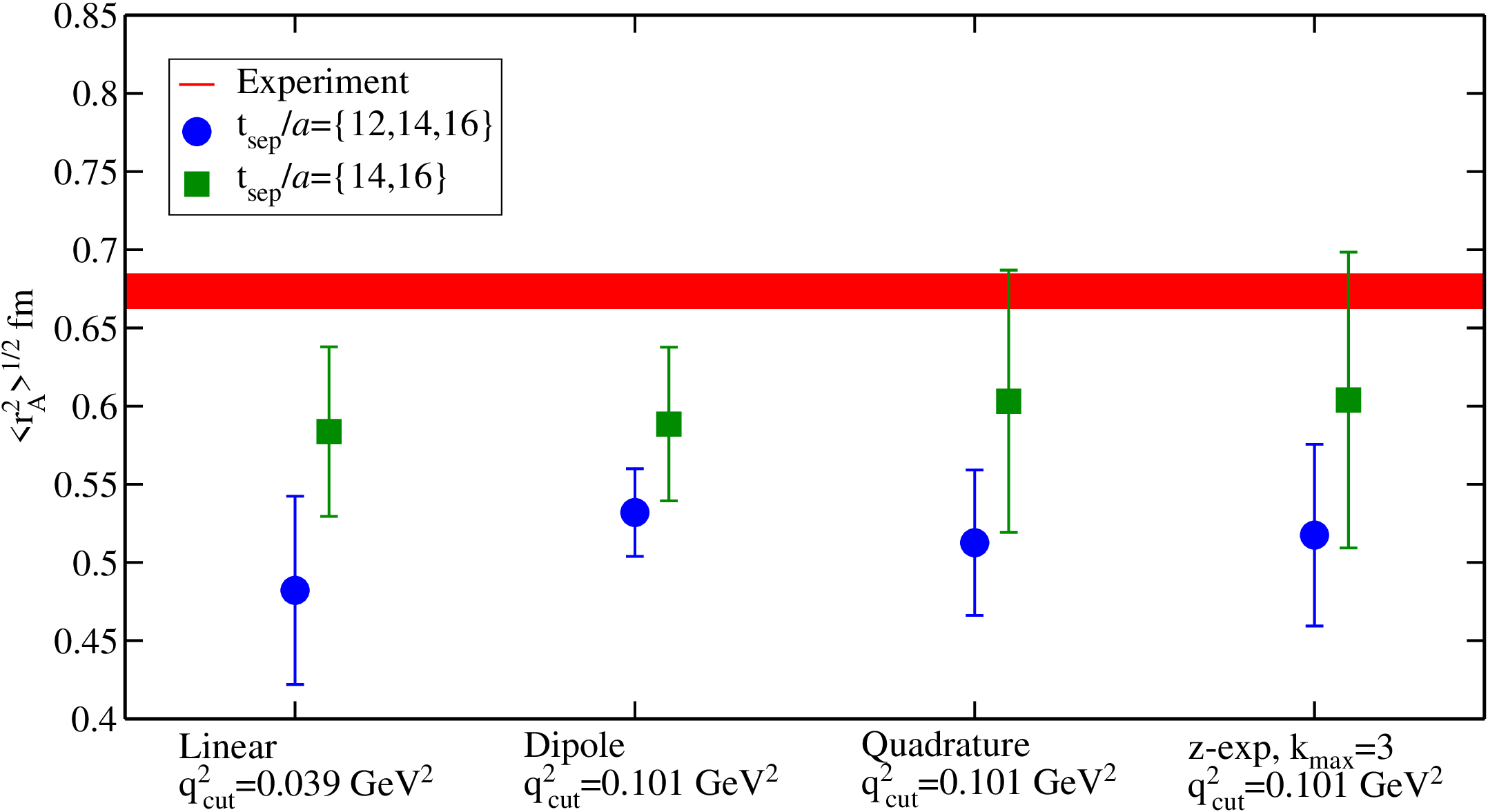}
\caption{The axial-vector RMS radius $\sqrt{\la r^2_A\ra}$ obtained by the fit for the combined data with the linear, dipole, quadratic forms and the $z$-expansion method. Horizontal bands represent the experimental values.
This figure replaces the right panel of Fig.~18 in the paper.}
\label{fig_new:ra_fit}
\end{figure*}

The effect of this correction increases as $q^2$ increases.
As a result, $G_E(q^2)$ and $F_A(q^2)$ are overestimated from
the experiments in the large $q^2$ region as presented in 
Figs.~\ref{fig_new:ge_q2}, \ref{fig_new:ge_q2_p_n}, and \ref{fig_new:fa_q2}.
The root-mean-square (RMS) charge radius 
and the axial-vector RMS radius become smaller than the ones quoted 
in the paper, and are underestimated from the experiments as shown in 
Figs.~\ref{fig_new:re_fitdep} and \ref{fig_new:ra_fit}, respectively.
In contrast to these RMS radii, other results obtained from the magnetic, 
induced pseudoscalar, and pseudoscalar form factors
are reasonably consistent with the ones that appeared in the paper.

\begin{figure*}[ht!]
  \includegraphics[width=16cm,keepaspectratio,clip]{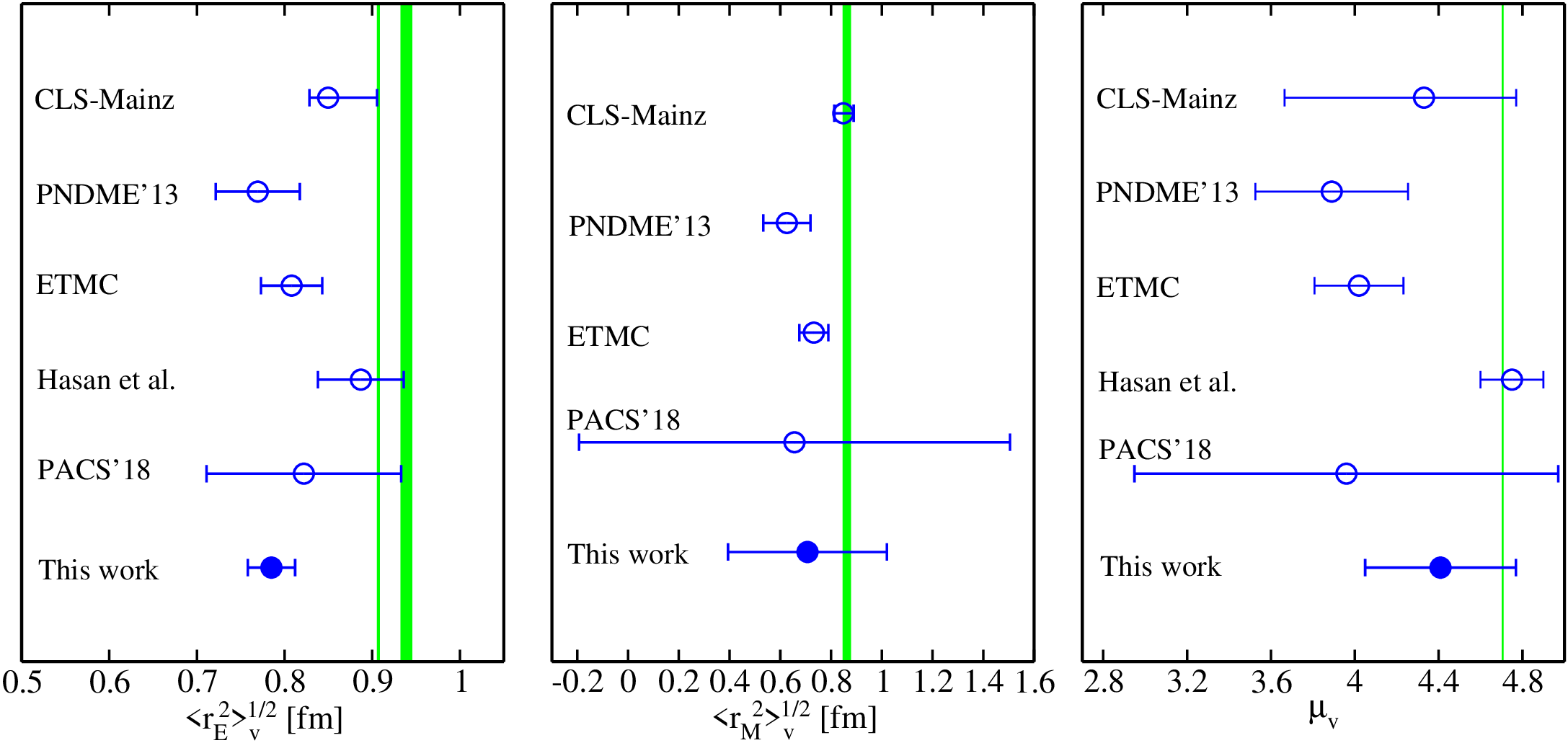}
  \vskip 3mm
  \includegraphics[width=5cm,keepaspectratio,clip]{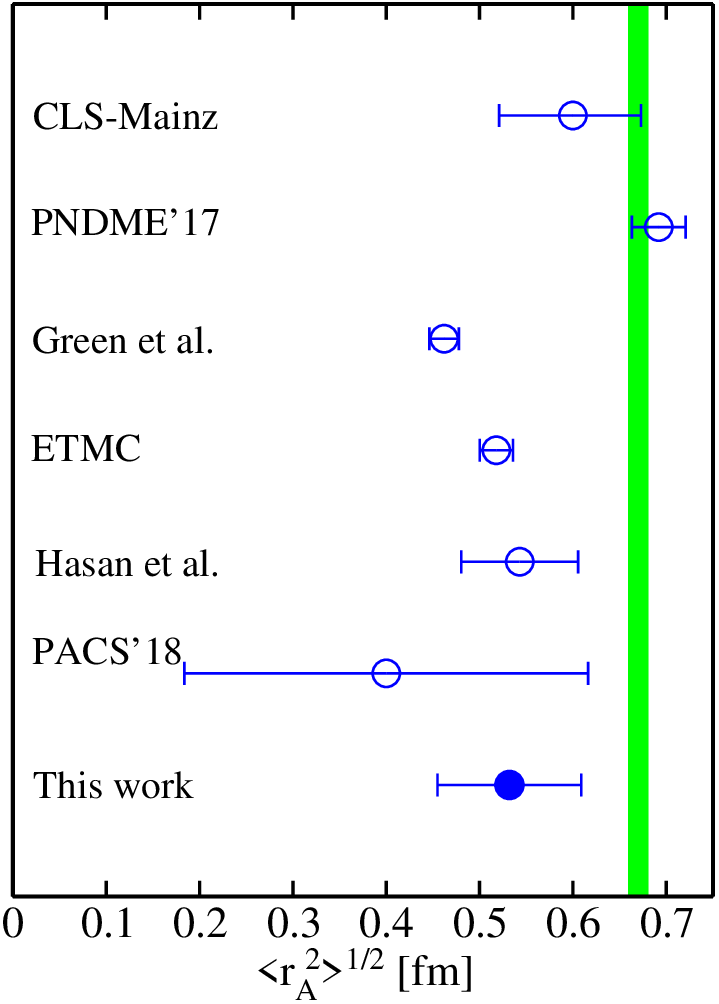}
\caption{(Top) Comparison of recent LQCD results for the isovector nucleon electric and the magnetic RMS radii, and magnetic moment obtained by CLS-Mainz~\cite{Capitani:2015sba}, PNDME~\cite{Bhattacharya:2013ehc}, ETMC~\cite{Alexandrou:2017ypw}, Hasan et al.~\cite{Hasan:2017wwt} and PACS~\cite{Ishikawa:2018rew}. (Bottom) Same as top panels for the axial RMS radius and the axial-vector coupling obtained by CLS-Mainz~\cite{Capitani:2017qpc}, Green et al.~\cite{Green:2017keo}, PNDME~\cite{Rajan:2017lxk,Gupta:2018qil}, ETMC~\cite{Alexandrou:2017hac}, CalLat~\cite{Chang:2018uxx}, and PACS~\cite{Ishikawa:2018rew}. Those errors are total one combined with statistical and systematic errors in the quadrature. Vertical bands denote experimental values. This figure replaces Fig.~26 in the paper, except for the figure of $g_A$.}
\label{fig_new:summary_iso}
\end{figure*}

Figure~\ref{fig_new:summary_iso} presents that the RMS charge radius becomes 
reasonably consistent with the recent lattice results~\cite{Capitani:2015sba,Bhattacharya:2013ehc,Alexandrou:2017ypw,Hasan:2017wwt}.
Thus, the finite volume effect could be small in this radius,
although we discussed the possibility of its existence in the paper. 
For the charge radius, it is important to understand reason of the discrepancy
between the lattice results and the experiments for solving the proton size
puzzle.
Towards understanding of the discrepancy, we plan to measure the RMS charge 
radius with a finer lattice spacing in the next calculation to estimate 
discretization error, which was not estimated in the paper.

Our best estimated values for the RMS radii and magnetic moments 
in Eqs.~(29)-(32) of the paper are replaced by the following values,
\ben
&&\begin{array}{cc}
\sqrt{\la r^2_{E}\ra}=0.785(17)(21)\,[{\rm fm}]& \textrm{(isovector)}\\
\sqrt{\la r^2_{E}\ra}=0.766(14)(32)\,[{\rm fm}]& \textrm{(proton)}\\
\la r^2_{E}\ra=\,-0.047(20)(13)[{\rm fm^2}]& \textrm{(neutron)}\\
\end{array}
,\label{Eq_new:final_rns_ge}\\
&&\begin{array}{cc}
\sqrt{\la r^2_{M}\ra}=0.707(36)(311)\,[{\rm fm}]& \textrm{(isovector)}\\
\sqrt{\la r^2_{M}\ra}=0.710(39)(211)\,[{\rm fm}]& \textrm{(proton)}\\
\sqrt{\la r^2_{M}\ra}=0.758(33)(286)\,[{\rm fm}]& \textrm{(neutron)}\\
\end{array}
,\label{Eq_new:final_rns_gm}\\
&&\begin{array}{cc}
  \mu_v = 4.409(136)(333)& \textrm{(isovector)}\\
  \mu_p = 2.743(92)(148)& \textrm{(proton)}\\
  \mu_n = -1.697(63)(147)& \textrm{(neutron)}\\
\end{array}
,\label{Eq_new:final_mm}\\
&&
\sqrt{\la r^2_{A}\ra}=  0.532(28)(72)\,[{\rm fm}].
\een 
The corresponding figures and tables related to the corrected form factors 
are shown below.

\begin{figure*}[ht!]
  \includegraphics[width=8.0cm,keepaspectratio,clip]{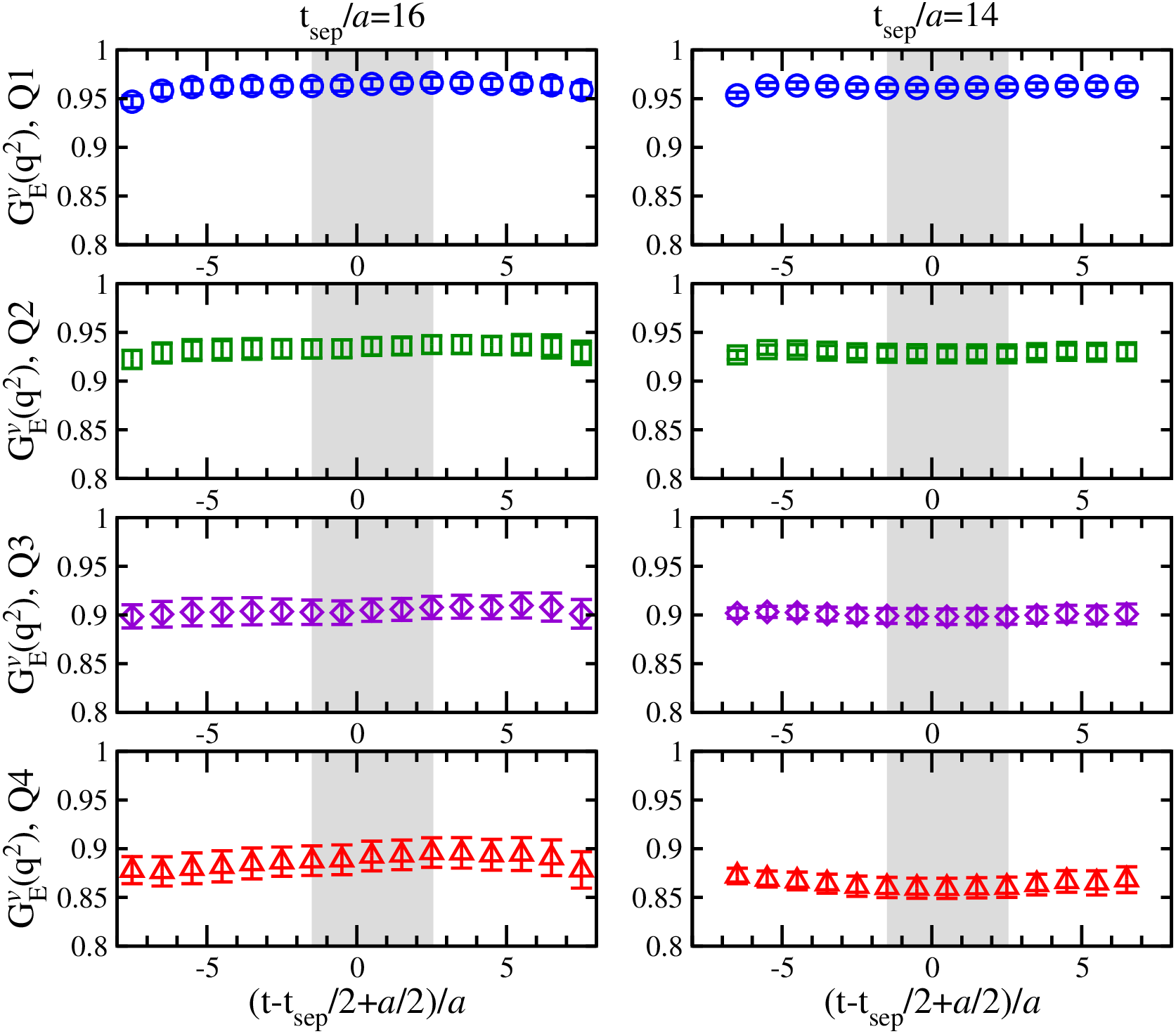}
  \hspace{3mm}
  \includegraphics[width=8.0cm,keepaspectratio,clip]{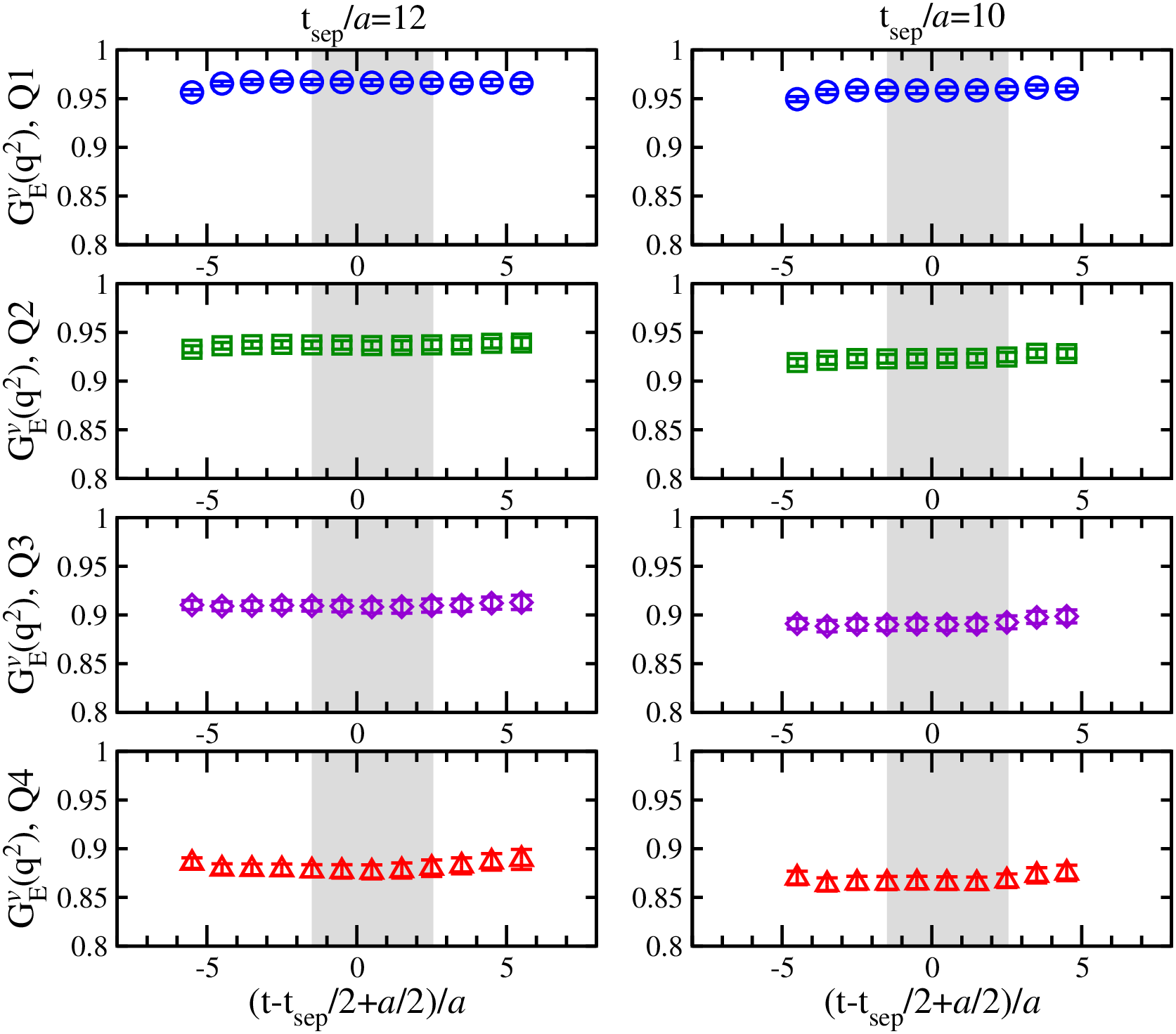}
\caption{Isovector electric form factor $G_E^v(q^2)$, which is extracted from the ratios of three- to two-point functions of Eq.~(20) in the paper, for $t_{\rm sep}/a=10,12,14,16$ with four lowest nonzero momentum transfers. Gray-shaded area denotes the fit range in each panel. This figure replaces Fig.~2 in the paper.}
\label{fig_new:ratio_ge}
\end{figure*}

\begin{figure}[ht!]
  \includegraphics[width=8.0cm,keepaspectratio,clip]{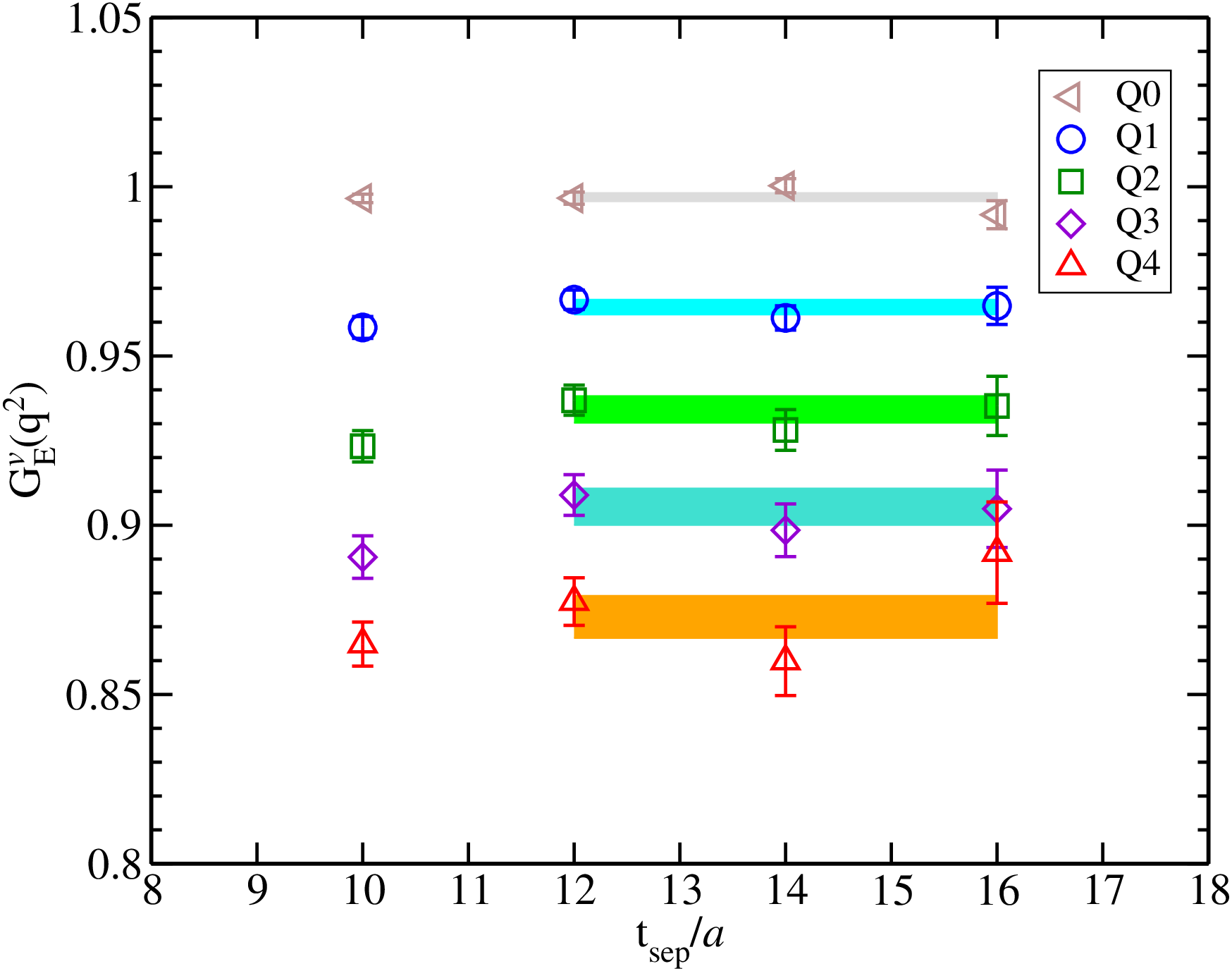}
\caption{$t_{\rm sep}$ dependence of the isovector electric form factor $G_E^v(q^2)$ with five lowest momentum transfers. Horizontal band represents the fit result of $G_E^v(q^2)$ at $t_{\rm sep}/a=12,14,16$ for each $q^2$.
This figure replaces Fig.~3 in the paper.}
\label{fig_new:ge_ts}
\end{figure}

\begin{table*}[ht!]
\begin{ruledtabular}
\caption{
  Results for the electric RMS charge radius $\sqrt{\la r^2_{E}\ra}$ in the isovector, proton and neutron channels. In the row of ``This work'' we present our best estimates, where the first error is statistical and the second one is systematic as explained in the text. Results for the proton and neutron are obtained without the disconnected diagram. Our previous work was performed on a $96^4$ lattice at $m_\pi = 146$ MeV in Ref.~\cite{Ishikawa:2018rew}, where only the statistical errors are presented. This table replaces Table~III in the paper.
\label{tab_new:re}}
\begin{tabular}{ccccccccccc}
  & & & \multicolumn{2}{c}{Isovector} & \multicolumn{2}{c}{Proton} & \multicolumn{2}{c}{Neutron}\\
  \hline
  Fit type & $q^2_{\rm cut}$ [GeV$^2$] & $t_{\rm sep}/a$ & $\sqrt{\la r^2_{E}\ra}$ [fm] & $\chi^2$/dof & $\sqrt{\la r^2_{E}\ra}$ [fm] & $\chi^2$/dof & $\la r^2_{E}\ra$ [fm$^2$] & $\chi^2$/dof\cr
  \hline
  \multirow{2}{*}{Linear} & \multirow{2}{*}{0.013} & $\{12,14,16\}$&  0.764(26)  &  $-$&  0.765(22)  &  $-$&  $-$&  $-$\\
 & & $\{14,16\}$&  0.806(35)  &  $-$&  0.791(28)  &  $-$ &  $-$&  $-$ \\
  \hline
  \multirow{2}{*}{Dipole} & \multirow{2}{*}{0.102} & $\{12,14,16\}$&  0.785(17)  &  1.2(8)&  0.766(14)  &  0.8(7)&  $-$&  $-$\\
  & & $\{14,16\}$&  0.806(26)  &  0.6(6)&  0.788(20)  &  0.6(6)&  $-$&  $-$\\
  \hline
  \multirow{2}{*}{Quadrature} & \multirow{2}{*}{0.102} & $\{12,14,16\}$&  0.785(19)  &  1.0(8)&  0.773(16)  &  0.4(5)&  $-$0.038(18) & 2.2(1.9)\\
  & & $\{14,16\}$&  0.783(30)  &  0.7(7)&  0.780(19)  &  0.8(8)&  $-$0.029(23)& 2.6(2.2)\\
  \hline
  z-exp & \multirow{2}{*}{0.102} & $\{12,14,16\}$&  0.776(28)  &  1.2(9)&  0.781(24)  &  0.6(7)&  $-$0.047(20) & 1.8(1.7)\\
  ($k_{\rm max}=3$)& & $\{14,16\}$&  0.796(37)  &  0.8(8)&  0.798(25)  &  0.5(6)& $-$0.034(25) & 2.4(2.1)\\
  \hline
  \multicolumn{2}{ c }{This work} & & 0.785(17)(21) & & 0.766(14)(32) & & $-$0.047(20)(13)\\
  \hline\hline
  \multicolumn{2}{ c }{PACS'18~\cite{Ishikawa:2018rew}} \cr
  \hline
  Dipole    & 0.215 & 15 & 0.822(63) & $-$ & $-$ & $-$ & $-$ & $-$ \cr
  \hline
  Quadratic & 0.215 & 15 & 0.851(62) & $-$ & $-$ & $-$ & $-$ & $-$  \cr
  \hline
  z-exp     & \multirow{2}{*}{0.215} & \multirow{2}{*}{15} & \multirow{2}{*}{0.914(101)} & \multirow{2}{*}{$-$} & \multirow{2}{*}{$-$} & \multirow{2}{*}{$-$}  & \multirow{2}{*}{$-$} & \multirow{2}{*}{$-$} \cr
  ($k_{\rm max}=3$)\\
  \hline\hline
  \multicolumn{2}{ c }{Experimental value}  &  &  \cr
  \hline
  \multicolumn{2}{ c }{$ep$ scattering} & & 0.939(6) & & 0.875(6) & & $-$0.1161(22) \cr
  \multicolumn{2}{ c }{$\mu$H atom} & & 0.907(1) & & 0.8409(4) & & $-$\cr
\end{tabular}
\end{ruledtabular}
\end{table*}

\begin{figure*}[ht!]
  \includegraphics[width=8.0cm,keepaspectratio,clip]{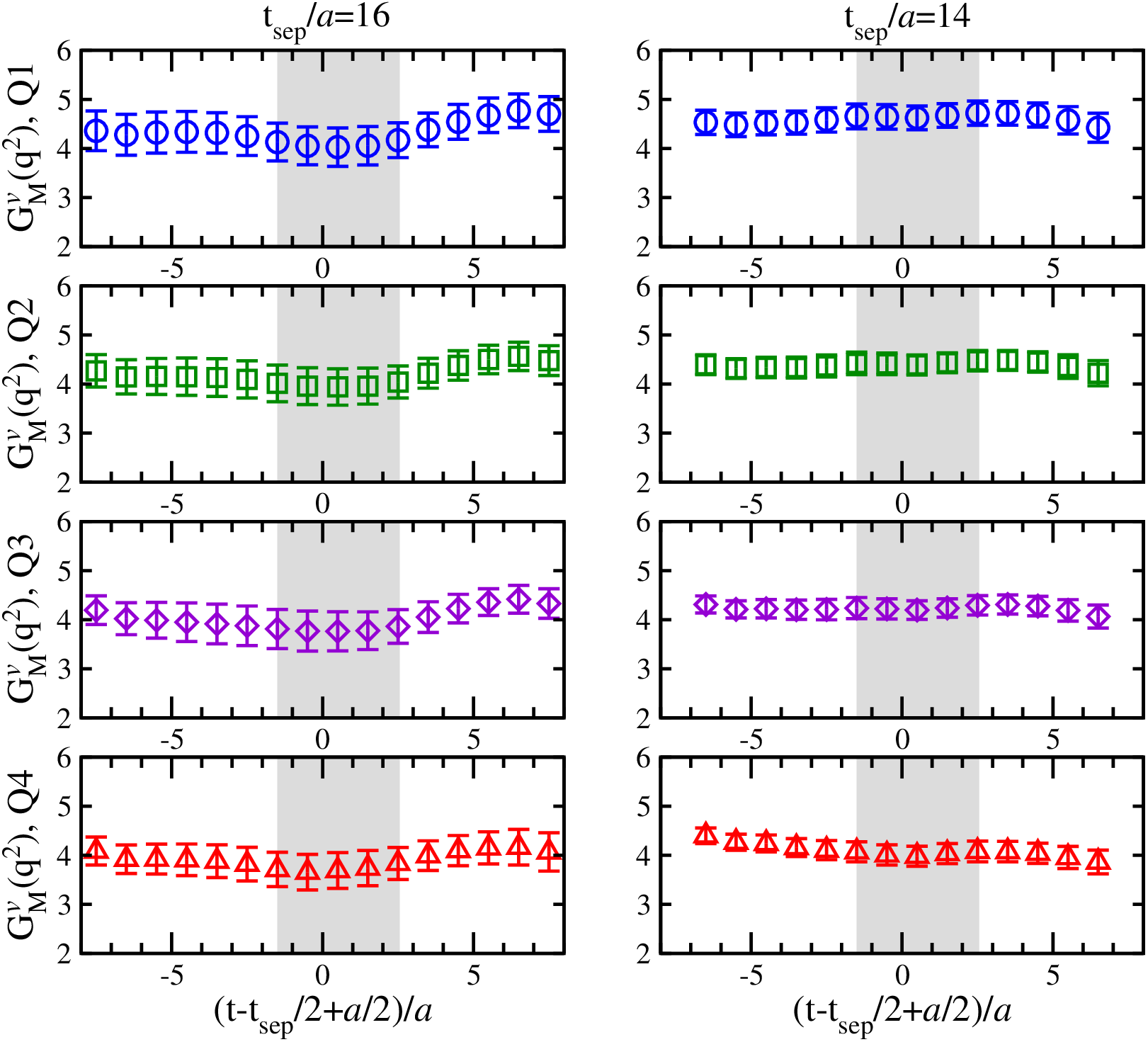}
  \hspace{3mm}
  \includegraphics[width=8.0cm,keepaspectratio,clip]{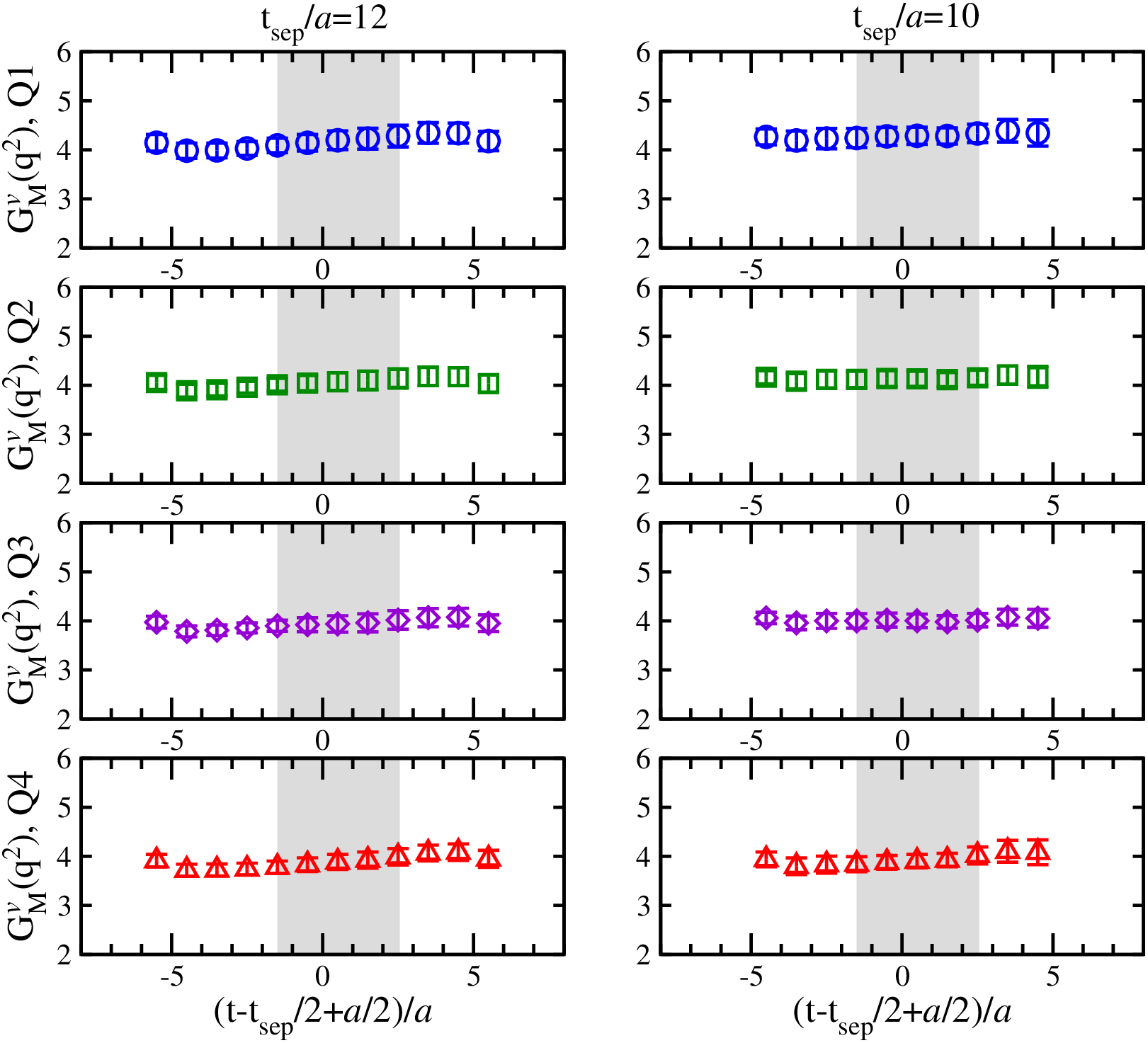}
\caption{Same as Fig.~\ref{fig_new:ratio_ge} for the ratio of Eq.~(21) in the paper to extract the isovector magnetic form factor $G_M^v(q^2)$. This figure replaces Fig.~7 in the paper.}
\label{fig_new:ratio_gm}
\end{figure*}

\begin{figure}[ht!]
  \includegraphics[width=8.0cm,keepaspectratio,clip]{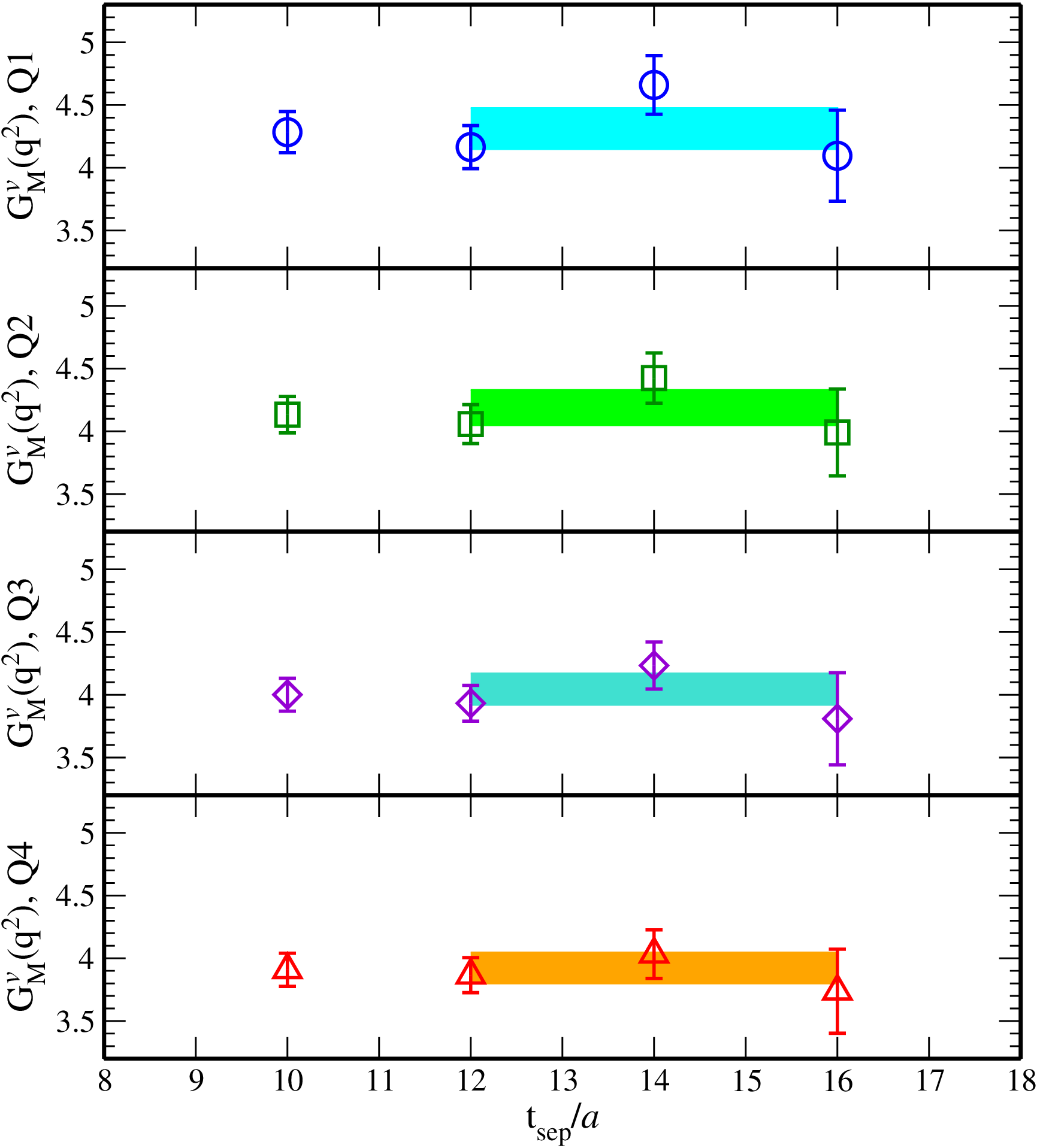}
  \caption{Same as Fig.~\ref{fig_new:ge_ts} for the isovector magnetic form factor $G_M^v(q^2)$ with four lowest nonzero momentum transfers. This figure replaces Fig.~8 in the paper.}
\label{fig_new:gm_ts}
\end{figure}

\begin{figure}[ht!]
  \includegraphics[width=8.0cm,keepaspectratio,clip]{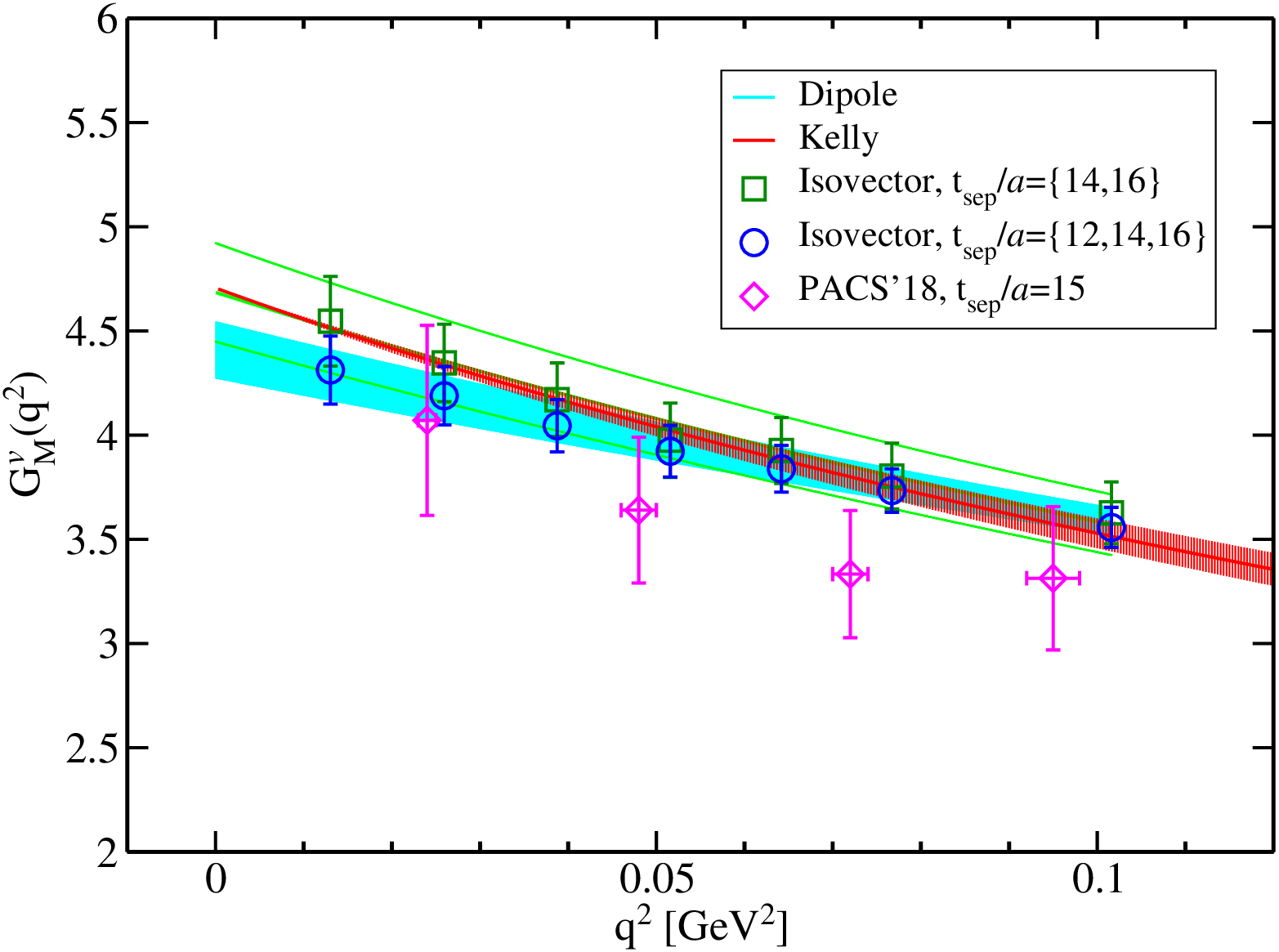}
\caption{Same as Fig.~\ref{fig_new:ge_q2} for the isovector magnetic form factor $G_M^v(q^2)$.
This figure replaces Fig.~9 in the paper.}
\label{fig_new:gm_q2}
\end{figure}

\begin{figure*}[ht!]
  \includegraphics[width=8.0cm,keepaspectratio,clip]{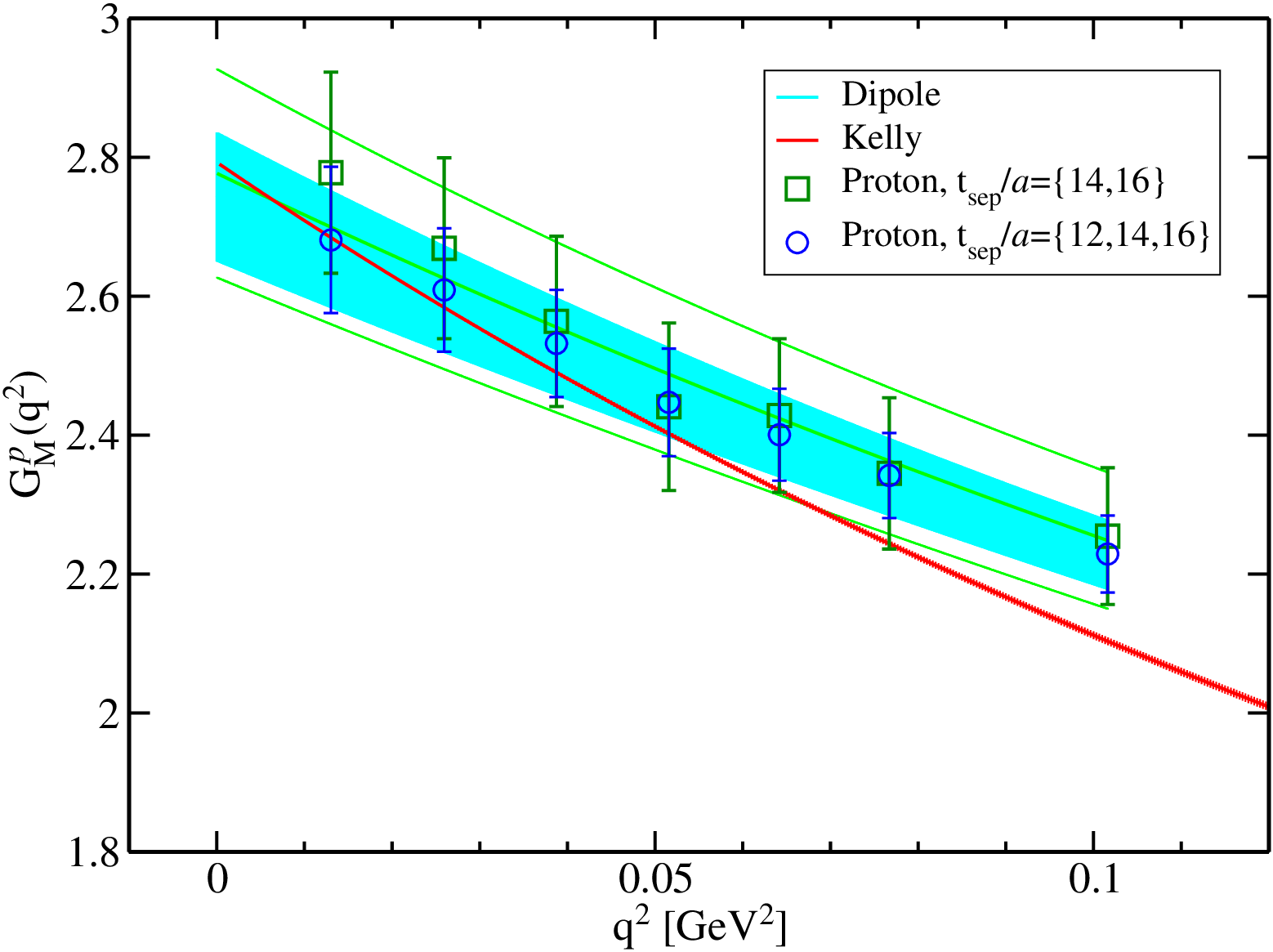}
  \hspace{3mm}
  \includegraphics[width=8.0cm,keepaspectratio,clip]{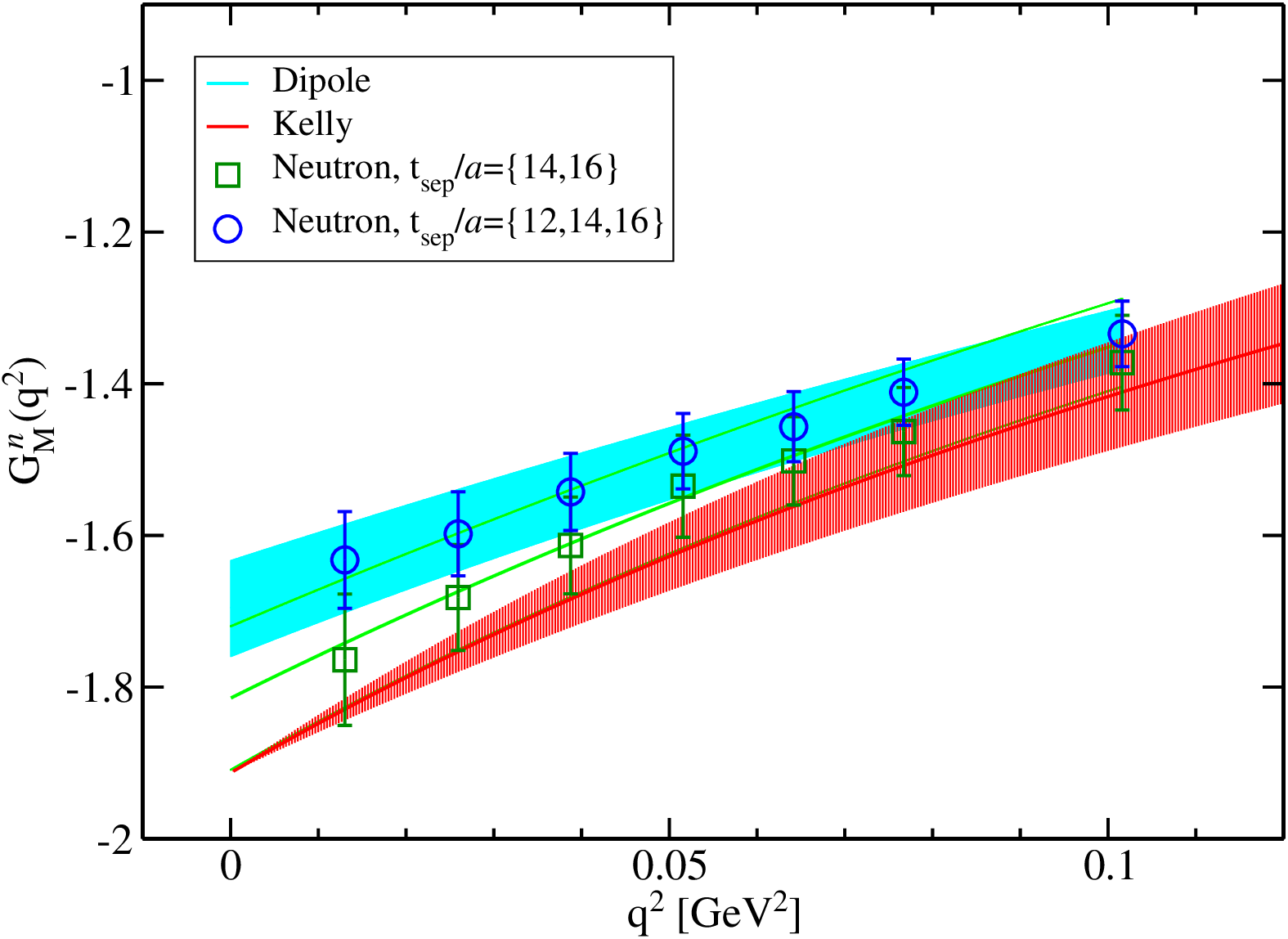}
\caption{Same as Fig.~\ref{fig_new:ge_q2} for the proton (left) and neutron (right) magnetic form factor $G_M(q^2)$. The
results are obtained without the disconnected diagram. This figure replaces Fig.~10 in the paper.}
\label{fig_new:gm_q2_p_n}
\end{figure*}

\begin{figure}[ht!]
  \includegraphics[width=7.0cm,keepaspectratio,clip]{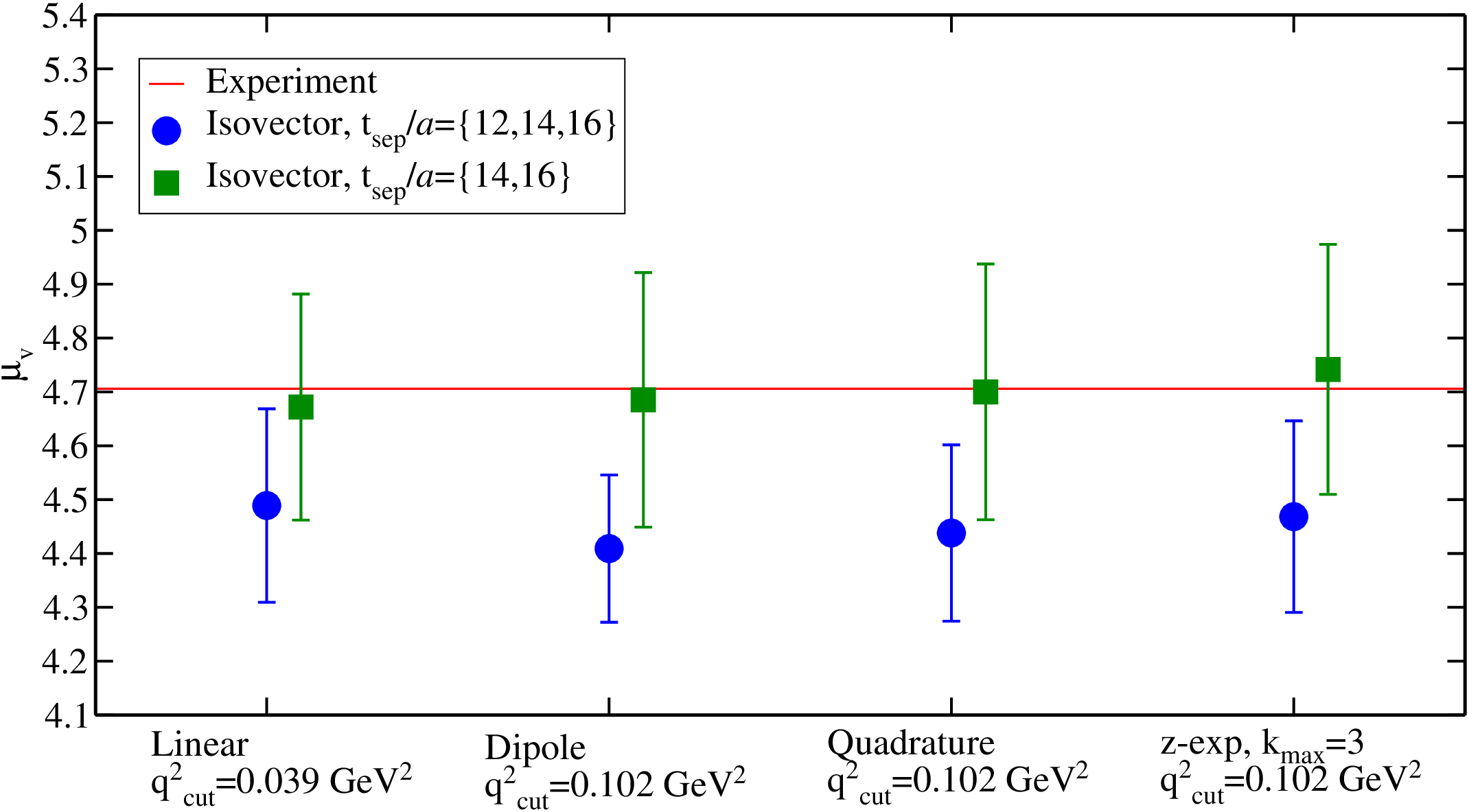}
  \vskip 3mm
  \includegraphics[width=7.0cm,keepaspectratio,clip]{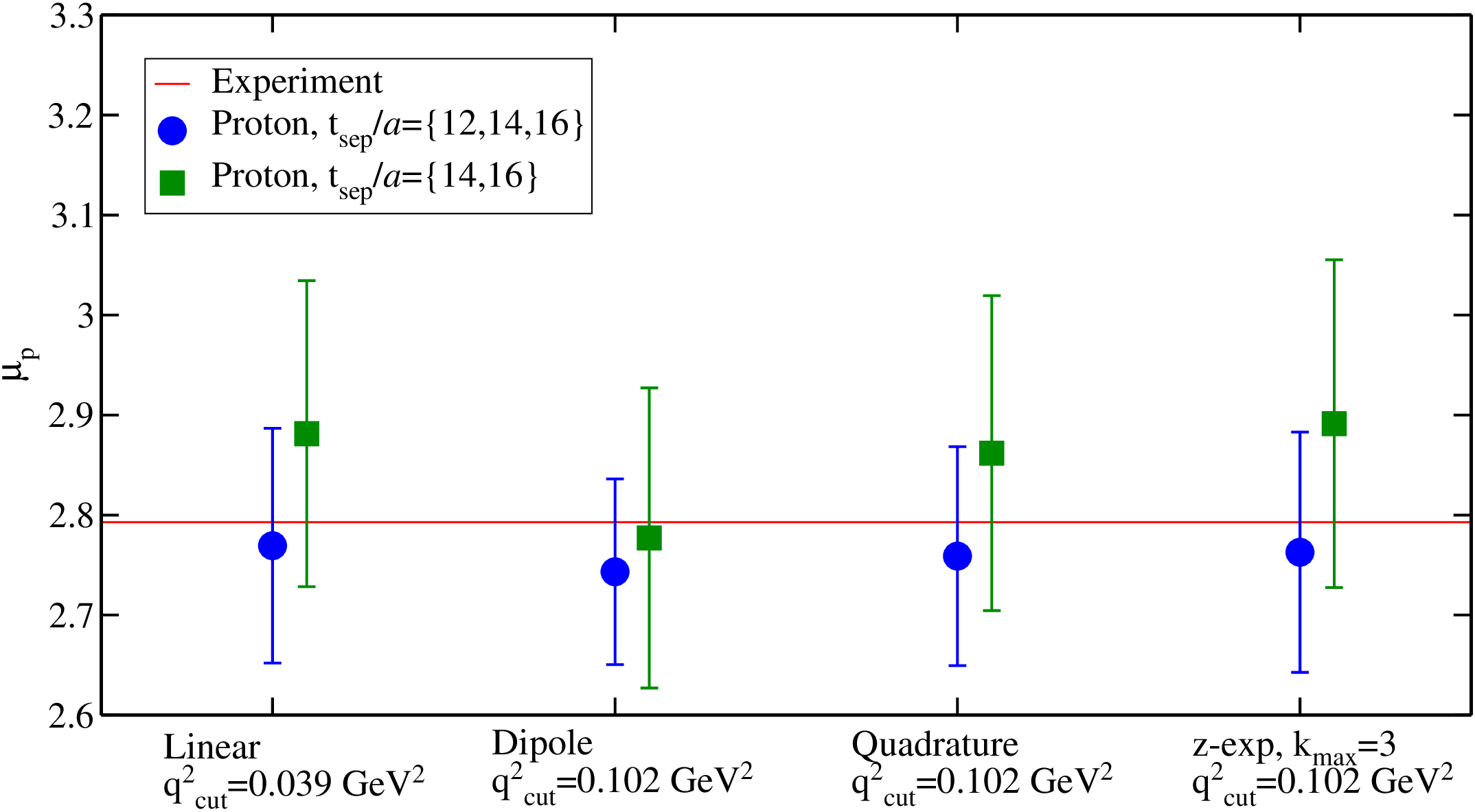}
  \vskip 3mm
  \includegraphics[width=7.0cm,keepaspectratio,clip]{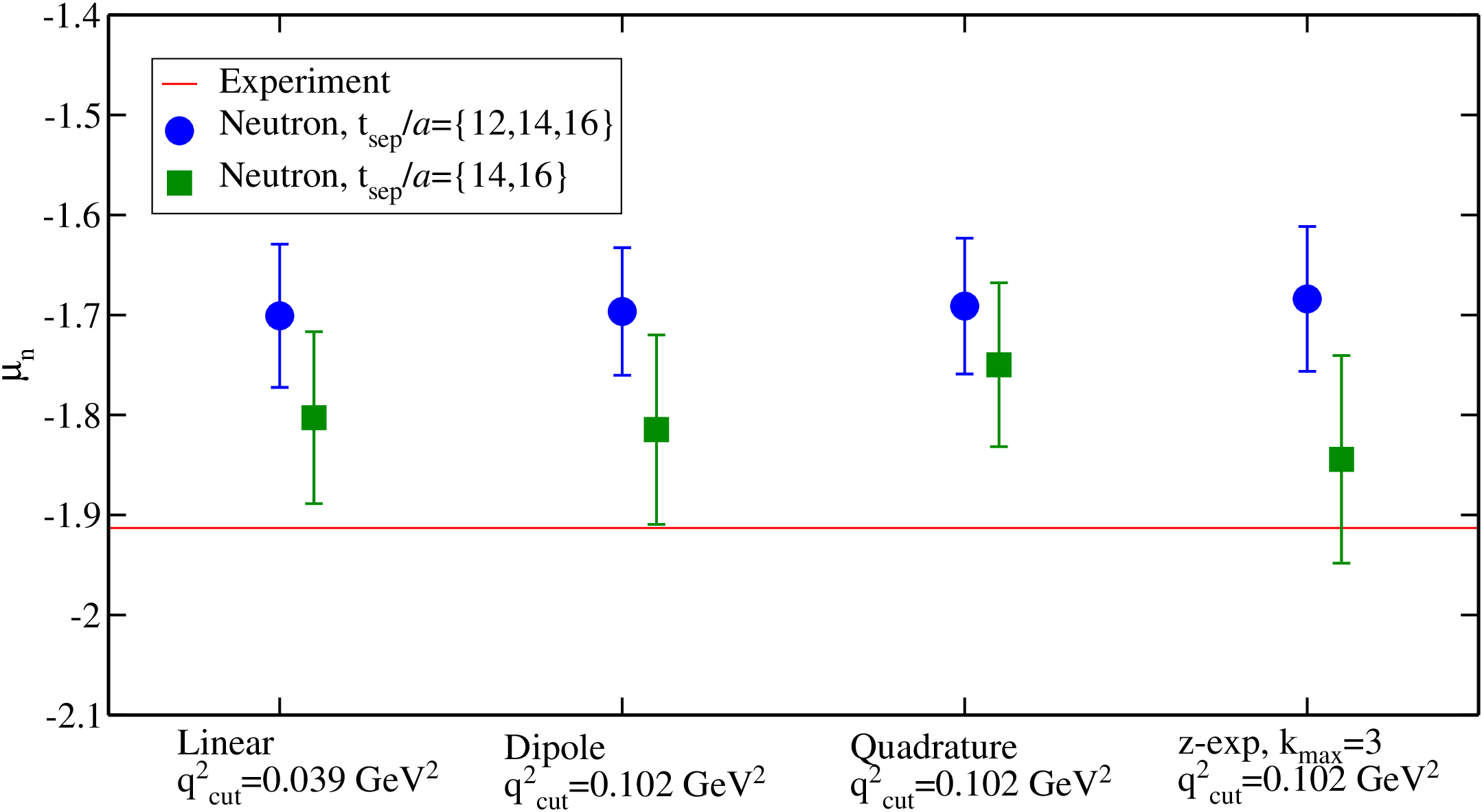}  
\caption{Magnetic moment $\mu$ for the isovector (top), proton (middle) and neutron (bottom) channels obtained by the fitting with the linear, dipole, quadratic forms and the $z$-expansion method for the combined data. Horizontal bands represent the experimental results. Two types of symbols denote the results with two choices of the combined $t_{\rm sep}$ ranges. Results for the proton and neutron channels are obtained without the disconnected diagram. This figure replaces Fig.~11 in the paper.}
\label{fig_new:mm_fit}
\end{figure}

\begin{figure}[ht!]
  \includegraphics[width=7.0cm,keepaspectratio,clip]{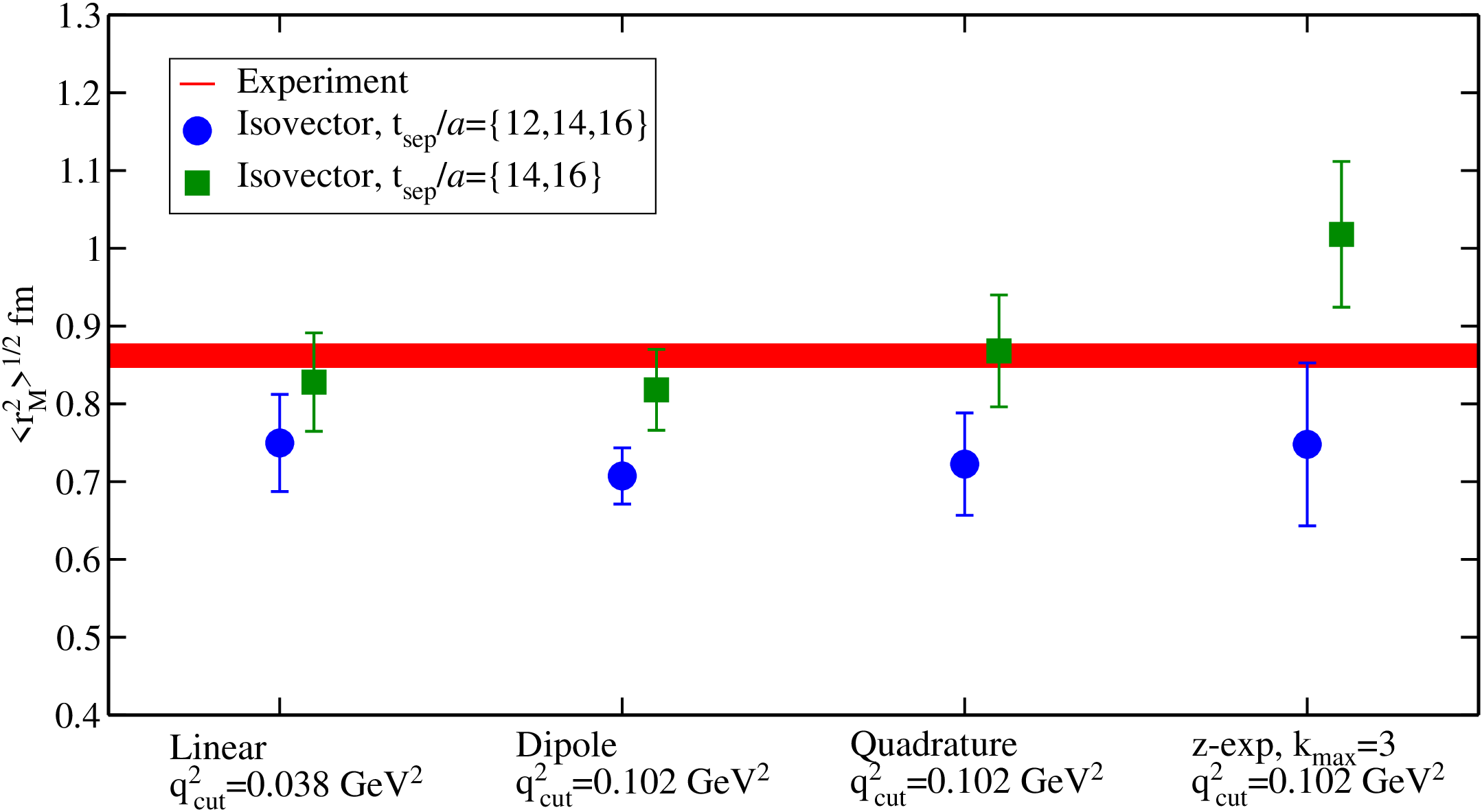}
  \vskip 3mm
  \includegraphics[width=7.0cm,keepaspectratio,clip]{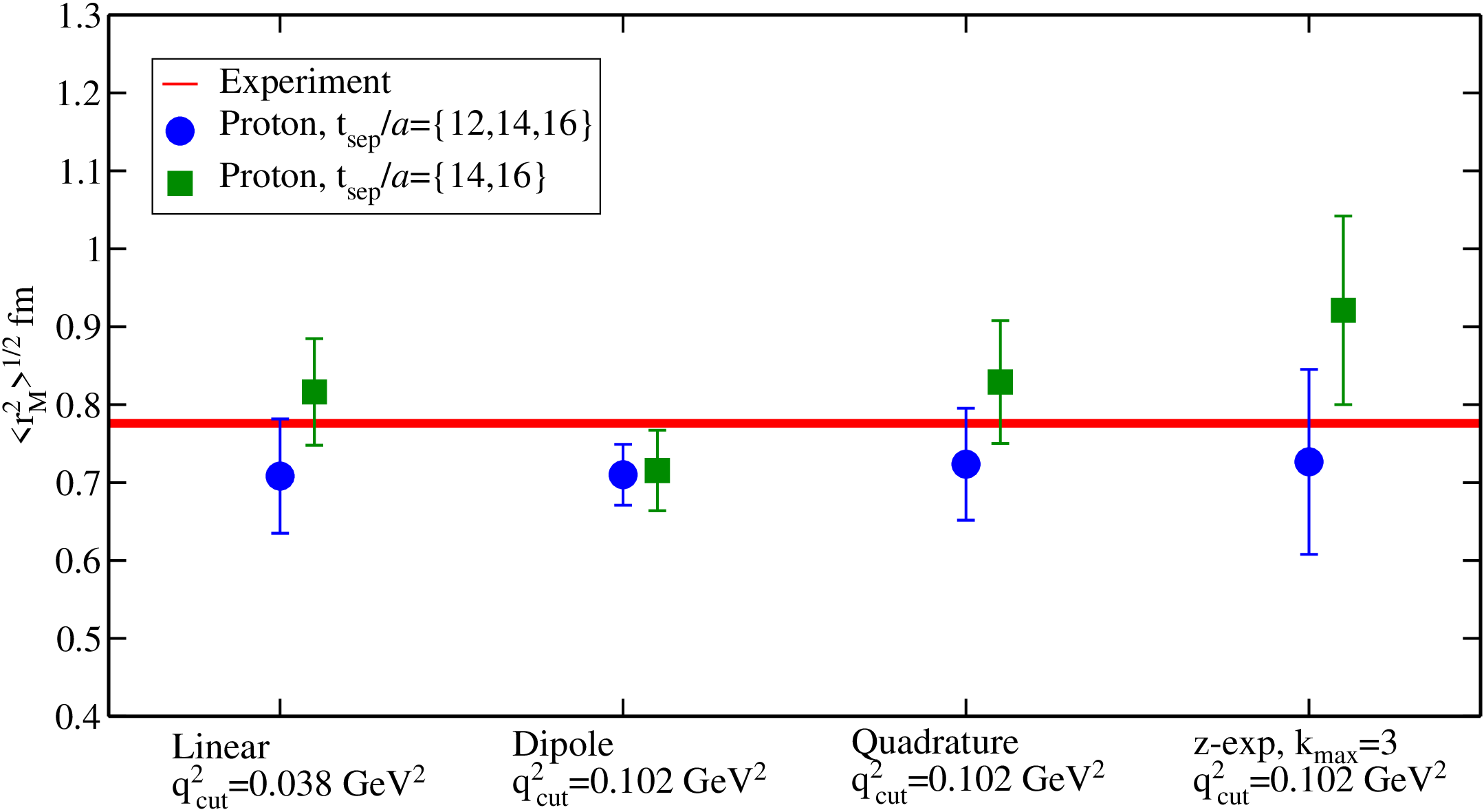}
    \vskip 3mm
  \includegraphics[width=7.0cm,keepaspectratio,clip]{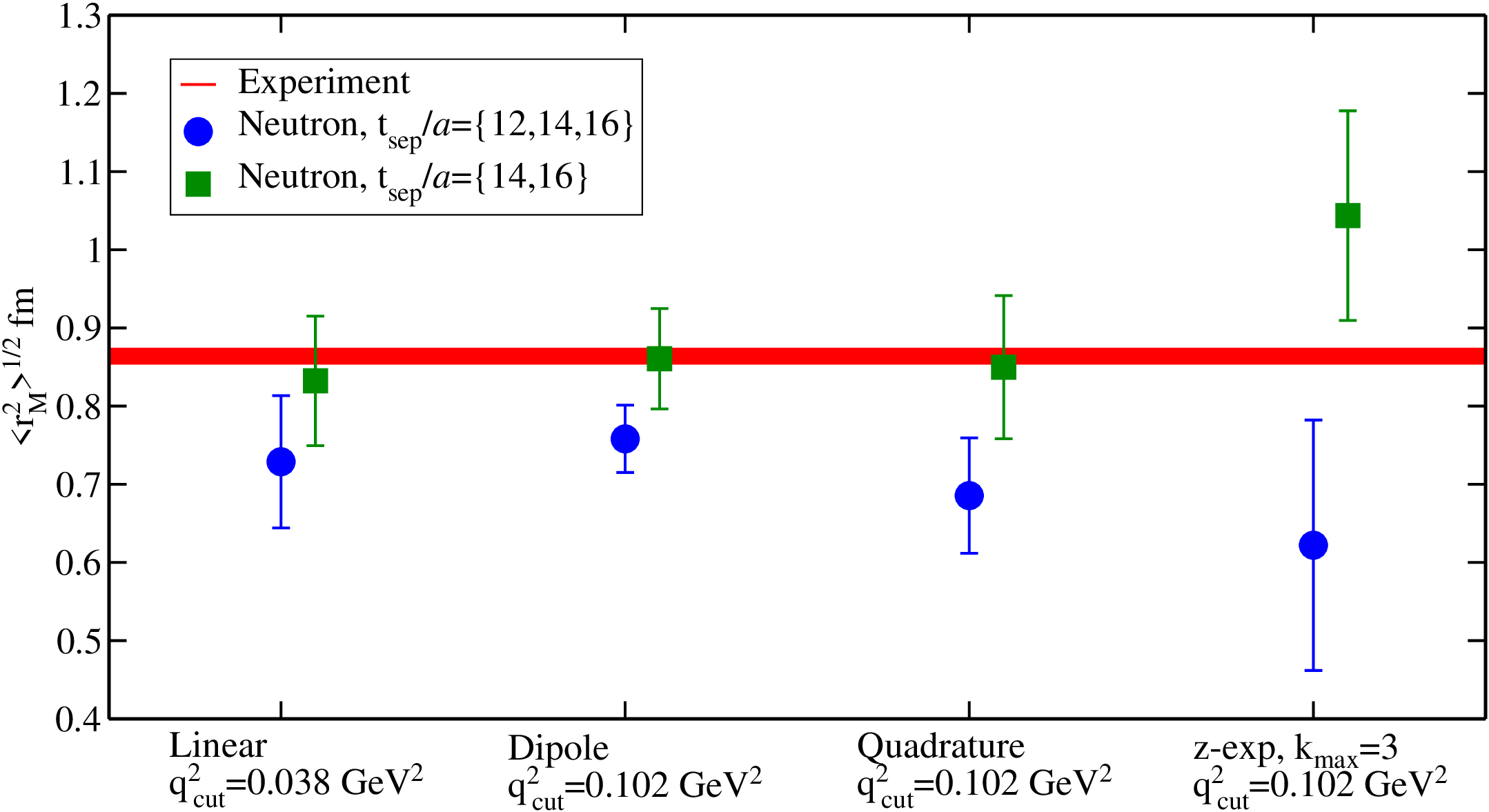}
\caption{Same as Fig.~\ref{fig_new:mm_fit} for the magnetic RMS radius $\sqrt{\la r^2_M\ra}$. This figure replaces Fig.~12 in the paper.}
\label{fig_new:rm_fit}
\end{figure}

\begin{table*}[ht!]
\begin{ruledtabular}
\caption{
Results for the magnetic moments $\mu$ and magnetic RMS radius
$\sqrt{\la r^2_{M}\ra}$ for the isovector, proton and neutron channels. In the row of ``This work'' we present our best estimates, where the first error is statistical and the second one is systematic as explained in the text. Results for the proton and neutron are obtained without the disconnected diagram. Previous work was performed on a $96^4$ lattice at $m_\pi = 146$ MeV in Ref.~\cite{Ishikawa:2018rew}, where only the statistical errors are presented. This table replaces Table~IV in the paper.
\label{tab_new:murm}}
\begin{tabular}{ccccccccccccc}
  & & & \multicolumn{3}{c}{Isovector}\\
  \hline
  Fit type & $q^2_{\rm cut}$ [GeV$^2$] & $t_{\rm sep}/a$ & $\mu_v$ & $\sqrt{\la r^2_{M}\ra}$ [fm]& $\chi^2$/dof \\
  \hline
  \multirow{2}{*}{Linear} & \multirow{2}{*}{0.039} & $\{12,14,16\}$&  4.489(179)  &  0.750(62)  &  0.8(1.9)\\
  & & $\{14,16\}$&  4.672(209)  &  0.828(63)  &  0.4(1.2)\\
  \hline  
  \multirow{2}{*}{Dipole} & \multirow{2}{*}{0.102} & $\{12,14,16\}$&  4.409(136)  &  0.707(36)  &  0.8(7)\\
  & & $\{14,16\}$&  4.685(236)  &  0.818(51)  &  3.6(3.8)\\
  \hline
  \multirow{2}{*}{Quadrature} & \multirow{2}{*}{0.102} & $\{12,14,16\}$&  4.438(163)  &  0.723(65)  &  1.0(9)\\
  & & $\{14,16\}$&  4.700(237)  &  0.868(71)  &  0.9(1.8)\\
  \hline
  z-exp & \multirow{2}{*}{0.102} & $\{12,14,16\}$&  4.468(177)  &  0.748(104)  &  0.9(9)\\
  ($k_{\rm max}=3$)& & $\{14,16\}$&  4.742(232)  &  1.018(93)  &  0.9(9)\\
  \hline
  \multicolumn{2}{ c }{This work} & & 4.409(136)(333) & 0.707(36)(311) & \\
  \hline\hline
  \multicolumn{2}{ c }{PACS'18~\cite{Ishikawa:2018rew}} \cr
  \hline
  Dipole    & 0.215 & 15 & 3.96(46) & 0.656(133) & $-$ &  \cr
  \hline
  Quadratic & 0.215 & 15 & 4.24(52) & 0.852(130) & $-$ \cr
  \hline
  z-exp     & \multirow{2}{*}{0.215} & \multirow{2}{*}{15} & \multirow{2}{*}{4.86(82)} & \multirow{2}{*}{1.495(437)} & $-$ \cr
  ($k_{\rm max}=3$)\\
  \hline\hline
  \multicolumn{2}{ c }{Experimental value} \cr
  \hline
  & & & 4.70589 & 0.862(14) \\
  \hline\hline\\
 \hline\hline
  & & & \multicolumn{3}{c}{Proton} & \multicolumn{3}{c}{Neutron}\\
  \hline
  Fit type & $q^2_{\rm cut}$ [GeV$^2$] & $t_{\rm sep}/a$ & $\mu_p$ & $\sqrt{\la r^2_{M}\ra}$ [fm]& $\chi^2$/dof & $\mu_n$ & $\sqrt{\la r^2_{M}\ra}$ [fm]& $\chi^2$/dof \\
  \hline
  \multirow{2}{*}{Linear} & \multirow{2}{*}{0.039} & $\{12,14,16\}$&  2.769(117)  &  0.708(73)  &  0.2(9)&  $-$1.701(71)  &  0.729(84)  &  2.4(3.0)\\
  & & $\{14,16\}$&  2.881(153)  &  0.816(68)  &  0.1(4)&  -1.803(85)  &  0.832(82)  &  0.5(1.6)\\
  \hline
  \multirow{2}{*}{Dipole} & \multirow{2}{*}{0.102} & $\{12,14,16\}$&  2.743(92)  &  0.710(39)  &  0.2(4)&  $-$1.697(63)  &  0.758(43)  &  0.3(6)\\
  & & $\{14,16\}$&  2.777(150)  &  0.715(51)  &  1.6(1.7)&  -1.815(94)  &  0.861(64)  &  1.3(2.5)\\
  \hline
  \multirow{2}{*}{Quadrature} & \multirow{2}{*}{0.102} & $\{12,14,16\}$&  2.759(109)  &  0.724(71)  &  0.3(4)&  $-$1.691(67)  &  0.686(73)  &  1.0(9)\\
  & & $\{14,16\}$&  2.862(157)  &  0.829(78)  &  0.8(1.7)&  -1.750(81)  &  0.850(91)  &  1.3(1.1)\\
  \hline
  z-exp & \multirow{2}{*}{0.102} & $\{12,14,16\}$&  2.763(120)  &  0.727(118)  &  0.3(4)&  $-$1.684(72)  &  0.622(160)  &  0.9(8)\\
  ($k_{\rm max}=3$)& & $\{14,16\}$&  2.891(163)  &  0.921(121)  &  0.5(1.3)&  -1.844(103)  &  1.044(134)  &  0.7(8)\\
  \hline
  \multicolumn{2}{ c }{This work} & & 2.743(92)(148) & 0.710(39)(211) & & $-$1.697(63)(147) & 0.758(33)(286)\\
  \hline\hline
  \multicolumn{2}{ c }{Experimental value} \cr
  \hline
  & & & 2.79285 & 0.776(38) & & $-$1.91304 & 0.864(9)\cr  
\end{tabular}
\end{ruledtabular}
\end{table*}

\begin{figure*}[ht!]
  \includegraphics[width=8.0cm,keepaspectratio,clip]{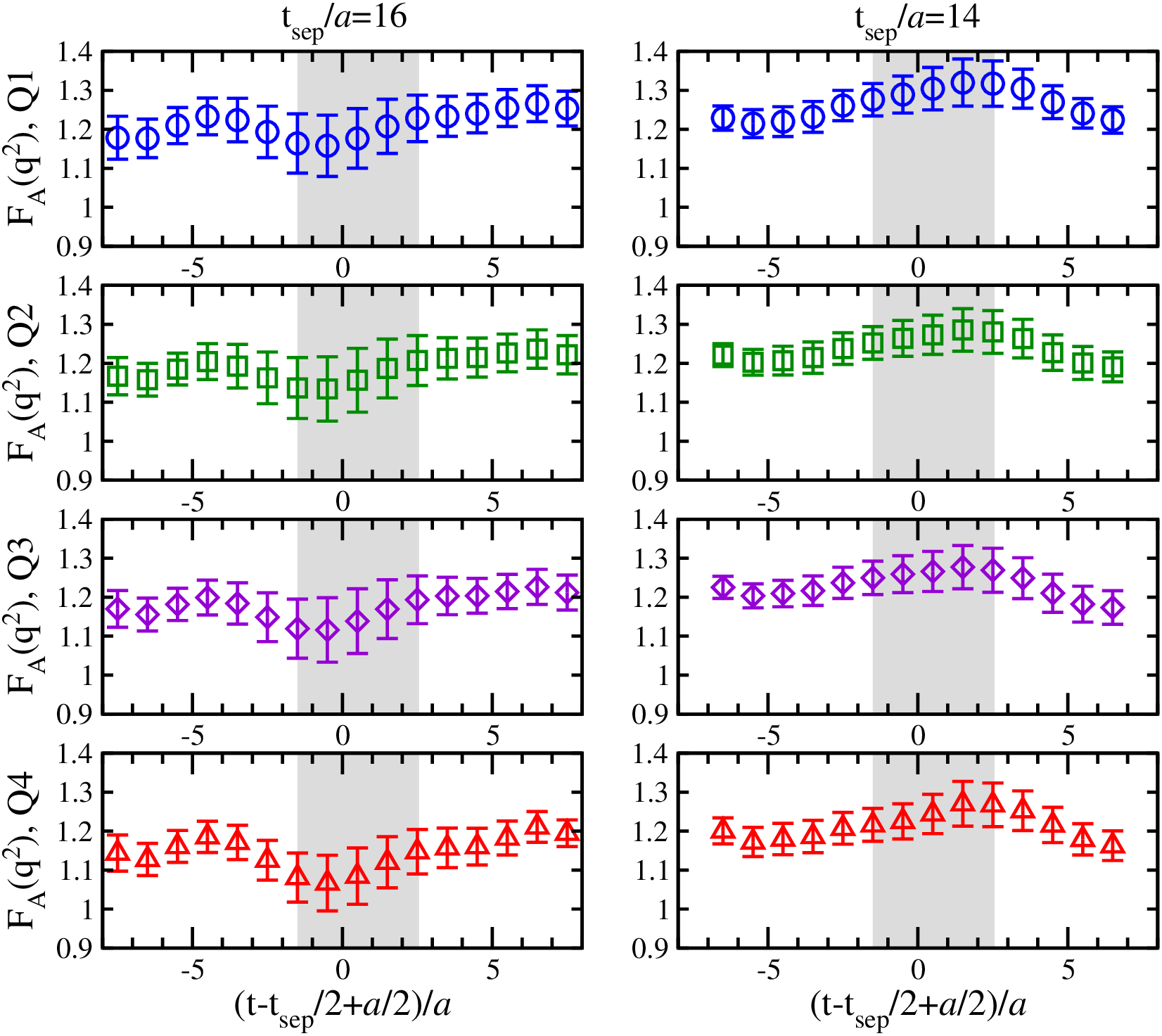}
  \hspace{3mm}
  \includegraphics[width=8.0cm,keepaspectratio,clip]{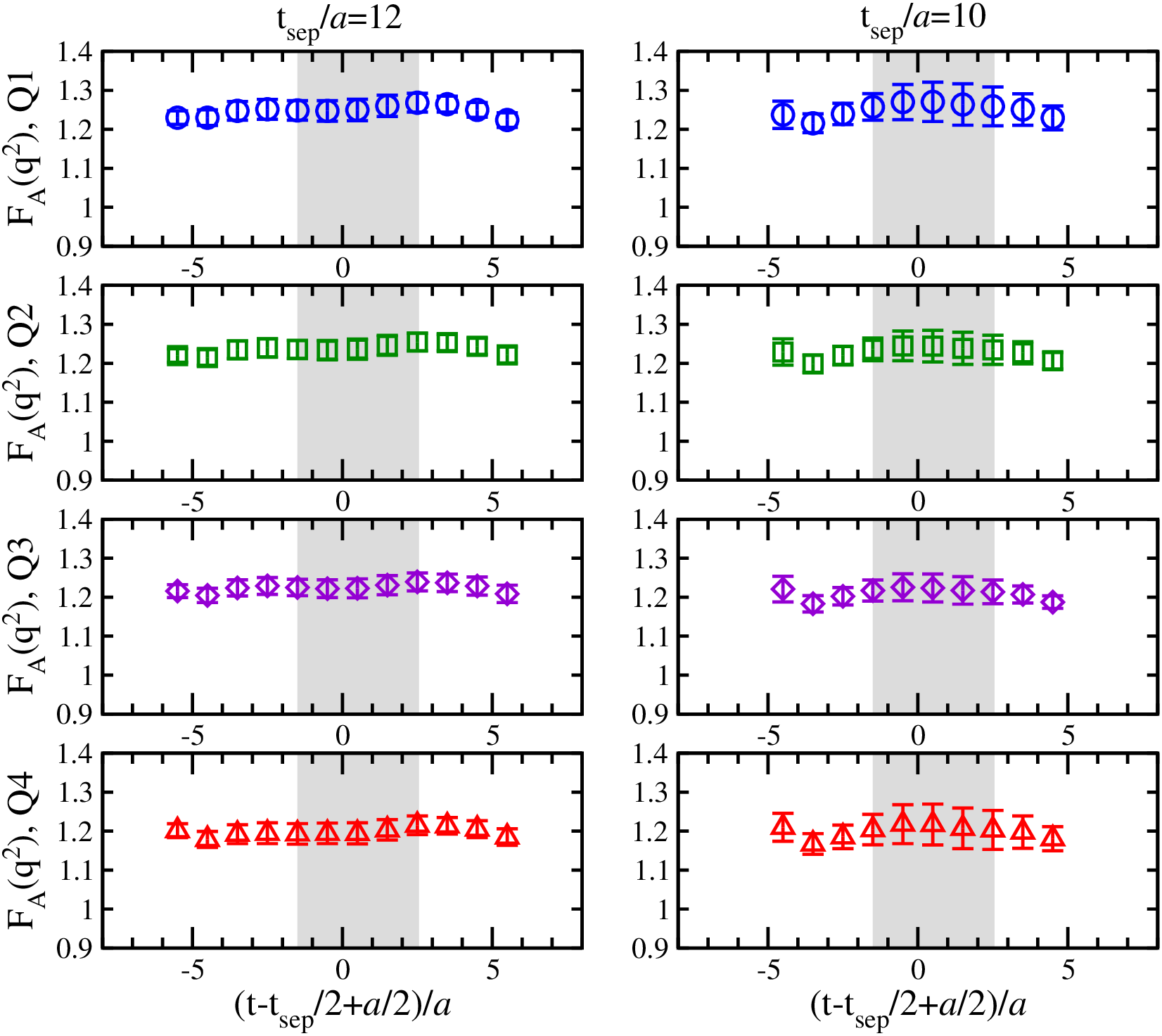}
\caption{Same as Fig.~\ref{fig_new:ratio_ge} for the axial-vector form factor $F_A(q^2)$ extracted from the ratio of Eq.~(22) in the paper. This figure replaces Fig.~15 in the paper.}
\label{fig_new:ratio_fa}
\end{figure*}

\begin{figure}[ht!]
\includegraphics[width=8.0cm,keepaspectratio,clip]{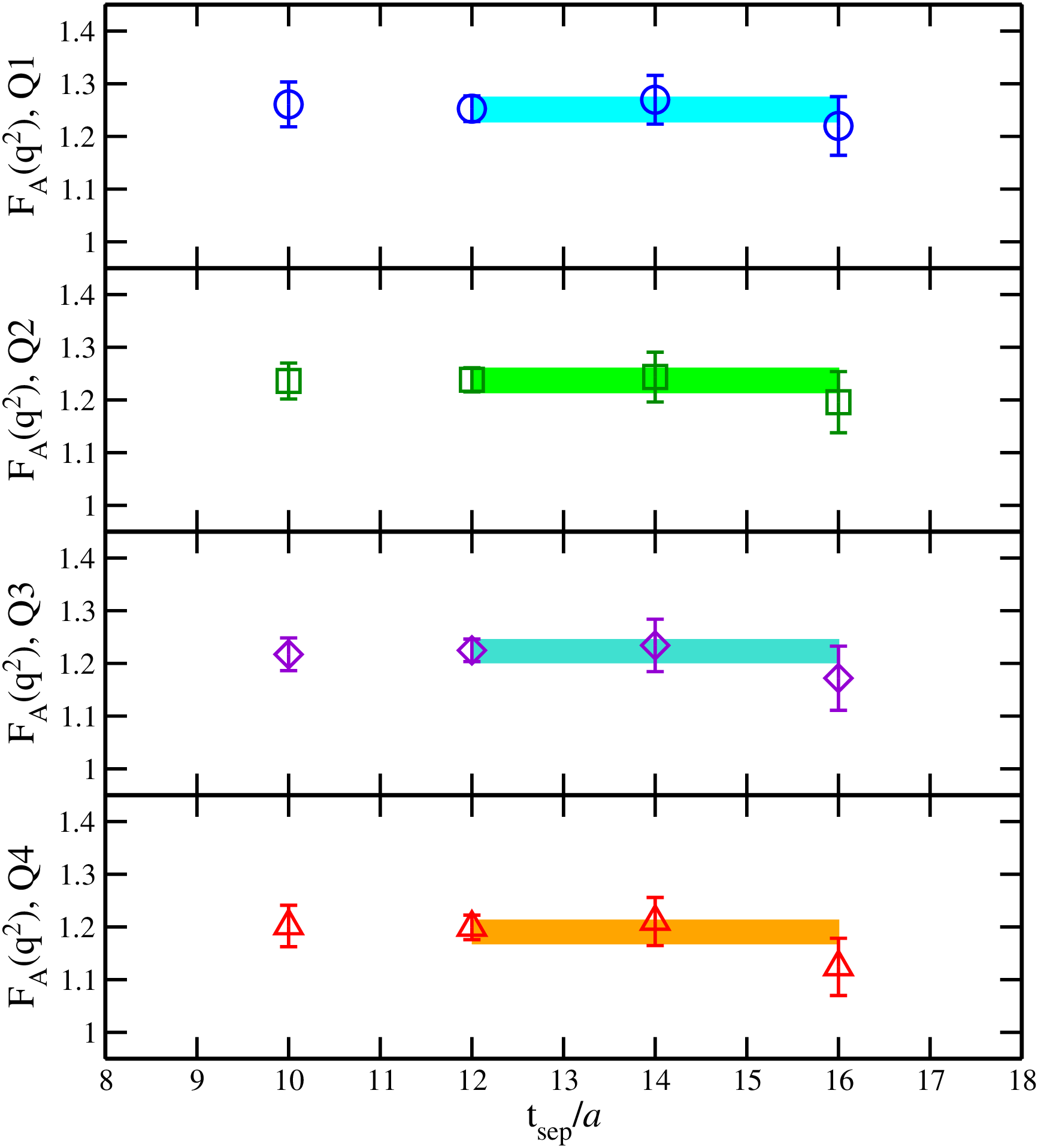}
\caption{Same as Fig.~\ref{fig_new:ge_ts} for $F_A(q^2)$ with four lowest nonzero momentum transfers.
This figure replaces Fig.~16 in the paper.}
\label{fig_new:fa_ts}
\end{figure}

\begin{table*}[ht!]
\begin{ruledtabular}
\caption{Results for the axial-vector coupling $g_A=F_A(0)$ and axial-vector RMS radius $\sqrt{\la r^2_A\ra}$.  In the row of ``This work'' we present our best estimates, where the first error is statistical and the second one is systematic as explained in the text. Results for the proton and neutron are obtained without the disconnected diagram. Previous work was performed on a $96^4$ lattice at $m_\pi = 146$ MeV in Ref.~\cite{Ishikawa:2018rew}, where only the statistical errors are presented. This table replaces Table~V in the paper.
\label{tab_new:gara}}
\begin{tabular}{ccccccc}
  Fit type & $q^2_{\rm cut}$ GeV$^2$ & $t_{\rm sep}/a$ & $F_A(0)$ & $\sqrt{\la r^2_A\ra}$ [fm]& $\chi^2$/dof   \cr
  \hline
  \multirow{2}{*}{linear} & 0.039 & $\{12,14,16\}$&  1.278(23)&  0.482(60)  &  1.1(1.5)\\
  & 0.052 & $\{14,16\}$&  1.262(29) &  0.584(54)  &  0.2(1.2)\\
  \hline  
  \multirow{2}{*}{dipole} & 0.102 & $\{12,14,16\}$&  1.288(20)&  0.532(28)  &  1.2(8)\\
  & 0.077 & $\{14,16\}$&  1.256(28) &  0.589(49) & 0.2(7)\\
  \hline
  \multirow{2}{*}{quadrature} & 0.102 & $\{12,14,16\}$&  1.287(20)&  0.513(46)  &  1.4(1.0)\\
  & 0.077 & $\{14,16\}$&  1.258(28)&  0.603(83) & 0.4(1.5)\\
  \hline
  z-exp & 0.102 & $\{12,14,16\}$&  1.287(20)&  0.517(58)  &  1.5(1.1)\\
  ($k_{\rm max}=3$)& 0.077 & $\{14,16\}$&  1.257(28) &   0.604(94) & 0.4(1.4)\\
  \hline
  \multicolumn{2}{ c }{This work} & & & 0.532(28)(72)\\
  \hline\hline
  \multicolumn{2}{ c }{PACS'18~\cite{Ishikawa:2018rew}} \cr
  \hline
  Dipole    & 0.215 & 15 & $-$ & 0.40(12) & $-$ \cr
  \hline
  Quadratic & 0.215 & 15 & $-$ & 0.22(49) & $-$ \cr
  \hline
  z-exp     & \multirow{2}{*}{0.215} & \multirow{2}{*}{15} & \multirow{2}{*}{$-$} & \multirow{2}{*}{0.46(11)} & \multirow{2}{*}{$-$} \cr
  ($k_{\rm max}=3$)\\
  \hline\hline
  \multicolumn{2}{ c }{Experimental value}  & & & \cr
  \hline
  & & & & 0.67(1) &\cr
\end{tabular}
\end{ruledtabular}
\end{table*}

\begin{figure*}[ht!]
  \includegraphics[width=8.0cm,keepaspectratio,clip]{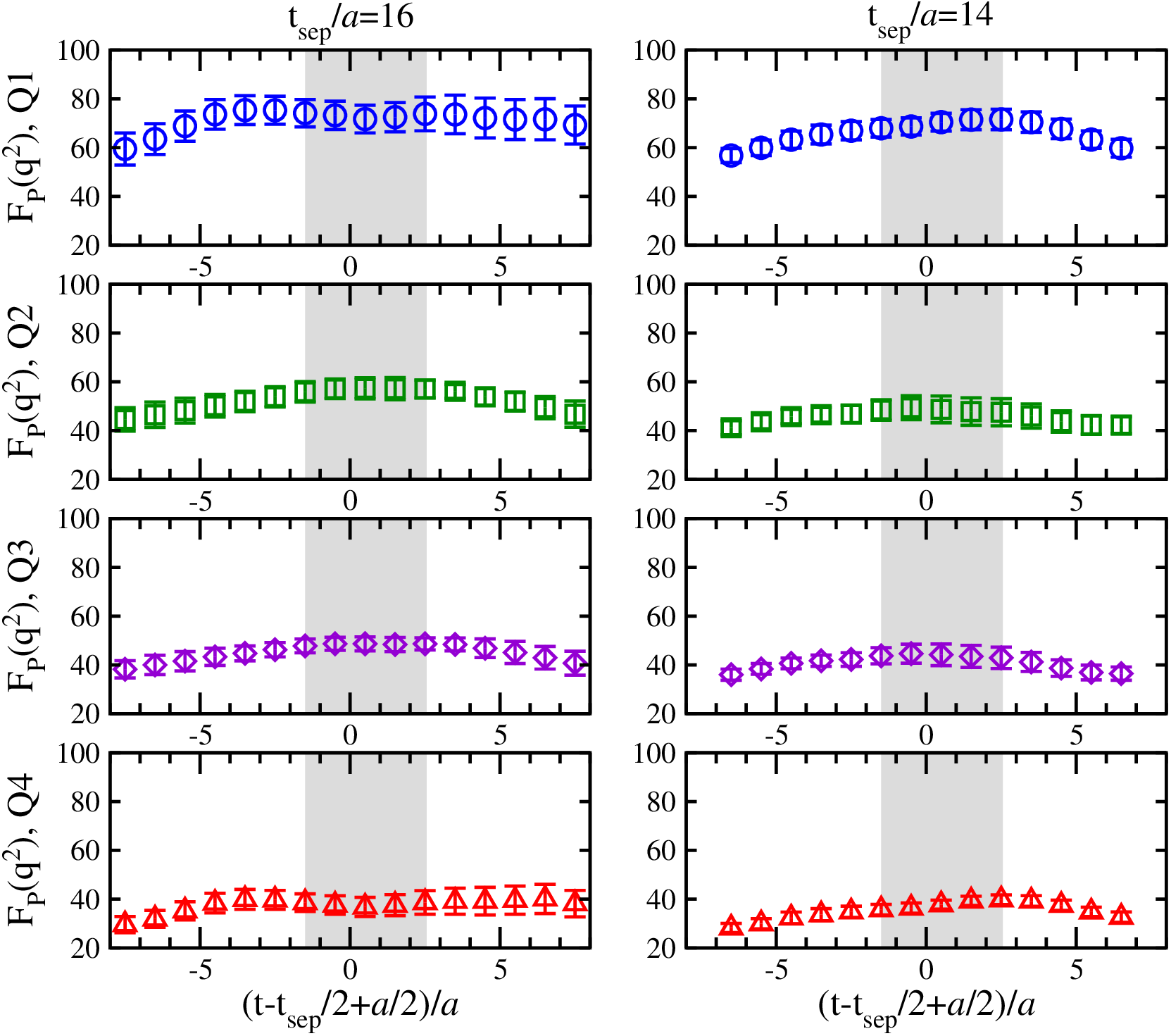}
  \hspace{3mm}
  \includegraphics[width=8.0cm,keepaspectratio,clip]{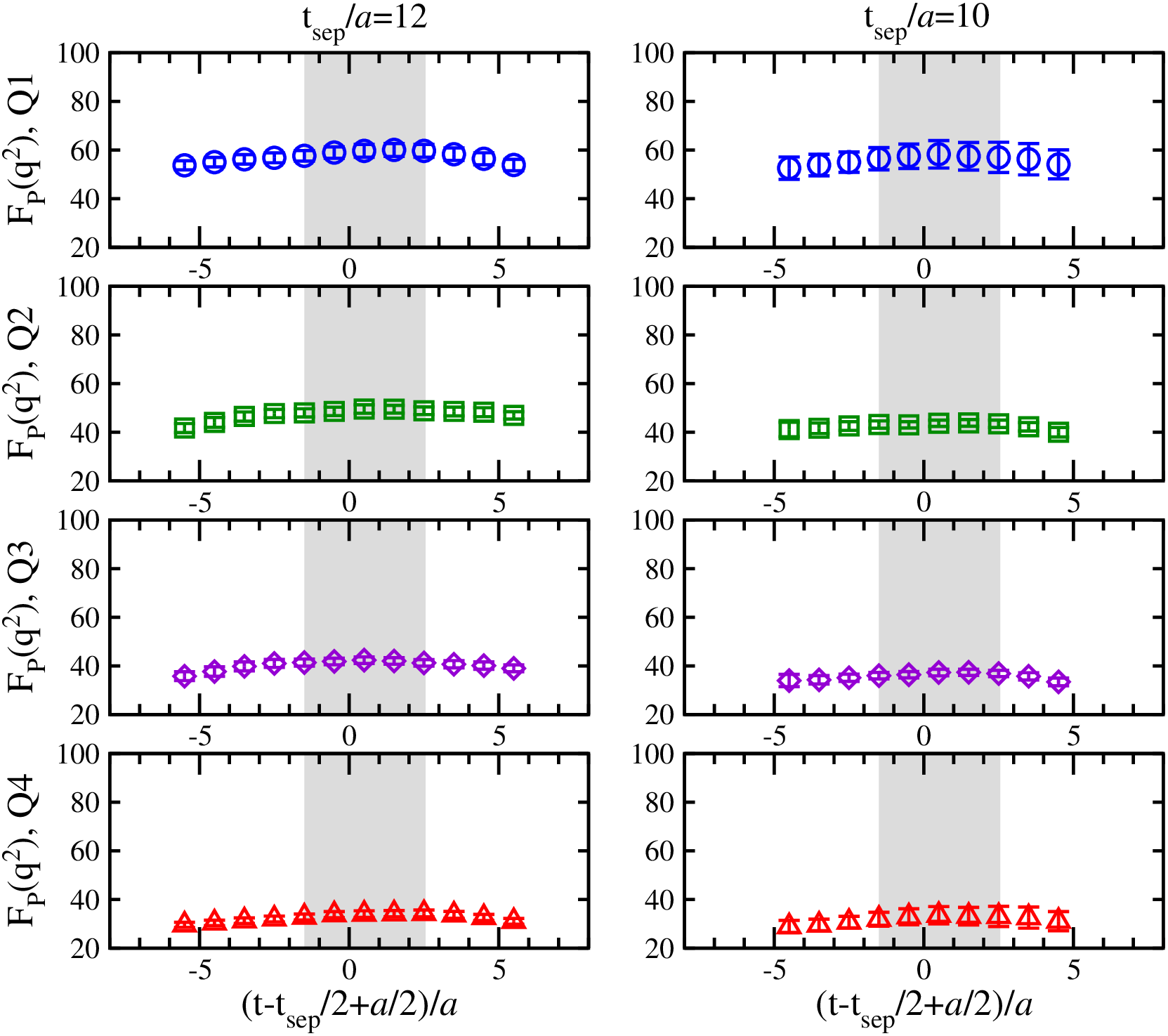}
\caption{Same as Fig.~\ref{fig_new:ratio_ge} for the induced pseudoscalar form factor $F_P(q^2)$ extracted from the ratio of Eq.~(22) in the paper. This figure replaces Fig.~19 in the paper.}
\label{fig_new:ratio_fp}
\end{figure*}

\begin{figure}[ht!]
  \includegraphics[width=8.0cm,keepaspectratio,clip]{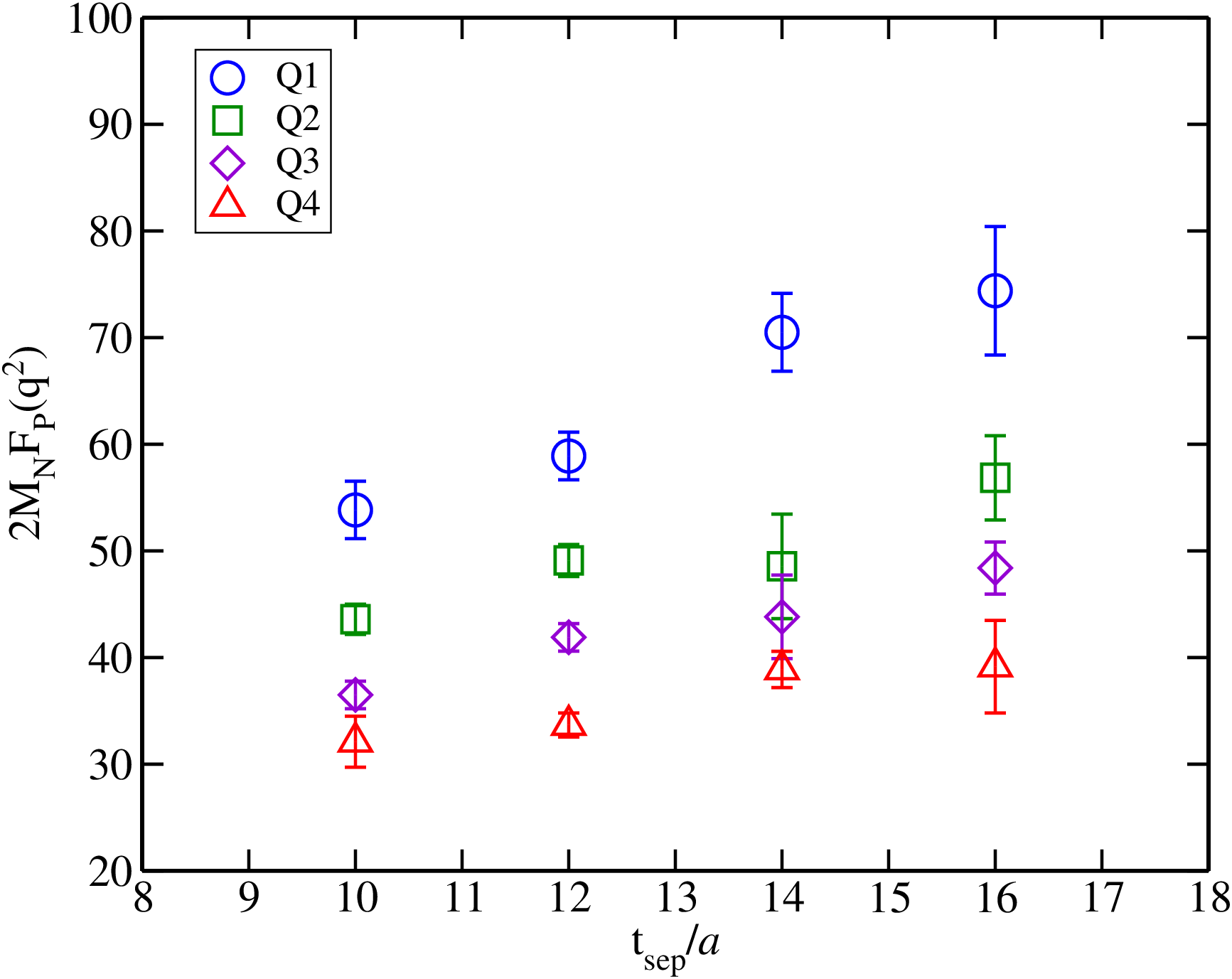}
\caption{Same as Fig.~\ref{fig_new:ge_ts} for the normalized induced pseudoscalar form factor $2M_NF_P(q^2)$ with four lowest nonzero momentum transfers. This figure replaces Fig.~20 in the paper.}
\label{fig_new:fp_ts}
\end{figure}

\begin{figure}[ht!]
  \includegraphics[width=8.0cm,keepaspectratio,clip]{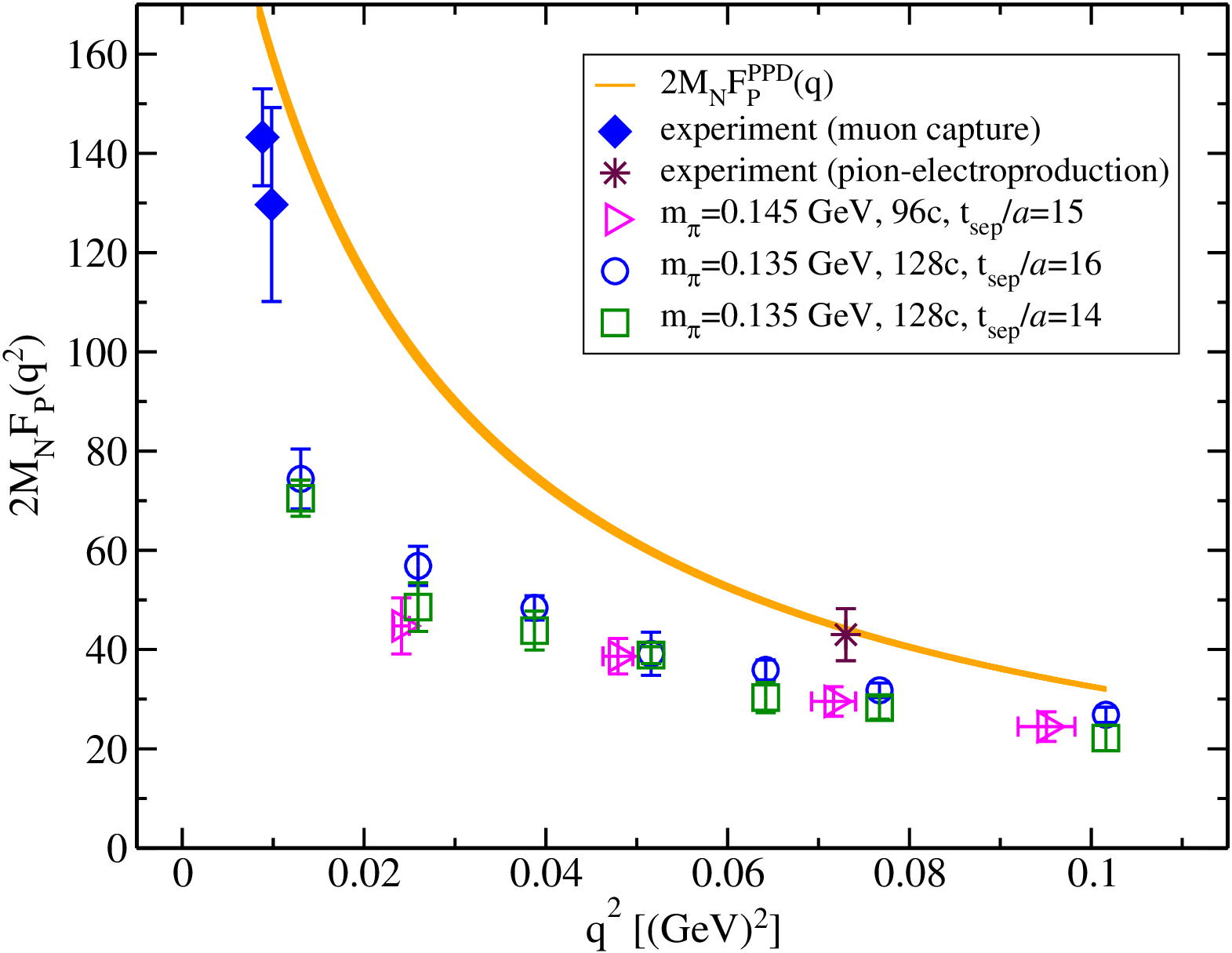}
  \caption{$q^2$ dependence of the normalized induced pseudoscalar form factor $2M_N F_P(q^2)$. We also plot the results in the previous work~\cite{Ishikawa:2018rew} for comparison. The colored band shows the function of the pion pole dominance model using the fit result of $F_A(q^2)$ with the dipole form in Fig.~\ref{fig_new:fa_q2}.
This figure replaces Fig.~21 in the paper.}
\label{fig_new:fp_q2}
\end{figure}

\begin{figure*}[ht!]
  \includegraphics[width=8.0cm,keepaspectratio,clip]{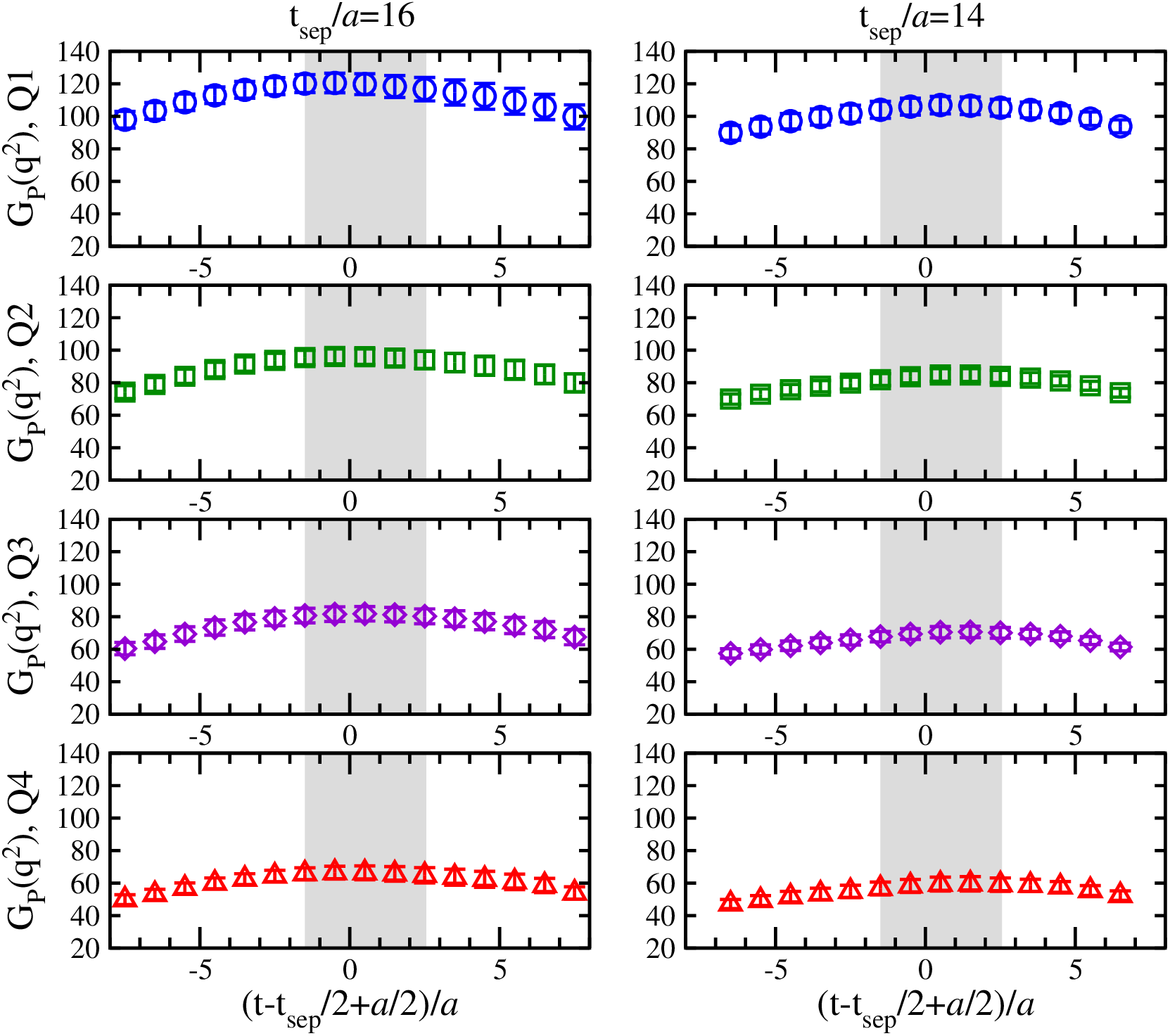}
  \hspace{3mm}
  \includegraphics[width=8.0cm,keepaspectratio,clip]{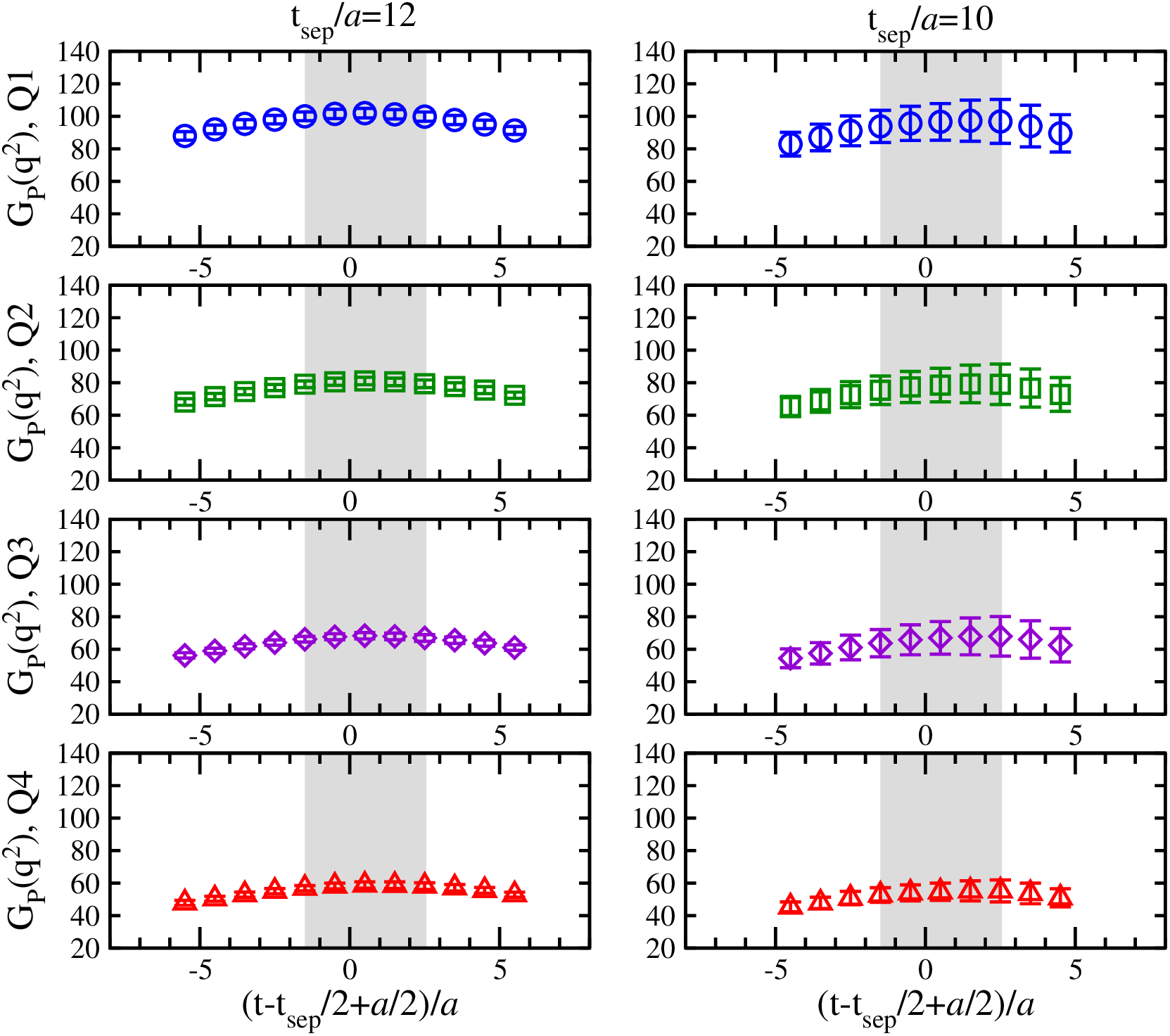}
\caption{Same as Fig.~\ref{fig_new:ratio_ge} for the pseudoscalar form factor $G_P(q^2)$ extracted from the ratio of Eq.~(23) in the paper. This figure replaces Fig.~22 in the paper.}
\label{fig_new:ratio_gp}
\end{figure*}

\begin{figure}[ht!]
  \includegraphics[width=8.0cm,keepaspectratio,clip]{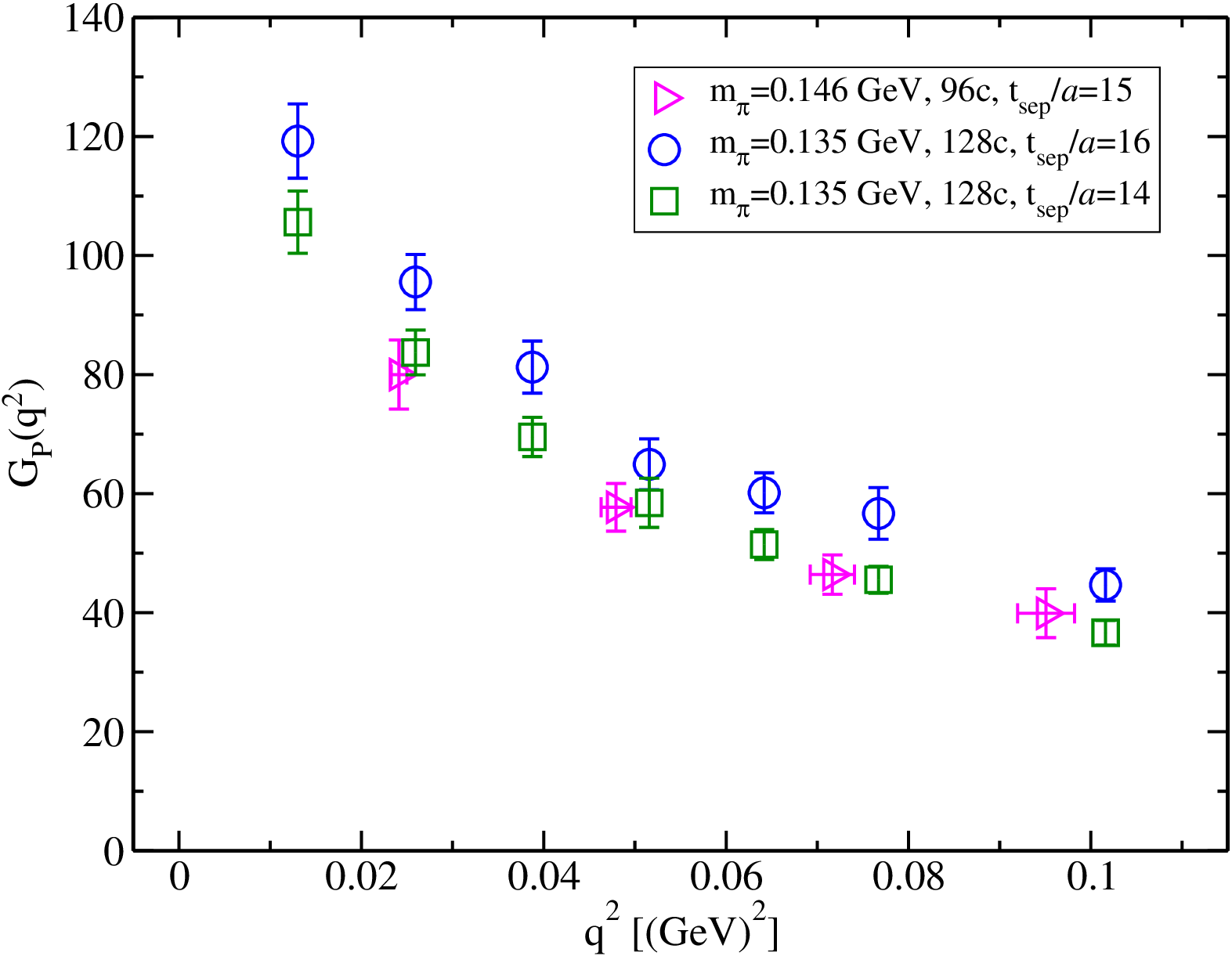}
\caption{$q^2$ dependence for the pseudoscalar form factor $G_P(q^2)$ at $t_{\rm sep}/a=14,\,16$. We also plot the results in the previous work~\cite{Ishikawa:2018rew} for comparison. This figure replaces Fig.~23 in the paper.}
\label{fig_new:gp_q2}
\end{figure}

\begin{figure}[ht!]
  \includegraphics[width=8.0cm,keepaspectratio,clip]{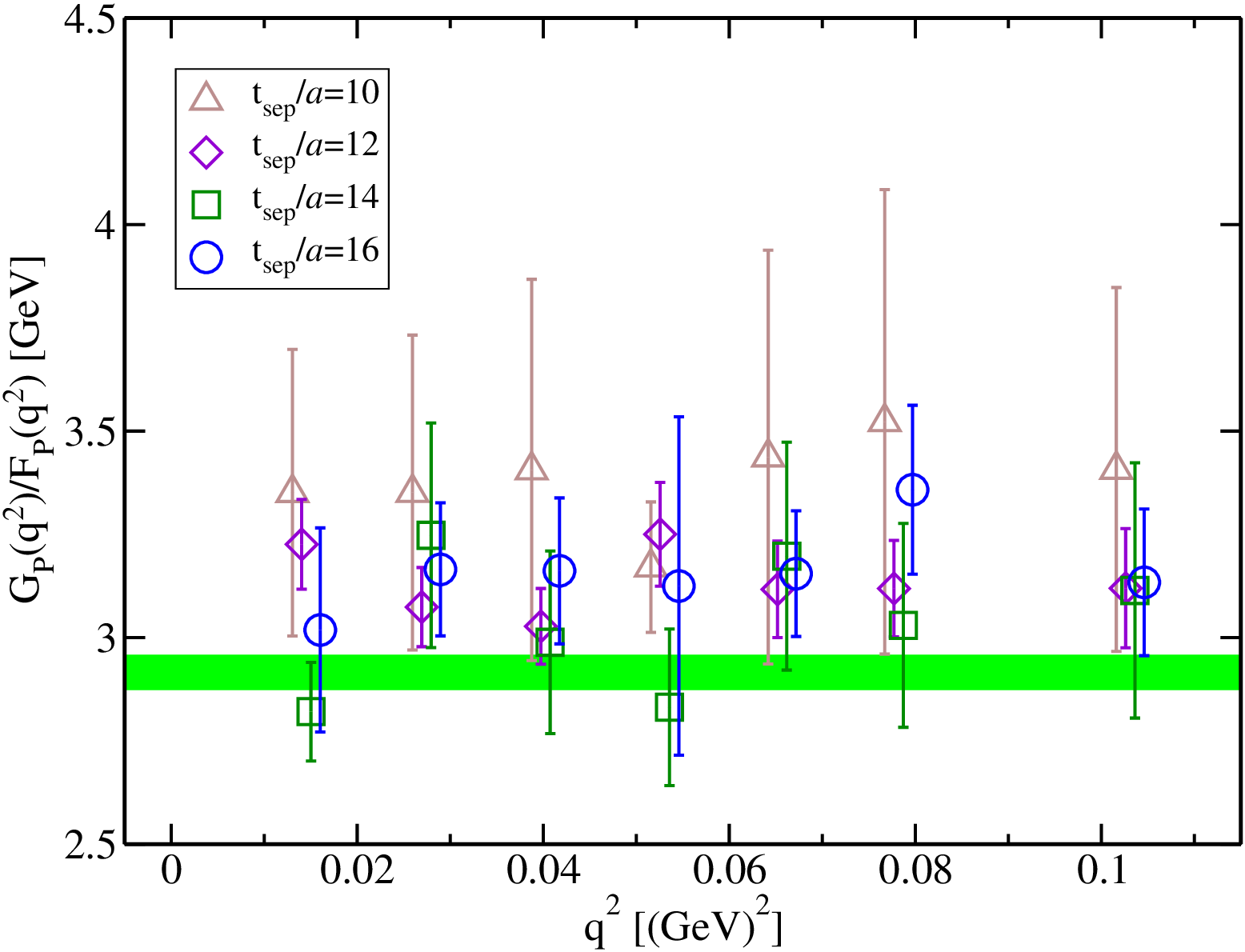}
\caption{Ratio of the pseudoscalar form factor $G_P(q^2)$ over the induced one $F_P(q^2)$ as a function of $q^2$ at each $t_{\rm sep}$. This figure replaces Fig.~24 in the paper.}
\label{fig_new:gp_q2_ratio}
\end{figure}
 
\begin{figure*}[ht!]
  \includegraphics[width=7.0cm,keepaspectratio,clip]{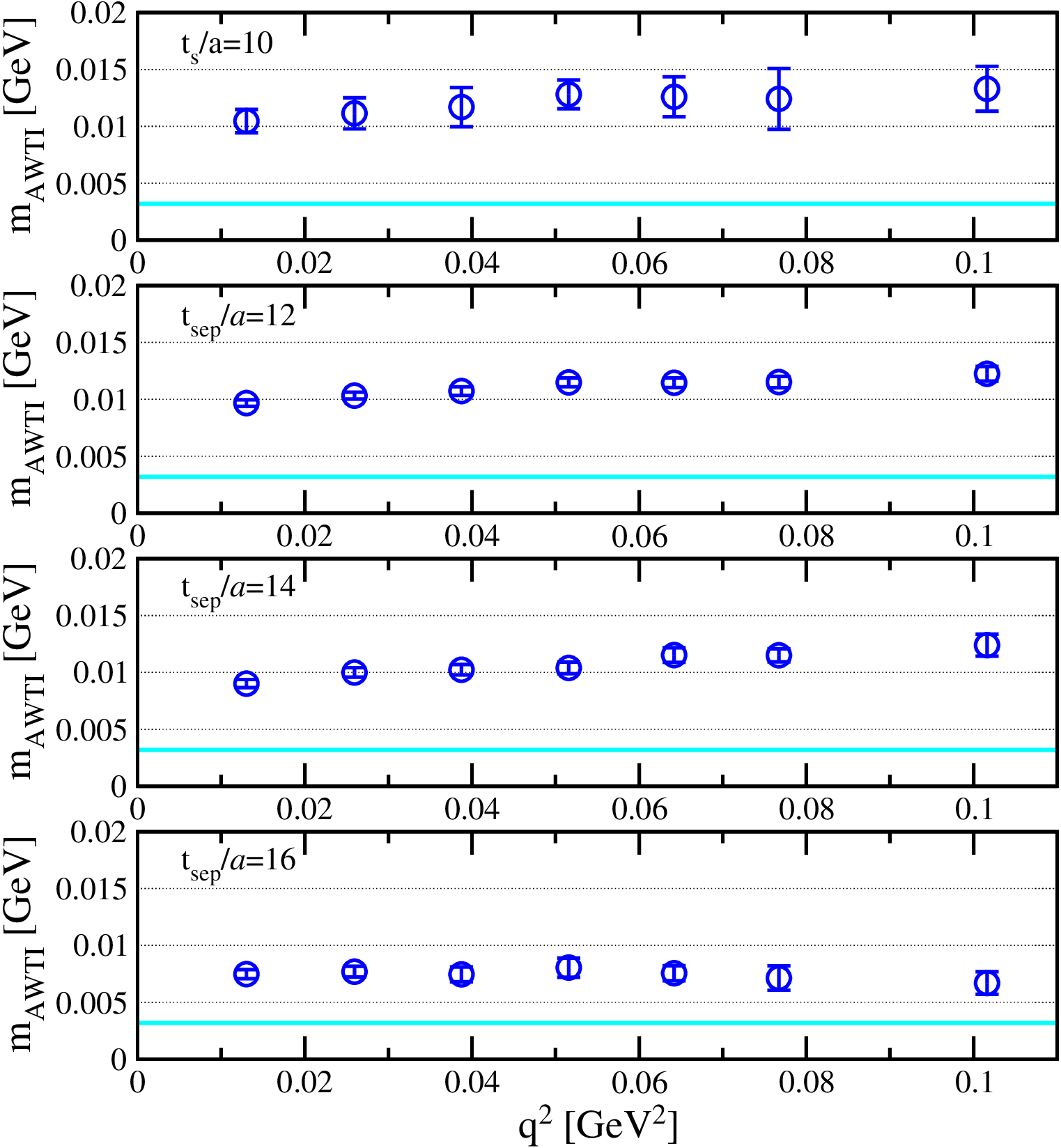}
  \hspace{3mm}
  \includegraphics[width=8.0cm,keepaspectratio,clip]{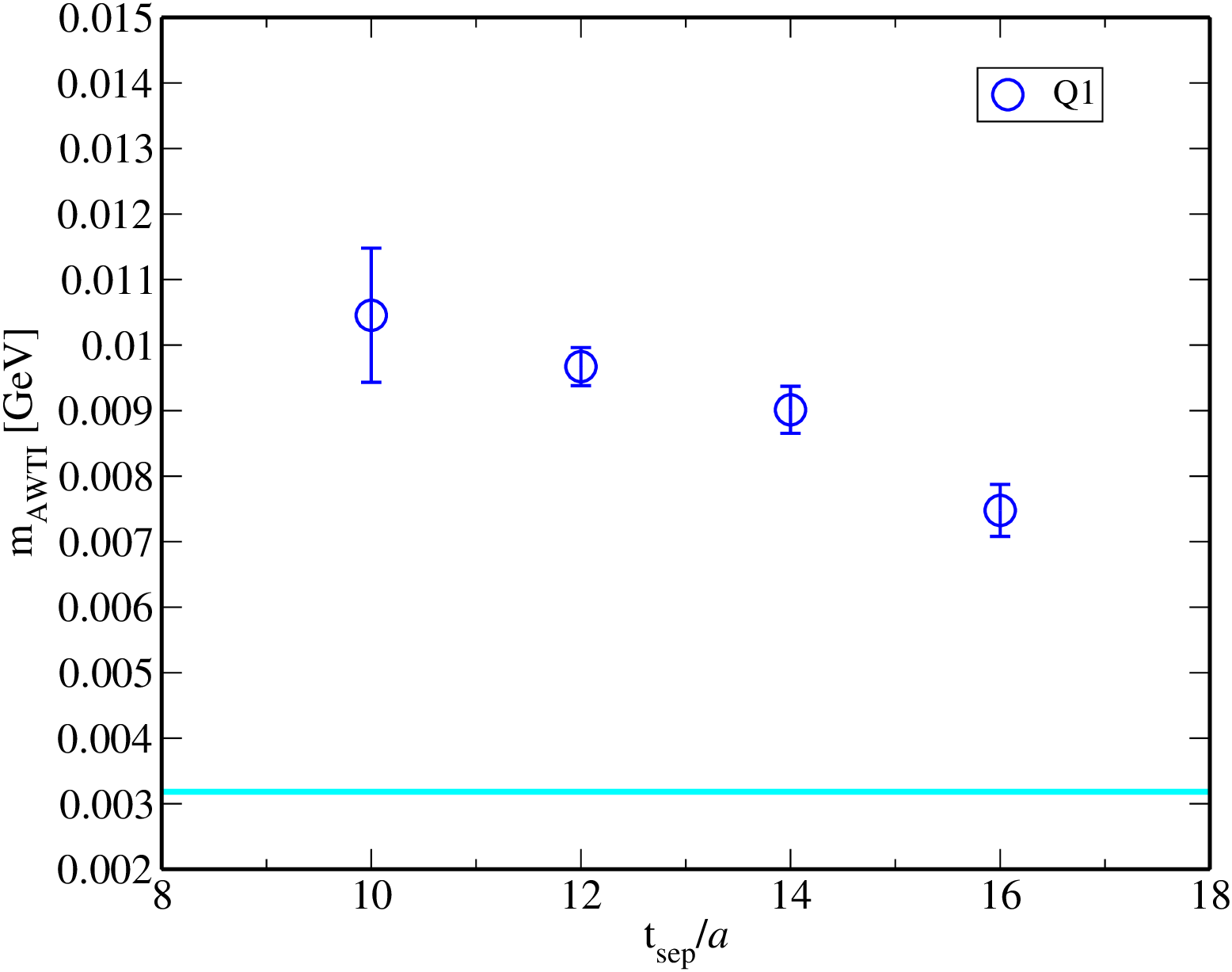}
\caption{$q^2$ dependence (left) and $t_{\rm sep}$ dependence (right) of the quark mass defined by the generalized GT relation of Eq.~(9) in the paper. The data in the right panel is obtained at Q1 corresponding to $|{\bm n}|^2=1$. Horizontal band denotes the quark mass obtained from the PCAC relation in Ref.~\cite{Ishikawa:2018jee}. This figure replaces Fig.~25 in the paper.}
\label{fig_new:mq_awti}
\end{figure*}

\begin{figure*}[ht!]
  \includegraphics[width=16cm,keepaspectratio,clip]{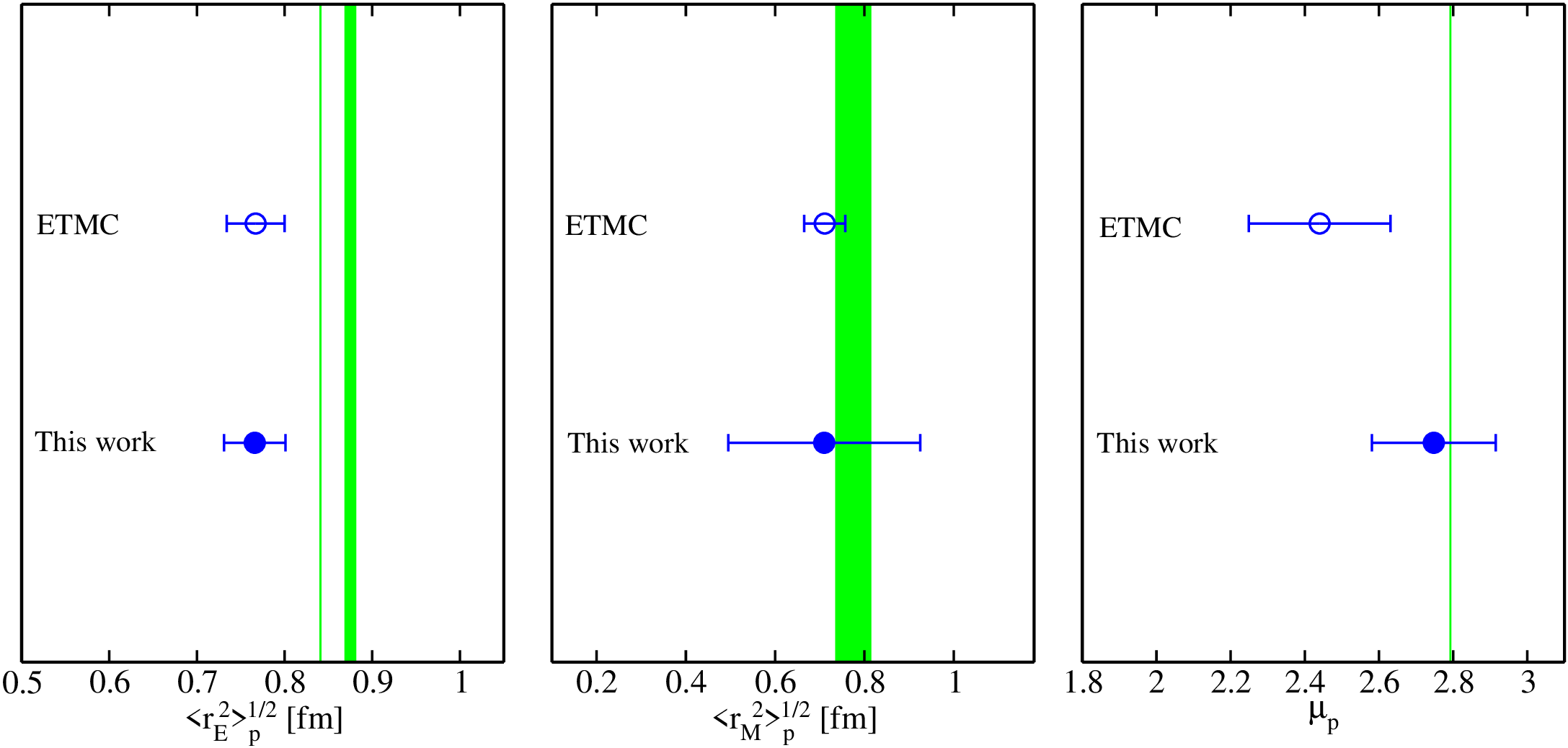}
  \vskip 3mm
  \includegraphics[width=11cm,keepaspectratio,clip]{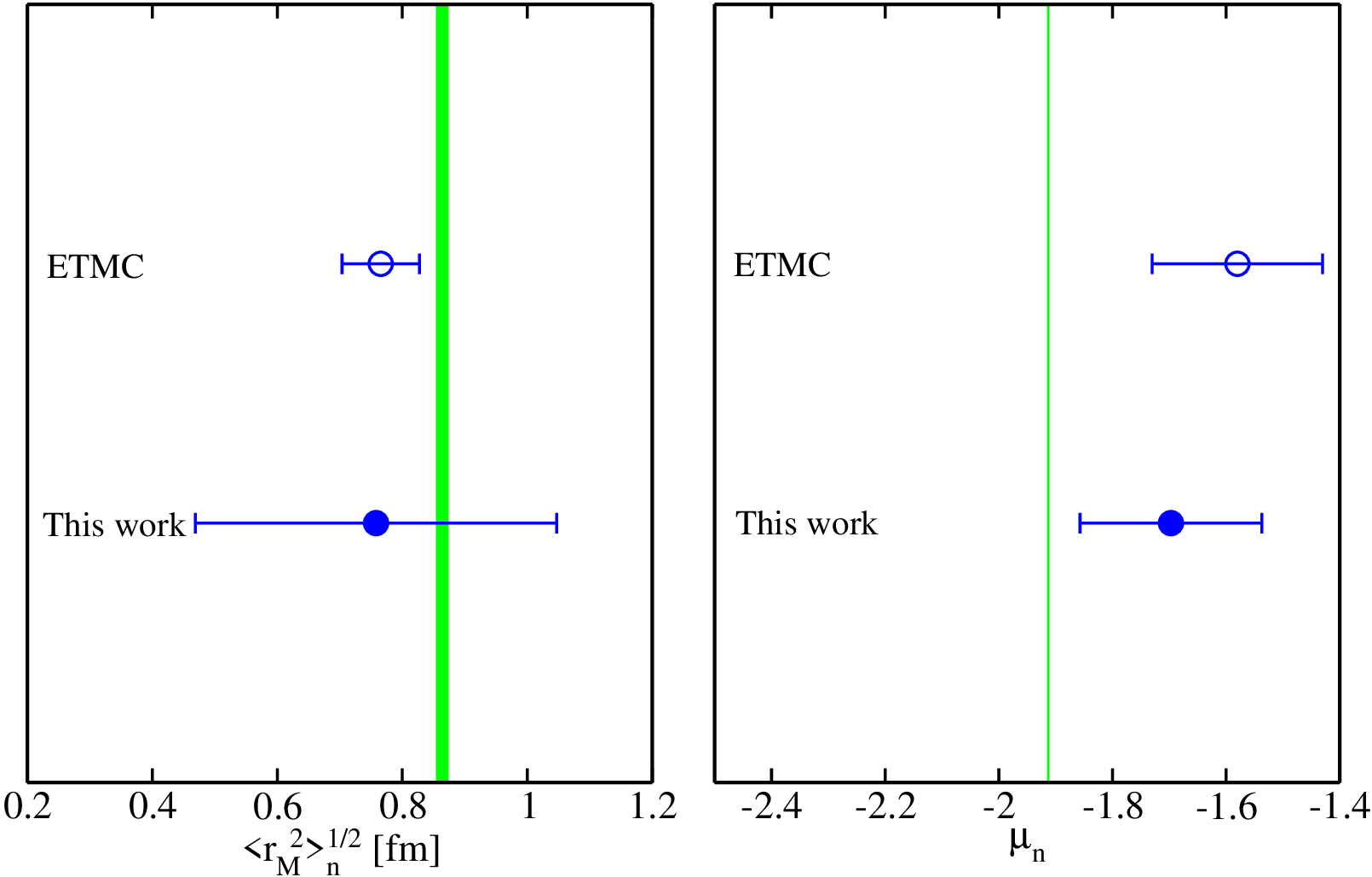}
\caption{Summary plot for the proton electric and magnetic RMS radii and magnetic moment (top) and the neutron electric and magnetic RMS radii and magnetic moment (bottom). We also plot the results of the ETM Collaboration~\cite{Alexandrou:2017ypw} for comparison. Vertical bands denote experimental values. Our results are obtained without the disconnected diagram. This figure replaces Fig.~27 in the paper.}
\label{fig_new:summary_p_n}
\end{figure*}

\begin{table*}[hb!]
\caption{$q^2$ dependence of the isovector form factors obtained by the constant fit for $t_{\rm sep}/a=\{12,14,16\}$ and $t_{\rm sep}/a=\{14,16\}$. In the previous work~\cite{Ishikawa:2018rew} the results are obtained with $t_{\rm sep}/a=15$ on a $96^4$ lattice at $m_\pi = 146$ MeV.
This table replaces Table~VII in the paper.
\label{tab_new:ff_q2_comb}
}
\begin{ruledtabular}
\begin{tabular}{c|cccccc} 
\multirow{2}{*}{$q^2\,[{\rm GeV}^2]$}  & \multicolumn{3}{c}{$t_{\rm sep}/a=\{12,14,16\}$} & \multicolumn{3}{c}{$t_{\rm sep}/a=\{14,16\}$} \\
  \cline{2-4}\cline{5-7}
  & $G_E^v(q^2)$ & $G_M^v(q^2)$ & $F_A(q^2)$ & $G_E^v(q^2)$ & $G_M^v(q^2)$ & $F_A(q^2)$\\  
  \hline
  0.000 & 0.997(1) & --- & 1.273(24) &  0.999(1) & --- & 1.269(34)\\
  0.013 & 0.964(2)&4.312(163)&1.251(22)& 0.962(2)& 4.546(215)& 1.249(33)\\
  0.026 & 0.934(3)&4.189(140)&1.237(22)& 0.930(5)& 4.347(185)& 1.226(39)\\
  0.039 & 0.905(5)&4.045(125)&1.223(21)& 0.900(6)& 4.170(176)& 1.213(42)\\
  0.052 & 0.873(6)&3.923(123)&1.190(21)& 0.870(8)& 3.976(178)& 1.174(30)\\
  0.064 & 0.848(6)&3.839(112)&1.189(22)& 0.843(9)& 3.926(158)& 1.157(42)\\
  0.077 & 0.825(7)&3.734(104)&1.176(22)& 0.817(10)& 3.804(158)& 1.144(43)\\
  0.102 & 0.778(8)&3.557(96)&1.146(23)& 0.766(12)& 3.627(148)& 1.103(51)\\
\hline\hline
\multirow{2}{*}{$q^2\,[{\rm GeV}^2]$} & \multicolumn{3}{c}{PACS'18~\cite{Ishikawa:2018rew} $t_{\rm sep}/a=15$}  \\
  \cline{2-4}
& $G_E^v(q^2)$ & $G_M^v(q^2)$ & $F_A(q^2)$\\  
  \hline
0.000 & 1.000(4) & --- & 1.163(75)\\
0.024 & 0.924(11) & 4.071(456) & 1.121(68)\\
0.048 & 0.861(19) & 3.640(350) & 1.137(69)\\
0.072 & 0.804(27) & 3.333(305) & 1.112(64)\\
0.095 & 0.774(30) & 3.313(344) & 1.118(72)\\
\end{tabular}
\end{ruledtabular} 
\end{table*}

\begin{table*}[hb!]
\caption{$q^2$ dependence of the proton and neutron form factors obtained by the constant fit for $t_{\rm sep}/a=\{12,14,16\}$ and $t_{\rm sep}/a=\{14,16\}$. 
This table replaces Table~VIII in the paper.
\label{tab_new:ff_p_n_q2_comb}
}
\begin{ruledtabular}
  \begin{tabular}{c|cccccccc}
\multirow{3}{*}{$q^2\,[{\rm GeV}^2]$}  & \multicolumn{4}{c}{$t_{\rm sep}/a=\{12,14,16\}$} & \multicolumn{4}{c}{$t_{\rm sep}/a=\{14,16\}$} \\
\cline{2-5}\cline{6-9}
  & \multicolumn{2}{c}{Proton} & \multicolumn{2}{c}{Neutron} & \multicolumn{2}{c}{Proton} & \multicolumn{2}{c}{Neutron} \\
  \cline{2-5}\cline{6-9}
  & $G_E^p(q^2)$ & $G_M^p(q^2)$ & $G_E^n(q^2)$ & $G_M^n(q^2)$ & $G_E^p(q^2)$ & $G_M^p(q^2)$ & $G_E^n(q^2)$ & $G_M^n(q^2)$\\  
  \hline
  0.000 & 0.9988(8) & --- &  0.0016(7) & --- & 1.000(0)&---& 0.002(1)&---\\
  0.013&0.966(1)& 2.681(105)& 0.002(1)& $-$1.632(63)& 0.965(2)& 2.778(144)& 0.003(1)& $-$1.764(86)\\
  0.026&0.936(3)& 2.609(88)& 0.003(1)& $-$1.598(55)& 0.933(3)& 2.669(130)& 0.004(2)& $-$1.682(69)\\
  0.039&0.907(4)& 2.532(77)& 0.004(2)& $-$1.543(50)& 0.904(5)& 2.564(122)& 0.005(4)& $-$1.613(63)\\
  0.052&0.879(5)& 2.447(77)& 0.008(3)& $-$1.489(49)& 0.877(6)& 2.441(120)& 0.014(11)& $-$1.535(67)\\
  0.064&0.854(5)& 2.401(66)& 0.008(3)& $-$1.457(46)& 0.850(7)& 2.428(110)& 0.013(9)& $-$1.502(58)\\
  0.077&0.830(6)& 2.342(61)& 0.008(3)& $-$1.411(43)& 0.825(7)& 2.345(108)& 0.012(8)& $-$1.463(58)\\
  0.102&0.786(7)& 2.229(55)& 0.010(4)& $-$1.334(43)& 0.778(9)& 2.255(98)& 0.015(8)& $-$1.372(62)\\
\end{tabular}
\end{ruledtabular} 
\end{table*}

\end{document}